\documentstyle[epsfig,psfig]{report}
\begin {document}
%
\newcommand{\be}{\begin{equation}}
\newcommand{\ee}{\end{equation}}
\newcommand{\ba}{\begin{eqnarray}}
\newcommand{\ea}{\end{eqnarray}}
\newcommand{\beq}{\begin{equation}}
\newcommand{\eeq}{\end{equation}}
\newcommand{\bea}{\begin{eqnarray}}
\newcommand{\eea}{\end{eqnarray}}
\newcommand{\ie}{{\it i.e.}}
\newcommand{\eg}{{\it e.g.}}
\newcommand{\etal}{{\it et al.}}
\newcommand{\mn}{{\mu\nu}}
\newcommand{\tr}{{\rm tr}}
\newcommand{\ave}[1]{\langle {#1} \rangle}
\newcommand{\tave}[1]{\langle\!\langle{#1}\rangle\!\rangle}
\newcommand{\ket}[1]{| {#1} \rangle}
\newcommand{\bra}[1]{\langle {#1} |}
\newcommand{\pkt}{\, .}
\newcommand{\bce}{\begin{center}}
\newcommand{\ece}{\end{center}}
\newcommand{\integ}{\int\!d}
\def\lsim{\mathrel{\rlap{\lower4pt\hbox{\hskip1pt$\sim$}}
    \raise1pt\hbox{$<$}}}         
\def\gsim{\mathrel{\rlap{\lower4pt\hbox{\hskip1pt$\sim$}}
    \raise1pt\hbox{$>$}}}         
\title
{Chiral Symmetry Restoration and Dileptons in Relativistic Heavy-Ion 
Collisions}
 
\author
{R. Rapp$^{1}$ and J. Wambach$^{2}$
\\[5mm]
1) Department of Physics and Astronomy, State University of New York,\\ 
    Stony Brook, NY 11794-3800, U.S.A.\\
2) Institut f\"ur Kernphysik, Technische Universit\"at Darmstadt,\\ 
   Schlo{\ss}gartenstr.~9, D-64289 Darmstadt, Germany}
 
\date{}
\maketitle
 
\begin{abstract}
The current theoretical status in the analysis and interpretation 
of low-mass dilepton measurements in (ultra-) relativistic heavy-ion 
experiments is reviewed. Special emphasis is put on potential signals 
of (partial) restoration of dynamically broken chiral symmetry in 
a hot and dense hadronic medium. It follows from chiral symmetry 
alone that parity partners of hadronic correlation functions must 
become identical when the symmetry is restored. The
assessment of medium effects in the vector channel, which governs the 
dilepton production,  thus necessitates a simultaneous
treatment of the vector and axialvector degrees of freedom. 
While significant progress in this respect has been made some open 
questions remain in establishing a rigorous link in the mass region below
1~GeV. From the present calculations a suggestive
'quark-hadron duality' emerges near the phase boundary. It implies 
substantial medium effects in the dilepton signal from the hadronic phase  
which smoothly matches a perturbative description within the plasma phase. 

\end{abstract}

\tableofcontents

\chapter{Introduction}

In recent years substantial experimental and theoretical efforts have 
been undertaken to investigate the versatile physics issues involved 
in (ultra-) relativistic heavy-ion collisions, \ie, 
collisions of atomic nuclei in which the center-of-mass ($cms$) 
energy per nucleon is (much) larger than the nucleon rest 
mass~\cite{quarkmatter,Wong}. 
The principal goal of this initiative 
is to explore the phase structure of the underlying
theory of strong interactions -- Quantum Chromodynamics (QCD) -- by creating
in the laboratory new states of matter. Through varying  
experimental conditions such as the collision energy or impact
parameter, one aims at covering
as broad a regime as possible in temperature and baryon density 
of the excited nuclear system. 
In nature such new states are believed to have existed and still
may be encountered on large scales in at least two astrophysical contexts:  
in the evolution of the early universe where a few tens of 
microseconds after the 'big bang' a transient stage of 
strongly interacting matter prevailed at temperatures a few times $10^{12}$~K
($\sim 200$~ MeV) with very small net baryon excess;  
in the interior of neutron stars where
mass densities are likely to exceed $10^{15}$~g/cm$^3$ 
-- about four times the central density of nuclei --  
while surface temperatures are as low as $10^5$~K or less.
Experiments have been performed until recently at the Alternating Gradient 
Synchrotron (AGS) in Brookhaven (BNL) with $cms$ energies around  
$\sqrt{s}\sim 5$~AGeV and are currently underway 
at the CERN Super-proton-Synchrotron (SpS) at $\sqrt{s}\sim 20$~AGeV. 
In the near future the Relativistic Heavy-Ion Collider (RHIC) at BNL
will start data-taking at $\sqrt{s}\sim 200$~AGeV, and eventually heavy ions 
will also be injected into the CERN Large Hadron Collider (LHC) 
reaching $\sqrt{s}\sim 10$~ATeV. 

\begin{figure}[!ht]
\bce
\epsfig{figure=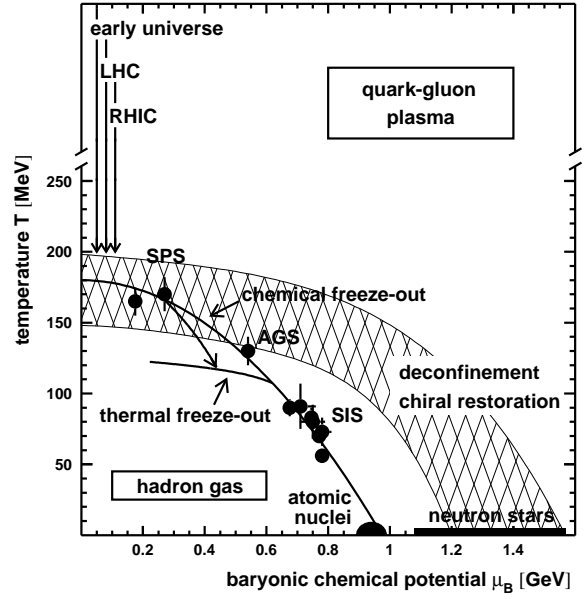,height=8cm}
\ece
\caption{QCD phase diagram~\protect\cite{pbm98,St99} in the temperature
and baryon chemical potential plane 
with freezeout points extracted from hadrochemical analyses at SIS, 
AGS and SpS energies.}
\label{fig_freeze}
\end{figure}

As nuclear matter is heated and compressed hadrons 
occupy more and more of the available space.
Eventually they start to overlap and the initially confined quarks and gluons 
begin to 'percolate' between the hadrons thus being 'liberated'. 
This simple picture has originally provided the basis for models
of the quark-hadron transition and has been essentially 
confirmed by ab-initio    
numerical QCD lattice calculations at finite temperature. The latter  
demonstrate that strongly interacting matter exhibits a rapid change 
in energy- and entropy-density (possibly constituting a true phase 
transition) within a narrow temperature interval indicating a change-over 
from confined hadrons to a 'quark-gluon plasma' (QGP). 
At the same time the quarks -- most notably up ($u$) and down ($d$) quarks, 
carrying an effective mass of a few hundred MeV in the confined 
phase -- lose their 'constituent mass' leading to the restoration
of 'chiral symmetry', an approximate symmetry of
QCD in the sector of 'light' quarks. Once massless, left- and 
right-handed quarks decouple leading to a degeneracy in (hadronic) 
states of opposite parity. 
The expected phase diagram of hadronic matter is shown in 
Fig.~\ref{fig_freeze}. 
The 'confined phase' consists of an interacting gas of hadrons 
(a 'resonance gas')
while the 'deconfined phase' is comprised of a (non-ideal) gas of quarks
and gluons. The hatched 'phase boundary' reflects the present uncertainties
from lattice QCD extrapolated  to finite baryochemical potential, $\mu_B$.
Also shown are locations realized in heavy-ion systems for
various laboratory bombarding energies at the point where {\it inelastic} 
collisions between the particles in the fireball cease -- the so-called
'chemical freezeout' -- characterizing the stage where the
fireball acquires its final particle composition.  
The 'thermal freezeout' refers to the stage where {\it elastic}
collisions are no longer supported as the mean-free paths of the hadrons 
exceed the size of the fireball. Their momentum distributions do 
no longer change and they stream freely to the detector. 
Based on the assumption of thermodynamic equilibrium the chemical 
freezeout is determined from a hadrochemical analysis
of the measured abundances of particle species~\cite{PSWX,pbm98} with 
the conclusion that at the highest presently available energies the 
produced systems must be close to the phase boundary. However, subsequent
rescattering still maintains local thermal equilibrium for about
10~fm/c, cooling  the system to appreciably lower temperatures until the  
thermal freezeout is reached. The latter is reconstructed from a 
combination of (comoving) thermal distributions and 
a collective transverse and longitudinal expansion of the  fireball.


If the quark-hadron transition can indeed be induced in heavy-ion collisions 
the challenge is how to detect it in the laboratory and 
isolate observable signals (see, \eg, Refs.~\cite{CB99,KKL97} for  recent 
reviews). Because of their negligible final-state 
interactions with the hadronic environment, dileptons (correlated 
lepton pairs, $l^+l^-=e^+e^-$ or $\mu^+\mu^-$) as well as photons are
considered ideal probes for the high-density/-temperature regions formed
in the early stages of the collision~\cite{Fein76,shur78}. 
However, as they are emitted continuously, 
they sense in fact the entire space-time history of the 
reaction~\cite{shur80,kkmm86}.  Because of an additional 
variable, the invariant pair mass $M_{ll}$, dileptons have the 
experimental advantage of a superior signal to background ratio as 
compared to real photons~\cite{itz95}.
Consequently, dileptons provide the most promising electromagnetic data
up to date. 

The finally measured dilepton spectra can be chronologically divided  
into several phases.  
Before the nuclear surfaces actually touch dileptons are produced 
through coherent Bremsstrahlung~\cite{Rueckel} in the decelerating 
Coulomb field of 
the approaching nuclei. Their contribution seems to be negligible as 
compared to subsequent sources~\cite{JaKo}. 
Within the first 
1~fm/c or so of the nuclear overlap, the 
excited hadronic system is far from thermal equilibrium and the
corresponding 'pre-equilibrium' dilepton radiation mostly consists 
of hard processes such as Drell-Yan annihilation leaving its trace 
mainly at large invariant masses $M_{ll}\gsim 3$~GeV.  A rapid 
thermalization~\cite{Geiger} is expected to subsequently establish  
the QGP phase, sometimes also called the 'partonic phase', where 
dilepton production proceeds predominantly via (perturbative) 
quark-antiquark annihilation. It should  
reflect a thermal spectrum even though, towards smaller masses, 
radiative corrections from gluons as well as thermal loop effects 
are likely to become important.
At later stages when, upon expansion and cooling, the QGP has converted 
into a hot hadron gas,  dileptons are preferentially radiated 
from pion and kaon annihilation processes as well as other 
collisions between various hadrons. The two-body annihilation processes 
are dynamically enhanced through the formation of (light) vector 
meson resonances, such as the $\rho$, $\omega$ and $\phi$ mesons, 
which directly couple to $l^+l^-$ pairs. 
Thus the invariant mass of the lepton pair directly reflects the mass 
distribution of the vector meson at the moment of decay! This
explains the distinguished role that vector mesons -- in conjunction with 
their in-medium modifications -- play for dilepton measurements in 
heavy-ion reactions. The situation is somewhat different for the heavy 
quarkonium states such as the $J/\Psi$ or $\Upsilon$: 
in contrast to the light vector mesons, their lifetime is substantially
longer than the typical one of the hadronic fireball such that they will
predominantly decay after freezeout. Therefore, as detailed
further below, the importance of the
corresponding dilepton signal largely resides in its magnitude and 
not so much in the spectral shape.   
Finally, when the freezeout stage is reached, the dominant sources are
hadronic resonance as well as Dalitz decays, mostly from 
$\pi^0$, $\eta$ and $\omega$ mesons, all feeding into the low-mass
region, $M_{ll}\lsim 1$~GeV.  

\begin{figure}[!ht]
\vspace{1cm}
\hspace{3.5cm}\epsfig{figure=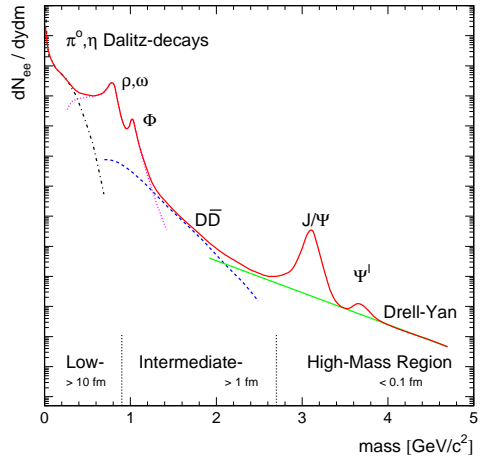,height=6cm}
\caption{Expected sources for dilepton production as a function of invariant 
mass in ultrarelativistic heavy-ion collisions\protect\cite{axdrees}.}
\label{fig_dlscheme}
\end{figure}

A schematic view of characteristic dilepton sources in ultrarelativistic 
heavy-ion collisions (URHIC's) is given in Fig.~\ref{fig_dlscheme}.
With respect to invariant mass one can roughly distinguish  
three regions. Let us try to draw some qualitative connections with  
the basic properties of strong interactions that might be addressed 
in the respective regimes. 
The low-mass region below and around the $\phi$ meson is governed by the 
light-quark sector of $u$, $d$ and $s$ quarks. In fact, it is known 
that in the limit of vanishing (current) quark masses the order
parameter of the QCD finite-temperature phase transition 
is associated with chiral symmetry restoration (\eg, the quark condensate)
and is most likely of first order for three flavors. 
Thus, in the low-mass region 
signals of chiral restoration should turn up, transmitted in terms of 
medium modifications of light hadrons. 
On the other hand, in the limit of very large current quark masses,
the order parameter of the QCD finite-$T$ transition is associated 
with deconfinement (the so-called 'Wilson line'), again realized in a strong
first-order transition. Thus, for heavy quarks one might hope  
to become sensitive to features of deconfinement. This seems indeed
to be the case:  the confining potential within heavy quarkonium 
states ($J/\Psi$, $\Upsilon$) will be Debye-screened due to 
freely moving color charges in a QGP   
leading to a dissolution of the bound states~\cite{satz86}.
As a consequence the final abundance of, \eg, $J/\Psi$ mesons -- and thus their 
contribution to the dilepton spectrum -- is suppressed, signaling
(the onset of) the deconfinement transition.   
This very important topic will not be covered in the present review, see
Refs.~\cite{jpsisum} for the recent exciting developments. 
Finally, the intermediate-mass region (IMR) might allow insights into
aspects of quark-hadron 'duality'. 
As is evident from the saturation of the vacuum annihilation cross section
$e^+e^- \to hadrons$ by perturbative QCD above $\sim 1.5$~GeV, the essentially 
structureless thermal 'continuum' 
up to the  $J/\Psi$ can be equally well described by either hadronic or 
quark-gluon degrees of freedom. However,  
as a QGP can only be formed at higher temperatures than a hadronic gas, the 
intermediate mass region might be suitable to observe   
a thermal signal from plasma radiation~\cite{shur78,mt85} in terms
of absolute yield. The most severe 'background' in this regime is 
arising from decays of 'open-charm' mesons,
\ie, pairwise produced $D \bar D$ mesons followed by individual
semileptonic decays. Although an enhanced charm production is interesting
in itself -- probably related to the very early collision stages -- it
may easily mask a thermal plasma signal. To a somewhat lesser extent, 
this also holds true for the lower-mass tail of Drell-Yan production.

Until today, the measurement of dilepton spectra 
in URHIC's has mainly been carried out
at the CERN-SpS by three collaborations: CERES/NA45
is dedicated to dielectron measurements in the 
low-mass region~\cite{ceres95,ceres96a,ceres96d,ceres96}, 
HELIOS-3~\cite{helios3} 
has measured dimuon spectra from threshold up to the $J/\Psi$ region, 
and NA38/NA50~\cite{na38,na50psi,na50int,na3850low} 
measures dimuon spectra from threshold to very high masses of
about 8~GeV, with emphasis on $J/\Psi$ suppression
(for a summary of low- and intermediate-mass dilepton
measurements see Refs.~\cite{drees98,itz98,wurm97}).
In the near future, with RHIC coming on line pushing the collision energies 
to new frontiers, high resolution dilepton spectra will be measured
by the PHENIX collaboration~\cite{phenix}. 
At much lower bombarding energies dilepton data have also been
taken by the DLS collaboration at the BEVALAC~\cite{DLS1,DLS2} and will 
soon become available from the high-precision detector HADES 
at SIS (GSI)~\cite{hades}. Here 
only the low-mass region up to the kinematical limit of around 1~GeV
is accessible and the focus is on the role of high baryon density.
Unfortunately, no dilepton measurements have been performed at the 
AGS where presumably the highest baryon densities were attained. 
However, the already commissioned low-energy run at the CERN-SpS at
a projectile energy of 40~AGeV is believed to close this gap 
penetrating into the regime of extreme baryon density.

\begin{figure}[!htb]
\bce
\epsfig{figure=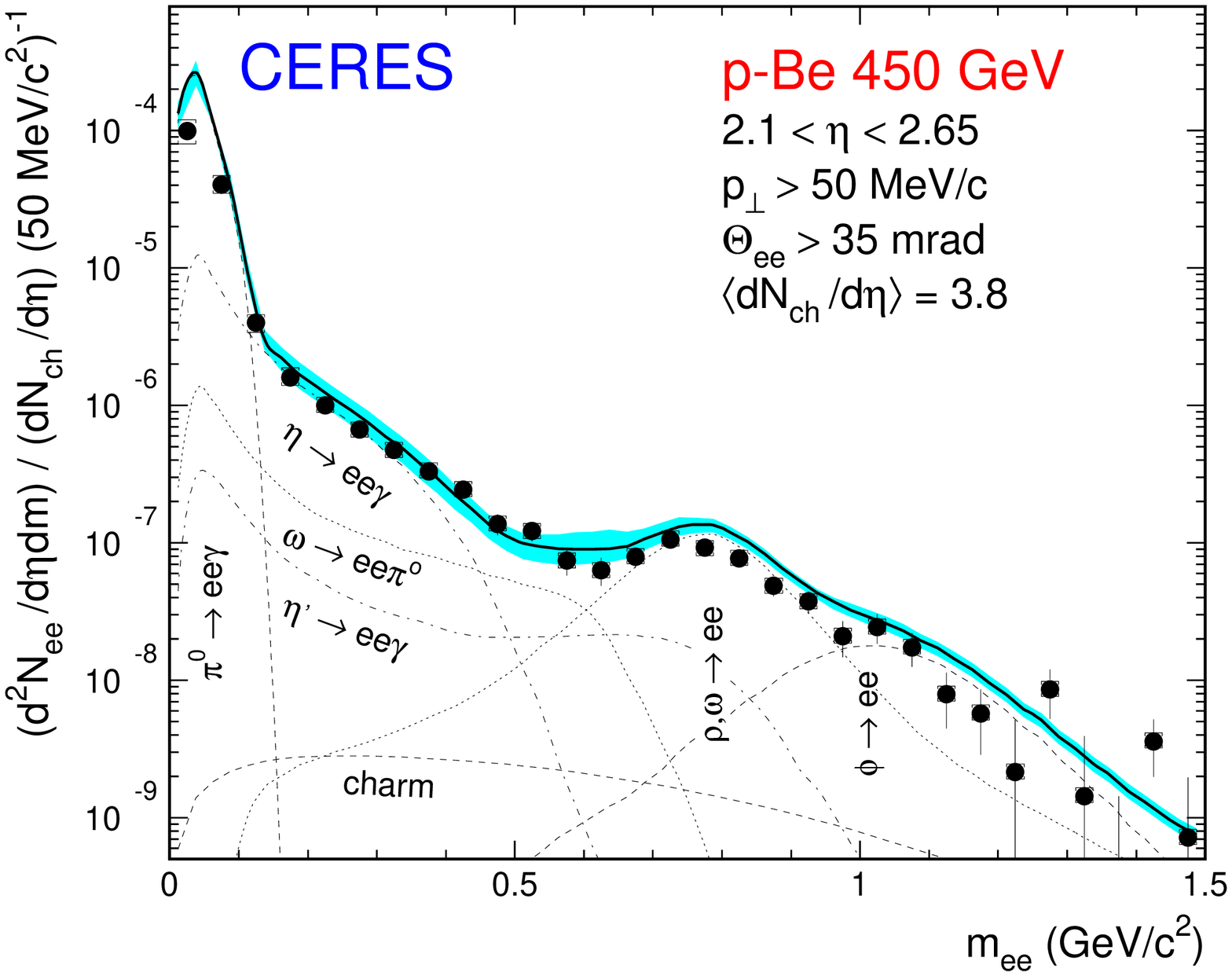,height=6.7cm}
\epsfig{figure=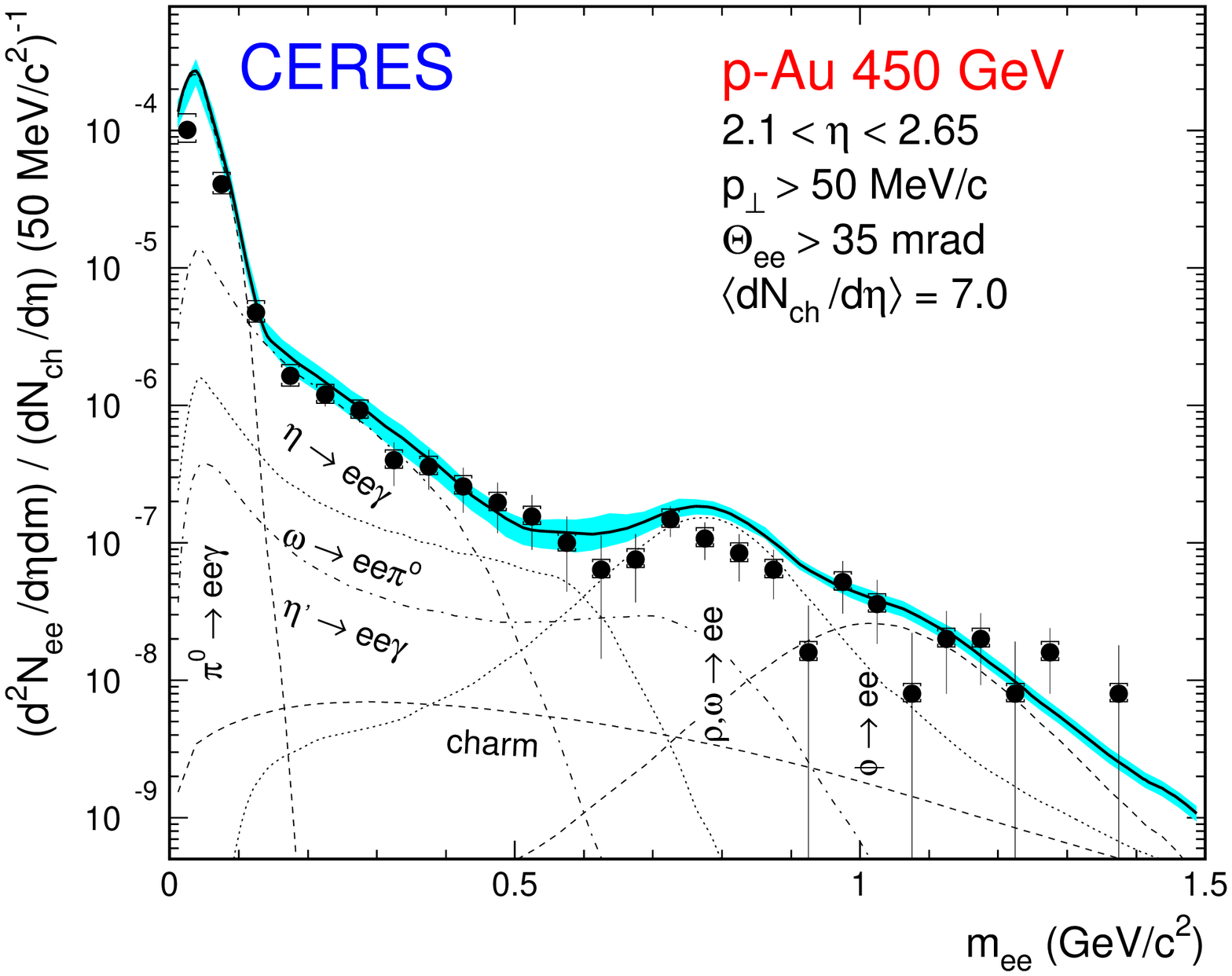,height=6.7cm}
\ece
\vspace{-1cm}
\caption{Invariant mass spectra of dileptons as measured by the
CERES collaboration~\protect\cite{ceres95} in 450~GeV proton-induced
collisions on
Beryllium (left panel) and Gold targets (right panel). The data are compared
to expectations from various hadron decay channels (labeled explicitly)
based on measured hadron multiplicities. The
bands indicate the systematic uncertainty in the cocktail.}
\label{fig_dlpa}
\end{figure}

\begin{figure}[!htb]
\vspace{1cm}
\bce
\epsfig{figure=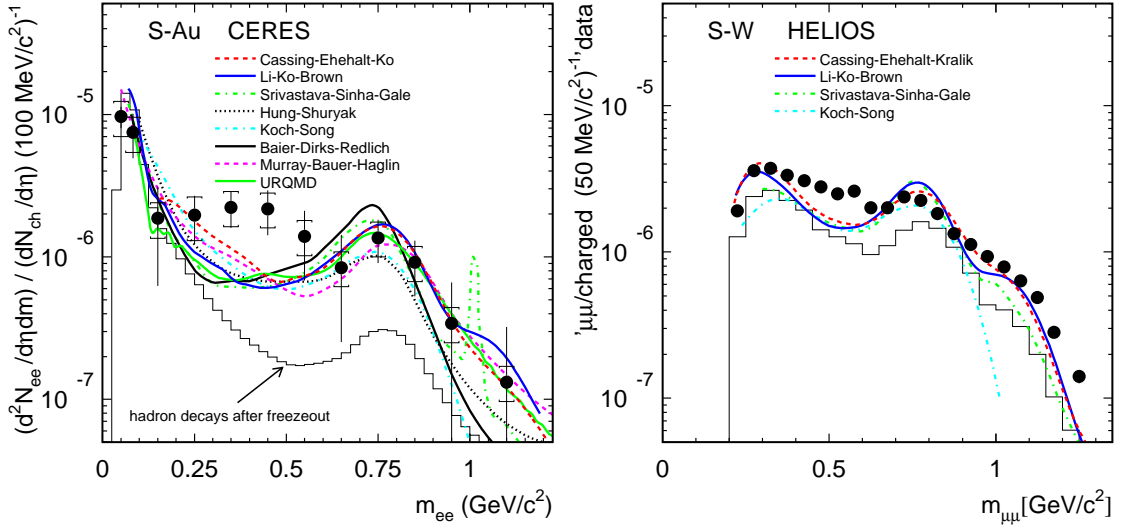,height=7cm}
\ece
\caption{Dilepton invariant mass spectra as measured in central collisions of 
200~AGeV sulfur nuclei with Au targets (left panel: $e^+e^-$ spectra from
CERES/NA45~\protect\cite{ceres95}) and $W$ targets (right panel: 
$\mu^+\mu^-$ spectra from HELIOS-3~\protect\cite{helios3}, no systematic
errors included) as compared to
a compilation~\protect\cite{ceres96d} of theoretical calculations
using free meson properties  performed by Koch/Song~\protect\cite{Koch96}, 
Li \etal~\protect\cite{LKB}, Cassing \etal~\protect\cite{CEK95,CEK96}, 
Srivastava \etal~\protect\cite{SSG96}, Baier \etal~\protect\cite{BDR971},
the Frankfurt group~\protect\cite{Winck96}, 
Hung/Shuryak~\protect\cite{HuSh97} as well as 
Murray \etal~\protect\cite{MBH98}. The histograms indicate
the hadron 'cocktail' of final state decays which gives a good description
of the $p$-$A$ data as shown in Fig.~\ref{fig_dlpa}.}
\label{fig_dlexth}
\end{figure}

The objective of the present article is to review the theoretical efforts
in understanding the experimental results in the low-mass region which 
-- as discussed above -- is intimately connected to the question of chiral
symmetry restoration. A great deal of theoretical activity has been 
triggered by recent observations from the CERES~\cite{ceres95,ceres96} and
the HELIOS-3~\cite{helios3} collaborations that central nucleus-nucleus 
($A$-$A$) collisions exhibit a strong  enhancement of low-mass dilepton  
production as compared to proton-nucleus reactions\footnote{One should note 
that similar effects have {\em not} been found by NA38/50~\cite{na3850low}, 
which, however, is most likely related to a rather large 
$M_T$-cut applied in their analysis. On the other hand, the 
DLS-data~\cite{DLS2}, taken at much lower incident energies, 
{\it do} show a very strong enhancement.}.  
Whereas the $p$-$A$ data can be well reproduced by 
final-state hadron decays with known abundances -- the so-called
hadronic 'cocktail' -- (Fig.~\ref{fig_dlpa}),
the latter strongly underestimate the $A$-$A$ spectra. 
As several hundreds of pions
are produced in $A$-$A$ collisions, the observed increase 
of dilepton pairs has been attributed to
$\pi^+\pi^-\to l^+l^-$ annihilation during the interacting phase of the 
hadronic fireball. Using vacuum meson properties many theoretical groups  
have included this process within different models for the space-time
evolution of $A$-$A$ reactions. Their results are   
in reasonable agreement with each other, but in 
disagreement with the data: the experimental spectra in the 
mass region from 0.3--0.6~GeV are significantly underestimated as seen
from Fig.~\ref{fig_dlexth} (see also Ref.~\cite{drees98}).
This has led to the suggestion of various medium effects that might be 
responsible for the observed enhancement. Among these, the 'dropping' 
vector meson mass scenario~\cite{LKB,BR91,HatsLee}, as will be detailed
below, provides an interesting possibility since it conjectures
a direct link between hadron masses and the quark condensate and
thus to the restoration of chiral symmetry towards the phase  
transition. When incorporated within a transport-theoretical treatment 
of the collision dynamics it is found to provide a unified 
description of both the CERES and HELIOS-3 data. Based on an interacting 
hadron gas many-body approaches also seem to be able to describe the 
observed phenomena~\cite{RCW,CBRW,Morio98,RW99}. 
Here the restoration of chiral symmetry 
manifests itself in more subtle ways which will be one of the key  
issues in our review. 

More explicitly, the article is organized as follows:  
In Chapter 2 we will first recall some basic properties of strong 
interactions in the light-quark sector, with emphasis on the underlying
symmetries of the QCD Lagrangian and their consequences in the 
nonperturbative regime, such as the emergence of quark and 
gluon condensates and the non-degeneracy of chiral partners in the 
hadronic spectrum. Special attention is paid to the vector and 
axialvector mesons as the former directly couple to dileptons. We then 
present model-independent methods to evaluate medium modifications of 
their spectral properties. Being based on virial
expansions the results are in general 
restricted to low density and low temperature. 

In Chapter 3 we move on to more specific model calculations 
that have been performed to assess medium modifications 
of vector and axialvector mesons. The presentation here concentrates 
on hadronic Lagrangians which can be 
classified into two categories, namely purely mesonic ones addressing 
temperature effects and those including the impact of finite 
baryon densities.   

Chapter 4 starts out by discussing how hadronic models can be 
(and have been) subjected to empirical constraints that
do {\em not} involve dilepton data in URHIC's. 
The philosophy here is to essentially fix the underlying model
parameters to enable reliable
predictions for thermal dilepton rates in hot and dense matter.  
In addition we will also discuss rate calculations within 
'non-standard' (or non-hadronic) scenarios for highly excited 
strong-interaction matter  that cannot be reliably founded on 
empirical information, such as  Disoriented Chiral Condensates 
(DCC's) or thermal quark-antiquark annihilation. 
The  main part of Chapter 4 is then devoted to a confrontation of 
the various model results within a detailed analysis 
of recent dilepton data taken in heavy-ion experiments at the BEVALAC
and CERN-SpS. An additional crucial ingredient needed  
to do so is the description of the space-time collision dynamics, which
thermal rates have to be convoluted over. We will briefly discuss 
three approaches, \ie, relativistic hydrodynamics, transport simulations
and more simplistic fireball expansions.  Chapter 4 will end with a critical
reassessment of the different mechanisms that have been invoked and outline 
possible theoretical implications for the nature 
of chiral symmetry restoration as indicative from the present status of
the observed spectra.    

Chapter 5 tries to summarize the major theoretical achievements 
in the field over the past five years or so. Based on these we will attempt  
to draw conclusions on our current understanding of the QCD
phase transition in hot/dense matter as evidenced from  
the interplay of vigorous experimental and theoretical efforts
in low-mass dilepton production.

\chapter{'Strong QCD' and Vector Mesons}
\label{chap_QCDlight}

In the Standard Model (SM) of particle physics strong interactions are 
described by Quantum Chromodynamics (QCD), a local $SU(3)$ gauge theory  
with quarks and gluons as elementary degrees of freedom~\cite{qcd}. 
The dynamics is governed by the QCD Lagrangian 
\beq
{\cal L_{\rm QCD}} = \bar \psi(i\gamma^\mu D_\mu-{\cal M}^\circ)\psi 
-{1\over 4} G_{\mu\nu}^a G^{\mu\nu}_a 
\label{LQCD}  
\eeq
with the non-abelian gluonic field-strength tensor given as
\beq 
G^a_{\mu\nu}=\partial_\mu A^a_\nu-\partial_\nu A^a_\mu+ig f^{abc}
A^b_\mu A^c_\nu \ , 
\eeq
where $A^a_\mu$ represents the spin-1 gauge field with
color index $a$ ($a=1,8$). The gauge covariant derivative
\beq
D_\mu=\partial_\mu-ig(\lambda_a/2)A^a_\mu
\label{Covder}
\eeq
induces a coupling between the spin-$1/2$, colored matter fields $\psi$
of $N_f$ flavors and the gauge fields $A^a_\mu$ (with $\lambda_a$ 
denoting the usual $SU(3)$ Gell-Mann matrices).

In Eq.~(\ref{LQCD}) ${\cal M}^\circ$ represents the diagonal matrix of 
current quark masses,  
\beq
{\cal M}^\circ =
\left( \begin{array}{cccc}  m_u &  &    &\\
                               &m_d&   &\\ 
                               &   &m_s&\\
                               &   &   & \ddots 
\end{array} \right) \ , 
\eeq
which are parameters of the SM. With $m_u,m_d,m_s\simeq 4,7,150$ MeV
and $m_c,m_b,m_t\simeq 1.5,4.5,175$ GeV there is an obvious separation
into sectors of 'light' and 'heavy' quarks. For the discussion in
the present article only the light-quark sector will be relevant. 

\begin{figure}[!ht]
\hspace{4cm}\epsfig{figure=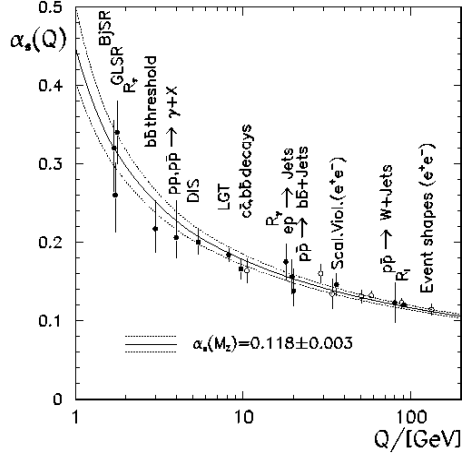,height=6cm,width=6cm}
\caption{The 'running' of the strong coupling constant $\alpha_s(Q)$
from various measurements compared to theory~\protect\cite{Schmell}.}
\label{fig_asrun}
\end{figure}

To fully specify QCD we need to take account of the fact that
-- due to quantum-loop effects -- the 'fine-structure constant'
$\alpha_s\equiv g^2/4\pi$ depends on the space-time distance or,
equivalently, the four-momentum transfer $Q$ of a given strong process,  
\beq
\alpha_s(Q)={\alpha_s(\Lambda)\over 1+\alpha_s(\Lambda){33-2N_f\over 12\pi}
\ln \bigl ({Q^2\over\Lambda^2}\bigr )} \ , 
\label{asrun}
\eeq
where $\Lambda$ is a scale at which the coupling constant is to be fixed by 
experiment,\eg, $\alpha_s(M_Z)=0.118$ at the $Z$ boson mass $M_Z=91$~GeV 
(see Fig.~\ref{fig_asrun}).
As $Q$ increases $\alpha_s$ decreases logarithmically ('asymptotic
freedom') and perturbation theory can be applied. In this regime QCD
is well tested. At momentum scales of $Q\simeq 1$ GeV -- a typical mass 
scale for 'light' hadrons -- standard perturbation theory, however, breaks
down due to a rapid increase of $\alpha_s$. 
This is the realm of 'strong QCD' where not only effective couplings
but also the relevant degrees of freedom change with scale. It is this
phenomenon that poses the most challenging intellectual problem in strong
interaction theory and is subject to intense study via ab-initio
lattice calculations and model building. At large distance scales 
the QCD degrees of freedom become encoded in colorless ('white') light 
mesons, low-mass baryons 
and glueballs rather than quarks and gluons. In dealing with their mutual 
interactions the underlying symmetries and anomalies of QCD are of utmost 
importance.  

\section{Symmetries and Anomalies of QCD}
\label{sec_sym-ano}

The structure of the lowest-mass hadrons -- involving the light-quark sector 
comprised of $u,d,s$ quarks -- is largely determined by chiral 
symmetry~\cite{chisym} and its dynamical breaking in the physical vacuum 
with confinement presumably playing a much
lesser role. This is evidenced from the dominant role of instantons 
-- believed to be responsible for chiral symmetry breaking and other 
nonperturbative phenomena -- in hadronic 
correlators~\cite{SS98}. Recent lattice QCD calculations with cooling 
algorithms corroborate such a picture~\cite{Negele}.

Apart from invariance under local $SU(3)_{\rm color}$ transformations
and a global $U(1)$ symmetry, \ie, multiplication of the matter fields
$\psi$ by a phase -- entailing baryon number conservation -- the QCD Lagrangian
(\ref{LQCD}) has additional symmetries for vanishing quark
masses. In this limit -- well justified in the up- and down-quark sector 
and to a somewhat lesser extent in the strange-quark sector -- the theory is 
invariant under global vector and axialvector transformations in 
$SU(3)$-flavor 
space
\beq
\psi\to e^{-i\alpha_V^i{\lambda^i\over 2}}\psi\ ,\qquad
\psi\to e^{-i\alpha_A^i{\lambda^i\over 2}\gamma_5}\psi
\label{VAtr} 
\eeq
with conserved vector and axialvector Noether currents
\beq
j_{V,i}^\mu=\bar\psi\gamma^\mu{\lambda_i\over 2}\psi \ ,\qquad
j_{A,i}^\mu=\bar\psi\gamma^\mu\gamma_5{\lambda_i\over 2}\psi\pkt
\eeq
As a consequence, the corresponding charges 
\beq
Q^V_i=\int\!d^3x\,\psi^\dagger{\lambda_i\over 2}\psi \ , \qquad
Q^A_i=\int\!d^3x\,\psi^\dagger{\lambda_i\over 2}\gamma_5\psi
\eeq
commute with the QCD Hamiltonian,  $[Q_i^{V,A},H_{\rm QCD}]=0$.

When decomposing the quark fields into left and right chirality 
components, $\psi_{L,R}=\frac{1}{2}(1\mp \gamma_5)\psi$, the 
Lagrangian (\ref{LQCD}) takes the form
\beq
{\cal L_{\rm QCD}} =\bar\psi_Li\gamma^\mu D_\mu\psi_L+ 
\bar\psi_Ri\gamma^\mu D_\mu\psi_R 
-{1\over 4} G_{\mu\nu}^aG^{\mu\nu}_a-(\bar\psi_L {\cal M}^\circ\psi_R+
\bar\psi_R {\cal M}^\circ\psi_L)  
\eeq
and the transformations (\ref{VAtr}) translate to
\bea
\psi_L &\to& e^{-i\alpha_L^i{\lambda^i\over 2}}\psi_L \ ,
\qquad\psi_R \to \psi_R\\
\psi_R &\to& e^{-i\alpha_R^i{\lambda^i\over 2}}\psi_R \ ,
\qquad\psi_L \to \psi_L \ , 
\eea
which -- in the limit of vanishing quark masses -- constitutes a global 
$SU(3)_L\times SU(3)_R$ {\em chiral symmetry} in flavor space. Left-
and right-handed quarks are not mixed dynamically and thus preserve their
'handedness' in strong interactions. The associated conserved
charges
\bea
Q^L_i &=& \int\!d^3x\,\psi_L^\dagger{\lambda_i\over 2}\psi_L
={1\over 2}(Q^V_i-Q^A_i)\\
Q^R_i &=& \int\!d^3x\,\psi_R^\dagger{\lambda_i\over 2}\psi_R
={1\over 2}(Q^V_i+Q^A_i)
\eea
again commute with the QCD Hamiltonian.

For $m_q=0$, the Lagrangian of QCD contains yet another 
symmetry. Under global $U_A(1)$ (axial) transformations
\beq
\psi\to e^{-i\alpha\gamma_5}\psi
\eeq
(\ref{LQCD}) is also invariant
implying a conserved (singlet) Noether current of the
form
\beq
j_{A,0}^\mu=\bar\psi\gamma^\mu\gamma_5\psi\pkt
\eeq 
However, in the full quantum theory the divergence of $j_{A,0}^\mu$ 
has an anomaly
\beq
\partial_\mu j_{A,0}^\mu={3\over 8}\alpha_s G^a_{\mu\nu}\tilde G_a^{\mu\nu}
\quad , 
\quad \tilde G^a_{\mu\nu}=\epsilon^{\mu\nu\alpha\beta}G^a_{\alpha\beta} \ ,
\eeq
-- the $U(1)_A$ {\em axial anomaly} -- such that finally QCD is symmetric only 
under the group 
\beq
SU(3)_L \times SU(3)_R \times U(1)_V \ .
\label{Sgroup}
\eeq
This implies the conservation of the baryonic as well as the 
vector and axialvector currents.

Moreover, at the classical level,  massless QCD is {\it scale invariant}. 
To see this consider a scale transformation in Minkowski space, 
\beq
x^\mu\to x'^\mu=\lambda x^\mu\pkt
\label{Scaletr}
\eeq
Quantum fields scale as 
\beq
\phi(x)\to \phi'(x)=\lambda^d\phi(\lambda x) \ , 
\eeq
where $d=3/2$ for fermions and $d=1$ for vector bosons. A field theory is
called scale invariant if the action $S$ remains invariant under
scale transformations (\ref{Scaletr}):
\beq
S'=\integ ^4x\,{\cal L}'(x)=\integ ^4x \lambda^4{\cal L}(\lambda x)=S \ , 
\eeq
\ie, ${\cal L}$ scales as $\lambda^4$. Associated with scale invariance
is another conserved current -- the dilation (or scale) current --, 
\beq
j_D^\mu=x_\nu T^{\mu\nu} \ , \quad \partial_\mu j_D^{\mu}=T_\mu^\mu=0 \ , 
\label{Jdil}
\eeq
where $T^{\mu\nu}$ denotes the energy-momentum tensor of the theory. 
Considering classical quark and gluon fields and explicitly 
constructing $T^\mn$ it is easily verified that the dilation current of
QCD is conserved.

As in the case the of the axial anomaly, scale invariance is broken in the
full quantum theory. Renormalization requires the introduction of a
scale $\Lambda$ resulting in a running coupling constant, Eq.~(\ref{asrun}). 
As a consequence the
dilation current is no longer conserved. Including finite quark masses
the full expression becomes
\beq
\partial_\mu j_D^\mu={\beta(\alpha_s)\over 4\alpha_s}
G_\mn^aG_a^\mn+(1+\gamma)\bar\psi{\cal M}^\circ\psi \ , 
\label{trano}
\eeq
where $\beta(\alpha_s)=\Lambda d\alpha_s(\Lambda)/d\Lambda$ is the 
Gell-Mann-Low $\beta$-function of QCD and $\gamma=
d(\ln m_q(\Lambda))/d(\ln\Lambda)$ the anomalous dimension. Expanding the 
$\beta$-function in powers of $\alpha_s$, 
\beq
\beta(\alpha_s)=-(33-2N_f){\alpha_s^2\over 6\pi}+O(\alpha_s^3) \ , 
\eeq
and keeping only the lowest term results in the operator 
identity~\cite{DrukLev}
\beq
T_\mu^\mu=-{9\over 8}G^2+\bar\psi{\cal M}^\circ\psi \ , 
\label{tmumu}
\eeq
where $G^2\equiv{\alpha_s\over\pi}G_\mn^aG_a^\mn$ and the (small) 
anomalous dimension $\gamma$ has been neglected.

\section{Vacuum Condensates}
In the physical vacuum quarks and gluons condense giving rise 
to nonvanishing vacuum expectation values 
$\ave{\bar\psi\psi}$~\cite{Nambu,GOR} and $\ave{G^2}$~\cite{SVZ}. 
The physical mechanism is believed to be provided by instantons
-- semiclassical configurations of the gluon fields in 4-dimensional
euclidean space~\cite{SS98}. The gluon condensate may be viewed as a
strength parameter associated with nonperturbative scale breaking effects.   
A finite quark condensate implies that chiral symmetry is
spontaneously broken. In mathematical terms the symmetry group (\ref{Sgroup}) 
is broken down to
\beq
SU(3)_V \times U(1)_V\pkt
\eeq
The baryon and vector current remain conserved but the QCD vacuum 
is no longer symmetric under axialvector transformations (\ref{VAtr}).
While the axial charges $Q^A_k$ still commute with the QCD Hamiltonian
the axial charge of the vacuum is nonvanishing: $Q^A_k\ket{0}\neq 0$.
The situation is analogous to a ferromagnet which consists of separate 
domains of aligned spins. For a given domain rotational symmetry is 
partially broken in the ground state, although the Hamiltonian is 
rotationally invariant.

\begin{figure}[!ht]
\hspace{3.5cm}\epsfig{figure=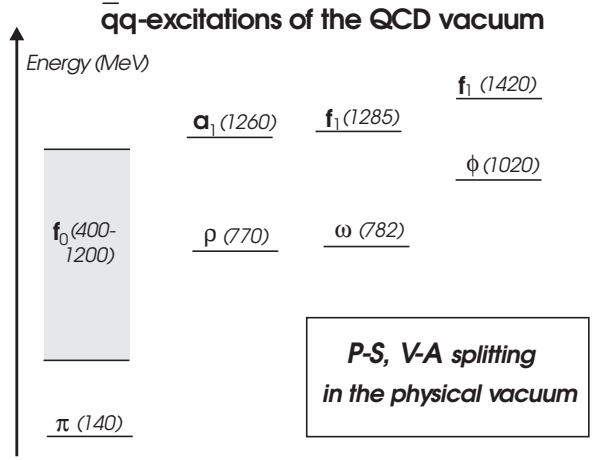,height=6cm}
\caption{Experimentally observed spectrum of low-mass mesons.}
\label{fig_nodub}
\end{figure}

For the light meson spectrum spontaneous 
chiral symmetry breaking manifests itself in two ways:

\begin{itemize}
\item[(i)] the appearance of eight (nearly) massless Goldstone bosons
          (pions, kaons, eta) which interact weakly at low energies.
          The ferromagnetic analogy of Goldstone particles is the occurrence 
          of a spin wave. For large wavelengths the spin configuration 
          begins to resemble a           
          uniform rotation of all the spins. In the limit of infinite 
          wavelength this does not cost any energy, thus yielding a massless 
          Goldstone mode.
\item[(ii)] the absence of parity doublets, \ie, the splitting of 
          scalar and pseudoscalar, as well as vector and axialvector 
          mesons. For massless fermions helicity eigenstates are also
          parity eigenstates. Were chiral symmetry unbroken
          one would expect degenerate hadronic isospin multiplets of
          opposite parity which is clearly not observed in nature 
          as apparent from Fig.~\ref{fig_nodub}.
\end{itemize}
As a further consequence of chiral symmetry breaking the axial-current
matrix element between the vacuum and a Goldstone boson is nonvanishing.
For pions one has for instance:
\beq
\bra{0}j_{A,k}^\mu(x)\ket{\pi_j(p)}=-i\delta_{jk}f_\pi p^\mu e^{-ipx} \ ,
\eeq
and the pion decay constant $f_\pi=93$ MeV serves as an order parameter
which measures the strength of the symmetry breaking. A second order 
parameter is the quark condensate
\beq
\ave{\bar\psi\psi}\equiv\bra{0}\bar\psi_L\psi_R+\bar\psi_R\psi_L\ket{0} \ , 
\eeq
which exhibits the explicit mixing of left- and right-handed quarks in
the QCD vacuum. The order parameters $f_\pi$ and $\ave{\bar\psi\psi}$ are
related. To see this one makes use of the operator identity
\beq
[Q^A_i,[Q^A_j,H_{QCD}]]
=\delta_{ij}\int\!d^3x\,\bar\psi(x){\cal M}^\circ\psi(x)\pkt
\label{OpId}
\eeq
Taking the vacuum matrix element and inserting a complete set of excited states
$\ket{n}$ one obtains the 'energy-weighted sum rule'~\cite{Chanfr}
\beq
\sum_n 2E_n|\bra{n}Q^A_i\ket{0}|^2=-\ave{m_u\bar uu+m_d\bar dd }=
-2\bar m\ave{\bar qq}
\eeq
where $\bar m$ denotes the average of $m_u$ and $m_d$ and $\ave{\bar qq}
\equiv\ave{\bar uu}=\ave{\bar dd}$.
Upon saturating $\ket{n}$ by single-pion states the 
Gell-Mann-Oakes-Renner relation (GOR)~\cite{GOR}  
\beq
m_\pi^2f_\pi^2=-2\bar m\ave{\bar qq}
\label{GOR}
\eeq
is obtained.
Taking $\bar m=6$ MeV yields a value for the
quark condensate, $\ave{\bar qq}=-(240\,{\rm MeV})^3=-1.8\,{\rm fm}^{-3}$.
Focusing on vector mesons a further order parameter can be 
specified as the difference between the vector and axialvector current 
correlators, $\ave{j_{V,k}^\mu (x) j_{V,k}^\mu (0)}
-\ave{j_{A,k}^\mu (x) j_{A,k}^\mu (0)}$. It provides a direct
link between chiral symmetry breaking and the spectral properties of
vector and axialvector mesons and will be of most relevance in 
connection with dilepton production in heavy-ion experiments. 
Also this order parameter is
related to $f_\pi$ via 'Weinberg sum rules' as will be discussed later.

For the medium modification of the quark condensate it will be important 
to consider matrix elements
of the operator identity (\ref{OpId}) for a given hadron $h$. This
defines the hadronic sigma commutator, or '$\sigma$-term'
\beq
\Sigma_h=\bra{h}[Q^A_i,[Q^A_j,H_{\rm QCD}]\ket{h}=\bra{h}
\bar\psi{\cal M}^\circ\psi\ket{h} \ , 
\eeq
and the ratio $\Sigma_h/\bar m$ has the physical interpretation of the scalar
quark density inside the hadron $h$. By using the Feynman-Hellmann Theorem
$\Sigma_h$ can also be expressed in terms of the hadron mass as
\be
\Sigma_h=m_q{\partial m_h\over\partial m_q}\pkt
\label{Sigh}
\ee 
From this relation and the GOR one immediately derives for the pion that 
$\Sigma_\pi=m_\pi/2=69$~MeV, \ie, the
scalar quark density inside the pion is $\Sigma_\pi/\bar m\simeq 12$. This
large value reflects the Goldstone nature of the pion in which
quark-antiquark pairs are highly correlated. The sigma commutator
for the nucleon can be inferred phenomenologically from low-energy
pion-nucleon scattering~\cite{GaLeSa} and has a value 
$\Sigma_N\simeq 45\pm 8$ MeV. In models such as the 'cloudy bag model'
about 2/3 of this value originates from the virtual pion cloud around
the nucleon~\cite{JTC,BMcG}, 
again reflecting the collective nature of the pion.

\section{In-Medium Condensates via Low-Density Expansions}
\label{sec_medcond}
It is to be expected that the quark and gluon condensates are modified 
at finite temperature, $T$, and quark chemical potential, 
$\mu_q$. To discuss these modifications we first recall some basic 
thermodynamics. The equilibrium 
properties of a given system of 3-volume V in contact with a 
reservoir are specified by the grand canonical partition function
\beq 
{\cal Z}(V,T,\mu_q)={\rm Tr}\{e^{-(\hat{H}-\mu_q \hat{N})/T}\} \ , 
\label{Partf}
\eeq
where $\hat{H}$ is the Hamiltonian and $\hat{N}$ the quark number
operator. The thermal average of any operator ${\cal O}$
is then given as
\be
\tave{{\cal O}}
={\cal Z}^{-1}\sum_n\bra{n}{\cal O}\ket{n}e^{-(E_n -\mu_q)/T} \ ,
\ee
where the sum extends over a complete set of eigenstates of $H$ and 
$E_n$ are the corresponding eigenvalues. When applied to the QCD 
condensates one has
\be
\tave{\bar\psi\psi}
={\cal Z}^{-1}\sum_n\bra{n}\bar\psi\psi\ket{n}e^{-(E_n -\mu_q)/T}
\label{thquark}
\ee
and
\be
\tave{G^2}={\cal Z}^{-1}\sum_n\bra{n}G^2\ket{n}e^{-(E_n -\mu_q)/T} \ , 
\label{thgluon}
\ee
where $\ket{n}$ and $E_n$ are now the exact QCD eigenstates and energies.
The equation of state (EoS) can be directly obtained from the logarithm of 
${\cal Z}$. The free energy density is given as
\be
\Omega  = - {T \over V} \ln {\cal Z} \ ,  
\label{freeenergy}
\ee
while the energy density and pressure are derivatives of $\ln{\cal Z}$ with 
respect to $T$ and $V$,
\begin{eqnarray}
\epsilon&=&{T^2 \over V}\biggl 
({\partial {\rm ln}{\cal Z} \over \partial T}\biggr )_{V,N}
+\mu_q\ \ {N\over V} \ 
\label{energy} \\
p&=&T \biggl ({\partial {\rm ln}{\cal Z} \over \partial V}\biggr )_{T,N} \ . 
\label{pressure}
\end{eqnarray}
In the thermodynamic limit ($V\to\infty ;\, N/V={\rm const}$) 
the pressure is directly given by the free energy density as
\be
p = -\Omega \ .    
\ee
In fact, the state variables of the EoS determine directly the condensates. 
The quark condensate is obtained as the derivative of the free energy 
density (or pressure)
with respect to the current quark mass,
\be
\ave{\bar\psi\psi}={\partial \Omega\over \partial m_q}
=-{\partial p\over \partial m_q}\pkt
\label{thcc}
\ee
This is analogous to a
spin system in an external magnetic field, where, in QCD,  the role of the 
latter is played by the quark mass.  
To relate the gluon condensate to the thermodynamic 
variables one should notice that the 
thermal average of the trace of the energy-momentum tensor is given by
\be
\tave{T^\mu_\mu}=\epsilon-3p \ . 
\ee
From Eq.~(\ref{tmumu}) we then have 
\be
\tave{G^2}=-{8\over 9}\biggl 
[(\epsilon-3p)+m_q{\partial p\over\partial m_q}\biggr ]  \  .
\label{thgc}
\ee 

Model-independent results for the changes of the quark and gluon condensate
can be obtained at low temperatures and small baryon densities. 
In both cases one deals with a low-density gas of confined hadrons. It is 
therefore appropriate to evaluate the thermal averages (\ref{thquark})
and (\ref{thgluon}) in a hadronic basis including the
vacuum as well as the lowest-mass mesons and baryons. 
As the temperature increases pions are thermally excited first since 
they represent the lightest hadrons. Considering
a dilute, non-interacting pion gas, the leading correction to the vacuum 
condensate $\ave{\bar\psi\psi}$ is therefore given by the matrix element
$\bra{\pi}\bar\psi\psi\ket{\pi}$, and from (\ref{GOR}) and (\ref{Sigh})
it is easily worked out that the condensate ratio becomes
\be
{\tave{\bar\psi\psi}\over\ave{\bar\psi\psi}}\simeq 1-
{\Sigma_\pi\varrho^s_\pi(T)\over f^2_\pi m^2_\pi } \ , 
\label{Dilute1}
\ee
where $\Sigma_\pi$ denotes the pion $\sigma$-term and $\varrho^s_\pi$
the pion scalar density at given temperature. This expression is consistent
with Eq.~(\ref{thcc}) when using the free energy density of a 
non-interacting Bose gas and the $\sigma$-term as the quark-mass
derivative (\ref{Sigh}). A similar argument holds
for finite baryon density and vanishing temperature. In this case nucleons 
give the dominant correction to $\ave{\bar\psi\psi}$ and
\be
{\tave{\bar\psi\psi}\over\ave{\bar \psi\psi}}\simeq 1-
{\Sigma_N\varrho^s_N(\mu_N)\over f^2_\pi m^2_\pi } \ , 
\label{Dilute2}
\ee
where $\Sigma_N$ is the nucleon $\sigma$-term and $\varrho^s_N$
the nucleon scalar density at given $\mu_q$ (the nucleon chemical potential
is related to the one of quarks by simply $\mu_N=3\mu_q$). Since nucleons
are heavy the scalar density is nearly equal to the number
density, $\varrho_N$. The underlying physical picture that emerges
from the low-density expansion is very simple. Whenever a 
hadron is created in the vacuum, the  condensate is changed locally
since the $\bar \psi\psi$-expectation value inside a hadron is different 
from that of the vacuum.
 
Let us now turn to the low-density expansion of the gluon condensate.
At finite temperature the leading correction to the vacuum condensate
should again be given by non-interacting pions. Since $G^2$ is a chiral
singlet its one-pion matrix element vanishes and there
is no contribution from the ideal pion gas. At finite baryon density
the situation is different. First we note that the nucleon mass is
determined by $T_\mu^\mu$ as
\be
\bra{N}T_\mu^\mu\ket{N}=m_N\bar\psi_N\psi_N=
\biggl [-{9\over 8}\bra{N}G^2\ket{N}+\bra{N}\bar\psi {\cal M}^\circ\psi\ket{N}
\biggr ] 
\bar\psi_N\psi_N \ , 
\ee
where $\psi_N$ represents the nucleon spinor and the operator identity
(\ref{tmumu}) has been used. According to (\ref{Sigh}) the term proportional 
the quark mass, $m_q$, is nothing but the nucleon $\sigma$-term which is about
45~MeV. Thus the bulk of the nucleon mass is in fact generated by the 
gluon field. Putting things together the leading correction to the
gluon condensate arises from finite baryon density such that
\be
\tave{G^2}-\ave{G^2}=-{8\over 9}m_N^{(0)}\varrho^s_N(\mu_N) \ , 
\label{Diluteg}
\ee   
where $m_N^{(0)}$ denotes the contribution of the nucleon mass from the 
$G^2$ matrix element.

One important conclusion which can be drawn from the above arguments  
is that, in spite of changing condensates, the properties of 
the involved hadrons by definition remain unchanged to lowest order.

At vanishing baryon density interactions in the pion gas can be included 
via chiral perturbation theory leading to a rigorous low-temperature
expansion of the condensates~\cite{GaLe,GeLe,Leut}. The starting point
are the thermodynamic relations (\ref{thcc}) and (\ref{thgc}). For
simplicity we restrict ourselves to the chiral limit, $m_q\to 0$.
In this case the leading contribution to $p$ or $\epsilon$ is from the
massless ideal Bose gas which is of order $O(T^4)$. The interactions among
the Goldstone bosons only show up at order $O(T^8)$ such that
\be
p={\pi^2\over 90}(N_f^2-1)T^4\biggl [1+N_f^2\bigl ({T^2\over 12 f_G^2}\bigr )^2
\ln\bigl ({\Lambda_p\over T}\bigr )\biggr]+O(T^{10}) \ , 
\label{pchpt}
\ee
where $f_G$ denotes the weak-decay constant of the Goldstone boson  
in the chiral limit and $\Lambda_p$
appearing in the chiral logarithm is the regularization scale.
There is no $T^6$-term since -- in the chiral limit -- the forward scattering
amplitude for two Goldstone bosons vanishes. The quark-mass derivative
occurring in (\ref{thcc}) can be rewritten in terms of a derivative 
{\it w.r.t.} the mass of the Goldstone boson, $m_{\rm G}$, and 
by means of the Gell-Mann-Oakes-Renner relation,
\be  
{\tave{\bar\psi\psi}\over\ave{\bar\psi\psi}}= 1+
{1\over f_G^2}{\partial p\over\partial m_{\rm G}^2} \ . 
\label{ccrchpt}
\ee
Injecting the expression (\ref{pchpt}) for the pressure one arrives at
\beq
{\tave{\bar\psi\psi}\over\ave{\bar\psi\psi}}=1-{(N_f^2-1)\over N_f}
{T^2\over 12 f_G^2}+{(N_f^2-1)\over 2N_f}
\biggl({T^2\over 12 f_G^2}\biggr)^2-N_f(N_f^2-1)
\biggl({T^2\over 12 f_G^2}\biggr)^3
\ln\bigl ({\Lambda_q\over T}\bigr )+O(T^8)\pkt
\label{qqbarT}
\eeq
While the $T^2$- and $T^4$-terms are model-independent,
model dependence enters at order $O(T^6)$ through the regularization scale 
$\Lambda_q$ which is related to $\Lambda_p$ as~\cite{Leut}
\be
\ln\bigl ({\Lambda_q\over\Lambda_p}\bigr)={N_f^2+1\over 6N_f^4}+0.491 \ .
\ee
The numerical value $\Lambda_q\simeq 470$ MeV ($N_f=2$) is determined by the
isoscalar $D$-wave $\pi\pi$ scattering length. For two flavors the temperature 
scale is set by $\sqrt{8}f_\pi\simeq 260$~MeV. 
To derive a low-temperature expansion for the gluon condensate from
(\ref{thgc}) one uses the fact that
\be
\tave{T_\mu^\mu}=\epsilon-3p=T^5{d\over dT}\biggl ({p\over T^4}\biggr) \ , 
\label{Tmmp}
\ee
from which $\tave{G^2}$ can be expressed in terms of the 
Bose gas pressure alone.
The use of Eq.~(\ref{pchpt}) leads to
\be
\tave{G^2}-\ave{G^2}=-{\pi^2\over 3240}N_f^2(N_f^2-1){T^8\over f_G^4}
\biggl [ {\Lambda_p\over T}-{1\over 4}\biggr ]+\cdots \ . 
\label{gcchpt}
\ee
The leading $O(T^8)$-behavior  is easily 
understood from the observation that to order $O(T^4)$ one has a massless 
ideal Bose gas for which $\epsilon=3p$ and hence
$\tave{T_\mu^\mu}=0$. This is in agreement with the fact that a free gas 
of massless particles is scale invariant. The change in the gluon condensate 
arises  solely on account of the interaction of Goldstone bosons which 
is not scale invariant. The
high power of $T$ implies that the gluon condensate 'melts' much more 
slowly than the quark condensate.

\section{Lattice Results}
\label{sec_lattice}
Obviously the low-density expansion of $\tave{\bar\psi\psi}$ and $\tave{G^2}$
discussed in the previous section
is of limited validity and cannot address the nature of the QCD phase 
transition. 
The low-temperature expansion is restricted to below $T\simeq 120$~MeV,
 mostly because at this point heavier mesons start to enter~\cite{GeLe}.
For finite $\varrho_N$ and vanishing $T$ the dilute gas expression 
(\ref{Dilute2}) predicts a
decrease of the chiral condensate ratio which is linear in the number density. 
At nuclear saturation density, $\varrho_0=0.16$~fm$^{-3}$, 
this yields a $\sim$~30~\% drop and a naive extrapolation would 
indicate chiral restoration at $\varrho_c\simeq 3\varrho_0$. 
This clearly cannot be trusted, since the EoS of nuclear
matter greatly differs from that of a free Fermi gas at such high
densities. Going beyond the dilute gas limit by using a realistic EoS it
has been found~\cite{Brockm} that deviations from the dilute gas set in
slightly above $\varrho_0$. 

In the vicinity of the phase boundary nonperturbative methods are needed.
Even though many-body approaches~\cite{Aouissat} and renormalization-group
techniques \cite{Schaefer} are quite promising  
the most stringent framework is lattice QCD. Here the aim is an ab-initio
understanding of the quark-hadron transition by evaluating the 
partition function (\ref{Partf}) of QCD numerically. Because of technical
difficulties this has been achieved, so far, only at vanishing baryon density.
 
In a theory of interacting boson fields $\phi$ and fermion fields 
$\psi$ in contact with a heat bath the partition function ${\cal Z}$ at
vanishing $\mu_q$ is given by the finite-temperature path integral
\be
{\cal Z}(V,T)=\int {\cal D}\phi{\cal D}\psi {\cal D}\bar\psi
e^{-S^E(\phi,\psi,\bar\psi)}
\ee
involving the euclidean action 
\be
S^E(\phi,\psi,\bar\psi)=\int _0^{T^{-1}}\!\!\!\!d\tau\int_Vd^3x {\cal L}^E(x)
\ , 
\ee 
where $x=(\vec x,\tau)$ with $\tau =it$ and ${\cal L}^E$ denotes 
the imaginary-time Lagrange density.
The boson (fermion) fields obey periodic (antiperiodic)
boundary conditions \cite{Kapusta}.
The thermal expectation value of a given operator $\cal O$
is given as an ensemble average
\ba
\ll\!{\cal O}\!\gg&=&{1\over {\cal Z}}
\tr\{e^{-(\hat H-\mu_q\hat N)/T} {\cal O}\}
\nonumber\\
    &=&{1\over {\cal Z}}\int\! {\cal D}\phi{\cal D}\psi {\cal D}\bar\psi
     \,{\cal O}(\phi,\psi,\bar\psi)
           e^{-S^E(\phi,\psi,\bar\psi)} \ , 
\ea
and the field theory has been turned into a statistical
mechanics problem. 

For QCD the euclidean Lagrange density is given by (cf.~Eq.~(\ref{LQCD}))
\be
{\cal L}^E_{QCD}(x)=
\bar \psi(x)(-i\gamma_\mu D_\mu-i{\cal M}^\circ)\psi(x)
+{1\over 4}G^a_\mn(x) G^a_\mn(x)
\ee
with the euclidean Dirac matrices obeying  
$\{\gamma_\mu,\gamma_\nu\}=2\delta_\mn$. The corresponding
euclidean action is obtained as 
\be
S^E_{QCD}(A_\mu,\psi,\bar\psi)=\int _0^{T^{-1}}\!\!\!\!d\tau
\int_Vd^3x {\cal L}^E_{QCD}(x) \ . 
\label{Seucl}
\ee 
For a numerical evaluation of the partition function ${\cal Z}_{QCD}$
via the path integral a hypercubical lattice 
of spacing $a$ with $N_s$ lattice points in each spatial direction and 
$N_\tau$ points in the temporal direction is introduced.
Temperature and volume are related to the lattice size, 
$N_s^3 \times N_\tau$, as
\be
T^{-1} = N_\tau a \ , \qquad V = (N_s a)^3 \ , 
\label{TandV}
\ee
and the temperature and volume derivatives are replaced according to
\ba
{\partial\over\partial T}&\to& {1\over N_\tau}{\partial\over\partial a}
\nonumber\\
{\partial\over\partial V}&\to& {1\over 3a^2N_s^3}{\partial\over\partial a}
 \  . 
\ea
Because of the scale dependence of $\alpha_s$, Eq.~(\ref{asrun}), 
the lattice spacing becomes a function of the bare gauge coupling 
$\beta \equiv 6/g^2$ 
which fixes the temperature and the physical volume at a given coupling.
The next step is to discretize the euclidean QCD action (\ref{Seucl}).
Two requirements have to be met. The first is
the correct continuum form of the action in the limit
$a\to 0$. The second is local gauge invariance. 
To proceed one considers the 'Schwinger line integral'
\be
U^\mu(x)=e^{ig\int_x^{x+a\hat\mu}dy A_\mu(x+y)}
\ee
rather than the gauge field $A_\mu$. In terms of
$U^\mu$ the QCD partition function is expressed as
\be
{\cal Z}_{QCD}=\int\!{\cal D}U{\cal D}\psi{\cal D}\bar\psi
e^{-S^E_{QCD}(U,\psi,\bar\psi)}\pkt
\label{Partc}
\ee
The discretization of the field variables $U^\mu(x)\to U^\mu_n$
and $\psi(x)\to \psi_n$ turns the line integral into a 'link
variable' which connects lattice site $n$ to its neighbor $n+\hat \mu$.
The quark fields are defined on the lattice sites $n$.
  
In the Wilson formulation~\cite{Wilson} the gluonic part of the 
action, $S_G=\int \frac{1}{4}(G_{\mu\nu}^a)^2$, is expressed in 
terms of elementary plaquettes
\be 
U_{\Box}\equiv U^\mu_nU^\nu_{n+\hat\mu}U^{\mu\dagger}_{n+\hat\nu}
U^{\nu\dagger}_n \ ,  
\ee
which constitute the smallest closed path starting from lattice site $n$.
In terms of these plaquettes one has
\be
S^E_G=\beta\sum_\Box (1-{1\over 3}{\rm Re}{\rm Tr} U_{\Box}) \ , 
\ee
where the sum runs over all possible plaquettes. For small $a$
\be
S^E_G={\beta g^2\over 12}\sum_na^4({\rm Tr}G_\mn(n) G_{\mn}(n)+{\cal O}(a^2))
 \ , 
\ee
which extrapolates to the proper continuum limit with
$\beta=6/g^2$ and hence satisfies both requirements of the lattice
action.

To find an appropriate form for the fermionic part is more difficult.
A naive discretization of the continuum Dirac action,
\be
S^E_F=\int _0^{T^{-1}}\!\!\!\!\!\!d\tau\int_Vd^3x
\bar \psi(x)(\gamma_\mu D_\mu+{\cal M}^\circ)\psi(x) \ , 
\ee
leads to a lattice action 
\be
S^E_F=\sum_{n,m}a^4\bar\psi_nK_{nm}[U]\psi_m \ , 
\ee
where
\ba
K_{nm}&=&\gamma_\mu D_{\mu,nm}+{\cal M}^\circ\delta_{nm}
\nonumber\\
D_{\mu,nm}&=&{1\over 2a}\{U^\mu_n\delta_{n+\hat\mu,m}-
U^{\mu\dagger}_{n-\hat\mu}\delta_{n-\hat\mu,m}\}
\label{Kmat}\pkt
\ea
In the continuum limit it describes $2^4=16$ fermion species rather 
then one. This fermion doubling per field component has
its origin in the first derivative occurring in the Dirac equation.
To remove the spurious degeneracy two methods have been
proposed: Wilson fermions~\cite{Wilson} in which the naive lattice action
is supplemented by an extra term which ensures that in the
continuum limit the extra 15 species are removed. It has
the disadvantage that the numerical realization of the chiral 
limit $m_q\to 0$ is extremely time consuming. In studies which
focus on the chiral phase transition the method of 
staggered fermions~\cite{KogutS} is more suitable. Here the
unwanted fermions are removed by doubling the effective
lattice spacing: two separate fermion fields are introduced 
for even and odd lattice sites, which in the contimuum limit 
are associated with  the upper-two and lower-two components of the 
4-component Dirac spinor, \ie, the quark fields $\psi$ can be 
reconstructed by suitable linear combination on hypercubes of length 
$2a$. More recently, another promising method called 'domain wall
fermions' has been developed to treat fermions in discretized 
vector gauge theories~\cite{dwf}. Here, a fifth dimension 
(in addition to 4-dimensional space-time) is introduced, and the low-lying
(zero mode) fermion states are localized on  'domain walls' in this extra
dimensions. This method has the attractive feature
that -- contrary to staggered or Wilson fermions -- the full chiral
symmetry of QCD is preserved to a high accuracy in its discretized version.   

On a finite lattice the QCD partition function finally takes the form
\be
{\cal Z}_{QCD}=\int\! \prod_{n,\mu}dU^\mu_n \prod_{n_1} d\bar \psi_{n_1} 
\prod_{n_2} d\psi_{n_2}
e^{-S^E_{QCD}(U^\mu_n,\bar \psi_{n_1},\psi_{n_2})}\pkt
\ee
In contrast to the symbolic notation in Eq.~(\ref{Partc}) the  measure
${\cal D}U{\cal D}\bar\psi{\cal D}\psi\equiv \Pi dU\Pi d\bar\psi
\Pi\psi$ now has a well-defined meaning: $dU^\mu_n$ refers
to the measure on the $SU(3)$ gauge group while $d\psi$,  
$d\bar\psi$ are the usual measures over Grassmann variables on site $n$.

The discretized fermion fields appear quadratic in the
action and can therefore be integrated out to yield
\be
\int\!\prod_{n_1} d\bar \psi_{n_1} \prod_{n_2} d\psi_{n_2}
e^{S^E_F[U^\mu_n,\bar \psi_{n_1} \psi_{n_2}]}={\rm det} K[U]\pkt
\label{Grass}
\ee
Thus the QCD partition function is given by a path integral
solely over gauge fields $U$
\be
{\cal Z}_{QCD}=\int\! \prod_{n,\mu}dU^\mu_ne^{-S^E_G(U^\mu_n)}{\rm det} K[U]\pkt
\ee
When factorizing out the quark mass dependence in the matrix
$K_{nm}$ (\ref{Kmat})
as
\be 
K_{nm}=m_q({1\over m_q}\gamma_\mu D^\mu_{nm}+1\delta_{n,m}) \ , 
\ee
one sees that in the limit of large quark masses ($m_q\to\infty$) 
the term involving
the covariant derivative gives a negligible contribution and
the Grassmann integration (\ref{Grass}) becomes a pure Gaussian
integral. This results in a constant multiplicative factor in
the partition function which cancels out in the expectation
values of operators, $\tave{\cal O}$. Thus the limit of large $m_q$ reduces
to a pure gauge theory and is referred to as the 
'quenched approximation'. Physically it corresponds to the
omission of vacuum polarization effects via quark loops.

Although in principle the condensates and the EoS can be derived 
from the free energy density as discussed in Sect.~\ref{sec_medcond},
in practice a direct
computation of the partition function is
rather difficult. Instead one calculates the expectation value
of the action by taking the derivative of $\ln {\cal Z}_{QCD}$ with respect 
to the bare gauge coupling $\beta$ and the bare quark masses $m_q$. 
In this way the pressure, the energy density and the condensates
are obtained by proper extrapolation to the continuum limit $a\to 0$ 
via a renormalization group analysis on the lattice.

\begin{figure}[!htb]
\epsfig{figure=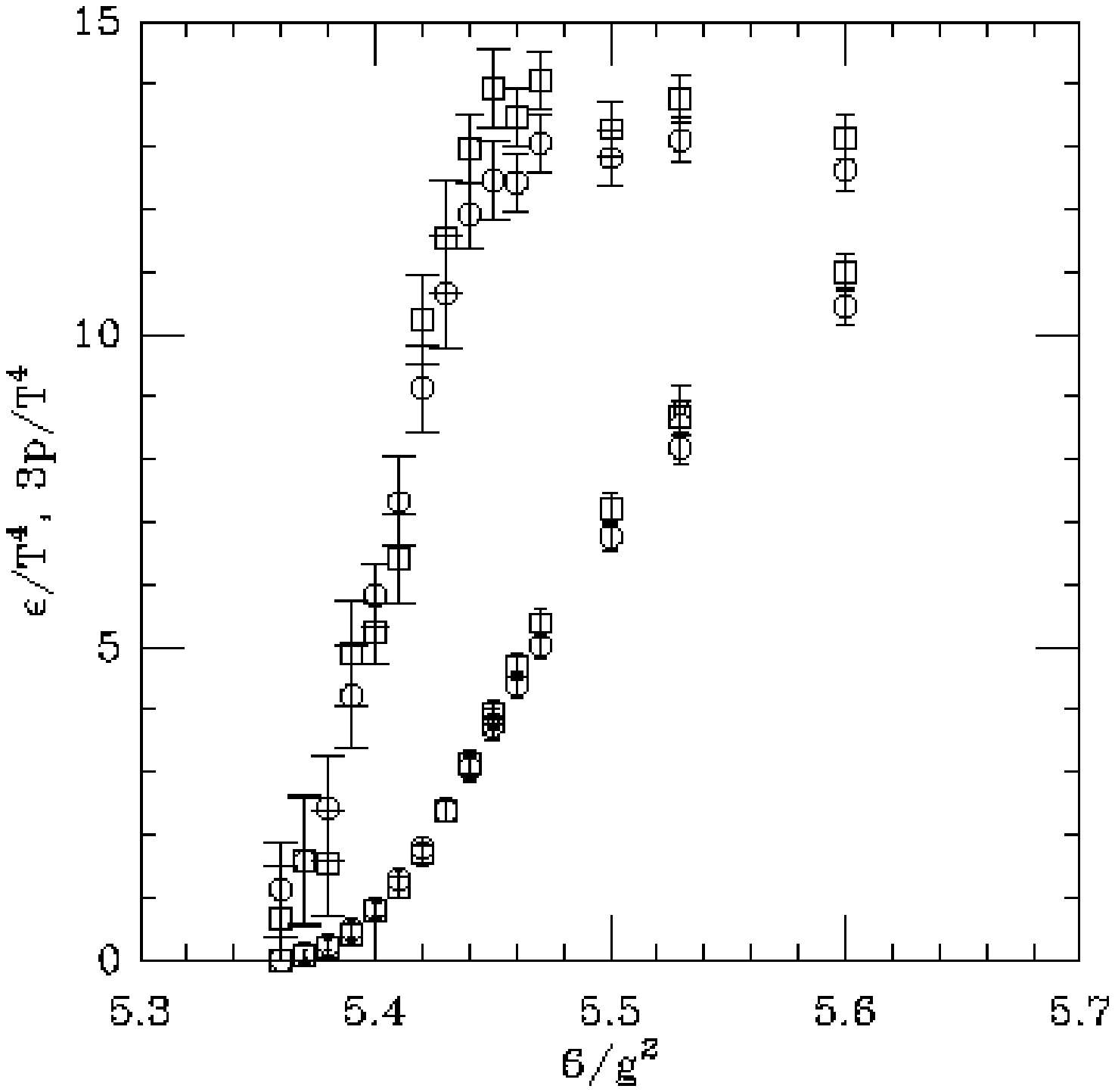,height=5.8cm}
\epsfig{figure=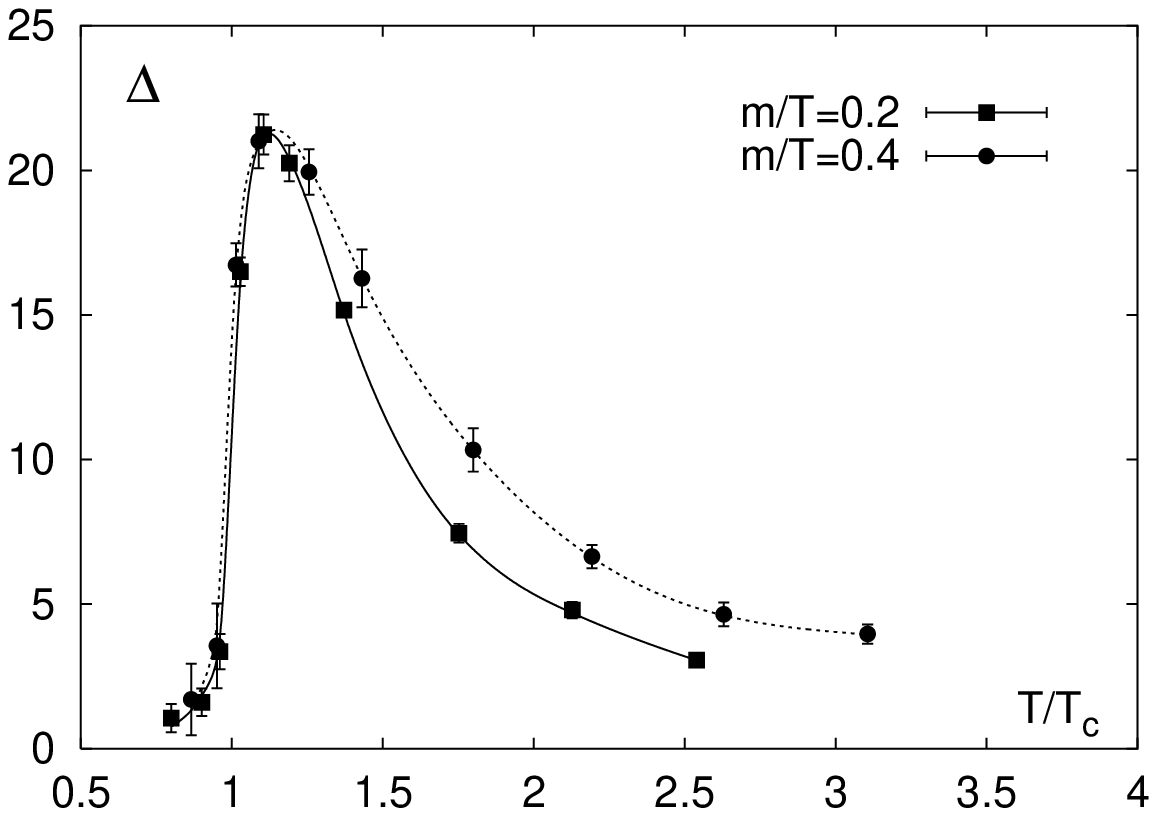,height=5.2cm}
\caption{Left panel: the energy density $\epsilon/T^4$ and $3p/T^4$  
        as a function of $\beta=6/g^2$~\protect\cite{Bernard}.
        The lattice data are for $am_q=0.0125$.
        Right panel:  the interaction measure $\Delta$ as a 
        function of $T/T_c$~\protect\cite{BoydM}.}
\label{fig_LatEoS}
\end{figure}

Recent two-flavor results for $\epsilon$, $p$ and the so-called 
'interaction measure' $\Delta\equiv(\epsilon-3p)/T^4=\tave{T_\mu^\mu}/T^4$ 
are shown in Fig.~\ref{fig_LatEoS}.
One observes a rapid rise in $\epsilon$ (left panel) above a critical
coupling of $\beta_c=5.36$ roughly reaching the continuum Stephan-Boltzmann 
limit at the last data point. The critical coupling of $\beta_c=5.36$
translates to a transition temperature $T_c= 140$ MeV. On the other hand,  
the pressure rises much more slowly such that $\Delta$ remains large
above the transition (right panel). This implies that above $T_c$ the system  
is far from being ideal. However, the entropy density quickly approaches that
of an ideal gas of quarks and gluons. A natural interpretation of this 
feature is that the dominant part of the nonperturbative pressure is 
provided by the remnant of the (vacuum) gluon condensate~\cite{Satzpr}, 
which persists across the transition to a substantial extent (see below). 
Thus, above $T_c$ one might indeed have a deconfined plasma of 
weakly interacting 
quarks and gluons in the background of a residual 'bag' pressure. 

\begin{figure}[!htb]
\vspace{1.5cm}
\begin{minipage}[t]{6.4cm}
{\makebox{\epsfig{file=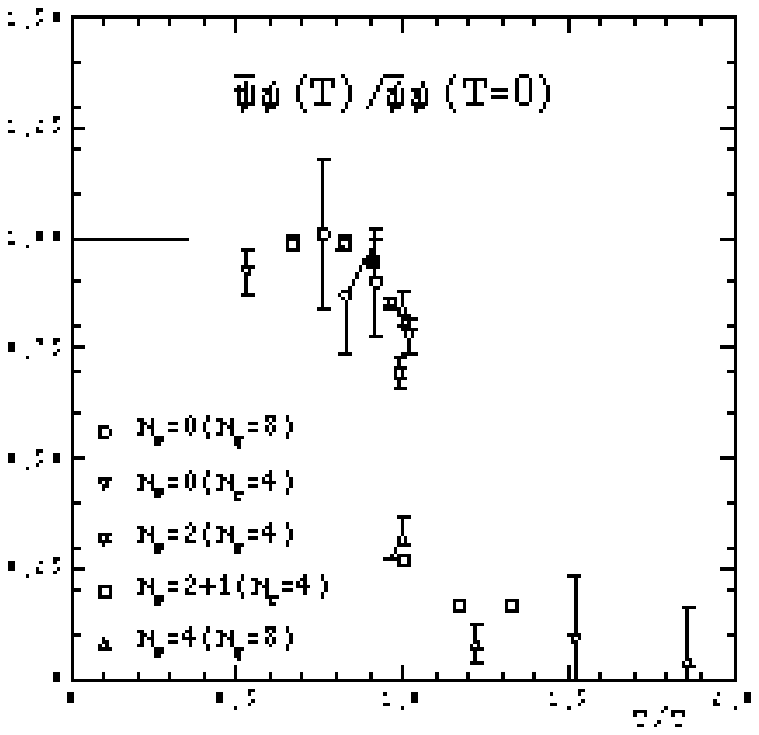,width=6.4cm}}}
\end{minipage}
\begin{minipage}[t]{6.9cm}
{\makebox{\epsfig{file=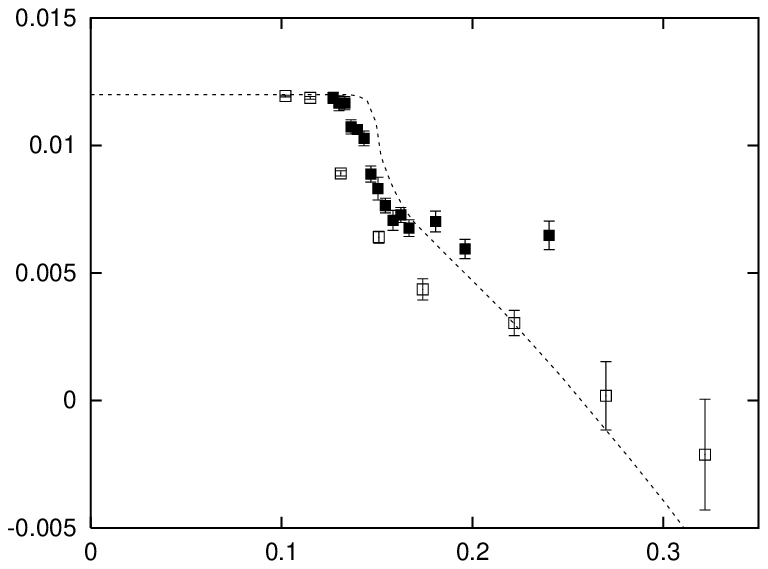,width=6.9cm}}}
\end{minipage}
\vspace{-1cm}
\caption{Left panel: the quark condensate ratio 
$\tave{\bar\psi\psi}/\ave{\bar\psi\psi}$
as a function of $T/T_c$~\protect\cite{Karsch};
right panel:  the temperature dependence of the gluon 
condensate~\protect\cite{BoydM}.} 
\label{fig_Latcon}
\end{figure}

Lattice results for the condensates $\tave{\bar\psi\psi}$ and $\tave{G^2}$ 
are shown in Fig.~\ref{fig_Latcon}.
The condensate ratio stays basically flat up to about 0.9 $T_c$ 
after which it rapidly decreases. At first sight the relative 
constancy at small temperatures is at variance
with the rigorous result from chiral perturbation theory 
which predicts a $T^2$-dependence
of the condensate ratio. It should, however, be realized that the 
lattice results are
not in the chiral limit. Instead the simulations implicitly contain a rather 
'heavy pion' with a mass
of roughly twice the physical mass. This explains the apparent differences with
chiral perturbation theory. The temperature dependence of the gluon 
condensate can be inferred from the interaction measure 
$\Delta$ and the quark condensate via the relations
(\ref{thgc}) and (\ref{thcc}). The right panel of Fig.~\ref{fig_Latcon} 
displays results
for two flavors (solid squares) and for four flavors (open squares) while the
dashed line shows the condensate in the 
quenched approximation rescaled by the number
of degrees of freedom. While the condensate remains essentially unchanged 
below $T_c$ which is consistent with the $T^8$-dependence 
predicted from chiral perturbation
theory, there is a rapid decrease slightly above 
$T_c$ and the condensate eventually
turns negative around $T=200$~MeV. Asymptotic freedom implies  
that the effects of nonperturbative scale breaking disappear at 
high temperature where perturbation theory
should become reliable. Intuitively one might therefore expect 
the gluon condensate to
vanish at high temperature in accordance with the notion of an ideal 
quark-gluon plasma. As the lattice results (as well as chiral 
perturbation theory) show, this is
however not the case. Scale invariance remains broken also in the 
high-temperature phase. This is related to the 
scale dependence of the running coupling constant (\ref{asrun}). 
Identifying the scale with $T$ introduces $1/\ln(T)$ corrections 
to the pressure
such that at high temperatures 
$\tave{T_\mu^\mu}$ is of order $T^4/(\ln T)^2\propto g^4T^4$~\cite{Leut}. 
While small compared to the energy density of the plasma,
 $\tave{T_\mu^\mu}$ itself grows without bounds as the temperature 
rises such that the gluon condensate
does not dissappear but becomes negative and large.

As a further result of lattice QCD we discuss the temperature dependence of 
'screening masses' which will become relevant for the later discussion of
axial-/vector correlation functions.
In general screening masses are extracted from the ensemble averaged
current-current
correlation functions $\tave{j_\mu(x)j_\mu(0)}$
where the appropriate currents are denoted generically by $j_\mu(x)$.
\begin{figure}[!htb]
\bce
\epsfig{figure=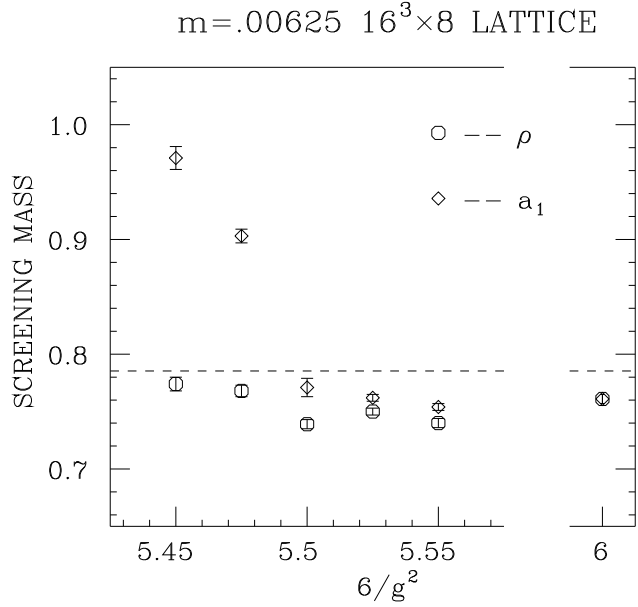,height=8cm}
\ece
\caption{The temperature dependence of the $\rho$ and $a_1$
screening masses. The results are
taken from Ref.~\protect\cite{Gottl}.}
\label{fig_latvec}
\end{figure}
In practical calculations these correlation functions are evaluated for spatial
separations $r$. For large $r$ they show exponential behavior
\beq
\tave{j_\mu(x)j_\mu(0)}\to e^{-m_h r} \ ,
\label{Tcorr}
\eeq
from which the (screening) masses of the 
lowest-energy hadron in the appropriate channel
can be extracted numerically. Fig.~\ref{fig_latvec} displays 
results for the screening masses of the $\rho$ and $a_1$ meson. 
As dictated by spontaneous chiral symmetry breaking  they are 
different in the
physical vacuum. At the chiral transition they are found 
to be degenerate. The point of 
degeneracy coincides with
the transition temperature for the melting of the condensate. 
The screening masses
need not necessarily correspond to the 'pole masses' in the propagators 
of the $\rho$ and $a_1$ meson but rather represent 
the 'centroids' of the in-medium
vector and axialvector spectral functions. The spectral distributions 
themselves are likely to be quite complicated with significant 
reshaping, \eg, a strong  broadening as emerging in models of an 
interacting hadron gas, to be discussed later.

\section{Dilepton Production and Vector Mesons}
\label{sec_DPR+VM}
As mentioned in the Introduction dileptons probe all stages in 
the course of a heavy-ion reaction. The most
interesting hot and dense phase emits low-mass dileptons predominantly 
from thermal annihilation processes such as quark-antiquark or
pion and kaon annihilation.

The thermal rate for the production of dileptons at four-momentum $q$ from a 
heat bath at temperature $T$ is given by~\cite{Fein76,mt85}
\beq
{d^8N_{l^+l^-}\over d^4xd^4q}\equiv{d^4R\over d^4q}=L^{\mu\nu}(q) 
W_{\mu\nu}(q) \ ,  
\label{Dlrate}
\eeq
where to lowest order in the electromagnetic coupling, $\alpha$=1/137, the
lepton tensor is obtained as
\begin{eqnarray}
L_{\mu\nu}(q) &=& \frac{(4\pi\alpha)^2}{M^4} \int
\frac{d^3p_+}{(2\pi)^3 2p_{0,+}} \ \frac{d^3p_-}{(2\pi)^3 2p_{0,-}}
 \ \tr\left[(\not\!{p_+}-m)\gamma_\mu (\not\!{p_-}+m)\gamma_\nu\right]
\delta^{(4)}(q-p_+-p_-) 
\nonumber\\
 &=&-\frac{\alpha^2}{6\pi^3 M^2} \left( g_{\mu\nu} -
\frac{q_\mu q_\nu}{M^2} \right) 
\label{Lmunu}
\end{eqnarray}
with $p_{0,\pm}=(m_{l^\pm}^2+{\vec p}_\pm^2)^{1/2}$.
For simplicity we will focus on the $e^+e^-$ case and thus neglect
the rest mass of the leptons as compared to their
individual 3-momenta $|\vec p_+|$, $|\vec p_-|$.
$M^2=(p_++p_-)^2$ is the total four-momentum
of the pair in the heat bath. The effect of the hadronic medium is 
encoded in the hadron tensor $W_{\mu\nu}(q)$. 
It is obtained from the thermal average of the electromagnetic 
current-current correlation function as~\cite{Fein76,mt85}
\beq
W_{\mu\nu}(q)=\int\!\!d^4x\,e^{-iqx}\tave{j^{\rm em}_\mu(x)
j^{\rm em}_\nu(0)} \ ,  
\label{Wmunu}
\eeq
where the average is taken in the grand canonical ensemble. For invariant
masses below the charm threshold $(M< 2m_c\simeq 3\,{\rm GeV})$ the current
can be decomposed as
\be
j^{\rm em}_\mu={2\over 3}\bar u\gamma_\mu u-{1\over 3}\bar d\gamma_\mu d
-{1\over 3}\bar s\gamma_\mu s\pkt
\ee
With the identification
\bea
j^{\rm em}_\mu&=&j^\rho_\mu + j^\omega_\mu + j^\phi_\mu
\\
j^\rho_\mu &=& {1\over 2}(\bar u\gamma_\mu u-\bar d\gamma_\mu d)
\\
j^\omega_\mu &=& {1\over 6}(\bar u\gamma_\mu u+\bar d\gamma_\mu d)
\\
j^\phi_\mu &=& -{1\over 3}(\bar s\gamma_\mu s) \ , 
\eea
its flavor content can be expressed in physical channels with the
quantum numbers of the $\rho$, $\omega$ and $\phi$ meson. 

\begin{figure}[!ht]
\vspace{1cm}
\bce
\epsfig{figure=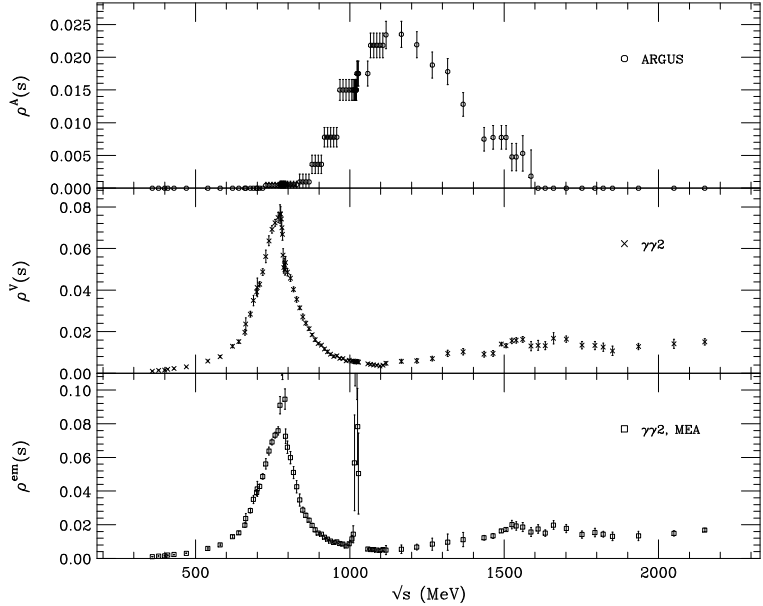,angle=90,height=8cm}
\ece
\caption{Lower panel: experimental cross section ratio 
$R^h(s)/12\pi^2$ according to Eq.~(\ref{Rh}) from a recent data 
compilation~\protect\cite{Huang} of various experiments on $e^+e^-$ 
annihilation~\protect\cite{gg2-79,ww79,olya,dm1-80,vepp2m}.  
The middle and upper panel show the individual experimental 
information on vector and axialvector spectral densities, Eqs.~(\ref{rhoV0})
and (\ref{rhoA0}), as extracted from $e^+e^-\to 2n\pi$ and $\tau$-decay
data, respectively.}
\label{fig_rhoAVem}
\end{figure}

For an ideal plasma of quarks and gluons at finite temperature and
vanishing chemical potential the rate (\ref{Dlrate}) is readily
evaluated. Applying lowest-order perturbation theory and 
integrating over the three-momentum of the dilepton pair
one obtains the familiar expression 
\be
{dR^q\over dM^2}=R^q{\alpha^2\over 6\pi^2}MTK_1(M/T)
\label{Dlrqp}
\ee
where $K_1(M/T)$ denotes a modified Bessel function and $R^q$
involves the sum over squared quark charges and the number
of colors,   
\be
R^q=N_c\sum_fe^2_f=3({4\over 9}+{1\over 9}+{1\over 9}) \ . 
\label{Rq}
\ee
A similar expression can be derived for an ideal resonance
gas at finite temperature:
\be
{dR^h\over dM^2}=R^h(M){\alpha^2\over 6\pi^2}MTK_1(M/T) \ , 
\label{Dlrhg}
\ee
where 
\be
R^h={\sigma(e^+e^-\to hadrons)\over
\sigma(e^+e^-\to \mu^+\mu^-)}
\label{Rh}
\ee
is accessible from experiment as indicated in the lower panel of 
Fig.~\ref{fig_rhoAVem}. 

\begin{figure}[!ht]
\hspace{3cm}\epsfig{figure=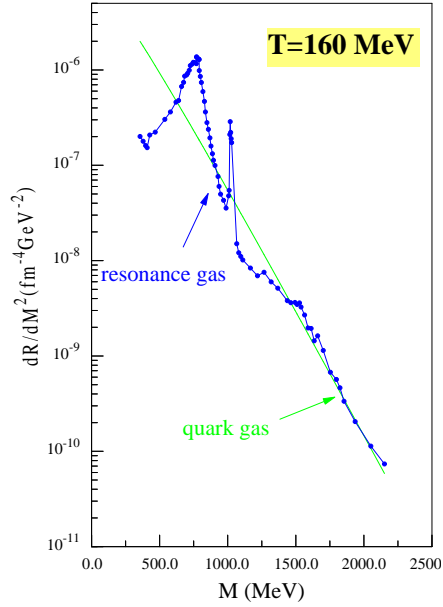,height=8cm}
\caption{Three-momentum integrated dilepton production rate at a temperature
$T=160$~MeV using either the resonance gas approximation, Eq.~(\ref{Dlrhg}), 
or the perturbative $q\bar q$ prediction, Eq.~(\ref{Dlrqp}).} 
\label{fig_dlgas}
\end{figure}

From the rates (\ref{Dlrqp}) and (\ref{Dlrhg}) a simple estimate can be 
made for the expected dilepton signal when the hadronic fireball is close to
the phase boundary (Fig.~\ref{fig_freeze}). As displayed in Fig.~\ref{fig_dlgas}
for $T=160$ MeV and $\mu_B=0$ the predicted rates coincide above $\sim$~1.5~GeV
but differ greatly below due to the $\rho$, $\omega$ and $\phi$
resonance structures in the electromagnetic spectral function (lower 
panel of Fig.~\ref{fig_rhoAVem}) 

\section{Vector-Axialvector Mixing}
\label{sec_vamix}
  
As has been discussed the quark-hadron phase transition is 
accompanied by the restoration of chiral symmetry, \ie, a 
'melting' of the quark condensate at the transition temperature,
$T_c\simeq 160$~MeV. The change of $\tave{\bar\psi\psi}$ is not
an experimental observable, however. On the other hand it follows from 
chiral symmetry alone that, at the phase boundary, vector and 
axialvector correlators must become identical (in the chiral limit). This 
is evidenced by the temperature dependence of the lattice screening 
masses (Fig.~\ref{fig_latvec}) as will be discussed more generally 
in the following.

First we consider the isovector ($I=1$( vacuum vector and 
axialvector correlators for
two flavors. The former clearly dominates the electromagnetic spectral 
function (Fig.~\ref{fig_rhoAVem}) and hence the dilepton rate. One has
\bea
\Pi^{\circ\mu\nu}_V(q) &=& -i\int\!d^4x \ e^{iq\cdot x}\bra{0}{\cal T}j^\mu_V(x)
j^\nu_V(0)\ket{0}
\nonumber\\
\Pi^{\circ\mu\nu}_A(q) &=&-i\int\!d^4x \ e^{iq\cdot x}\bra{0}{\cal T}j^\mu_A(x)
j^\nu_A(0)\ket{0}  \ , 
\label{VAcorrv}
\eea
where 
\bea
j^\mu_V &=& {1\over 2}(\bar u\gamma_\mu u-\bar d\gamma_\mu d)
\\
j^\mu_A &=& {1\over 2}(\bar u\gamma_\mu\gamma_5 u-\bar d\gamma_\mu\gamma_5 d)
\eea
carry the quantum numbers of the $\rho$ and $a_1$ meson, respectively.
The imaginary parts of (\ref{VAcorrv}) can be expressed in 
terms of the vector and axialvector spectral densities, 
$\rho_V$ and $\rho_A$, as
\bea
{1\over\pi}{\rm Im}\Pi^{\circ\mu\nu}_V(q) &=& 
(q^2g^{\mu\nu}-q^\mu q^\nu)\rho^\circ_V(q^2)
\nonumber\\
{1\over\pi}{\rm Im}\Pi^{\circ\mu\nu}_A(q) &=& 
(q^2g^{\mu\nu}-q^\mu q^\nu)\rho^\circ_A(q^2)
-q^\mu q^\nu f_\pi^2\delta(q^2-m_\pi^2)\pkt
\eea
Due to spontaneous symmetry breaking in the physical vacuum 
and the resulting Goldstone nature of the pion the axialvector correlator
contains an additional pion pole term. Chiral symmetry also
dictates a relationship between the vector and axialvector
sector which is encoded in two sum rules~\cite{Wein67}:
\bea
\int\!ds
\bigg (\rho^\circ_V(s)-\rho^\circ_A(s)\biggr ) &=& f_\pi^2
\\
\int\!dss\, \biggl (\rho^\circ_V(s)-\rho^\circ_A(s)\biggr ) &=&0 \ . 
\label{Weinsrv}
\eea
The first one directly links the spectral functions to $f_\pi$,
one of the order parameters of spontaneous symmetry breaking, while
the second one is a well-known consequence of the conservation of
vector and axialvector currents in the chiral limit.
 
The vacuum spectral functions $\rho^\circ_V(s)$ and $\rho^\circ_A(s)$ are 
related to physical 
processes. The vector spectral function can be obtained from the 
$e^+e^-$-annihilation into an even number of pions: 
\beq
\rho^\circ_V(s)=-{s\over 16\pi^3\alpha^2}
\sum_{n=1}{\sigma(e^+e^-\to {\rm 2n}\pi}) \ ,
\label{rhoV0}
\eeq
while $\rho^\circ_A(s)$ can be extracted from data on $\tau$-decay into a 
$\nu_\tau$-neutrino and an odd number of pions:
\be
\rho^\circ_A(s)=\frac{8\pi m_\tau^3}{G_F^2\cos\theta_c(m_\tau^2+2s)
(m_\tau^2-s)^2} \sum_{n=1}{d\Gamma(\tau\to\nu_\tau{\rm(2n+1)}\pi)\over 2s}
 \ . 
\label{rhoA0}
\ee
An available data compilation~\cite{Huang} is displayed in the upper and 
middle panel of Fig.~\ref{fig_rhoAVem}.
The two spectral functions are clearly different which is
one of the experimental signatures that chiral symmetry is spontaneously
broken. Replacing the spectral functions by a simplifying pole ansatz
\beq
\rho^\circ_V(s)={m^4_\rho\over g^2_\rho}{1\over s}\delta(s-m^2_\rho) 
\ ,\quad  
\rho^\circ_A(s)={m^4_{a_1}\over g^2_{a_1}}{1\over s}\delta(s-m^2_{a_1}) \ , 
\label{Polv}
\eeq
one immediately derives from the Weinberg sum rules (\ref{Weinsrv}) that
\beq
{m^4_\rho\over g^2_\rho}={m^4_{a_1}\over g^2_{a_1}} \ ,\quad
m^2_\rho=ag^2_\rho f_\pi^2~~{\rm with~} 
a=\biggl(1-{m^2_\rho\over m^2_{a_1}}\biggr)^{-1}\pkt
\label{Polvv}
\eeq
For $m_{a_1}=\sqrt{2}m_\rho$ (\ie, $a=2$), which is not too far from reality,
the KSFR relation~\cite{KSFR} is recovered.

Turning to the hot hadronic medium the current-current correlation functions
-- in analogy to Eq.~(\ref{VAcorrv}) -- are given as the thermal averages
\bea
\Pi^{\mu\nu}_V(q)&=&-i\int\!d^4x e^{iq\cdot x}\tave{j^\mu_V(x)
j^\nu_V(0)}
\\
\Pi^{\mu\nu}_A(q)&=&-i\int\!d^4x e^{iq\cdot x}\tave{j^\mu_A(x)
j^\nu_A(0)} \ . 
\label{Vcorr}
\eea
Since the thermal medium specifies a preferred frame and thus explicitly
breaks Lorentz invariance the tensor structure
is now more complicated giving rise to a separate dependence on energy
$q_0$ and three-momentum $\vec q$ as well as a splitting into  
longitudinal and transverse components. One has
\be
\Pi^{\mu\nu}_{V,A}(q_0,\vec q)=\Pi_{V,A}^L(q_0,\vec q) \ P_L^{\mu\nu}
+\Pi_{V,A}^T(q_0,\vec q) \ P_T^{\mu\nu}
\label{Plt}
\ee
where $P_L$ and $P_T$ are the usual longitudinal and transverse 
projection operators:
\bea
P_L^{\mu\nu} & = & \frac{q^\mu q^\nu}{M^2}-g^{\mu\nu}-P_T^{\mu\nu}
\nonumber\\
P_T^{\mu\nu} & = &  \left\{ \begin{array}{l}
 \quad~~ 0 \qquad , \ \mu=0 \ {\rm or} \ \nu=0
 \\
\delta^{ij}-\frac{q^iq^j}{{\vec q}^2} \ , \ \mu,\nu \in \lbrace 1,2,3
\rbrace
\end{array}   \right. \  
\label{PLT}
\eea
(the space-like components of $\mu$ and $\nu$ are denoted by
$i$ and $j$, respectively),
and $\Pi_{V,A}^{L,T}(q_0,\vec q)$ denote the longitudinal and transverse
polarization functions. In general they are different and
only coincide for vanishing three-momentum $\vec q=0$, \ie, excitations
which are at rest relative to the medium. In-medium Weinberg sum rules have
been derived in Ref.~\cite{KapuSh}. Introducing the in-medium spectral 
distributions for vanishing three-momentum as
\be
\rho_{V,A}(q_0)=-{1\over q_0^2\pi}\Pi_{V,A}^L(q_0,0) \ , 
\label{specmed}
\ee
these sum rules are given by
\bea
\int\!dq_0^2
\bigg (\rho_V(q_0)-\rho_A(q_0)\biggr ) &=& 0
\label{WSRm1}
\\
\int\!dq_0^2 q_0^2\,
\biggl (\rho_V(q_0)-\rho_A(q_0)\biggr ) &=&0 \  . 
\label{WSRm2}
\eea

\begin{figure}[!ht]
\vspace{1cm}
\begin{center}
\epsfig{figure=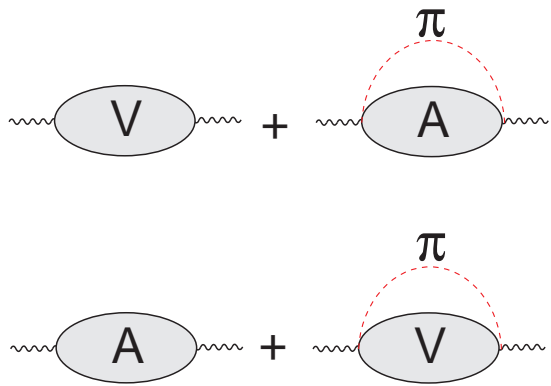,height=5cm}
\end{center}
\hspace{1cm}
\caption{Diagrammatic representation of the mixing of vector and axialvector
correlators in a heat bath of pions.}
\label{fig_vamix}
\end{figure}

\begin{figure}[!ht]
\vspace{1cm}
\hspace{1cm}\epsfig{figure=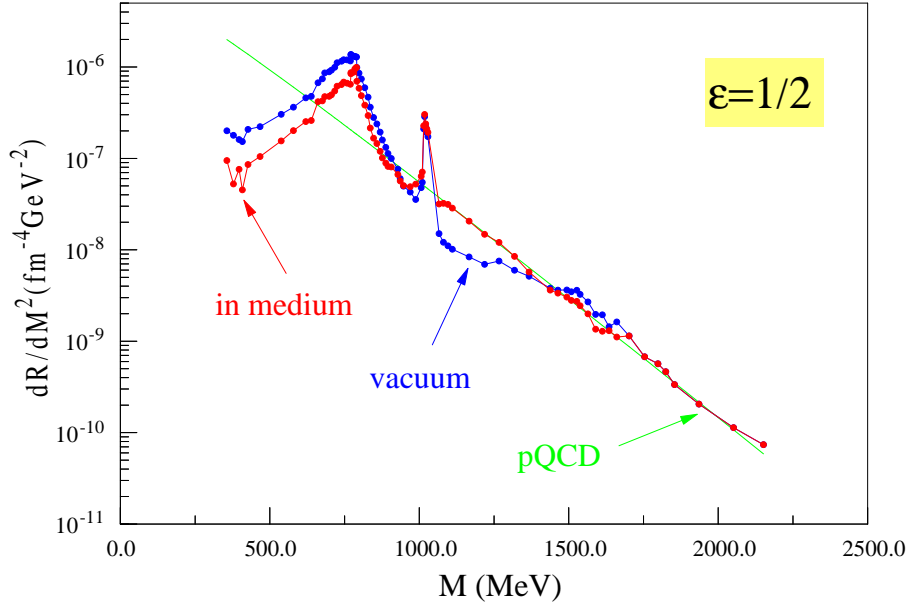,height=8cm}
\caption{Three-momentum integrated dilepton production rates at fixed 
temperature $T=160$~MeV using the free electromagnetic correlator
(labeled 'vacuum'), the fully mixed one from Eq.~(\ref{VAmixing}) 
('in-medium') and the perturbative quark-gluon one according to 
Eq.~(\ref{Dlrqp}) ('pQCD').}
\label{fig_remm}
\end{figure}

Due to the presence of pions in the thermal heat bath the 
vector and axialvector
correlators mix (Fig.~\ref{fig_vamix}). 
At low temperature this mixing
can be calculated in a virial expansion. To lowest order in 
temperature one obtains~\cite{DEI90} the model-independent 
'mixing theorem' of vacuum correlators: 
\bea
\Pi^{\mu\nu}_V(q) &=& (1-\epsilon) \ \Pi^{\circ\mu\nu}_V(q)
+ \epsilon \ \Pi^{\circ\mu\nu}_A(q)
\nonumber\\
\Pi^{\mu\nu}_A(q) &=& (1-\epsilon) \ \Pi^{\circ\mu\nu}_A(q)
+\epsilon \ \Pi^{\circ\mu\nu}_V(q) \ . 
\label{VAmixing}
\eea
The mixing coefficient $\epsilon$ is given by the thermal pion loop
\beq
\epsilon={2\over f^2_\pi}\int\!\!{d^3k\over (2\pi)^3}
{f^\pi(\omega_\pi(k))\over \omega_\pi(k)} \  
\eeq 
($f^\pi(\omega)=1/(\exp[\omega/T]-1)$: pion Bose distribution, 
$\omega_\pi(k)^2=m_\pi^2+k^2$). 
In the chiral limit ($m_\pi\to 0$) this reduces to
\be
\epsilon={T^2\over 6f_\pi^2} \ . 
\label{epscl}
\ee
To lowest order there is no change in the spectral shapes themselves  
but rather a temperature-dependent coupling between the free vector and
axialvector correlators. As is easily verified the mixing theorem
fulfills the in-medium Weinberg sum rules and according to (\ref{VAmixing})
chiral symmetry is restored for $\epsilon=1/2$. Eq.~(\ref{epscl}) thus implies
a transition temperature $T_c=\sqrt{3}f_\pi\simeq 160$ MeV which coincides with
the transition temperature $T_c=150\pm 20$ MeV from lattice QCD. Clearly the
low-temperature expansion cannot be trusted to such high temperature. 
Nonetheless it is instructive to see what the consequences for the dilepton 
spectrum are. This is displayed in Fig.~\ref{fig_remm}.
By comparing with the results from a quark-gluon plasma 
one observes that both are
indistinguishable down to invariant masses of $\sim$~1~GeV. This is
significantly lower than without mixing in which case 
 the two rates only coincide
above $\sim$~1.5~GeV (the latter is conceivable from the fact that
chiral symmetry breaking is a long distance phenomenon which does not
impact the short-distance behavior of the correlators). In other words:
the vector-axialvector mixing at finite temperature entails that 
the 'duality threshold' -- where hadronic and quark-gluon based descriptions
start to agree -- is reduced from its vacuum location at $M\simeq 1.5$~GeV 
to about $M\simeq 1$~GeV in the medium. At this level, the $\phi$ and
$\rho$/$\omega$ resonance structures (which themselves are not affected by 
the simple mixing mechanism) inhibit a further penetration of the duality
threshold into the low-mass region. However, let us note already at this 
point that it is precisely the broadening (or 'melting')
of the resonances in the medium (as predicted in various hadronic model
calculations) that -- at the same time -- flattens the ($\rho$) resonance
structure and generates low-mass dilepton enhancement below the
free $\rho$ mass. These important issues will be reiterated in some detail
throughout this article.    

Further interesting conclusions about the in-medium correlators can be drawn
by using a pole approximation~\cite{KapuSh} similar to (\ref{Polv}):
\be
\rho_V(q_0)={m^4_\rho\over g^2_\rho}Z_\rho{1\over q_0^2}
\delta(q_0^2-m^{*2}_\rho)\ , \quad  
\rho_A(q_0)={m^4_{a_1}\over g^2_{a_1}}Z_{a_1}
{1\over q_0^2}\delta(q_0^2-m^{*2}_{a_1})
+f_\pi^{*2}\delta(q_0^2) \ , 
\label{Polm}
\ee
where $m^*_\rho$, $m^*_{a_1}$ and $f_\pi^*$ denote the in-medium masses 
and pion decay constant while $Z_\rho$ and $Z_{a_1}$ are 
the residues at the quasiparticle pole.
Inserting (\ref{Polm}) into the Weinberg sum rules (\ref{WSRm1}),
(\ref{WSRm2}) yields $Z_\rho=Z_{a_1}$ from the second one, and thus
from the first one
\be
{f_\pi^{*2}\over f_\pi^2}=aZ_\rho\biggl ({m_\rho^2\over m^{*2}_\rho}
-{m_\rho^2\over m^{*2}_{a_1}}\biggr ) \ , 
\ee
where $a$ is given by (\ref{Polvv}). Since $f^*_\pi$ is an order parameter of
chiral symmetry the latter is restored when $m^*_\rho=m^*_{a_1}$. Such an 
approach to mass degeneracy is indeed observed in lattice QCD calculations.

In close reminiscence to the finite temperature case, Krippa~\cite{Kr98}
derived an analogous mixing theorem for a zero-temperature gas of 
noninteracting nucleons using soft pion theorems and current algebra, 
\bea
\Pi^{\mu\nu}_V(q) &=& (1-\xi) \ \Pi^{\circ\mu\nu}_V(q)
+\xi \ \Pi^{\circ\mu\nu}_A(q)
\nonumber\\
\Pi^{\mu\nu}_A(q) &=& (1-\xi) \ \Pi^{\circ\mu\nu}_A(q)
+\xi \ \Pi^{\circ\mu\nu}_V(q) \ .
\label{VAmixnuc}
\eea
Here, the mixing parameter 
\bea
\xi\equiv\frac{4\varrho_N \bar{\sigma}_{\pi N}} {3f_\pi^2 m_\pi^2} 
\eea
appears in terms of the leading nonanalytic term in the current quark mass
($\propto m_\pi^3$) in the chiral expansion of the nucleon $\sigma$-term,
given by  
\bea
\bar{\sigma}_{\pi N} &=& 4\pi^3 m_\pi^2 \langle N|\pi^2|N \rangle \ 
\nonumber\\
&\simeq& 20~{\rm MeV} \ . 
\eea
It is related to the pion mass contribution
to the in-medium nucleon mass 
and arises from the long-distance physics encoded in the nuclear pion cloud 
being governed by chiral symmetry~\cite{Bi96}. 
Naive extrapolation of the mixing to
chiral restoration (\ie, $\xi=1/2$) yields $\rho_c\simeq 2.5\varrho_0$, 
which again is not unreasonable.

\section{QCD Sum rules}
\label{sec_qcdsr}
The QCD sum rule approach~\cite{SVZ} aims at an understanding of 
physical current-current correlation functions in terms of QCD 
by relating the observed hadron spectrum to the 
nonperturbative QCD vacuum structure. This is achieved
by a separation of short- and long-distance scales.  
The principal tool is the 'operator product expansion' (OPE) 
which evaluates  the time-ordered product of the light-quark QCD 
currents at large space-like momenta $Q^2\equiv -q^2>0$. In this case the
current product can be related to a series of gauge invariant 
local operators ${\cal O}_n$ as
\be
-i\!\int\! d^4x\, e^{iq\cdot x}{\cal T}j^\mu(x)j^\nu(0)=
-\bigl (g^{\mu\nu}-{q^\mu q^\nu\over q^2}\bigr )
\sum_n c_n(Q^2,\Lambda^2){\cal O}_n(\Lambda^2) \ , 
\label{OPE}
\ee
where $\Lambda$ is the renormalization scale and $c_n$ denote c-number 
functions (the Wilson coefficients) which contain the short-distance 
physics and can be calculated reliably. The $Q^2$-independent operators 
${\cal O}_n$, on the other hand, encompass the long-distance properties of QCD, 
manifest in the appearance of various condensates such as $\ave{\bar\psi\psi}$
or $\ave{G^2}$. These operators
have various dimensions, $d$, such that at large $Q^2$ (\ref{OPE}) can 
be considered as an expansion in inverse powers of $Q^2$ (corrected
by logarithms due to renormalization). An increase in dimension implies
extra inverse powers of $Q^2$ such that operators of higher
dimension are suppressed. 

The general strategy is to take the vacuum matrix elements 
of ({\ref{OPE}) and to consider a dispersion relation (possibly subtracted)
of the type
\be
\Pi^\circ(Q^2=-q^2)=\Pi^\circ(0)-{Q^2\over\pi}\int_0^\infty\! {ds\over s}\,
{{\rm Im}\Pi^\circ(s)\over Q^2+s}=
\Pi^\circ(0)+Q^2\int_0^\infty\! ds\,{\rho^\circ(s)\over Q^2+s} \ , 
\label{Disp}
\ee
where $\rho^\circ(s)=-(1/\pi s){\rm Im}\Pi^\circ(s)$ denotes the 
vacuum spectral function
of the current-current correlator in question. For the electromagnetic
current $\Pi^\circ(0)=0$, since the photon is massless in the vacuum.
The {\it l.h.s.}~of Eq.~(\ref{Disp}) is evaluated in the OPE by 
determining
the Wilson coefficients up to a certain dimension $d$ while the
{\it r.h.s.}~is taken from measured cross sections (or 
simple parameterizations thereof) in the time-like region. For the 
{\it l.h.s.}~one obtains
\be
{12\pi\over Q^2}\Pi^\circ(Q^2)={d\over \pi}\biggl [
-c_0{\rm ln}(Q^2/\Lambda^2)+{c_1\over Q^2}+{c_2\over Q^4}
+{c_3\over Q^6}+\cdots \biggr ] \ , 
\label{OPE1} 
\ee
which exhibits the power series expansion in $Q^2$. In case 
of the $\rho$ meson~\cite{KKW97}
\bea 
c_0^\rho&=&1+{\alpha_s\over \pi}
\nonumber\\
c_1^\rho&=&-3(m_u^2+m_d^2)
\nonumber\\
c_2^\rho&=&{\pi^2\over 3}\ave{G^2} +4\pi^2\ave{m_u\bar uu+m_d\bar dd}
\nonumber\\
c_3^\rho&\propto&\alpha_s\ave{(\bar qq)^2} \ , 
\eea
where $\alpha_s(Q^2)$ is the running QCD coupling constant given in 
Eq.~(\ref{asrun}). 
Explicit expressions for the coefficients $c_i$ of other mesons can be 
found in Ref.~\cite{KKW97}. While for $c_2$ the quark and gluon condensates
enter, which are rather well known, $c_3$ contains the four-quark
condensate, \ie, $\ave{(\bar qq)^2}$, which is quite uncertain. 
Based on the assumption of vacuum saturation it is usually
approximated in factorized form, $\ave{\bar qq}^2$, and higher
meson states, especially pions, are incorporated by a phenomenological
factor $\kappa>1$ such that
\beq
c_3=\kappa\alpha_s\ave{\bar qq}^2
\eeq 
The parameter $\kappa$ typically varies between 
1 and 6~\cite{HatsLee,Lein97,KKW97}.

Rather than working with (\ref{OPE1}) the convergence of the hadronic 
side can be improved by observing that the physical spectrum is
dominated by low-lying resonances (Fig.~\ref{fig_rhoAVem}). To enhance 
their weight  one employs the fact that dispersion theory implies
the following relation for derivatives of $\Pi^\circ$:
\be
{1\over n!}\biggl (-{d\over d Q^2}\biggr )^n\Pi^\circ (Q^2)_{|Q^2=Q_0^2}
={1\over\pi}\integ s{{\rm Im}\Pi^\circ (s)\over 
(s+Q_0^2)^{n+1}} \ . 
\ee
This leads to the introduction of the 'Borel transformation'
\be
\hat L_M=\lim_{\stackrel{Q^2\to\infty, n\to\infty}{Q^2/n=M^2}}
{1\over (n-1)!}(Q^2)^n\biggl (-{d\over dQ^2}\biggr )^n \ , 
\ee
where $M$ is the so-called Borel mass. When applied to both
sides of Eq.~(\ref{Disp}) and by using general properties
of $\hat L_M$ the result of the transformation is
\be
{1\over\pi M^2}\integ s \rho^\circ_V(s)e^{-s/M^2}=
{d_V\over 12\pi^2}\biggl [c_0+{c_1\over M^2}+{c_2\over M^4}+{c_3\over 2M^6}
+\cdots\biggr ]\pkt
\label{Bort}
\ee
Note the appearance of the exponential factor in the integrand of the 
{\it l.h.s.} which suppresses the contribution from higher resonances. The
{\it r.h.s.} converges rapidly if $M$ is sufficiently large such that the
few lowest terms in the OPE suffice. Typically the minimum value of the
Borel mass to achieve rapid convergence is around 1 GeV.

The QCD sum rule analysis in the vacuum can now be performed in 
two ways. Either the phenomenological side is experimentally
accessible in which case values for the
various condensates can be extracted, or, by using 'known' values
for the condensates the properties of the physical spectrum, \ie, 
masses and coupling constants, can be inferred. The latter procedure
forms the basis of QCD sum rule applications in hadronic matter. The most 
simple ansatz for the vacuum spectral density consists of a 'delta function'
parameterization of the resonance part supplemented by a 'continuum
step function' (cf.~Fig.~\ref{fig_rs}):
\beq
\rho^\circ(s)={{\cal Z}_V\over 12\pi^2}\delta(s-m_V^2)+
{d_V\over 12\pi^2}\bigl (1+{\alpha_s\over\pi}\bigr )\Theta(s-s_V) \ , 
\label{parsimp}
\ee
\begin{figure}[!ht]
\bce
\epsfig{figure=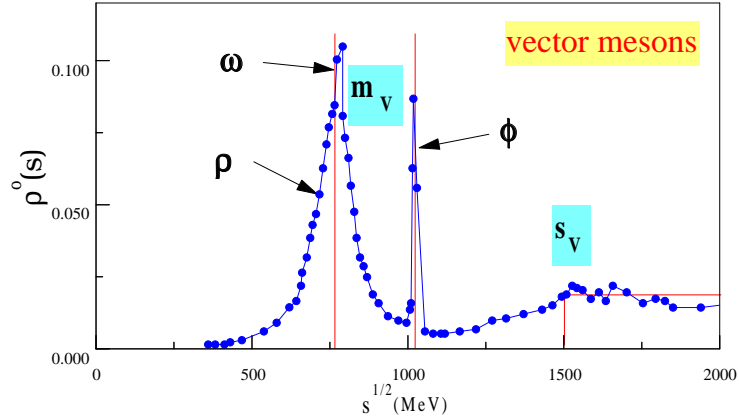,height=6cm}
\ece
\caption{QCD sum rule parameterization of the strength function.}
\label{fig_rs}
\end{figure}
where ${\cal Z}_V$ denotes the pole strength, $m_V$ the vector
meson mass and $s_V$ the continuum threshold. The continuum
strength $d_V$ is determined by perturbative QCD with
$d_\rho=3/2,\, d_\omega=1/6,\, d_\phi=1/3$. Applying this
parameterization to (\ref{Bort}) with $\alpha_s=0.36$,
$\ave{\bar uu}\simeq \ave{\bar dd}
\simeq \ave{\bar ss}=(-250\,{\rm MeV})^3$, $\ave{{\alpha_s\over \pi}
G^a_{\mn}G_a^{\mn}}=(330\,{\rm MeV})^4$ and $\kappa=2.36$ the {\it l.h.s.} and 
{\it r.h.s.} match for ${\cal Z}_\rho=9$~GeV$^2$, 
${\cal Z}_\omega=2.4$~GeV$^2$, ${\cal Z}_\phi=0.79$~GeV$^2$, 
$m_{\rho,\omega}=0.77\,{\rm GeV},\, m_\phi=1.02\,{\rm GeV}$
and $s_{\rho,\omega}=1.5\, {\rm GeV}^2, s_\phi=2.2\,{\rm GeV}^2$ in
a 'Borel window' of 0.8-1.5~GeV~\cite{KKW97}. Inclusion of the 
vector-meson decay widths (especially for the $\rho$ meson) does not 
affect these results appreciably.

The in-medium QCD sum rule analysis involves the vector current-current 
function in the interacting hadron gas (\ref{Vcorr}). 
In the following it will be focused on the case of $\vec q=0$ for which
$\Pi_V^{L,T}(q_0)$ from Eq.~(\ref{Plt}) coincide. 
In analogy to the vacuum 
case (\ref{Bort}) the in-medium sum rules at zero  
temperature are given by   
\beq
{1\over\pi M^2}\biggl [\Pi_V(0)+
\integ q_0^2 \rho_V(q_0)e^{-q_0^2/M^2} \biggr]=
{d_V\over 12\pi^2}\biggl [c_0+{c_1\over M^2}+{c_2(\varrho)\over M^4}
+{c_3(\varrho)\over 2M^6} +\cdots\biggr ] \ , 
\label{srmed}
\eeq
where $\rho_V(q_0)$ denotes the in-medium vector 
spectral function (\ref{specmed}).
Note that, in contrast to the vacuum, $\Pi_V(0)$ no longer vanishes for the
$\rho$ and $\omega$ meson. It is related to the $\rho/\omega$-nucleon 
forward-scattering amplitude~\cite{KKW97}. 
On the {\it r.h.s.} of Eq.~(\ref{srmed}) the
medium enters through the density-dependent Wilson coefficients $c_2(\varrho)$ and
$c_3(\varrho)$. These in turn are chiefly determined by the density-dependent
quark and gluon condensates 
\beq
\tave{\bar qq}=\ave{\bar qq}-{\Sigma_N\over 2\bar m}\varrho \ , \quad 
\tave{G^2}=\ave{G^2}-{8\over 9} m^{(0)}_N\varrho\pkt 
\eeq
which follow from the dilute gas
expressions (\ref{Dilute2}) and (\ref{Diluteg}) at zero temperature. 
As was pointed out in Ref.~\cite{HatsLee} additional contributions to
the Wilson coefficients arise from new condensates which involve mixed
quark and gluon fields, the latter entering through the gauge covariant
derivative $D_\mu$ (\ref{Covder}). These matrix elements are proportional
to moments of the quark and antiquark distribution functions
\beq
A^q_n=2\int^1_0 dx \, x^n[q(x)+\bar q(x)]
\eeq
in the nucleon. Restricting oneself to the lowest moments and leading
order in density one finally arrives for the $\rho$ meson at~\cite{HatsLee}
\beq
c^\rho_2(\varrho)
\simeq c^\rho_2(0)-({8\pi^2\over 27}m^{(0)}_N-2\pi^2A^{u+d}_1m_N)\varrho
\eeq
and
\beq 
c_3(\varrho)\simeq c_3(0)+({896\over 81}\kappa\pi^3\alpha_s
{\Sigma_N\over \bar m}\ave{\bar qq}-{10\over 3}\pi^2A^{u+d}_3m^3_N)\varrho
 \ , 
\eeq
where $m_N$ denotes the physical nucleon mass while $m_N^{(0)}$ represents
the nucleon mass in the chiral limit ($m_N^{(0)}\simeq 750$~MeV~\cite{BoMei96}).
Expressions for the density-dependent Wilson coefficient of other mesons can
be found in Ref.~\cite{KKW97}.

In keeping the simple parameterization (\ref{parsimp}) for the in-medium 
spectral function, 
\be
\rho_V(q_0)={{\cal Z}^*_V\over 12\pi^2}\delta(q_0^2-{m^*_V}^2)+
{d_V\over 12\pi^2}\bigl (1+{\alpha_s\over\pi}\bigr )\Theta(q_0^2-s^*_V) \ , 
\label{parsimpm}
\ee
Hatsuda and Lee~\cite{HatsLee} extracted the medium dependence of the 
non-strange vector-meson masses as
\be
{m^*_{\rho,\omega}\over m_{\rho,\omega}}=1-(0.18\pm 0.06){\varrho\over\varrho_0}\pkt
\ee
The fact that these masses decrease as density increases has initially 
been taken as an indication of the 'dropping mass scenario'
 of Brown and Rho~\cite{BR91} to which we will return below. 
Rather than using the parameterization (\ref{parsimpm}), it is, however, 
natural to take into account the fact that
the strength distributions might broaden significantly in the hadronic medium,
as will be discussed in detail later.
A first step in this direction was taken in Ref.~\cite{AK93} by including
effects of the $\Delta$-nucleonhole polarization in the pion cloud of 
the $\rho$ meson. Its net impact on the in-medium QCD sum rule, however, turned
out to be rather moderate, \ie, a strong decrease of the in-medium $\rho$ mass
very similar to the Hatsuda-Lee results was still needed to satisfy  
the sum rule.  
\begin{figure}[!htb]
\bce
\epsfig{figure=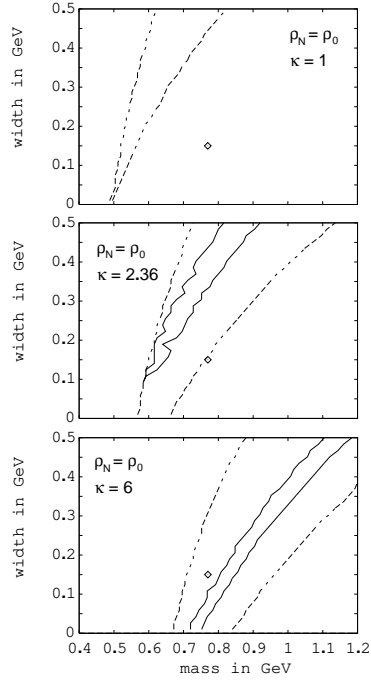,height=9cm}
\ece
\caption{Constraints on allowed values for the in-medium width and mass of
the $\rho$ meson from the QCD sum rule analysis of Ref.~\protect\cite{Leup}.
The full and dashed lines boarder allowed regions at 0.2\% and 1\% accuracy
level, respectively. The diamond marks the mass and width of the free
$\rho$ meson.}
\label{fig_qcdsrmed}
\end{figure}
A more general investigation of broadening effects was performed in 
in Ref.~\cite{Leup} by assuming a schematic  
Breit-Wigner spectral function 
\beq
A_V(q_0)={1\over \pi}{q_0\Gamma_V(q_0)
\over (q_0^2- m_V^2)^2+q_0^2\Gamma_V(q_0)^2}
\eeq
with
\beq
\Gamma_V(q_0)=\Gamma_0 \biggl ({1-(\omega^{\rm thr}_V/q_0)^2
\over (1-(\omega^{\rm thr}_V/m_V)^2}\biggr )^{1\over 2}\Theta(q_0^2-
(\omega^{\rm thr}_V)^2) \ , 
\eeq
$\omega^{\rm thr}$ denoting the appropriate in-medium threshold (for the 
$\rho$ meson $\omega^{\rm thr}_\rho=m_\pi$ was chosen which is correct
to leading order in the density) and $\Gamma_0$ being a constant width 
parameter.
One concludes from Fig.~\ref{fig_qcdsrmed} that QCD sum rules give no stringent 
prediction for a dropping of vector-meson masses. This result is corroborated
by the findings of Ref.~\cite{KKW97} where a microscopic spectral function
for vector mesons was used.

\section{Chiral Reduction Formalism}
\label{sec_chired}

Another model-independent approach that has been put forward to 
assess medium modifications of the vector correlation function
is the so-called chiral master formula framework
developed in Ref.~\cite{YZ96}. In the general case, it starts 
out from gauge covariant divergence equations 
(Veltmann-Bell equations~\cite{VB67}) including explicit chiral 
breaking in the presence of external sources, 
\bea
\nabla_\mu j_{V,a}^\mu + \epsilon_{abc} a_{\mu, b} j_{A,c}^\mu &=&
-f_\pi \epsilon_{abc} p_b \pi_c   
\\
\nabla_\mu  j_{A,a}^\mu +\epsilon_{abc} a_{\mu, b} j_{V,c}^\mu   &=&
f_\pi (m_\pi^2 +s) \pi_a-f_\pi p_a \sigma \ , 
\eea
where the (axial-) vector currents 
\beq
j_{V,a}^{\mu} (x)=\frac{\delta S}{\delta v_\mu^a(x)} \ ,  \qquad  
j_{A,a}^{\mu} (x)=\frac{\delta S}{\delta a_\mu^a(x)}
\eeq
and (pseudo-) scalar densities 
\beq
\sigma =-\frac{m_q}{f_\pi m_\pi^2} \ \bar qq  \ , \qquad  
\pi^a= \frac{m_q}{f_\pi m_\pi^2} \ \bar q i\gamma_5 \tau^a q \ 
\eeq
have been defined as functional derivatives of an action $S$  
{\it w.r.t.}~a pertinent set of auxiliary fields 
$\phi\equiv \{v^a_\mu,a^a_\mu,s,p^a\}$, respectively (here, $a,b,c$=1--3 
are isospin indices and the short-hand notation for the covariant derivative
is defined as $\nabla_\mu  j_{a}^\mu\equiv [\partial_\mu \delta_{ac} +
\epsilon_{abc} v_{\mu,b}] j_c^\mu$).  More specifically, the action
can be thought of as the QCD action plus an external source part, 
\bea
S &=& \int d^4x \left\{ {\cal L}_{QCD} + {\cal L}_{ext}  \right\} 
\nonumber\\
{\cal L}_{ext} &=& \bar q \left( \gamma^\mu 
[v_\mu^a + a_\mu^a \gamma_5] \frac{\tau^a}{2} -\frac{m_q}{m_\pi^2} [m_\pi^2+s-
i\gamma_5 \tau^a p^a] \right) q  \ .   
\eea
With the help of the Peierls-Dyson formula~\cite{Pei52}  for the 
$\cal S$-matrix, the Veltmann-Bell equations can be rewritten as 
\bea
\left( {X}_V^a  + \epsilon^{abc} p^b \frac{\delta}{\delta p^c} \right)  
 {\cal S} &=& 0 
\label{VB1}
\\
\left( {X}_A^a -(m_\pi^2+s) \frac{\delta}{\delta p^a} +p^a 
\frac{\delta}{\delta s}  \right)  {\cal S} &=& 0 \ ,  
\label{VB2} 
\eea
where $X_{V,A}$ are functional differential operators involving
the $v_\mu^a$ and $a_\mu^a$ fields. 
The feature of spontaneous chiral symmetry breaking is then imposed 
through  appropriate boundary conditions, namely that asymptotic stable
states are given in terms of (massive) pion fields,
\ie,  for $x^0\to \pm \infty$ one requires  
\bea
j_{A}^{\mu,a} (x) &=& -f_\pi \partial^\mu \pi^a_{in,out}(x) 
\\
\partial_\mu j_{A}^{\mu,a}(x) & \to & f_\pi m_\pi^2 \pi^a_{in,out}(x) \ . 
\eea
With these boundary conditions the Veltmann-Bell equations 
(\ref{VB1}),~(\ref{VB2}) can
be integrated to give the so-called master formula for the  
$SU(2)_L \times SU(2)_R$ symmetric  ${\cal S}$-matrix 
for massive pions, which can be found in Ref.~\cite{YZ96}. 
From the master formula one can derive 
expressions for pion Greens functions in terms of the (axial-) vector
currents and (pseudo-) scalar densities,  as well as corresponding 
Ward identities. 
Since the $\cal S$-matrix plays the role of a time evolution
operator (within the Heisenberg picture), the Fourier transform
of the master formula takes the form of a LSZ reduction formula, 
incorporating the proper chiral Ward identities, and thus
has been coined 'chiral reduction formula'. It allows to express hadronic
on-shell scattering reactions, determined by the {\em on-shell} 
$\cal S$-matrix (given by the limit of vanishing external fields 
$\phi\to 0$), through well-defined correlation functions 
and form factors. The latter have to be either inferred from other experimental 
information (in which case  the predictions for the processes under 
consideration are, in principle, exact, \ie, compatible with unitarity, 
crossing symmetry and broken chiral symmetry) or evaluated in an 
appropriate expansion scheme. 

Applications of this formalism to calculate medium modifications 
are readily performed using  
virial-type low-density approximations, as the processes have to be expressed 
in terms of stable final states within the hadronic matter, \ie, 
pions and nucleons. For electromagnetic thermal production rates, this
has been carried out in Refs.~\cite{SYZ1,SYZ2}. 
Starting from the usual eightfold-differential rate expression,   
\beq
\frac{d^8N}{d^4x d^4q}  =-\frac{\alpha^2}{6\pi^3 q^2} W(q) \ ,  
\eeq
the thermal correlator $W$ has been related to the time-ordered
(Feynman) one by 
\bea
W(q) &=& \frac{2}{1+{\rm e}^{q_0/T}} \ {\rm Im} W^F(q)
\nonumber\\
W^F(q)&=& i \int d^4x \ {\rm e}^{iqx} \  \frac{1}{\cal Z} \ 
\tr\left[{\rm e}^{(\hat{H}-\mu_N \hat{N})/T} 
{\cal T} j^\mu(x) j_\mu(0) \right] \ .  
\eea
Up to first order in the density of either pions or nucleons 
the imaginary parts of the latter become 
\bea
{\rm Im} W^F(q)&=&-3 \ {\rm Im} \Pi_V^\circ(q) 
+ \frac{1}{f_\pi^2} \int \frac{d^3k}{(2\pi)^3 2\omega_\pi(k)}  \ 
f^\pi(\omega_\pi(k);T) \ {\rm Im} W^F_\pi(q,k) 
\nonumber\\
 & & \qquad + \int \frac{d^3p}{(2\pi)^3 2E_N(p)} \  
f^N(E_N(p);\mu_N,T)  \ {\rm Im} W^F_N(q,p) \  
\eea 
($f^\pi$: pion Bose distribution, $f^N$: nucleon Fermi distribution; note
that our definition of the correlation functions $\Pi(q)$ differs from
the one in Refs.~\cite{SYZ1,SYZ2} by a factor of $q^2$).
The first term on the {\it r.h.s.} represents the vacuum part containing
the free electromagnetic
current correlator, whereas the second and third terms involve    
the forward scattering amplitudes of (real or time-like) photons on on-shell 
pions and nucleons from the medium, 
\bea
W_\pi^F(q,k) &=& i \int d^4x \ e^{iqx} \ 
\langle \pi(k)|{\cal T} j^\mu(x) j_\mu(0)|\pi(k)\rangle 
\nonumber\\
W_N^F(q,p) &=& i \int d^4x  \ e^{iqx} \  
\langle N(p)|{\cal T} j^\mu(x) j_\mu(0)|N(p)\rangle \ . 
\eea
The relevant expansion parameters for
both cases have been quoted as $\kappa_\pi= n_\pi / 2m_\pi f_\pi^2$ and
$\kappa_N=\varrho_N g_A^2 / 2m_N f_\pi^2$, which should provide reasonable
lowest-order results for $\kappa \lsim 0.3$, corresponding to 
temperatures $T\lsim 140$~MeV and nucleon
densities $\varrho_N\lsim 2.5\varrho_0$.  
However, especially for the nucleonic case, this counting scheme may be  
subject to large corrections if low-lying resonances or opening thresholds
are present~\cite{SYZ2}, see below.   

Applying the chiral reduction formulae to the scattering tensors, 
the relevant terms of the pionic piece take the form~\cite{SYZ1}  
\bea
{\rm Im} W_\pi^F (q,k) &\simeq& 12  \ {\rm Im} \Pi_V^\circ(q) - 
6 \ {\rm Im} \Pi_A^\circ(k+q) +  
6 \ {\rm Im} \Pi_A^\circ(k-q)
\nonumber\\ 
& & + 8[\frac{(k\cdot q)^2}{q^2}-m_\pi^2] \ {\rm Im} \Pi_V^\circ(q) \  
{\rm Re}[D_\pi^R(k+q)+D_\pi^R(k-q)] 
\eea
with $D_\pi^R$ denoting the retarded pion propagator and $\Pi_A^\circ$ the
free axialvector current correlator (extracted from $\tau$-decay data
as indicated in the upper part of Fig.~\ref{fig_rhoAVem}). 

The nucleonic piece is more difficult to assess; at the photon point 
it can be inferred via the optical theorem from the total 
$\gamma N$ cross section, 
\beq
e^2 \ {\rm Im} W^F_N(q,p)=-4(s-m_N) \ \sigma_{\gamma d}^{tot}(s) \   
\eeq
(here, the deuteron cross section has been taken as representative for the
isospin summed result).  
For time-like photons, an absorptive part starts to build up only  
from loop corrections for which a  one-pion loop expansion was 
performed for non-resonant $\pi N$ states (being related to on-shell
$\pi N$ scattering data). However, as is obvious from
the experimental $\gamma N$ cross section, the $\Delta$ resonance has to be 
included. Its contribution has been evaluated as
\beq
{\rm Im} W^F_\Delta(q,p)= {\rm Im}\left[\frac{4 m_N m_\Delta}{s-m_\Delta^2+
im_\Delta \Gamma_\Delta^0}\right] \overline{|{\cal M}_{N\Delta}|^2} 
+ (s\to u)   
\eeq 
with $\overline{|{\cal M}_{N\Delta}|^2}$ the 
spin-isospin summed modulus squared 
of the $N\Delta$ transition amplitude. The latter has been constructed 
compatible with current conservation and crossing symmetry, and its parameters 
are constrained by electric and magnetic polarizabilities as well as
the electromagnetic decay width $\Gamma_{\Delta\to N\gamma}\simeq 0.7$~MeV.
In subsequent work -- after the importance of the $N(1520)$ for dilepton 
production has been realized~\cite{PPLLM,RUBW,CBRW,BLRRW,Morio98} -- 
the $N(1520)$ contribution has been
included along similar lines as the $\Delta$ in Ref.~\cite{SZ99}. 
Higher order terms $\propto \kappa_\pi \kappa_\pi, \kappa_\pi \kappa_N, 
\kappa_N \kappa_N$ have also been estimated and claimed to be rather
small in the region of interest (\ie, above invariant masses of about 
200~MeV). 

The discussion of the numerical results for the dilepton and 
photon production rates in the chiral reduction formalism is deferred 
to Sect.~\ref{sec_dlrates},  where it will be put into context with 
other (model) approaches, most of which are elucidated in the following 
Chapter.

                               

\chapter{Modeling Vector Mesons in the Medium}
\label{Vmodels}
So far our assessment of medium effects in the current-current correlators
was of a rather general nature focusing on the QCD aspects and 
model independence.
We have repeatedly stressed the intimate 
relation between vector and axialvector channels, governed by chiral 
symmetry. The discussion now proceeds to various models 
that have been employed  to investigate the properties of vector and
axialvector 
mesons in hot and dense matter. On the one hand, this allows for much
more specific predictions, but also implies at least a  
partial loss of generality. However, a careful comparison of the 
underlying assumptions and associated characteristic features of the results 
should provide valuable information on the relevant mechanisms for chiral
symmetry restoration. 
Large efforts have been undertaken to investigate the in-medium vector meson 
properties, especially those of the $\rho$ meson, due to its prominent 
role in heavy-ion dilepton measurements, as will be extensively addressed 
in Chapter~\ref{chap_dilepton}. Much less has been done in the axialvector 
channel, dominated by the $a_1(1260)$ meson. 
This is mainly due to the fact that 
experimental information, in particular concerning medium effects, 
is and will be scarce (the hope is, of course, that ultimately QCD
lattice calculations will be able to overcome this 
unfortunate situation). Thus, in the following, the emphasis will 
inevitably be biased towards the vector channel, in particular
the $\rho$ meson.  

The various approaches can be roughly divided into two categories, namely
those which are based on purely mesonic Lagrangians, addressing
the impact of finite temperature, and those including baryonic 
fields to account for finite density effects. For each model we will
first briefly discuss its construction and vacuum properties,  
subsequently proceeding to the in-medium applications.   

\section{Effective Meson Lagrangians: Impact of Finite Temperature}
\label{sec_Vmodtemp}

\subsection{Gauged Linear $\sigma$-Model + VDM}
\label{sec_glsm}
Based on the presumption that for the properties of the 
lowest-lying meson multiplets near the phase boundary chiral 
symmetry restoration (rather than deconfinement) should be the prevailing 
feature of the QCD transition Pisarski proposed~\cite{Pisa95} to study 
the simplest version of an effective theory which incorporates 
the appropriate symmetry structure, \ie, the linear $\sigma$-model.
A low-lying genuine $\sigma$ meson might not be the most 
realistic description of the zero temperature situation, as it strongly
couples to two-pion states. As a result the width is of the order of 
the mass such that the $\sigma$ meson does not represent a well-defined degree
of freedom (or quasiparticle). He argued, however, that given the 
fact that at chiral restoration the pion and the sigma have to 
become degenerate there might well arise the situation in which the phase 
space for $\sigma\to \pi\pi$ is locked and the $\sigma$ field becomes a 
well-defined excitation~\cite{HK87a}.  

In the linear $\sigma$-model the pion and sigma fields are grouped  
into the standard four-dimensional vector 
\beq
\Phi =\sigma \ t^0 + i\vec{\pi} \cdot \vec{t} \ ,  
\eeq
where $\vec{t}=\vec{\tau}/2$ is defined via the standard Pauli matrices
and $t^0=1\!\!1/2$ is proportional to the unit-matrix in isospin space. 
The (axial-) vector fields are introduced via left- and right-handed 
combinations as 
\bea
A_L^\mu &=& 
(\omega^\mu+f_1^\mu) \ t^0+ (\vec{\rho}^\mu+\vec{a_1}^\mu) \cdot \vec{t} \  
\nonumber\\
A_R^\mu &=& 
(\omega^\mu-f_1^\mu) \ t^0 + (\vec{\rho}^\mu-\vec{a_1}^\mu) \cdot \vec{t} \  
\eea
in obvious notation. The crucial step is now to assume that the 
$SU(2)$ chiral transformations for vector fields are {\em local} 
ones, promoting it to a gauge symmetry. The basic motivation is
a natural emergence of conventional vector dominance. 
Once the field-strength tensor and covariant derivative on the scalar fields 
are accordingly defined, 
\bea
F^{\mu\nu}_{L,R} &=& \partial^\mu A_{L,R}^\nu - \partial^\nu A_{L,R}^\mu 
-ig \left[A_{L,R}^\mu,A_{L,R}^\nu\right] \  
\nonumber\\
D^\mu &=& \partial^\mu \Phi -i g \left(A_L^\mu \Phi - \Phi A_R^\mu \right) \ , 
\eea
the gauged linear $\sigma$-model Lagrangian takes the form  
\bea
{\cal L}_{{\rm gl}\sigma} &=& \tr |D_\mu\Phi|^2 - 2 \ h \ t^0 \ 
\tr (\Phi) - \mu^2 \ \tr |\Phi|^2 
+\frac{1}{2} \lambda \left(\tr |\Phi|^2\right)^2 
\nonumber\\
 & & +\frac{1}{4} \tr \left((F_L^{\mu\nu})^2+(F_R^{\mu\nu})^2\right) 
+\frac{1}{2} m_0^2 \ \tr \left((A_L^\mu)^2+(A_R^\mu)^2\right) \ . 
\label{Lgls}
\eea
The corresponding vector (Noether) current is then solely determined by 
the mass term, resulting in
\beq
j_{L,R}^\mu=\frac{m_0^2}{g} A_{L,R}^\mu \ . 
\label{field-curr}
\eeq
This is precisely  the desired current-field identity of the 
vector dominance model~\cite{Sa69} (VDM) where $g$ is the universal 
(dimensionless) vector coupling and $m_0$ the (bare) vector meson 
mass. Furthermore, in Eq.~(\ref{Lgls}), $\lambda$ denotes the 
(dimensionless) scalar coupling. The 'mass parameter' $\mu^2$ is taken 
positive to generate the spontaneous breakdown of chiral symmetry in 
the physical vacuum. Explicit chiral symmetry breaking 
through a 'magnetic background field' $h$ ensures that the resulting vacuum 
state is aligned in the $\sigma$ direction implying 
a non-vanishing expectation value $\ave{\sigma}\equiv 
\sigma_0=\mu/\sqrt{\lambda}$ and a finite pion mass. However, 
after the standard shift 
$\sigma\to \sigma_0+\sigma$ a 'spurious' mixing term of the form 
$g \vec{a_1}^\mu \cdot \partial_\mu \vec{\pi}$ between the pion 
and the $a_1$ field emerges
which has to be eliminated by a shift of the $a_1$ field,   
\beq
\vec{a_1}^\mu \to \vec{a_1}^\mu - \frac{g\sigma_0}{m_0^2+(g\sigma_0)^2} 
 \ \partial^\mu\vec\pi \ . 
\eeq
As a consequence the vector and axialvector meson masses, determined by the 
$\tr(A_{L,R})^2$ term, split according to  
\bea
m_\rho^2 &=& m_\omega^2 ~=~ m_0^2 
\nonumber\\
m_{a_1}^2 &=& m_{f_1}^2 ~=~ m_0^2 +(g\sigma_0)^2 \ ,  
\label{Masslsm}
\eea 
and the standard relations of the linear $\sigma$-model are modified as 
\bea
f_\pi&=&\frac{m_\rho}{m_{a_1}} \sigma_0
\nonumber\\
m_\pi^2&=&\left(\frac{m_\rho}{m_{a_1}}\right)^2 \frac{h}{\sigma_0}
\nonumber\\
m_\sigma^2&=&\frac{h}{\sigma_0} + 2 \lambda \sigma_0^2 \ . 
\eea
At the mean-field level, the parameter values are fixed as 
$\sigma_0=152$~MeV, $g=6.55$, $h=(102 {\rm MeV})^3$ by imposing
the experimental values for $f_\pi=93$~MeV, $m_\pi=138$~MeV,
$m_\rho=770$~MeV and $m_{a_1}=1260$~MeV. Some latitude arises
in the choice of $\lambda$ and $\mu$ 
due to the uncertainty in the $\sigma$ mass; when identifying 
it as $m_\sigma=600$~MeV, one has $\lambda=7.6$ and $\mu=412$~MeV.   
As emphasized in Ref.~\cite{Pisa95}, the virtue of complying with 
VDM through the requirement of locally gauge invariant couplings
greatly restricts the number of possible interaction terms (\ie, the mere
requirement of a global chiral symmetry would allow many more terms).

When moving to the finite-temperature modifications, Pisarski 
evaluated selfenergy corrections in terms of a thermal 
loop expansion to lowest order in $g$ for two limiting cases.  
At low temperatures and in the chiral limit, one can additionally
expand in small pion momenta $p\sim T\ll m_\rho, m_{a_1}$ to obtain
for the on-shell thermal pole masses of the $\rho$ and $a_1$ 
meson~\cite{Pisa95} 
\bea
m_\rho^2(T) &\simeq& m_\rho^2 -\frac{g^2 \pi^2 T^4}{45m_\rho^2} 
\left(\frac{4m_{a_1}^2(3m_\rho^2+4 q^2)}{(m_{a_1}^2-m_\rho^2)^2} 
-3 \right) + \dots   
\nonumber\\
m_{a_1}^2(T) &\simeq& m_{a_1}^2 + \frac{g^2 \pi^2 T^4}{45m_\rho^2} 
\left(\frac{4m_{a_1}^2(3m_\rho^2+4 q^2)}{(m_{a_1}^2-m_\rho^2)^2}
+\frac{2m_\rho^4}{m_{a_1}^2 (m_{a_1}^2-m_\sigma^2)}
-\frac{m_{a_1}^2}{m_\rho^2}\right) + \dots \ .  
\label{lslowt}
\eea 
This result is consistent with the model-independent mixing theorem 
(\ref{VAmixing}) 
of Dey et al.~\cite{DEI90} stating that there are no mass corrections 
to order $T^2$.  
However, as stressed in Ref.~\cite{Pisa95}, this only holds strictly 
on the mass shell, \ie, for $q^2=m_\rho^2$, but not away from it.  
Moreover, for explicitly broken chiral symmetry, \ie, for $h\ne 0$, the 
on-shell $a_1$ pole mass does pick up a $T^2$-term 
\beq
m_{a_1}^2(T)\simeq m_{a_1}^2 +\frac{g^2m_\pi^2 T^2}{4m_\sigma^2} 
+ \dots \ . 
\eeq
One should also note that in Ref.~\cite{EI95}, where the $O(T^4)$
corrections have been assessed using the OPE
in connection with deep-inelastic scattering amplitudes on the pion, 
{\em both} the $\rho$ and $a_1$ masses have been found to decrease. 

As a second limit Pisarski considered the behavior of the masses
at the critical temperature for chiral restoration.
For the gauged linear $\sigma$-model the latter is  given in terms of 
the zero temperature $\sigma$ expectation value by $T_c^\chi=\sqrt{2}
\sigma_0\simeq 215$~MeV (in the chiral limit). 
By definition, $\sigma_0\to 0$ for
$T\to T_c^\chi$, such that in the immediate vicinity of the transition 
several of the trilinear vertices $\propto \sigma_0$ encoded 
in the Lagrangian~(\ref{Lgls}) vanish. Neglecting furthermore
small terms of order $T^2/m_\rho^2$, the temperature corrections to
both the $\rho$ and $a_1$ selfenergies turn out to be
\beq
\Sigma_{\rho,a_1}^{T_c^\chi}=\frac{1}{6} g^2 T^2
\eeq   
resulting in $m_\rho^2(T_c^\chi)=m_{a_1}^2(T_c^\chi)
=(2m_\rho^2+m_{a_1}^2)/3=(962~{\rm MeV})^2$ (note that
the $(g\sigma_0)^2$-term in Eq.~(\ref{Masslsm}) vanishes at $T_c^\chi$). 
Since the rather large value of the coupling constant $g$ implies 
that a lowest-order calculation cannot be quantitatively trusted 
the emphasis here is not on the exact mass values but rather on the 
qualitative feature that $\rho$ and $a_1$ masses become degenerate
at a common value {\em in between} their vacuum masses. 
One should note the somewhat peculiar feature that in  
the low-temperature limit (\ref{lslowt}) the $\rho$ and $a_1$ masses 
start out by moving {\it apart}. As another striking
result the $\omega$ meson mass turns out not to be affected at all,
$m_\omega(T_c^\chi)=m_\omega$, thus lifting the (theoretically
not well-understood) zero temperature degeneracy with 
the $\rho$ meson. 

Pisarski also studied situations where vector meson dominance does
not hold, \eg, when replacing the local chiral symmetry-breaking 
vector mass term $\propto m_0^2$ in Eq.~(\ref{Lgls}) by
\beq
{\cal L}_\zeta=
\zeta \ \tr (|\Phi|)^2 \ \tr \left[ (A_L^\mu)^2+(A_R^\mu)^2\right] \ . 
\eeq
In this case the vacuum $\rho$ and $a_1$ masses are still split
by the spontaneous breaking term $(g\sigma_0)^2$, but an explicit
finite-temperature calculation shows that close to the 
transition the masses uniformly decrease
to $m_\rho(T_c^\chi)=m_{a_1}(T_c^\chi)=m_\omega(T_c^\chi)=629$~MeV.
Other possible terms outside the VDM might induce different behavior. 
Thus, within the gauged linear $\sigma$-model, the fate of
the (axial-) vector masses (especially for the $\rho$) crucially
depends on whether vector meson dominance, as represented by the 
field-current identity Eq.~(\ref{field-curr}),  continues to hold 
at finite temperature.  

\subsection{Massive Yang-Mills Approach} 
As a second variant of the chiral Lagrangian framework to study 
in-medium vector meson properties we discuss the massive 
Yang-Mills (MYM) approach. It is very similar in spirit to the
gauged linear $\sigma$-model,  only that the $\sigma$ degrees
of freedom have been eliminated using the non-linear realization  
of the $SU(3)_L\times SU(3)_R$ chiral symmetry~\cite{Meis88}. From 
a phenomenological point of view this might be the more appropriate
effective theory at zero and low temperatures. The Lagrangian is expressed 
through a matrix representation
\beq
U=\exp(i\sqrt{2}\phi/f_\pi) \ ,\qquad 
\phi\equiv \phi_a\frac{\tau_a}{\sqrt{2}} \ ,  
\label{ufield}
\eeq 
where, in the $SU(2)$ case, the isospin index $a$ of the pseudoscalar
fields $\phi_a$ runs from 1 to 3, contracted with the Pauli 
matrices $\tau_a$. 
The vector and axialvector fields are introduced in complete analogy 
to the linear $\sigma$-model of the previous Section, \ie, 
as massive gauge fields. Defining them via
\bea
V^\mu \equiv V^\mu_a\tau_a/\sqrt{2} \  \quad 
&,& \quad A^\mu \equiv A^\mu_a\tau_a/\sqrt{2}
\nonumber\\
A_L^\mu \equiv \frac{1}{2}(V^\mu+A^\mu) \ 
&,& \quad A_R^\mu \equiv \frac{1}{2}(V^\mu-A^\mu)
\label{vecmym}
\eea
leads to the Lagrangian
\bea
{\cal L}_{\rm mym} &=& \frac{1}{4} f_\pi^2 \ 
\tr \left[ D_\mu U D^\mu U^\dagger\right]
-\frac{1}{2} \tr \left[ (F_L^{\mu\nu})^2+(F_R^{\mu\nu})^2\right]
+m_0^2 \ \tr \left[(A_L^\mu)^2+(A_R^\mu)^2\right] 
\nonumber\\
 && -i\xi \ \tr \left[D_\mu U D^\mu U^\dagger F_L^{\mu\nu} + 
D_\mu U D^\mu U^\dagger F_R^{\mu\nu} \right] +  
\sigma \ \tr \left[F_L^{\mu\nu} U F_{R \mu\nu} U^\dagger \right] \ . 
\label{Lmym}
\eea
Note that the normalization convention chosen in Eq.~(\ref{vecmym})
(which differs from the previous Section by a factor of $1/\sqrt{2}$),
entails a factor of 2 in the terms bilinear in the (axial-) vector
fields. Also, the $\rho\pi\pi$ coupling constant picks up an additional
factor in its relation to the gauge coupling constant $g$, \ie, 
$g=\sqrt{2} g_{\rho\pi\pi}$. This is due to the definition of 
the covariant derivative which in (\ref{Lmym}) is taken as 
\beq
D^\mu U=\partial^\mu-ig (A_L^\mu U - U A_R^\mu) \ .  
\eeq
The last two terms in the Lagrangian (\ref{Lmym}) are so-called 
non-minimal coupling terms
(\ie, of higher order in the derivatives than the other ones), which 
are necessary for a realistic description of the 
vector and axialvector meson sector in vacuum. The four free parameters 
$(m_0, g, \sigma, \xi)$ are readily adjusted to reproduce 
the phenomenological masses and decay widths of $\rho$, $\omega$, 
$a_1$, etc..   

At finite temperature the calculation of the vector correlator 
-- saturated by the $\rho$ meson in VDM --  has been shown~\cite{LSY95}
 to obey the general low-energy theorem of Ref.~\cite{DEI90}.   
To lowest order in the 'mixing parameter' 
$\epsilon=T^2/6f_\pi^2$, the finite-temperature $\rho$ meson selfenergy 
receives two corrections from thermal one-pion loop diagrams. 
The relevant terms in the MYM Lagrangian (\ref{Lmym}),  
\beq
{\cal L}_{\rm mym}= \frac{1}{2} m_\rho^2 \vec{\rho}_\mu^2 
+\left[m_\rho^2+g^2 f_\pi^2\right] \vec{a_1}^2_\mu + 
g^2 f_\pi \vec{\pi} \times \vec{\rho}^\mu \cdot \vec{a_1}_\mu  
+\frac{1}{2} g^2 \left[ \vec{\rho}_\mu^2 \vec{\pi}^2 -\vec{\rho}^\mu 
\cdot \vec{\pi} \ \vec{\rho}_\mu \cdot \vec{\pi} \right] + \dots \ ,  
\label{Lmym2}
\eeq
induce a $\rho\rho\pi\pi$ 'tadpole' diagram (last term) as well as a 
$\pi a_1$ resonance loop (prelast term). The leading temperature 
dependence is driven by loops of pions from the heat bath. When evaluated
in the chiral limit the resulting vector correlator takes the form
\bea 
\Pi_V^{\mu\nu}(q) &=& 
\left(g^{\mu\nu}-\frac{q^\mu q^\nu}{m_\rho^2}\right) \ g_\rho^2 \  
\nonumber\\ 
 && \times \left[\frac{i}{m_\rho^2-q^2}+\frac{i}{m_\rho^2-q^2} i g^2f_\pi^2
\epsilon \frac{i}{m_\rho^2-q^2}+ \frac{i}{m_\rho^2-q^2} 
\frac{i^3 g^4 f_\pi^4 \epsilon}{m_{a_1}^2-q^2}\frac{i}{m_\rho^2-q^2}
\right] \ ,  
\eea
where the second and third terms arise from the interaction vertices
in Eq.~(\ref{Lmym2}) and $g_\rho=m_\rho^2/g$ is the VDM coupling constant.  
Making use of the Weinberg relation, $m_{a_1}^2=m_\rho^2+g^2f_\pi^2$, one 
finally obtains
\beq
\Pi_V^{\mu\nu}(q) = \left(g^{\mu\nu}-\frac{q^\mu q^\nu}{m_\rho^2}\right)
 \ g_\rho^2 \ \left[(1-\epsilon) \frac{i}{m_\rho^2-q^2} +
\epsilon \frac{i}{m_{a_1}^2-q^2} \right] 
\eeq
in accordance with the model-independent mixing theorem (\ref{VAmixing}). 
Deviations occur, \eg, through the finite
pion mass or when including the non-minimal coupling term with 
$\sigma\ne 0$ in the Lagrangian (\ref{Lmym}). As reported
by Song~\cite{song93}  the $\rho$ meson mass tends to increase 
whereas the $a_1$ mass decreases.
Quantitatively, however, the temperature dependencies seem to be very
weak. The masses only change by a few percent even 
at temperatures as high as $T=200$~MeV which already 
is clearly beyond the range of applicability of the low-temperature 
expansion. Much before that chiral restoration is likely to  be realized 
through the mixing effect.

\subsection{Hidden Local Symmetry}
The third chiral Lagrangian framework to incorporate vector mesons
is the so-called 'Hidden Local Symmetry' (HLS) approach  
proposed by Bando \etal~\cite{Bando}. It originated from the 
observation that the conventional $[SU(2)_L\times SU(2)_R]_{global}$-symmetric 
non-linear $\sigma$-model Lagrangian, 
\beq
{\cal L}_{{\rm nl}\sigma}= \frac{1}{4} f_\pi^2 \ 
\tr \left[\partial_\mu U \partial^\mu U^\dagger\right] 
\label{Lnls}
\eeq
($U$ as given in Eq.~(\ref{ufield})), can be recast
in a form that exhibits an additional $[SU(2)_V]_{local}$ symmetry. This can
be made explicit by rewriting the $U$-field as  
\beq
U\equiv \xi_L^\dagger \  \xi_R
\eeq    
in terms of new SU(2)-valued variables $\xi_L$ and $\xi_R$ (notice that this
implies the appearance of three additional, unphysical scalar degrees
of freedom).  Defining the usual covariant derivative  
\beq
{\cal D}^\mu= \partial^\mu -igV^\mu  
\label{Dmuhls1}
\eeq
with an auxiliary gauge field $V^\mu=V_a^\mu \tau_a/2$ and associated gauge
coupling $g$ allows to construct two invariants under 
$\left[SU(2)_L\otimes SU(2)_R\right]_{global} \times 
\left[SU(2)_V\right]_{local}$ transformations, namely  
\bea
{\cal L}_A &=& -\frac{1}{4} f_\pi^2 \ \tr \left[{\cal D}^\mu \xi_L\cdot
\xi_L^\dagger + {\cal D}^\mu \xi_R\cdot \xi_R^\dagger \right] 
\nonumber\\
{\cal L}_V &=& -\frac{1}{4} f_\pi^2 \ \tr \left[{\cal D}^\mu \xi_L\cdot
\xi_L^\dagger - {\cal D}^\mu \xi_R\cdot \xi_R^\dagger \right] \ . 
\eea
When imposing the so-called unitary gauge 
\beq
\xi_L^\dagger=\xi_R= \exp(-\pi/f_\pi) 
\eeq
(which eliminates the unphysical scalar degrees of freedom) 
it can be verified that, with arbitrary constant $a$, the 
combination  
\beq
{\cal L}={\cal L}_A + a {\cal L}_V
\eeq
is indeed equivalent to Eq.~(\ref{Lnls}). In fact, ${\cal L}_A$ is  
in one-to-one correspondence with the starting Lagrangian
${\cal L}_{{\rm nl}\sigma}$, whereas ${\cal L}_V$ identically vanishes by virtue
of the equations of motion for the gauge field. It can be assigned a physical 
significance by assuming that it develops its own dynamics, generating 
a kinetic energy term in the Lagrangian (usually attributed to
the underlying QCD-dynamics or quantum effects at the composite
level~\cite{Bando}). The physical vector field is then identified  
with the (isovector) $\rho$ meson, and the HLS Lagrangian becomes 
\bea
{\cal L}_{\rm hls} &=& {\cal L}_A + a {\cal L}_V - \frac{1}{4} 
(\vec\rho^{\mu\nu})^2   
\nonumber\\
 &=&   \frac{1}{4} f_\pi^2 
tr\left[\partial_\mu U \partial^\mu U^\dagger\right]
-\frac{1}{4} \vec\rho^{\mu\nu} \cdot \vec\rho_{\mu\nu} + \frac{a}{2} 
g^2 f_\pi^2 \vec\rho_\mu^2 + \frac{a}{2} g \vec\rho^\mu \cdot
(\vec\pi \times \partial_\mu \vec\pi) + {\cal O}(\vec\pi^4) \    
\label{Lhls}
\eea 
with the non-abelian field strength tensor 
\beq
\vec\rho^{\mu\nu}= \partial^\mu \vec\rho^\nu -\partial^\nu \vec\rho^\mu
+g  \ \vec\rho^\mu \times \vec\rho^\nu \ . 
\eeq
The second line in Eq.~(\ref{Lhls}) has been obtained using the weak-field
expansion for the $\xi$-fields. One reads off that  
\bea
g_{\rho\pi\pi}&=&\frac{1}{2} a g 
\label{grho}
\\
m_\rho^2&=&a g^2 f_\pi^2        \ ,  
\label{mrho}
\eea 
\ie, the gauge symmetry is spontaneously broken generating
a mass for the vector field via the Higgs mechanism. This is accompanied by the
disappearance of the scalar modes which have turned into the longitudinal
components of now massive vectors. A particular advantage of the HLS
framework is the unique way of introducing electromagnetic interactions.
Since the photon couples to the charge $Q=I_3^{(L)}+I_3^{(R)}$ 
corresponding to the global $\left[SU(2)_L\otimes SU(2)_R\right]$ 
isospin symmetry, the electromagnetic field can be introduced as a 
global gauge symmetry. Thus the covariant derivative 
(\ref{Dmuhls1}) can be simply extended to 
\beq
{\cal D}^\mu \xi_{L,R}= (\partial^\mu -igV^\mu)\xi_{L,R} 
+ e_0 \xi_{L,R} B^\mu \frac{\tau_3}{2} 
\label{Dmuhls2}
\eeq 
with the $U(1)_Q$ gauge field $B^\mu$ and associated coupling $e_0$. 
In addition, the corresponding kinetic term $-\frac{1}{4} B_{\mu\nu}^2$ 
($B_{\mu\nu}=\partial_\mu B_\nu-\partial_\nu B_\mu$) 
has to be added to the HLS Lagrangian in Eq.~(\ref{Lhls}). 
After rediagonalizing the fields one finds the following mass 
relations:
\beq
m_\gamma^2=0 \ , \quad m_{\rho^0}^2=a(g^2+e_0^2) f_\pi^2 \ , \quad
m_{\rho^\pm}^2=a g^2 f_\pi^2\pkt
\eeq
For the special case of  $a=2$, Eqs.~(\ref{grho}) and (\ref{mrho}) 
give the universality of the $\rho$ couplings as well as 
the KSFR relation~\cite{KSFR}. Moreover, vector dominance emerges due to
the vanishing of the direct $\gamma\pi\pi$ coupling and the
$\rho$-$\gamma$ coupling is given as $g_{\rho\gamma}=m_\rho^2/g$. 

Let us now turn to the finite temperature calculations performed 
within the HLS framework.   
The minimal version of the HLS Lagrangian does not involve the $a_1$ 
field. 
Nevertheless, the low-energy theorem, Eq.~(\ref{VAmixing}), ought to 
be satisfied for the vector correlator. This is indeed the case
and is realized through temperature-dependent corrections to the VDM 
coupling constant $g_{\rho\gamma}$. 
As has been shown in Ref.~\cite{LSY95}, to lowest
order in $\epsilon$, the thermal pion tadpole loop on the
$\rho-\gamma$ vertex, shown in Fig.~\ref{fig_grpitad}, leads
to correction factor $(1-\frac{1}{2}\epsilon)$. 
\begin{figure}[t]
\vspace{-2cm}
\begin{center}
\epsfig{figure=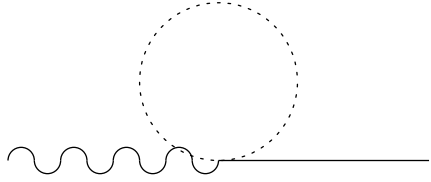,height=6cm}
\end{center}
\vspace{-2cm}
\caption{Finite temperature correction to the $\gamma\rho$ vertex 
through a thermal pion tadpole loop in the HLS approach (wavy line:
photon, solid line: $\rho$ meson, dotted line: pion).}
\label{fig_grpitad}
\end{figure}
Therefore  
\beq
g_{\rho\gamma}^2(T)=(1-\epsilon) g_{\rho\gamma}^2 + {\cal O}(\epsilon^2) \ .
\eeq 
Moreover, in the 'minimal' HLS there is no $\rho\rho\pi\pi$ contact 
interaction. Thus, to lowest thermal pion-loop order, the in-medium 
$\rho$ mass is only modified through temperature effects in the two-pion 
loop which do not pick up $T^2$-corrections due to the additional 
derivative in the $\rho\pi\pi$ coupling.    
  
For practical purposes the HLS approach has been mainly 
employed to study finite-temperature modifications
of the pion electromagnetic form factor, $F_\pi$, and 
dilepton/photon production 
rates. The latter will be discussed in Chap.~\ref{chap_dilepton}. 
In the thermal medium, the former 
can be defined through the total electromagnetic vertex 
for the $\pi\pi\to\gamma$ transition, 
\beq
\Gamma_\mu^{\gamma\pi\pi}(T)=q_\mu \ F_\pi(q_0,\vec q;T) \ .     
\eeq
For $a=2$ and in free space, it reduces to the well-known
VDM expression,
\beq
F_\pi^\circ(M)=\frac{g_{\rho\pi\pi} g_{\rho\gamma}}{M^2-(m_\rho^{(0)})^2+ 
\Sigma_{\rho\pi\pi}(M)} \ .
\eeq
In Ref.~\cite{SK96}, $F_\pi(T)$ has been evaluated in terms of
thermal one-pion loop corrections (using $a=2$). The total 
$\gamma\pi\pi$ vertex function can then be written as  
\beq
\Gamma_\mu(T)=\Gamma_\mu^{\rm mix}+\Gamma_\mu^{\rm vert}+
\Gamma_\mu^{\rm rho}+\Gamma_\mu^{\rm dir} \ ,
\label{vxppT}
\eeq  
where $\Gamma_\mu^{\rm mix}$ encodes the pion-tadpole diagram
(inducing $V$-$A$ mixing diagrammatically represented in
Fig.~\ref{fig_grpitad}), 
$\Gamma_\mu^{\rm vert}$ represents thermal-loop corrections of the 
$\pi\pi\rho$ vertex (left panel of Fig.~\ref{fig_vxppT}) and 
$\Gamma_\mu^{\rm rho}$ accounts for the temperature
dependence in the $\rho$ selfenergy $\Sigma_{\rho\pi\pi}$ (\ie, in 
the two-pion bubble). The appearance of a direct $\gamma\pi\pi$ 
vertex $\Gamma_\mu^{\rm dir}$ is solely due to finite-temperature
vertex modifications, induced by the diagrams shown in the
right panel of Fig.~\ref{fig_vxppT}. 
\begin{figure}[t]
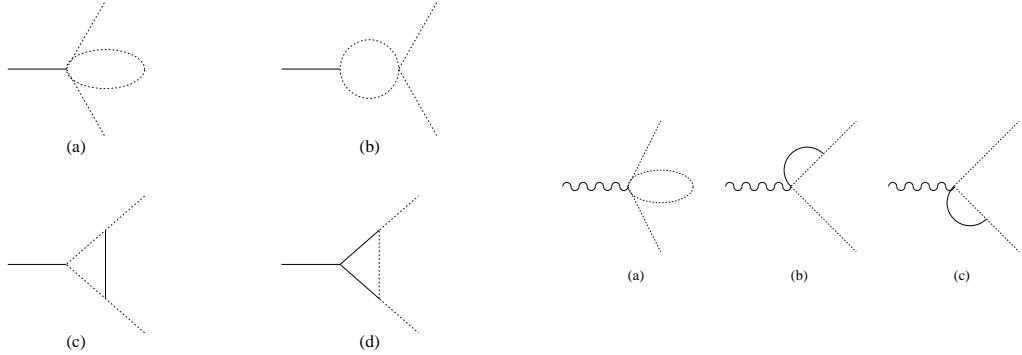

\epsfig{figure=vxrppT.eps,height=6.5cm}
\hspace{1cm}
\epsfig{figure=vxgppT.eps,height=6.5cm}
\vspace{-1cm}
\caption{Thermal loop corrections to the $\rho\pi\pi$ vertex (left
panel) as well as the 'direct' $\gamma\pi\pi$; the latter vanishes
at zero temperature in the VDM (\ie~ for $a=2$ in the HLS framework).
The figures are taken from Ref.~\protect\cite{SK96}.} 
\label{fig_vxppT}
\end{figure}
In the limit of vanishing three-momentum, $\vec q=0$, the in-medium 
form factor can be characterized by a single scalar function 
(see Eq.~(\ref{specmed})) 
depending on invariant mass $M$ only, according to 
\beq
F_\pi(T)=Z_\pi(T) \left[ \frac{g_{\rho\pi\pi}(T) \ g_{\rho\gamma}(T)}
{M^2-m_\rho^2+im_\rho \Gamma_\rho^\circ-\Sigma_\rho(T)} +F'_\pi(T) \ . 
\right]  
\label{fpiThls} 
\eeq
Here, $Z_\pi(T)$ is the pion wave function renormalization constant. 
It can be inferred from the relevant Ward-Takahashi identity,
\beq
(p^\mu-p'^\mu) \Gamma_\mu(p,p')=\Sigma_\pi(p)-\Sigma_\pi(p') \ ,  
\eeq
which ensures gauge invariance of the in-medium $\gamma\pi\pi$ 
transition with $\Sigma_\pi$ denoting the in-medium pion selfenergy
to be evaluated to thermal one-loop order. 
For on-shell pions of vanishing three momentum, as  
considered in Ref.~\cite{SK96}, the simple relation 
\beq
Z_\pi(T)= \left[ 1- \frac{\partial\Sigma_\pi}
{\partial p_0^2}(p_0=m_\pi,\vec p =0)\right]^{-1}  
\label{zpi}
\eeq
is obtained.  In Eq.~(\ref{fpiThls}), 
the temperature-dependent $\rho\gamma$ and $\rho\pi\pi$ couplings, 
the temperature part of the $\rho$ selfenergy, $\Sigma_{\rho\pi\pi}^T$,
as well as the direct $\gamma\pi\pi$ piece, $F'_\pi$,   
arise from the various vertex terms in Eq.~(\ref{vxppT}), respectively  
($\Gamma^\circ_{\rho\pi\pi}$ denotes the free $\rho$ decay width, and the
real part of the free $\rho$ selfenergy has been absorbed into the
physical $\rho$ meson mass, $m_\rho$).  
The resulting pion electromagnetic form factor is displayed
in Fig.~\ref{fig_fpiThls}. One observes a strong suppression 
with increasing temperature over the entire invariant mass range 
which, to a large extent, is driven by the reduction of the vector dominance
coupling $g_{\rho\gamma}(T)$, representing the vector-axialvector
mixing effect. 
\begin{figure}[!htb]
\vspace{-3cm}
\begin{center}
\epsfig{figure=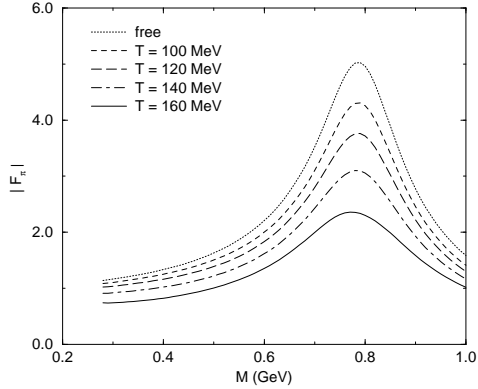,height=8cm}
\vspace{-0.5cm}
\end{center}
\caption{Pion electromagnetic form factor in a finite temperature pion gas
as calculated within the hidden local symmetry approach 
in Ref.~\protect\cite{SK96}.} 
\label{fig_fpiThls}
\end{figure}
One should note, however,  that the imaginary part of the vector correlator 
(which in VDM coincides with the imaginary part of the $\rho$ meson
propagator, \ie, the spectral function), involves an additional
factor from the imaginary part of the in-medium $\rho$ selfenergy;
\eg, in VDM the isovector correlator is related to the pion 
electromagnetic form factor through
\beq
{\rm Im} \Pi_V^{I=1}=\frac{{\rm Im} \Sigma_{\rho\pi\pi}}{g_{\rho\pi\pi}^2}
 \ |F_\pi(T)|^2 \ .  
\eeq
Within the quasiparticle approximation for pions, one obtains 
\beq
{\rm Im} \Sigma_{\rho\pi\pi}(q_0=M, \vec q=0)\propto 
\frac{k^4}{\omega_\pi(k)^2} 
|\frac{\partial\omega_\pi(k)}{\partial k}|^{-1}
\label{sigrppon}
\eeq
with the in-medium pion dispersion relation $\omega_\pi(k)$. 
In a pion gas, the latter is softened at small momenta due to the attractive
interaction with thermal pions in the $\pi\pi\to \rho$ channel, causing  
a reduction of the group velocity 
$v_k=|\partial\omega_\pi(k)/{\partial k}|^{-1}$. This  generates  
some enhancement in the vector correlator at low $M$,
which is not captured by the electromagnetic form factor
shown in Fig.~{\ref{fig_fpiThls} (the functional form of 
$\Sigma_{\rho\pi\pi}$ as quoted in Eq.~(\ref{sigrppon}) should be 
taken with care. It only leads to a gauge invariant vector correlator 
in connection with the $Z_\pi^2$ factor (\ref{zpi})   
which in Ref.~\cite{SK96} has been absorbed into the definition of 
$|F_\pi|^2$. In fact, one can show that  
$Z_\pi^2k^2/\omega_\pi(k)^2=v_k^2$, which reduces the effect of 
the pion softening, see, \eg, Ref.~\cite{ChSc} for a nice 
discussion on this point).   

Qualitatively similar features for the in-medium behavior of
the pion electromagnetic form factor have been found earlier within a  
schematic treatment in Ref.~\cite{Dom91}. Starting from the
on-shell expression for the free $\rho$ meson decay width (ignoring  
any pion mass), 
\beq
\Gamma_\rho^\circ = \frac{g_{\rho\pi\pi}^2}{4\pi} \frac{m_\rho}{12} \ , 
\eeq 
the finite temperature corrections have been estimated assuming 
the validitiy of the in-medium KSFR relation, 
$2 g_{\rho\pi\pi}^2 f_\pi^2 =m_\rho^2$, as well as an unmodified
$\rho$ mass (later on it has been realized that $m_\rho(T)$ does indeed 
not attain corrections to lowest order $O(T^2)$~\cite{EI93}, as 
said before). Then, using 
the lowest-order chiral perturbation theory result for the pion decay 
constant (for three massless flavors), 
\beq
f_\pi(T)=f_\pi (1-T^2/8f_\pi^2) \ 
\eeq
(for $N_f=2$ the coefficient 1/8 is to be replaced by 1/12),  
the temperature dependence of the $\rho$ decay width has been  cast in the
form
\beq
\Gamma_\rho(T)\simeq\frac{\Gamma_\rho^\circ}{1-T^2/4f_\pi^2} \ , 
\label{GrhoT}
\eeq
indicating a broadening with increasing $T$. 
Vertex corrections of the $\pi\pi\rho$ coupling 
as required by Ward identities to ensure the conservation
of the vector current have not been included in this estimate. 
When naively extrapolating Eq.~(\ref{GrhoT}) to high temperatures
one finds the $\rho$ width to diverge at $T=2f_\pi \simeq 185$~MeV (230~MeV
when using the two-flavor result for $f_\pi(T)$).  
In Refs.~\cite{Dom89,Dom91} this kind of resonance melting has been 
qualitatively associated with the  
approach towards color deconfinement. 

\subsection{Phenomenological Meson Lagrangians} 
\label{sec_phenlagrang}
The chiral Lagrangian frameworks discussed in the previous Sections
allow a for systematic investigation of the low-temperature chiral 
dynamics of the vector meson properties. The inclusion of higher 
resonances, however, can become a quite formidable task  due to the 
increasing number of interaction vertices. Furthermore, chiral symmetry
does not always give unique prescriptions for the latter, as we have seen 
above. Since the impact of certain meson resonances, which have not 
been incorporated via chiral Lagrangians so far, may be non-negligible 
more phenomenologically oriented approaches have been 
pursued~\cite{GaKa,GaLi,Ha95,RCW,EK99,Gao99,RG99}. They  
aim at including the empirically important interactions in a tractable 
way that also respects the relevant symmetries, such as vector current 
conservation or chiral symmetry. In the following, 
we will elaborate on two variants that have been employed in this
context, namely kinetic-theory and many-body type calculations.  

At low and moderate temperatures the thermal meson gas is dominated
by the light pseudoscalar Goldstone bosons $P=\pi,K,\bar K$. 
A rather extensive treatment of the possible scattering processes of 
on-shell vector mesons $V$ in such a system has been undertaken by  
Haglin~\cite{Ha95}. He specified the following interaction 
Lagrangians (isospin structure suppressed) 
\bea
{\cal L}_{VPP} &=& G_{VPP} \ V^\mu \ P \ {{\partial}}_\mu \ P 
\nonumber\\
{\cal L}_{VVP} &=& G_{VVP} \ \epsilon_{\mu\nu\alpha\beta} \ \partial^\mu
                  \ V^\nu \ \partial^\alpha \ V^\beta \ P  
\nonumber\\
{\cal L}_{\rm AVP} &=& G_{AVP} \ A_{\mu\nu} \ V^{\mu\nu}  
\eea 
for the exchange of pseudoscalar, 
vector ($V$) and axialvector ($A$) mesons, respectively ($V^{\mu\nu}$ and 
$A^{\mu\nu}$ denote the usual field strength tensors). 
The average collision rate for the vector mesons in binary collisions
with particles $h$ from the heat bath, $V h\to 3 4$ was then obtained 
from the kinetic theory expression 
\bea
\bar\Gamma_V^{\rm coll}(T) &=& \frac{g_V g_h}{n_V(T)} \int 
d^3\tilde p_V d^3\tilde p_h d^3\tilde p_3 d^3\tilde p_4 
|\bar {\cal M}_{\rho h\to 34}|^2 (2\pi)^4 \delta^{(4)}(p_V+p_h-p_3-p_4) 
\nonumber\\ 
 & & \qquad \qquad \times f^V(T) \  f^h(T) \  [1+f^3(T)] [1+f^4(T)] \ , 
\eea
where $d^3\tilde p_i\equiv d^3p_i/2\omega_i(p_i)(2\pi)^3$, etc., $g_i$ 
are the spin-isospin degeneracy factors, $f^i(T)$ thermal Bose-Einstein
distribution functions, and $n_V(T)$ is the number density of the vector meson $V$. 
The coherent sum of invariant amplitudes, 
\beq
{\cal M}_{\rho P}= \sum\limits_R {\cal M}_{\rho PR} \ , 
\eeq 
has been computed for both $s$- and $t$-channel exchanges of mesons $R$. For 
elastic $\rho\pi$ scattering, $R=\{\pi,\omega,\phi,a_1(1260),\omega(1390)\}$, 
for $\rho K$ scattering, $R=K_1(1270)$, and for $\omega \pi$ interactions,
$R=\{\rho,b_1(1235) \}$ were used.  For the $\phi$ meson the dominant processes
involve kaon exchange (including inelastic channels such as 
$\phi \pi\to K^* K$ or $\phi K\to K^* \pi$). 
The final results, displayed in Fig.~\ref{fig_collhag}, reveal a moderate
collisional broadening of about 40~MeV for the $\rho$ meson at $T=150$~MeV. 
A similar value of $\sim$~30~MeV has been found for the $\omega$ meson 
in which case, however, it amounts to a factor of four times its natural 
width. The effects for the $\phi$ meson are smaller. 
\begin{figure}
\vspace{1.5cm}
\hspace{3.6cm}
\epsfig{figure=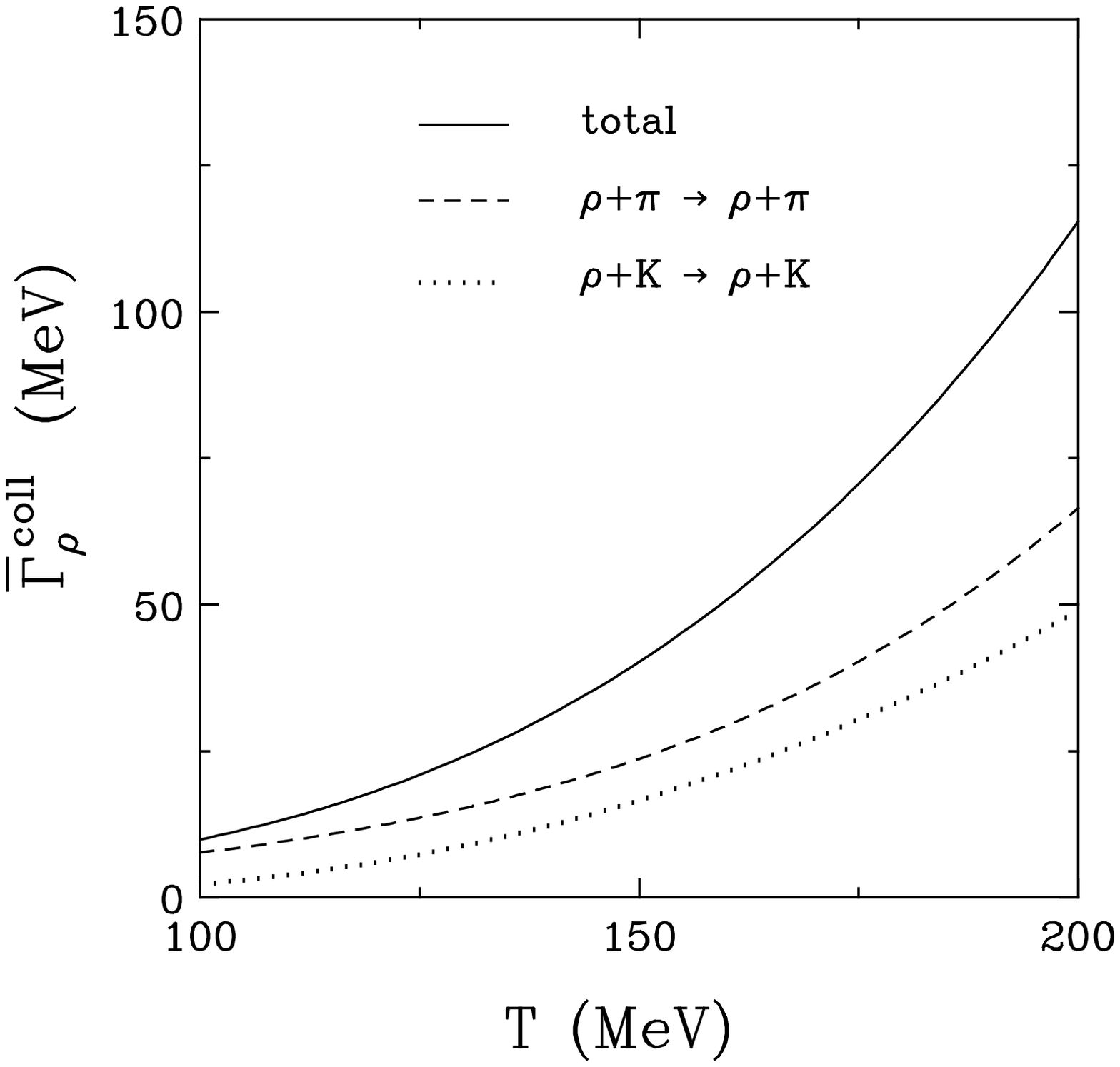,width=4cm,angle=90}
\hspace{3.1cm}
\epsfig{figure=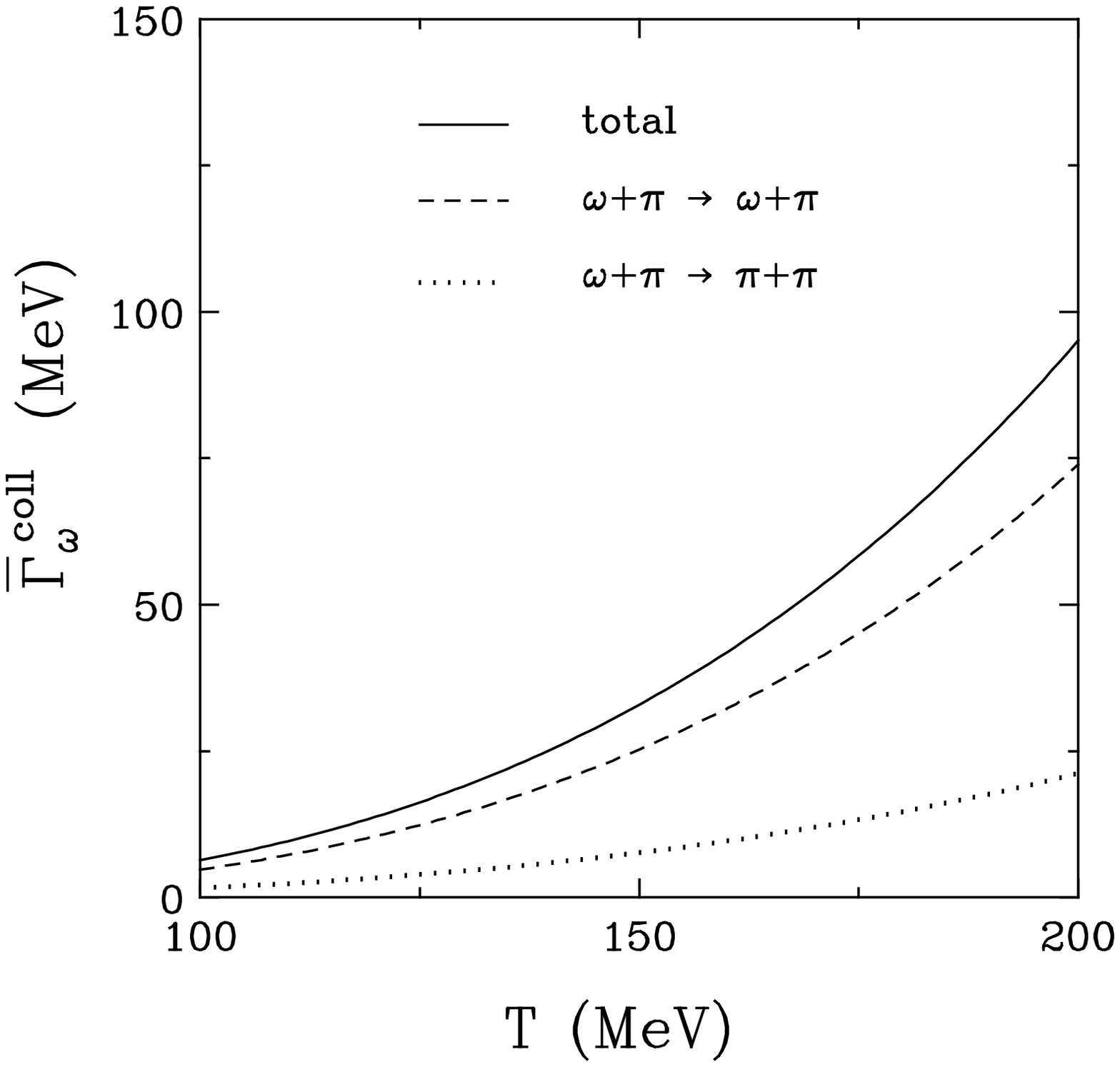,width=4cm,angle=90}
\hspace{3.1cm}
\epsfig{figure=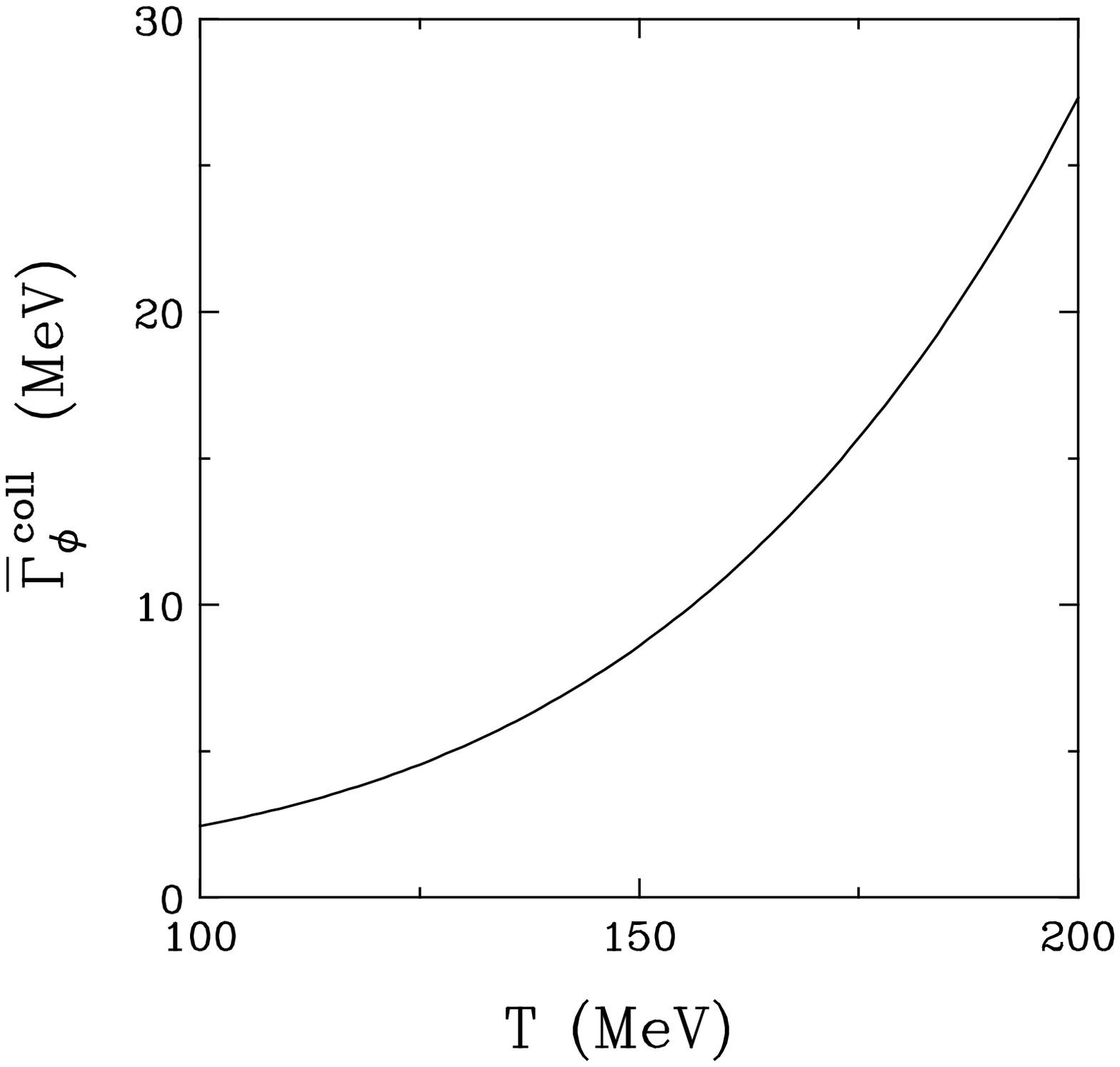,width=4cm,angle=90}
\caption{Temperature dependence of the collisional broadening of
$\rho$, $\omega$ and $\phi$ mesons (from left to right) as  
calculated in the kinetic
theory approach of Ref.~\protect\cite{Ha95}.}
\label{fig_collhag}
\end{figure}

Along similar lines, Gao \etal~\cite{Gao99} extended Haglin's analysis
for $\rho\pi$ scattering by including isospin-exchange interactions such 
as $\pi^+\rho^0\to \pi^0\rho^+$ and by using a different regularization 
method for the singularity in the $t$-channel pion-exchange diagram. 
He also employed a somewhat modified $\pi\rho a_1$ vertex which improves
the phenomenology of the $a_1\to\pi\rho$ decay~\cite{GG98}.  
In addition, the in-medium broadening of the $\rho\pi\pi$ decay width
through Bose-Einstein enhancement factors of the pions (which will be 
discussed in more detail below) were accounted for. The resulting $\rho$ 
meson spectral function, 
\beq
A_\rho(M)=-\frac{2 {\rm Im} \Sigma_\rho(M;T)}{[M^2-m_\rho^2-{\rm Re}
\Sigma_\rho(M;T)]^2+[{\rm Im} \Sigma_\rho(M;T)]^2} \ , 
\eeq
has been obtained in terms of a total in-medium $\rho$ selfenergy, 
\beq
\Sigma_\rho(M;T)=\Sigma_{\rho\pi\pi}(M;T)+\Sigma_\rho^{\rm coll}(T) \ ,
\eeq
consisting of a medium-modified  $\rho\to\pi\pi$ part (including the
Bose enhancement) and a collisional contribution that has been approximated
by its on-shell value (\ie, for $M=m_\rho$). 
\begin{figure}[!htb]
\vspace{-4cm}
\epsfig{figure=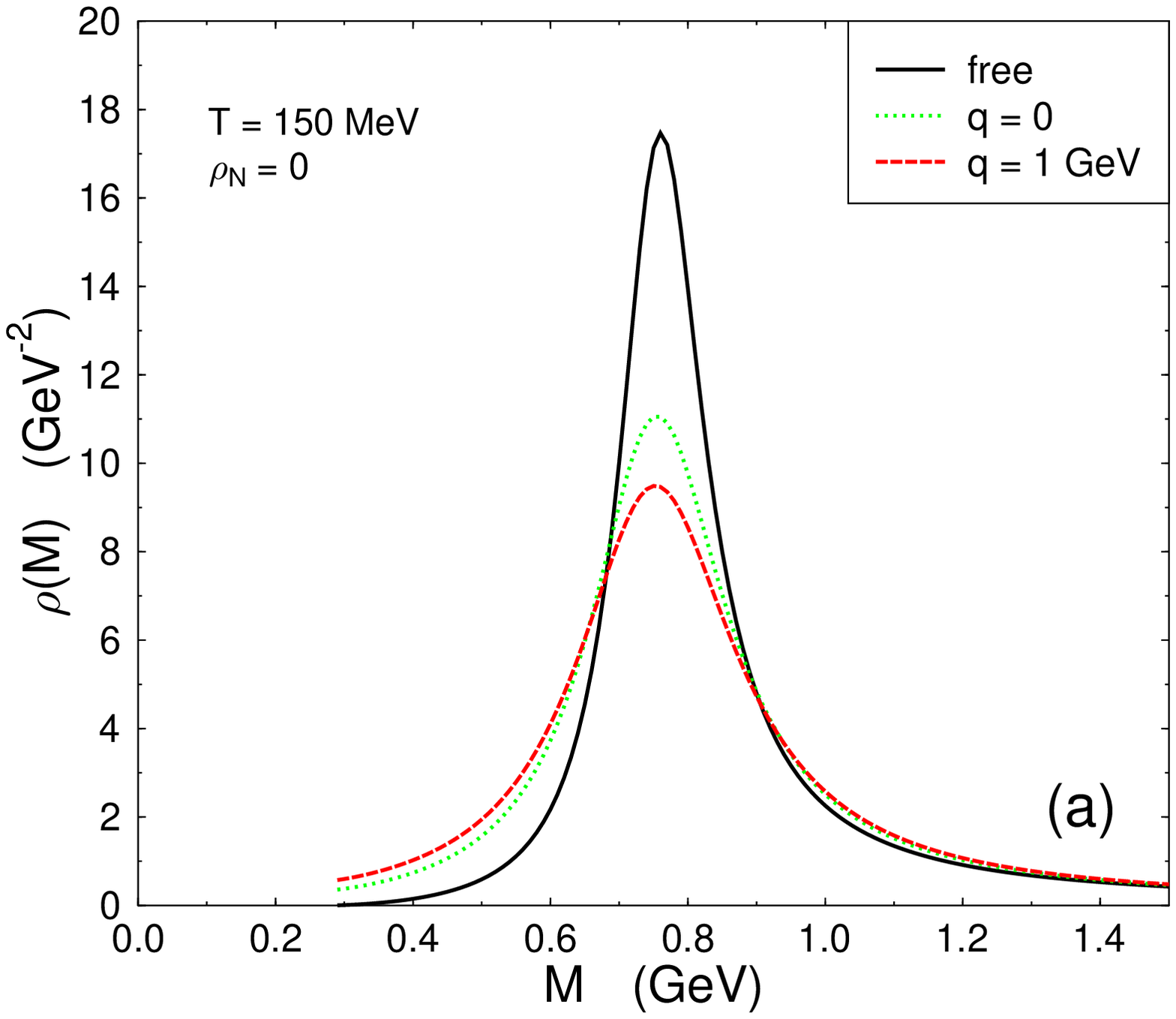,width=7.5cm,height=10cm}
\epsfig{figure=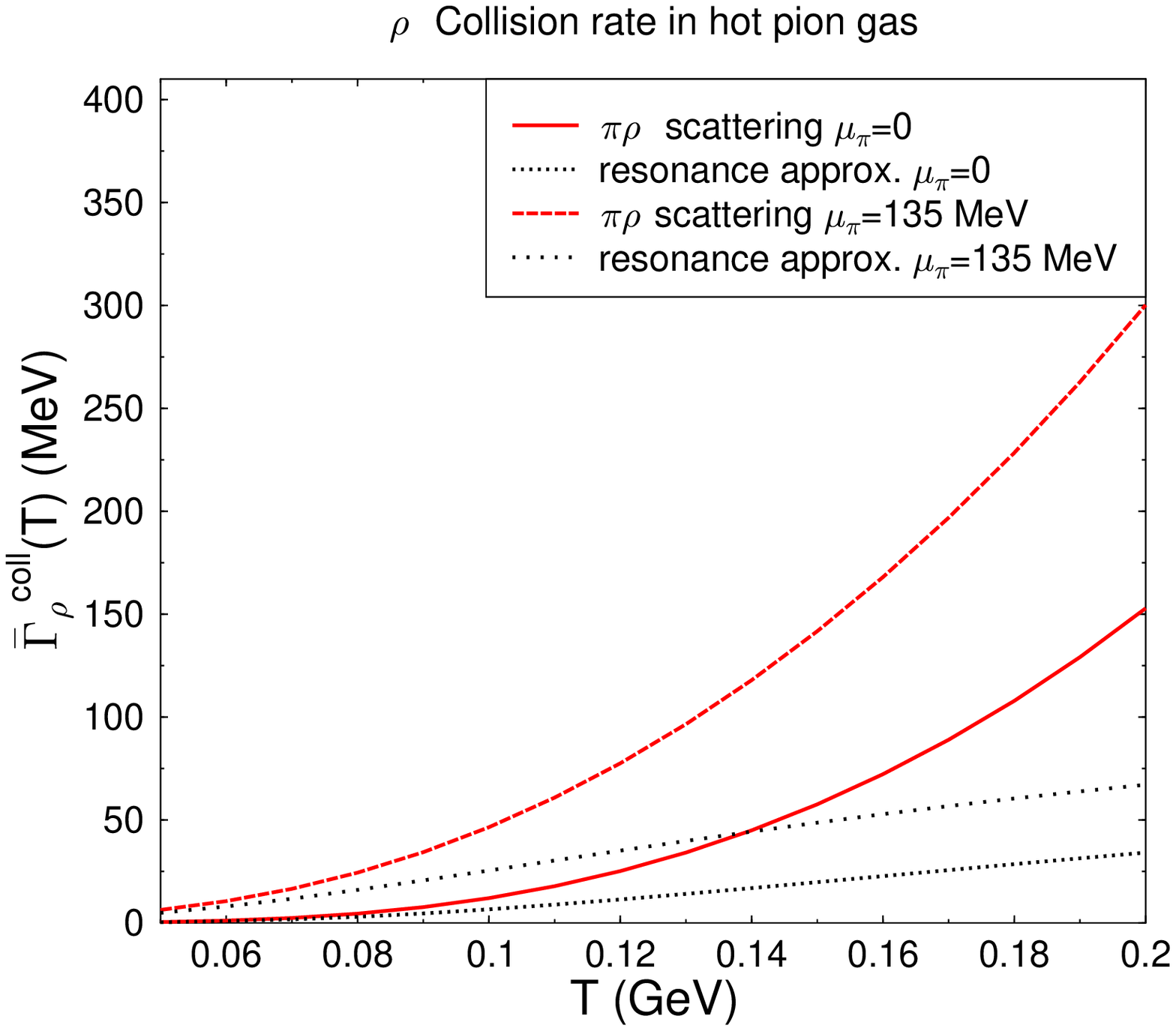,width=7cm,height=10.6cm}
\caption{$\rho$ meson spectral function at $T=150$~MeV (left panel) as 
evaluated in the kinetic 
theory calculations of Ref.~\protect\cite{Gao99} including the 
in-medium Bose-enhancement of the $\rho\to\pi\pi$ width. The right
panel shows the temperature dependence of the collisional part (using
two values for the pion chemical potential) for
the full result including $t$-channel processes (full curves, labeled 
'$\pi\rho$ scattering')  as well as for a calculation where only
$s$-channel resonances have been accounted for (dotted curves, labeled
'resonance approx.').}
\label{fig_gaorho}
\end{figure}
Therefore it depends on temperature only.
As apparent from Fig.~\ref{fig_gaorho}, the $\rho$ meson exhibits a 
thermal broadening which substantially exceeds the previous estimate 
by Haglin: at $T=150$~MeV, the scattering contribution 
from a pion gas turns out to be $\bar\Gamma_\rho^{\rm coll}=58$~MeV 
(see also right panel of Fig.~\ref{fig_gaorho}), whereas the 
in-medium Bose-enhancement of the $\rho\to\pi\pi$ decay width
amounts to 25~MeV. The three-momentum dependence of the spectral
function is rather weak.

The evaluation of the in-medium vector meson properties in 
the kinetic theory treatments of Refs.~\cite{Ha95,Gao99} 
was restricted to physical (on-shell) pole masses $m_V$. 
On the other hand, one might expect important effects from off-shell dynamics 
especially for the $\rho$ meson as it is characterized 
by an appreciable width already in free space (this, in turn, marks its 
distinguished role for low-mass dilepton yields, $M_{ll}<m_\rho$, in 
heavy-ion reactions, to be discussed in Chap.~\ref{chap_dilepton}).
Off-shell dynamics are naturally accounted for within a many-body treatment
of in-medium selfenergies, as we are going to discuss now.

The usual starting point is a microscopic model for the $\rho$ meson with
coupling to its 'pion cloud' via two-pion states. This not only renders the correct
decay width but also quantitatively describes its energy dependence over 
a broad range of invariant mass 
as encoded, \eg, in $\pi\pi$ $P$-wave scattering phase shifts or the pion 
electromagnetic form factor. Given the free $\pi +\rho$ Lagrangian, 
\beq
{\cal L}_{\pi+\rho}^{\rm free}
=\frac{1}{2} \ \tr \left[(\partial^\mu \pi)^2 -\phi^2\right]
-\frac{1}{2} \ \tr \left[(\rho_{\mu\nu}^2\right] 
+ (m_\rho^{(0)})^2 \ \tr \left[\rho_\mu^2\right] \ ,  
\eeq
Sakurai proposed to adopt the $\rho$ meson as the 
gauge boson of the conserved isospin~\cite{Sa60}, which can be realized 
by introducing the covariant derivative  
$\partial^\mu \to (\partial^\mu+ig_{\rho\pi\pi}\rho^\mu)$.  This is 
rather close in spirit to the massive Yang-Mills approach discussed
above. The resulting $\pi\rho$ interaction vertices are then given 
by 
\beq
{\cal L}_{\rho\pi}^{\rm int}=g_{\rho\pi\pi} (\vec\pi\times\partial^\mu\vec\pi) 
 \cdot \vec\rho_\mu  -\frac{1}{2} g_{\rho\pi\pi}^2 \ 
\vec\rho^\mu\cdot\vec\pi \ \vec\rho^\mu\cdot\vec\pi \ .  
\label{Lintrhopi}
\eeq
To lowest order in $g_{\rho\pi\pi}$ the corresponding selfenergy 
in vacuum reads 
\bea
\Sigma_{\rho\pi\pi}^{\circ,\mu\nu}(M) &=& 
 i g_{\rho\pi\pi}^2\int {d^4p\over{(2\pi)^4}}\,{(2p+q)_{\mu}(2p+q)_{\nu}\over
    {((p+q)^2-m_{\pi}^2+i\eta)(p^2-m_{\pi}^2+i\eta)}}
    \nonumber\\
 & &-i 2g_{\rho\pi\pi}^2 \ g^{\mu\nu}\int {d^4p\over{(2\pi)^4}}\,{1\over
    {p^2-m_{\pi}^2+i\eta}} \ .
\eea
The loop integrals have to be regularized. A symmetry conserving 
procedure is, \eg, provided by the Pauli-Villars scheme which applies  
subtractions to the divergent integrals according to
\beq
\Sigma^{\mu\nu}(q;m_\pi) \to \Sigma^{\mu\nu}(q;m_\pi) +\sum\limits_{i=1}^{2}
c_i \Sigma^{\mu\nu}(q;M_i) \ . 
\eeq
The required regulator masses $M_i$ can be related to a single
form factor cutoff $\Lambda_\rho$ with~\cite{UBRW}
\bea
c_1=-2 \ , \quad M_1&=&\sqrt{m_\pi^2+\Lambda_\rho^2} 
\nonumber\\
c_2= 1 \ \  , \quad M_2&=&\sqrt{m_\pi^2+2\Lambda_\rho^2} \ .  
\eea
Effectively, the same can also be achieved by writing the two-pion 
loop selfenergy in terms of a once-subtracted dispersion integral~\cite{ChSc}
as
\bea
\Sigma_{\rho\pi\pi}^\circ(M) & = & \bar{\Sigma}_{\rho\pi\pi}^\circ(M)
-\bar{\Sigma}_{\rho\pi\pi}^\circ(0) \ 
 \nonumber\\
\bar{\Sigma}_{\rho\pi\pi}^\circ(M) & = & \int \frac{p^2 dp}{(2\pi)^2} 
v_{\rho\pi\pi}(p)^2 \ G_{\pi\pi}^0(M,p)  \  
\label{sigrho0}
\eea
with the vacuum two-pion propagator
\begin{equation}
G_{\pi\pi}^\circ(M,p)=\frac{1}{\omega_\pi(p)} \
\frac{1}{M^2-(2\omega_\pi(p))^2+i\eta}
 \ , \quad \omega_\pi(p)=\sqrt{m_\pi^2+p^2} \
\label{Gpipi}
\end{equation}
and vertex functions
\begin{equation}
v_{\rho\pi\pi}(p) 
= \sqrt{\frac{2}{3}} \ g_{\rho\pi\pi} \ 2p \ F_{\rho\pi\pi}(p) \ 
\label{vrhopipi}
\end{equation}
involving a hadronic form factor $F_{\rho\pi\pi}$~\cite{RCW}. From gauge 
invariance it follows that $q_\mu \Sigma_{\rho\pi\pi}^{\circ,\mu\nu}=0$ 
and the $\rho$ meson selfenergy can be cast in the general form 
\beq
\Sigma^{\circ,\mu\nu}_{\rho\pi\pi}(q)=\biggl(-g^{\mu\nu}+
\frac{q^\mu q^\nu}{M^2} \biggr) \Sigma^\circ_{\rho\pi\pi}(M) \ . 
 \eeq
Iterating it to all orders by solving the Dyson equation for the  
propagator, one arrives at
\beq
D_\rho^{\circ,\mu\nu}(q)=\biggl(-g^{\mu\nu}+\frac{q^\mu q^\nu}{M^2}
\biggr) D_\rho^\circ(M) 
\eeq
with the scalar part
\beq
D^\circ_\rho(M)
=\left[M^2-(m_\rho^{(0)})^2-\Sigma^\circ_{\rho\pi\pi}(M)\right]^{-1} \ . 
\eeq
The three free parameters (coupling constant, bare mass and cutoff)
can be readily adjusted to the $P$-wave $\pi\pi$ phase shifts, 
\beq
\tan(\delta_{\pi\pi}^{JI=11}(M))
=\frac{{\rm Im} D_\rho^\circ(M)}{{\rm Re} D_\rho^\circ(M)} \ , 
\eeq  
and the pion electromagnetic form factor, which, imposing VDM, becomes 
\beq
|F_\pi^\circ(M)|^2=(m_\rho^{(0)})^4  \ |D_\rho^\circ(M)|^2 \ ,  
\eeq
cf.~Fig.~\ref{fig_fpidpp}.
\begin{figure}[!htb]
\begin{center}
\epsfig{figure=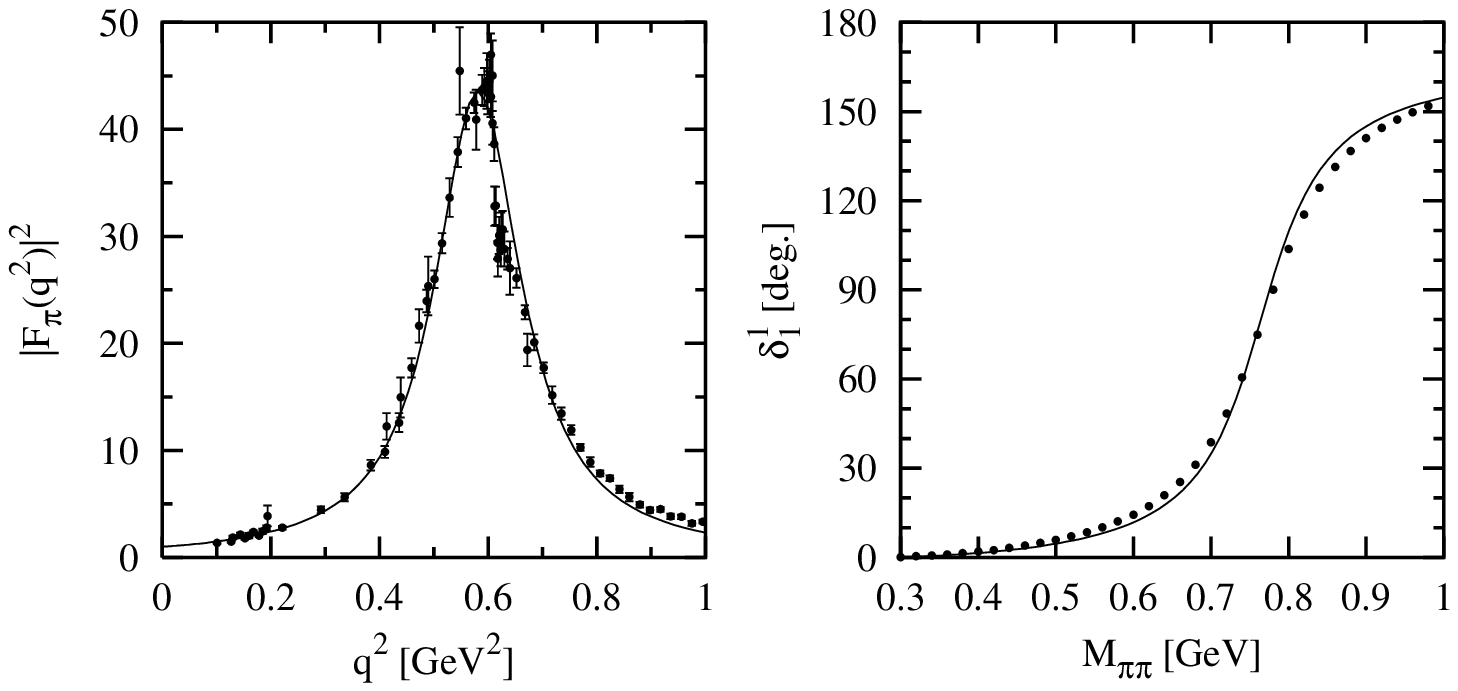,height=6.5cm}
\vspace{-0.5cm}
\end{center}
\caption{Pion electromagnetic form factor (left panel) and $P$-wave
$\pi\pi$ scattering phase shifts in free space as obtained in typical
fits from phenomenological models for the $\rho$ meson propagator.}
\label{fig_fpidpp}
\end{figure}
Typical values are  $g_{\rho\pi\pi}^2/4\pi=(2.7-2.9)$, 
$m_\rho^{(0)} =(0.82-0.85)$~GeV and $\Lambda_\rho=(1-3)$~GeV, 
depending on the regularization procedure.

Based on this standard description of the free $\rho$ meson one can 
distinguish two types of in-medium effects:
(i) modifications of the pion cloud, leading to a temperature-dependent
$\rho\pi\pi$ selfenergy $\Sigma_{\rho\pi\pi}$, and
(ii) scattering of the (bare) $\rho$ meson on surrounding matter
particles. Concerning (i) one needs to evaluate the 
in-medium properties of the pions. In a thermal pion gas they 
are only mildly affected chiefly because of their Goldstone nature. 
The corresponding pion 'optical' potentials amount to less than 
10\% corrections to the free pion dispersion relation even at 
temperatures as large as $T=200$~MeV~\cite{Sh91,RW9395}, 
cf.~Fig.~\ref{fig_UpionT}. 
\begin{figure}[t]
\begin{center}
\vspace{-2.5cm}
\epsfig{figure=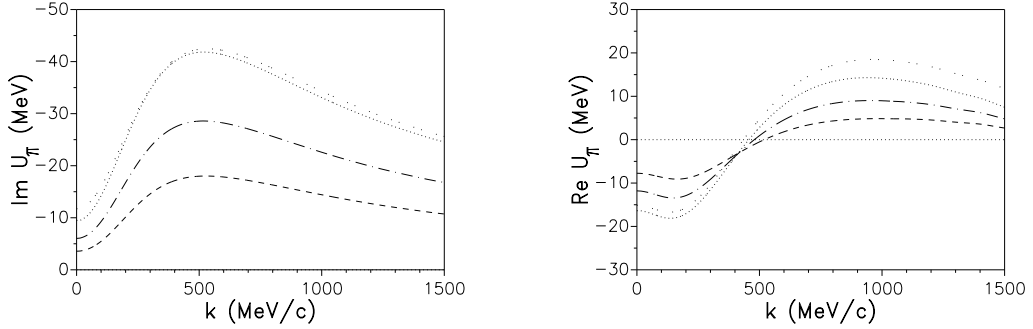,height=14cm,angle=90}
\vspace{-2.5cm}
\end{center}
\caption{Pion 'optical' potentials (left panel: imaginary part, right
panel: real part) in a hot pion gas as arising
from a selfconsistent Brueckner calculation~\protect\cite{RW9395} 
of the pion selfenergy $\Sigma_\pi$ and a chirally symmetric 
$\pi\pi$ interaction ('J\"ulich model'~\protect\cite{RPhD,RDW96})
within the Matsubara formalism. The on-shell potentials shown 
are defined as $U_\pi(k;T)=\Sigma_\pi(e_\pi(k),k;T)/2\omega_\pi(k)$,
where the quasiparticle energy $e_\pi(k)$ is determined by the solution of 
$e_\pi(k)^2=\omega_\pi(k)^2+\Sigma_\pi(e_\pi(k),k;T)$.
The  dashed, dashed-dotted and dotted lines correspond to temperatures 
$T=$150, 175 and 200~MeV, respectively (the light-dotted line is obtained
in first order of the selfconsistency iteration at $T$=200~MeV).}  
\label{fig_UpionT}
\end{figure}
In what follows, we shall therefore neglect the effects of a modified 
pion dispersion relation on the pion
cloud of the $\rho$ meson. A more important modification of 
$\Sigma_{\rho\pi\pi}$ 
stems from the Bose-Einstein enhancement factors of the (on-shell) 
intermediate two-pion states, representing an enhancement of the in-medium
$\rho\to\pi\pi$ width by 'stimulated emission'. In the Matsubara formalism 
the (retarded) two-pion propagator (\ref{Gpipi}) takes the form
\beq
G_{\pi\pi}(M,p;T)=\frac{1}{\omega_\pi(p)} \
\frac{\left[1+2f^\pi(\omega_\pi(p);T)\right]}{(M+i\eta)^2
-(2 \omega_\pi(p))^2} \ , 
\label{GpipiT}
\eeq 
which has been first quoted in Ref.~\cite{HK87b}.
Strictly speaking, the temperature factors are only exact for the 
imaginary part and the real part should be calculated from a dispersion
integral. It has been shown~\cite{RPhD}, however, that the latter 
is  well approximated by (\ref{GpipiT}) as long as the in-medium pion
dispersion relation is close to the free one. Furthermore, in 
Eq.~(\ref{GpipiT}) we have restricted ourselves to vanishing total 
three-momentum of the pion pair, the so-called 'back-to-back kinematics'. 
It has been verified in the model of Urban~\etal~\cite{UBRW} that the 
inclusion of finite three-momentum gives virtually identical results.  

Next we turn to the contributions from direct $\rho$ scattering off 
thermal mesons. In the many-body treatment of Refs.~\cite{RCW,RG99}  
it has been assumed that the interactions in each spin-isospin channel 
are saturated  by $s$-channel resonance formation 
('leading resonance approximation'). A clue as to which meson resonances 
might be of importance is provided by their branching ratios into
$\rho P$ states. This decay mode, however, is kinematically suppressed for  
resonance masses significantly below the naive kinematical 
threshold $m_\rho+m_P$. 
On the other hand, it should be emphasized that it is
just these subthreshold states which potentially generate substantial
strength for the in-medium $\rho$ meson spectral function at low invariant
masses (\eg, in $\rho\pi\to\omega(782)$ where, given a typical thermal
pion energy of 300--400~MeV, the appropriate $\rho$ meson mass would 
be $M\simeq 400-500$~MeV). Using VDM the subthreshold states can be 
largely identified through their radiative decays $R\to P\gamma$.      
The various resonances $R$ in $\rho P$ collisions can be grouped into
three major categories, namely vectors $V$,  
axialvectors $A$ and pseudoscalars $P'$. Following 
Ref.~\cite{RG99} we restrict our discussion to states 
with masses $m_R\le1.3$~GeV; higher ones are only relevant for invariant
$\rho$ masses beyond $M\simeq 1$~GeV. 

For $\rho P A$ vertices a suitable interaction 
Lagrangian, compatible with chiral symmetry and electromagnetic current
conservation, is given by
\begin{equation}
{\cal L}_{\rho P A}=G_{\rho PA} \ A_\mu \ (g^{\mu\nu} \ q_\alpha
p^\alpha - q^\mu p^\nu) \ \rho_\nu \ P \ ,
\label{LrhoPA}
\end{equation}
although other choices are possible~\cite{GG98}. The $\rho P$ scattering via
intermediate vector mesons $V$ is determined by  Wess-Zumino
anomaly terms which are of unnatural parity and involve
the four-dimensional antisymmetric Levi-Civita tensor
$\epsilon^{\mu\nu\sigma\tau}$:
\begin{equation}
{\cal L}_{\rho PV}=G_{\rho PV} \ \epsilon_{\mu\nu\sigma\tau} \ k^\mu
\ V^\nu \ q^\sigma \rho^\tau \ P \ .
\label{LrhoPV}
\end{equation}
In both Lagrangians (\ref{LrhoPA}) and~(\ref{LrhoPV}),  $p^\mu, 
q^\mu$ and $k^\mu$ denote the four-momenta of the pseudoscalar, $\rho$ and
(axial-) vector mesons, respectively. As a third possibility
$\rho P$ scattering can proceed via a pseudoscalar resonance. Here
an obvious candidate is $\rho\pi\to\pi'(1300)$ which
can be described by
\begin{equation}
{\cal L}_{\rho P P'}=G_{\rho PP'} \ P' \ ( k \cdot q \ p_\mu -
p \cdot q \ k_\mu) \ \rho^\mu \ P \ .
\label{LrhoPP}
\end{equation}
With increasing temperature the heat bath will consist of more and
more heavier resonances, especially those with high spin-isospin
degeneracy. After the light pseudoscalars the meson multiplet with the 
smallest masses are the vectors, most notably the $\rho$ meson with 
nine-fold degeneracy.   
Motivated by the observation that the $f_1(1285)$ resonance exhibits a 
large $\rho\gamma$ decay width (together with a predominant 4$\pi$ 
decay), it has been interpreted as a 'resonance' in $\rho\rho$ 
scattering~\cite{RG99}. 
The interaction vertex is also related to anomaly
terms~\cite{KM90} and has been chosen in the following form,
\begin{equation}
{\cal L}_{\rho VA}=G_{\rho VA} \ \epsilon_{\mu\nu\sigma\tau} \
\ p^\mu V^\nu \ \rho^{\sigma\alpha} \ k_\alpha A^\tau
- \frac{\lambda}{2} \ (k_\beta A^\beta)^2 \ ,
\label{LrhoVA}
\end{equation}
which again satisfies the appropriate conservation laws.
Here, the kinetic-energy term of the axialvector field has been
explicitly written to indicate a gauge freedom  
associated with the constant $\lambda$~\cite{ItZu}. For practical 
purposes -- following Ref.~\cite{RG99} -- $\lambda$ has been set to 1.  

The free parameters involved, which are the coupling constants
and the cutoffs for the hadronic vertex form factors, can be rather 
accurately determined from a simultaneous fit to both the
hadronic $R\to \rho P$ and radiative $R\to \rho\gamma$ branching
ratios as will be detailed in Sect.~\ref{sec_hadscat}. 

Within the imaginary time (Matsubara) formalism the $\rho$ meson
selfenergy tensor arising from binary collisions can now be calculated as
\begin{equation}
\Sigma_{\rho h}^{\mu\nu}(q)=\int \frac{d^3p}{(2\pi)^3}
\frac{1}{2\omega_h(p)} \left[f^h(\omega_h(p))-
f^{R}(k_0)\right] \ {\cal M}_{\rho h}^{\mu\nu}(p,q)
\label{sgrhoTmunu}
\end{equation}
with the thermal Bose-Einstein distribution function 
$f^h(\omega_h(p))=[\exp(\omega_h(p))/T-1]^{-1}$  of the corresponding
hadron species $h$ with on-shell energy $\omega_h(p)=\sqrt{m_h^2+\vec p^2}$.
${\cal M}_{\rho h}^{\mu\nu}$ denotes the isospin-averaged
forward scattering amplitude which, in the leading resonance 
approximation employed here, can be written as 
\beq
{\cal M}_{\rho hR}^{\mu\nu}(p,q)  = 
IF \ G_{\rho hR}^2 \ F_{\rho hR}(q_{cm})^2 \ D_R(s) \ v_R^{\mu\nu}(p,q) \ . 
\label{Mrhohmunu}
\eeq
The explicit expressions for the vertex functions $v_R^{\mu\nu}(p,q)$ can 
be derived from the above interaction Lagrangians as has been done in
Ref.~\cite{RG99}.  $IF$ denotes an isospin factor 
and the hadronic vertex form factor has been chosen of dipole 
type
\beq
F_{\rho hR}(q_{cm})=\left(\frac{2\Lambda_{\rho h R}^2+m_R^2}
{2\Lambda_{\rho h R}^2+\left[\omega_\rho(q_{cm})+\omega_P(q_{cm})\right]^2}
\right)^2 \ ,
\label{ffrhohR}
\eeq
normalized to 1 at the resonance mass $m_R$.
The scalar part of the intermediate resonance propagators is given by
\beq
D_R(s)=\frac{1}
{s-m_{R}^2+im_{R}\Gamma_{R}^{\rm tot}(s)}
\eeq
with $s=k^2=(p+q)^2$ and the total resonance width 
$\Gamma_R^{\rm tot}(s)$. Using the standard projection operators of
Eq.~(\ref{PLT}) the selfenergy tensors (\ref{sgrhoTmunu}}) can be
conveniently decomposed into longitudinal and transverse components, 
\beq
\Sigma_\rho^{\mu\nu}(q)= \Sigma_\rho^L(q_0,\vec q) \
P_L^{\mu\nu} + \Sigma_\rho^T(q_0,\vec q) \ P_T^{\mu\nu} \ ,  
\label{sgrhoLT}
\eeq 
which build up two independent modes of the in-medium $\rho$ propagator 
according to 
\begin{equation}
D_\rho^{\mu\nu}(q) = \frac{P_L^{\mu\nu}}{M^2-(m_\rho^{(0)})^2
-\Sigma_\rho^L(q_0,\vec q)}+\frac{P_T^{\mu\nu}}{M^2-(m_\rho^{(0)})^2
-\Sigma_\rho^T(q_0,\vec q)} + \frac{q^\mu q^\nu}{(m_\rho^{(0)})^2 M^2} \ . 
\label{drhomunu}
\end{equation}
Fig.~\ref{fig_sgrhoT} shows the individual contributions to the spin-averaged
selfenergy, 
\beq
\Sigma_{\rho hR}(M,\vec q)=\frac{1}{3}
\left[ \Sigma_{\rho hR}^L(M,\vec q)+2\Sigma_{\rho hR}^T(M,\vec q) \right] \ ,
\eeq
at fixed three-momentum modulus $|\vec q|=0.3$~GeV and temperature
$T=150$~MeV.  
\begin{figure}[htb]
\begin{center}
\epsfig{figure=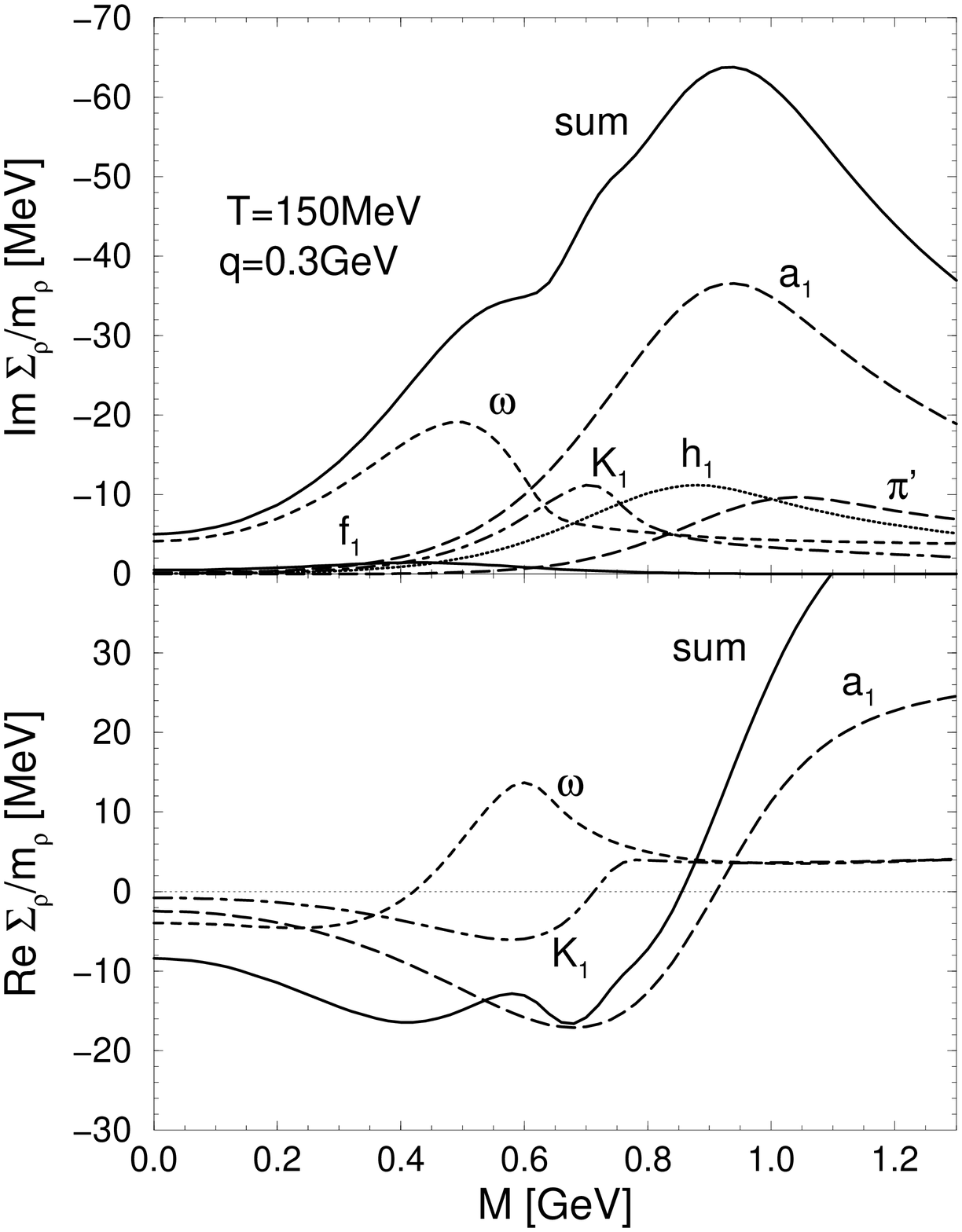,height=7.5cm}
\hspace{1cm} 
\epsfig{figure=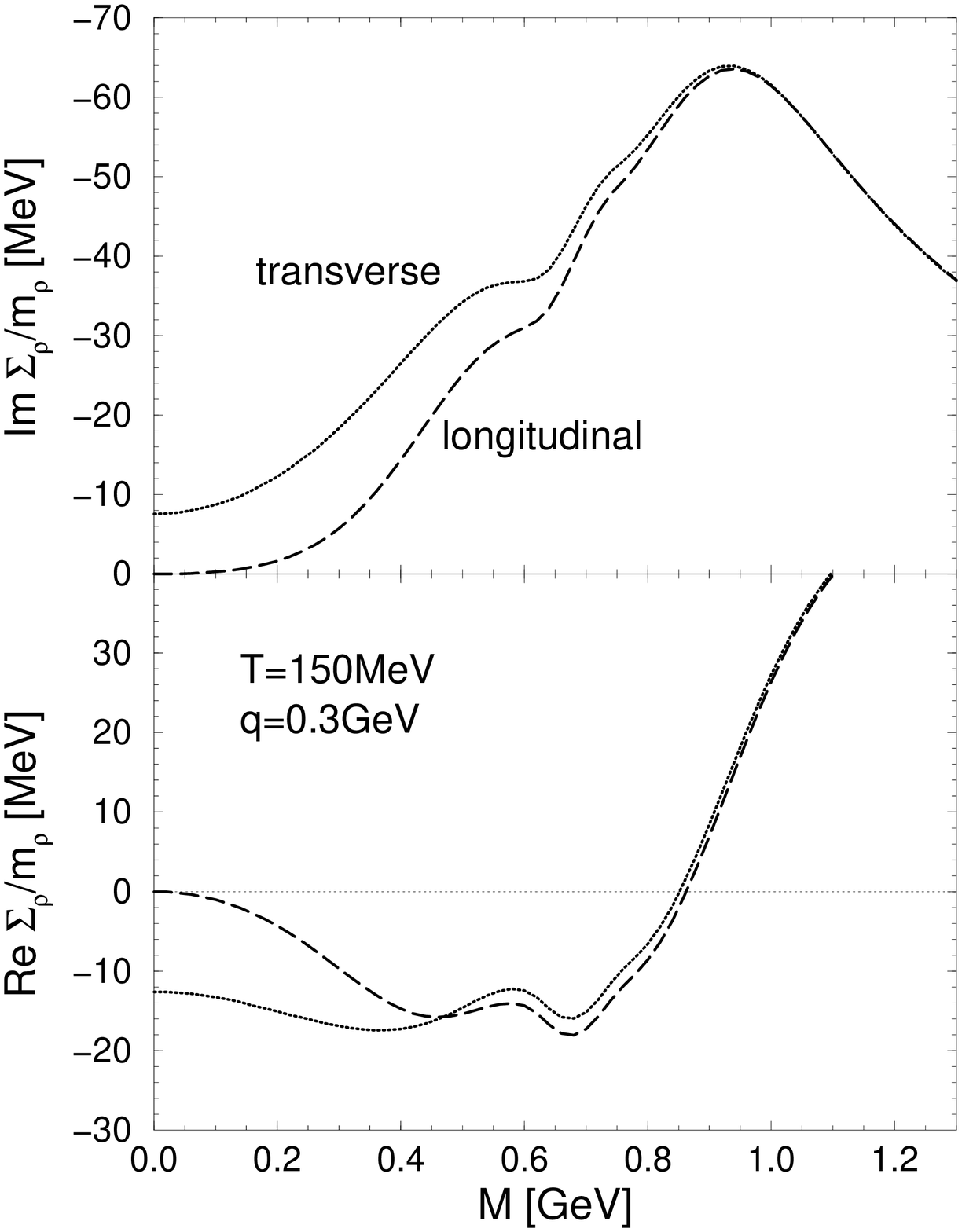,height=7.5cm}
\end{center}
\caption{The real and imaginary parts (lower and upper panels,
respectively) of the polarization-averaged
$\rho$ selfenergy (left panel) from
resonant scattering off thermal $\pi, K, \bar K$ and $\rho$ mesons
in a heat bath at $T=150$~MeV and $\mu_\pi=0$; the  
different channels are labeled by their intermediate resonances;
right panel: polarization decomposition of the 
 combined selfenergy contributions.}
\label{fig_sgrhoT}
\end{figure}
Around and above the free $\rho$ meson mass the strongest 
absorption is caused by
$a_1(1260)$ resonance formation, which is about as large as the sum of
all other channels, shared to roughly equal parts among 
$K_1(1270)$, $h_1(1170)$ and $\pi'(1300)$. The $K_1(1270)$ contribution
acquires its maximum at lower $M$ than the pion-resonances due to
the higher thermal energies of the kaons (including their rest mass).
In the low-mass region, $M\le 0.6$~GeV, the
prevailing contribution is due to the $\omega$ meson which, however,
leaves little trace in the resonance region.
It is also seen that the effect of the $f_1$ meson is very small.
In the real part of the
total selfenergy one observes appreciable cancellations  until eventually all
contributions turn repulsive (the latter feature is likely to be
modified when accounting for further higher resonances).
Such cancellations are typical for this kind of 
many-body calculations. They are the reason that one
usually encounters only moderate modifications
of the in-medium pole masses. On the other hand, the imaginary parts of
$\Sigma_\rho$ strictly add up, generating significant broadening.
Not shown here is the in-medium Bose enhancement of the $\rho\to\pi\pi$
selfenergy. It has a very smooth behavior with a broad maximum of
${\rm Im}(\Sigma_{\rho\pi\pi}(M;T)-\Sigma_{\rho\pi\pi}^\circ(M))/m_\rho\simeq 25$~MeV 
at about $M\simeq0.6$~GeV. We also note that the three-momentum
dependence of the selfenergies is rather weak, being most pronounced
at low $M$ where the transverse part is responsible for the build-up
of finite values. This can be seen more explicitly from the right 
panel of Fig.~\ref{fig_sgrhoT} where the summed selfenergy
contributions have been separated into the two polarization states. 
 
In Fig.~\ref{fig_spectral} the full spin-averaged imaginary part of the 
$\rho$ meson propagator,
\begin{equation}
{\rm Im} D_\rho(M,\vec q;T) =\frac{1}{3} \left[{\rm Im} D_\rho^L(M,\vec q;T)+
2{\rm Im} D_\rho^T(M,\vec q;T) \right] \ , 
\label{imdrho}
\end{equation}
in a thermal meson gas of temperatures $T=120$, 150 and 180~MeV (left panel)
as appropriate for the hadronic phase in ultrarelativistic heavy-ion
collisions is shown. 
\begin{figure}[!htb]
\begin{center}
\epsfig{figure=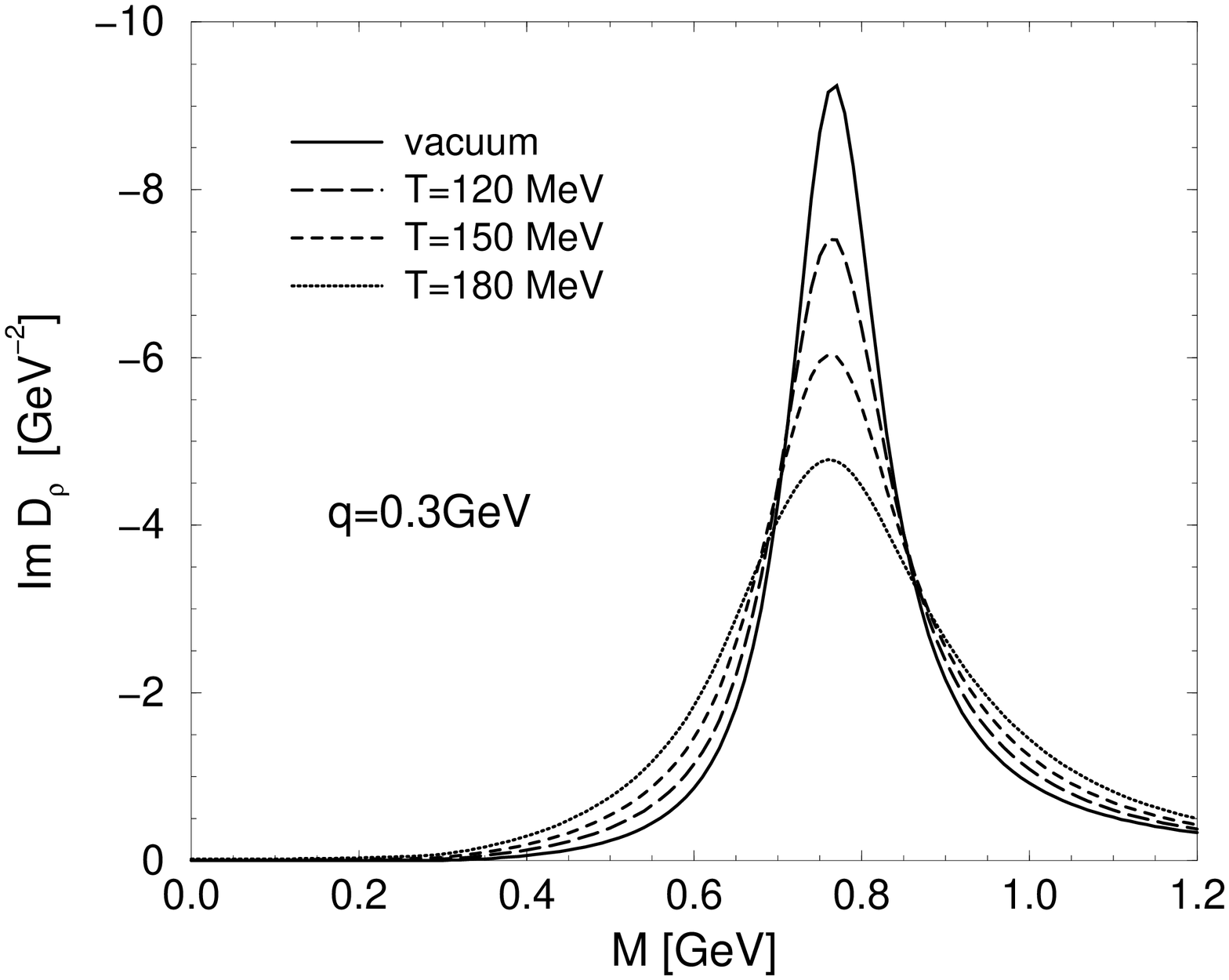,height=7cm,angle=-90}
\epsfig{figure=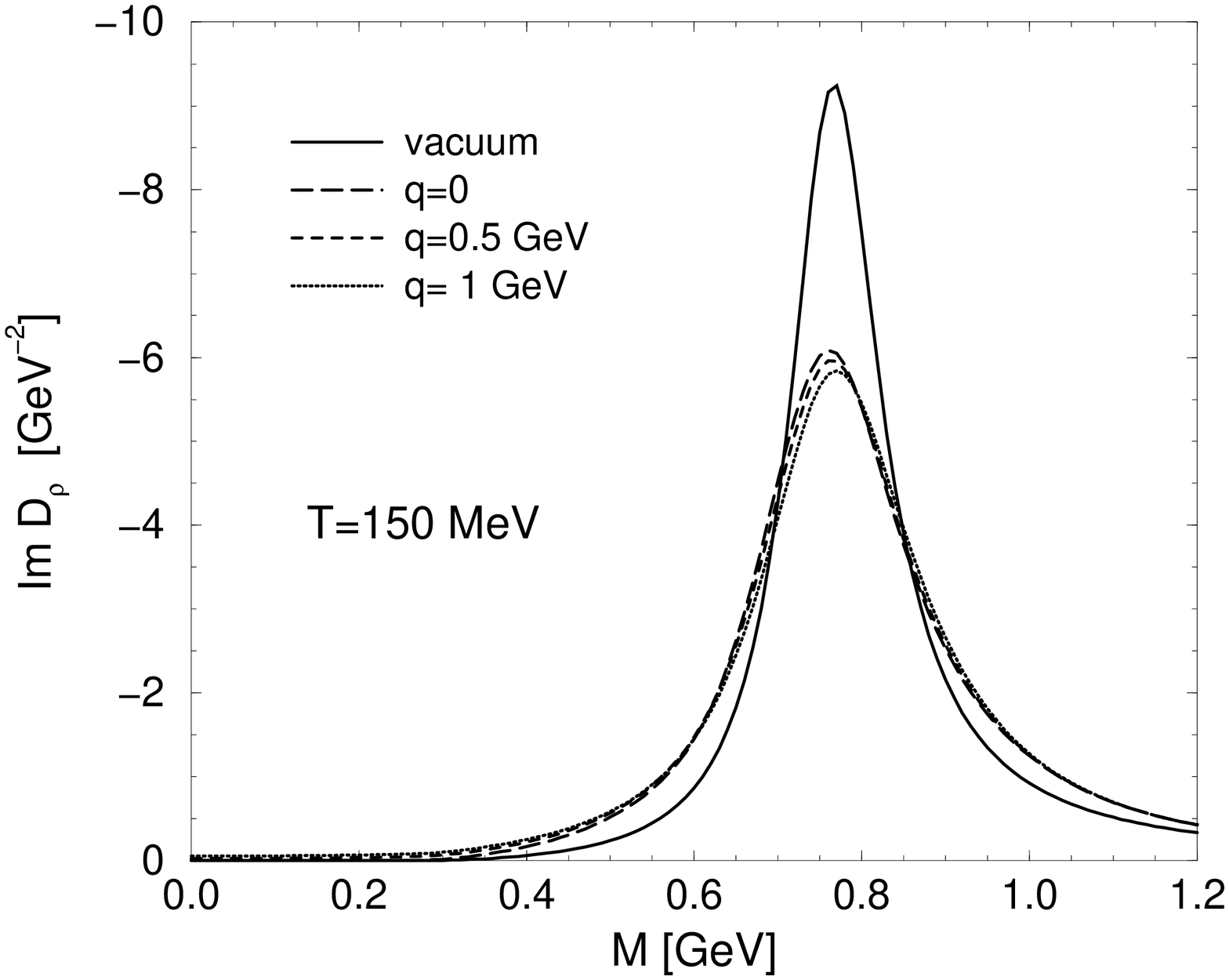,height=7cm,angle=-90}
\end{center} 
\caption{Imaginary part of the $\rho$ propagator (which, up to a factor
of (-2), coincides with the spectral function) in the vacuum (full curve) 
and in a thermal $\pi K \bar K \rho$ gas as calculated in the many-body 
framework of Ref.~\protect\cite{RCW,RG99}; left panel: for fixed
three-momentum $q=0.3$~GeV  at temperatures $T=120$~MeV (long-dashed curve),
$T=150$~MeV (dashed curve) and $T=180$~MeV (dotted curve); 
right panel: for fixed temperature $T=150$MeV at three-momenta
$q=0$ (long-dashed curve), $q=0.5$~GeV (dashed curve) and $q=1$~GeV 
(dotted curve). }
\label{fig_spectral}
\end{figure}
More explicitly, one has
\begin{equation}
{\rm Im}D_\rho^{L,T}(M,\vec q)=
\frac{{\rm Im} \Sigma_\rho^{L,T}(M,\vec q)}{|M^2-(m_\rho^{(0)})^2
-\Sigma_\rho^{L,T}(M,\vec q)|^2}  \
\label{imdrhoLT}
\end{equation}
with the longitudinal and transverse selfenergy parts
\beq
\Sigma_\rho^{L,T}  = \Sigma_{\rho\pi\pi} + \sum\limits_{\alpha}
\Sigma_{\rho\alpha}^{L,T} \ , 
\label{selfep}
\eeq
where the summation is over the mesonic excitation channels
$\alpha$=$\pi\omega$, $\pi h_1$, $\pi a_1$, $\pi\pi'$, $KK_1$,
$\bar K \bar K_1$, $\rho f_1$, as discussed,
and $\Sigma_{\rho\pi\pi}$ now contains the
Bose-Einstein factors through Eq.~(\ref{GpipiT}).
The thermal $\rho$ meson spectral function exhibits an additional 
broadening (defined as the full width at half maximum) of about
80~MeV at $T=150$~MeV which almost doubles to $\sim 155$~MeV at
$T=180$~MeV. On the other hand, the three-momentum dependence of the 
spectral function at fixed temperature is rather weak, cf.~right panel
of Fig.~\ref{fig_spectral}.

The results of the many-body approach are quite close to those obtained 
recently using kinetic theory~\cite{Gao99} (see Fig.~\ref{fig_gaorho}).  
The latter, however, attribute some portion of the collisional broadening
to $t$-channel meson exchanges. Those have not been included 
in the many-body calculations, where the net medium effect is 
entirely driven by large imaginary parts from resonant $s$-channel 
interactions (including subthreshold states). At tree level, the 
neglected $t$-channel 
exchanges do not generate an imaginary part. They would do so  once  
iterated in a Lippmann-Schwinger-type equation to construct a scattering
amplitude beyond tree level. In this case, however, care has to be taken 
in avoiding double counting when performing a combined treatment
of $s$- and $t$-channel graphs (in fact, the 'leading resonance
approximation' implies that non-resonant interactions are, at least
partially, subsumed in the resonance parameters).   
For the $\rho$ meson selfenergy in a hot pion gas
an  unambiguous way to disentangle $s$- and $t$-channel interactions
would require experimental  information on $\pi\rho$
scattering phase shifts which, owing to the short lifetime
of the $\rho$ meson, can only be inferred  by very indirect means.

\section{Finite Baryon Density}
\label{sec_Vmoddens}
The investigation of vector meson modifications in an environment of finite
nucleon density via effective hadronic models has mainly been
pursued within more phenomenologically oriented approaches. This
is partly due to the fact that the impact of chiral symmetry on 
$\rho$-nucleon interactions is much less obvious than in the purely
mesonic case (such as $\pi\rho a_1$ dynamics). 
Also, the extension of VDM to the baryonic sector is less accurate 
such that electromagnetic observables provide less direct access 
to vector meson interactions involving baryons. 
Pionic interactions with nucleons and nuclei, on the other hand, 
are much better known:  a wealth of pion-nucleus scattering data has 
provided a detailed understanding of the underlying physical mechanisms
for their modification in cold nuclear matter~\cite{ErWe} which are 
substantial. Since especially $\rho$
mesons (and, to a lesser extent, also $\omega$ mesons) exhibit a 
strong coupling to two- (three-) pion states, early analyses have focused 
on medium effects in the virtual pion cloud. These developments will be 
reviewed in the second part of this Section. Subsequently it has 
been realized that also
direct $\rho$-$N$ interactions can induce substantial modifications
of the $\rho$ meson spectral function which we will discuss in the 
third part of this Section.  
The first part, however, will be devoted to recall the mean-field based 
analysis of effective chiral Lagrangians by Brown and Rho~\cite{BR91} which 
culminated in the famous conjecture 
of 'Brown-Rho Scaling'. It is not exaggerated to say that, 
although (or just because) this conjecture is controversial, 
the associated hypothesis for a 'dropping' of vector meson masses  
has been one of the main triggers for an ensuing intense theoretical
(and experimental) activity.

\subsection{Mean-Field Approach: Brown-Rho Scaling} 
\label{sec_drop}
As we have already eluded to in Sect.~\ref{sec_sym-ano}, in the massless  
limit the QCD action is scale invariant on the classical level
implying that the QCD Lagrangian has scale dimension 4. 
Chiral meson Lagrangians, being constructed as effective low-energy 
theories of QCD, should in principle exhibit the same property. In the 
non-linear realization they are formulated in terms of the chiral field 
$U\equiv {\rm e}^{i\pi/f_\pi}$ ($\pi = \vec \pi \cdot \vec\tau$) which, 
when including up to fourth-order derivatives (involving the quartic 
'Skyrme term'), read
\beq
{\cal L}= \frac{f_\pi^2}{4} \ {\rm tr} (\partial_\mu U \partial^\mu U)
+ \frac{\eta^2}{4} \ 
\tr \left[U^\dagger\partial_\mu U, U^\dagger \partial_\nu U\right]^2
+ c \ \tr ({\cal M}^\circ U +{\rm h.c.}) \ .   
\label{Leff} 
\eeq
Here, also the explicit symmetry breaking term being proportional to the 
current quark mass $m_q$ in the quark mass matrix ${\cal M}^\circ$ 
has been incorporated.  
As has been argued in Ref.~\cite{Ellis} the $U$ field carries 
scale dimension zero which means that the three terms on the 
{\it r.h.s.} have scale dimensions 2,4 and 0, respectively.  

Following Ref.~\cite{BR91} the first 
step in the derivation of Brown-Rho scaling  
consists of modifying the effective Lagrangian (\ref{Leff}) to reflect 
the appropriate scaling behavior of the QCD Lagrangian.  
The simplest way to restore it has been proposed 
by Ellis \etal~\cite{Ellis} and is realized by introducing an effective  
'glueball' field $\chi$ with scale dimension 1 according to 
\bea
{\cal L_\chi} &=& \frac{f_\pi^2}{4} \ \left(\frac{\chi}{\chi_0}\right)^2
 \ {\rm tr} (\partial_\mu U \partial^\mu U)
+ \frac{\eta^2}{4} \ 
\tr \left[U^\dagger\partial_\mu U, U^\dagger \partial_\nu U\right]^2
+ \frac{1}{2} \ \partial_\mu\chi \partial^\mu\chi 
\nonumber\\ 
 & & \qquad + c \ \left(\frac{\chi}{\chi_0}\right)^3 \ 
\tr ({\cal M}^\circ U +{\rm h.c.})  + V(\chi) \ .
\label{lchi}
\eea
Besides a scale invariant (\ie, dimension-4) kinetic energy term 
for the glueball field a potential-energy term of the form
\beq
V(\chi)=B \left[ \frac{1}{4} \chi_0^4 
+\chi^4 \ln (\chi/{\rm e}^4 \chi_0)\right] \  
\eeq
has also been added. 
Minimizing $V$ in $\chi$ yields a nonzero ground state expectation value 
$\chi_0\equiv\langle 0|\chi|0\rangle$ signaling the spontaneous breakdown
of scale invariance which mimics the (quantum part of the) QCD scale anomaly 
on the effective Lagrangian level. The divergence of the corresponding 
dilation current (\ref{Jdil}) becomes~\cite{Ellis} 
\beq
\partial_\mu j_D^\mu = -B \chi^4 \ . 
\eeq
Thus, comparing to the trace anomaly of QCD (\ref{trano}) $\chi_0$
can be related to the gluon condensate $\ave{G^2}$. In addition, the QCD 
trace anomaly receives a contribution from explicit scale breaking 
through the quark mass term $(\bar\psi {\cal M}^\circ\psi)$ which 
has scale dimension 3 requiring the $\chi^3$ factor in the corresponding 
term in Eq.~(\ref{lchi}). 
  
As has been emphasized by Brown and Rho in subsequent work~\cite{BR96},
the introduced $\chi$-field is to be understood as consisting 
of a 'soft' (mean-field) and a 'hard' (fluctuation) component 
according to
\beq
\chi=\chi^* +\chi' \ . 
\eeq
It is the soft mean-field component $\chi^*$ that will govern the
medium modifications in the chiral effective Lagrangian, 
whereas the hard component $\chi'$ is to be associated with the glueball
mass scale of $>1$~GeV, well beyond the applicability range of 
low-energy effective theory. Indeed, as discussed in 
Sect.~\ref{sec_lattice}, lattice calculations show that in the chiral 
transition with light quarks only about half of the gluon condensate 
is 'melted', corresponding to the $\chi^*$ field (an analogous feature 
emerges within the instanton model~\cite{SS98}
where it has been identified  as a rearrangement of the 
chirally broken 'random' instanton liquid into a chirally restored 
phase with $I$-$A$ molecules, the latter characterizing the 'hard' 
component of the gluon fields that survive the transition).

In their second main step Brown and Rho postulate that, as the (quark
and gluon) condensates change in dense matter, the symmetries of the
Lagrangian remain intact such that the variation in the condensates can
be absorbed in a density-dependent change in masses and 
coupling constants of the effective theory. In line with the above
arguments, the change in the quark condensate at a given density can be 
expressed as
\beq
{\tave{\bar\psi\psi}\over\ave{\bar\psi\psi}}
=\left(\frac{\chi^*}{\tilde\chi_0}\right)^3 \ ,
\label{qqgl}  
\eeq
where $\tilde\chi_0$ denotes the vacuum expectation value of the 
$\chi^*$ field. 
This suggests to define an in-medium pion decay constant as
\beq
f_\pi^*=f_\pi \frac{\chi^*}{\tilde\chi_0}   
\eeq  
leading to the effective Lagrangian of the form
\bea
{\cal L}^* &=& \frac{{f_\pi^*}^2}{4} 
 {\rm tr} (\partial_\mu U \partial^\mu U)
+ \frac{\eta^2}{4} 
\tr \left[U^\dagger\partial_\mu U, U^\dagger \partial_\nu U\right]^2
+  c  \left(\frac{f_\pi^*}{f_\pi}\right)^3 
\tr ({\cal M}^\circ U +{\rm h.c.})  + \cdots \ , 
\label{lstar}
\eea
where the fields are now defined as ensemble averages
in cold nuclear matter, \ie, $\chi^*\equiv\tave{\chi}$ etc..
In particular, with 
\beq
U=\exp(i\pi^*/f_\pi^*) \ , 
\eeq
the pion field $\pi^*\equiv \pi \chi^*/\tilde\chi_0$ has picked up a scale 
dimension of one. Furthermore, it was argued that the fluctuating 
part of the glueball field strongly mixes with an effective low-lying
$\sigma$ meson (the quark-antiquark component of the scalar), and therefore 
the effective mass of the latter is inferred from
\beq
 f_\pi^*/f_\pi \approx \chi^*/\tilde\chi_0\approx m_\sigma^*/m_\sigma \ .  
\eeq
Moreover, when using the Goldberger-Treiman relation one has
\beq
m_N^*/m_N \approx (g_A^*/g_A)^{1/2} f_\pi^*/f_\pi \ . 
\eeq 
Since in the Skyrme model, the (scale invariant) coefficient $\eta^2$
of the quartic Skyrme term is directly related to the axialvector
coupling constant $g_A$, the latter
is not affected at the mean-field level. (The 'quenching' of $g_A$
from 1.26 to 1, observed in Gamow-Teller and magnetic transitions in nuclei,
has been argued to be due to loop effects, 
indicating an additional, lower scale induced in nuclei). 
Finally, making use of the KSFR relation~\cite{KSFR}, 
\beq
m_V^2=2g^2 f_\pi^2 \ ,  
\eeq
and the fact that within the Skyrme model the hidden-gauge
coupling $g^2=\frac{1}{8}\eta^2$ is scale invariant, 
the vector meson masses are conjectured to complete the (approximate) 
Brown-Rho (BR) scaling relation: 
\beq
\Phi(\varrho)\equiv\frac{f_\pi^*}{f_\pi}=\frac{m_\sigma^*}{m_\sigma}=
            \frac{m_N^*}{m_N} = \frac{m_\rho^*}{m_\rho} = 
            \frac{m_\omega^*}{m_\omega} \ .
\label{BRS}
\eeq
In fact, early QCD sum rules calculations of Hatsuda and 
Lee~\cite{HatsLee} have given support to this relation, as 
discussed in Sect.~\ref{sec_qcdsr}, with typical 
values for $\Phi(\varrho_0)=0.82\pm0.06$ at normal nuclear matter density.
  
There might be, however, some problems with the BR scaling hypothesis, mainly
in the finite temperature sector. On very general grounds
the chiral condensate is locally altered whenever a hadron is present as
discussed in Sect.~\ref{sec_medcond}. This leads to the dilute gas expressions 
(\ref{Dilute1}) and (\ref{Dilute2}) which are rigorously valid 
at low temperature and low density. Even though the quark condensate 
is decreased, by definition, nothing happens to the masses, the pion 
and the nucleon in this case. Secondly, when applied at low temperatures, 
the scaling relation (\ref{BRS}) is at variance
with the low-temperature expansion of the in-medium vector and axialvector
correlators (\ref{VAmixing}) and the chiral 
condensate ratio (\ref{qqbarT}). While the
latter is reduced already at order 
$O(T^2)$, the $T^2$-dependence of the correlators is governed by mixing
and corrections to the mass are of order $O(T^4)$ (in the chiral limit). 
This is implied by chiral 
symmetry~\cite{DEI90} and manifest in various effective models that we have
discussed in the previous Section.

\subsection{Pion Cloud Modifications} 
\label{sec_pioncloud} 
Let us first recall some basic features of the single-pion properties
in nuclear matter (for a review see, \eg, Ref.~\cite{MSTV}). 
The most prominent effect on (on-shell) pions propagating through 
the nuclear environment is generated through resonant $P$-wave
interactions exciting isobar-hole ($\Delta N^{-1}$) states. 
As is well-known from pion nuclear physics another important pionic  
excitation channel is represented by $P$-wave nucleon-hole 
($NN^{-1}$) states. On the other hand, $S$-wave $\pi N$ interactions  
are suppressed by about an order of magnitude in symmetric nuclear 
matter owing to an almost exact cancellation between the isospin 
1/2 and 3/2 partial waves~\cite{ErWe}. 
The standard Lagrangians for the $\pi NN$ and $\pi N\Delta$
$P$-wave interactions are given (in non-relativistic form) by
\bea
{\cal L}_{\pi NN}      &=& \frac{f_{\pi NN}}{m_\pi} \ \Psi_N^\dagger \  
\vec\sigma\cdot \vec k \ \vec\tau\cdot \vec\pi \ \Psi_N
\nonumber\\
{\cal L}_{\pi N\Delta} &=& \frac{f_{\pi N\Delta}}{m_\pi} \  
\Psi_\Delta^\dagger \ \vec S\cdot \vec k \ \vec T \cdot \vec\pi \ \Psi_N 
+ {\rm h.c.} \ ,  
\eea  
which leads to corresponding pion selfenergies in nuclear matter 
of the type 
\beq
\Sigma_{\pi\alpha}^{(0)}(k) 
= -\vec k^2 \ \chi_{\pi\alpha}^{(0)}(k) \  
\eeq
($k=(k_0,\vec k), \alpha=NN^{-1}, \Delta N^{-1}$). The so-called 
(pionic) susceptibilities are given by
\beq
\chi_{\pi\alpha}^{(0)}(k)  =  {\left
( \frac{f_{\pi\alpha} \  F_{\pi\alpha}(k)} {m_\pi}
\right )}^2  \ SI(\pi\alpha) \ \phi_\alpha(k) \ 
\label{chi0}
\eeq
in terms of spin-isospin factors  $SI(\pi\alpha)$, coupling constants
$f_{\pi\alpha}$ (cf.~Table~\ref{tab_pind})  and a hadronic vertex form 
factor usually chosen of monopole type,  
\begin{equation}
F_{\pi\alpha}(k)= \left( \frac{\Lambda_\pi^2-m_\pi^2}{\Lambda_\pi^2
+\vec k^2} \right) \ . 
\label{ffpinn}
\end{equation} 
\begin{table}
\bce
\begin{tabular}{c|cccc}
 $\pi\alpha$ & $\pi NN^{-1}$ & $\pi\Delta N^{-1}$ &
$\pi N\Delta^{-1}$ & $\pi\Delta\Delta^{-1}$ \\
\hline
$SI(\pi\alpha)$ & 4 & 16/9 & 16/9 & 400 \\
$f_{\pi\alpha}^2/4\pi$ & 0.08 & 0.32 & 0.32 & 0.0032 \\
\end{tabular}
\ece
\caption{\em Spin-isospin transition factors and coupling constants for
pion induced (longitudinal) $P$-wave particle-hole excitations in a hot
$N\Delta$ gas~\protect\cite{RW94}.}
\label{tab_pind}
\end{table}
In Eq.~(\ref{chi0}), $\phi_\alpha$ denote the Lindhard functions 
which, for the more general case of a $\alpha=ab^{-1}$ particle-hole 
excitation, read
\begin{equation}
\phi_\alpha(k)=-\int \frac{p^2 dp} {(2\pi)^2}
f^b[E_b(p)] \int\limits_{-1}^{+1}
 \ dx \ \sum_{\pm} \ \frac{1-f^a[E_a(\vec p +\vec k)]}
 {\pm k_0+E_b(p)-E_a(\vec p +\vec k) \pm \frac{i}{2}(\Gamma_a+\Gamma_b)}   \
\label{Lindhard1}
\end{equation}
including direct (sign '+') and exchange (sign '--') diagrams with
\begin{eqnarray}
E_a(p) & = & (m_a^2+p^2)^{1/2} \ , \nonumber\\
E_b(\vec p+\vec k) & = & (m_b^2+p^2+k^2+2pkx)^{1/2} \ 
\end{eqnarray} 
and Fermi-Dirac distribution functions $f^a$, $f^b$. $\Gamma_a$ and $\Gamma_b$
are the (energy- and density-dependent) total decay widths of 
particle $a$ and hole $b^{-1}$, respectively (notice that the holes 
carry the imaginary part with opposite sign).  
For a realistic description of the pion selfenergy short-range 
correlations between particle and hole have to be accounted for. These
are conveniently parameterized in terms of 'Migdal parameters'
${g'}_{\alpha\beta}$, which also induce a mixing between the channels.  
The resulting system of coupled equations, 
\begin{equation}
\chi_\alpha  =  \chi_\alpha^{(0)}-\sum_{\beta} \
\chi_\alpha^{(0)} \ {g'}_{\alpha\beta} \ \chi_\beta \ ,  
\label{chi}
\end{equation}
is solved by an elementary matrix inversion yielding  
\begin{equation}
\Sigma_{\pi}(k_0,\vec k)= -\vec k^2 \ \sum_\alpha 
\chi_{\alpha}(k_0,\vec k)  \ .
\label{Sigmapi}
\end{equation}
As an illustrative example we show in Fig.~\ref{fig_g2pic} the off-shell 
two-pion propagator (restricted to zero total momentum) in nuclear matter, 
\bea
G_{\pi\pi}(E,\vec k)&=&\int\frac{idk_0}{\pi} \ 
D_\pi(k_0,\vec k)
\ D_\pi(E-k_0,-\vec k)
\nonumber\\
D_\pi(k_0,\vec k)&=&\left[ k_0^2-\vec k^2-m_\pi^2-
\Sigma_{\pi}(k_0,\vec k) \right]^{-1} \ ,
\label{piprop}  
\eea
which directly enters into the two-pion selfenergy of the $\rho$ meson,  
cf.~Eq.~(\ref{sigrho0}). 
\begin{figure}[th]
\bce
\epsfig{figure=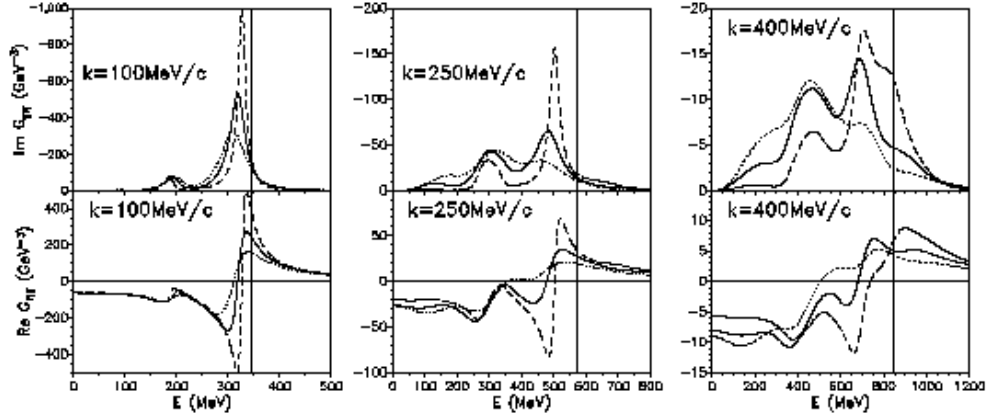,width=14.2cm}
\ece
\vspace{-0.5cm}
\caption{Two-pion propagator (upper panels: imaginary part, lower panels:
real part) in cold nuclear matter at densities
$\varrho_N/\varrho_0=$0.5, 1, and 1.5 represented by the long-dashed, full 
and dotted lines, respectively (the pion selfenergies have been evaluated 
using the $\pi\alpha$ form factor cutoff from the Bonn potential, 
$\Lambda_{\pi NN}=\Lambda_{\pi N\Delta}=1.2$~GeV, together with 
correspondingly large values for the Migdal parameters of 
$g'_{NN}=0.8, g'_{N\Delta}=0.5$). 
The vertical lines indicate the 
on-shell two-pion energy $E_{on}=2\omega_\pi(k)$ in free space.} 
\label{fig_g2pic}
\end{figure}
Clearly, the combination of $\Delta N^{-1}$ and $NN^{-1}$
excitations entails a rather rich structure with substantial shifts
of strength towards low energies (\ie, invariant $\rho$ masses). 

Besides
the modification of the intermediate pion propagators in the in-medium
$\rho$ meson selfenergy, the $\Delta N^{-1}$- and $NN^{-1}$-bubbles induce a 
number of corresponding vertex corrections for the  $\rho\pi\pi$ and
$\rho\rho\pi\pi$ couplings that have to be incorporated to ensure
transversality (\ie, gauge invariance) of the resulting vector propagator.
They can be systematically inferred from the
appropriate Ward-Takahashi identities~\cite{WaTa}, as will be
detailed below.
The first investigations  along these lines were performed by 
  Herrmann \etal~\cite{HeFN},
Chanfray and Schuck~\cite{ChSc} as well as Asakawa \etal~\cite{AKLQ}.   
Their calculations were restricted to the effects from the $\Delta N^{-1}$ 
excitation and  to vanishing total 3-momentum
$\vec q=0$ of the $\rho$ meson. Nevertheless, appreciable modifications 
of the spectral function were found, in particular an in-medium  
broadening as well as a rather pronounced peak structure at invariant
masses $M\simeq 3m_\pi$ stemming from transverse $\Delta N^{-1}$
excitations (corresponding to the vertex correction represented by the 
left diagram in the middle panel of Fig.~\ref{fig_vtxcorr}).   
In a next step, the additional effects from $P$-wave $NN^{-1}$
excitations were incorporated in Refs.~\cite{CRW,RCW}. Although
the $NN^{-1}$ channel predominantly populates the space-like 
momenta in the in-medium pion spectral function, it is nevertheless
of significance for the $\rho$ selfenergy, since the off-shell integration
over intermediate pion states does involve space-like pion kinematics, 
generating additional low-mass strength in the spectral function, 
as we have seen above. At the same time, additional broadening at the
$\rho$ meson resonance peak emerges.  

In subsequent work, Urban \etal~\cite{UBRW} could overcome the 
restriction to back-to-back kinematics of the previous analyses.
One starts from the Ward-Takahashi identities, which for the 
$\rho\pi\pi$ and $\rho\rho\pi\pi$ interaction vertices from
Eq.~(\ref{Lintrhopi}) read   
\bea
q^{\mu} \ \Gamma_{\mu ab}^{(3)}(k,q)
  &=& g_{\rho\pi\pi}  \ \epsilon_{3ab} \ 
\Big(D_{\pi}^{-1}(k+q)-D_{\pi}^{-1}(k)\Big)
  \label{WT3}
\\
q^{\mu} \ \Gamma_{\mu\nu ab}^{(4)}(k,k,q)
  &=&ig_{\rho\pi\pi} \ \Big(\epsilon_{3ca} \ \Gamma_{\nu bc}^{(3)}(k,-q)-
            \epsilon_{3bc} \ \Gamma_{\nu ca}^{(3)}(k+q,-q)\Big) \ ,  
\label{WT4}
\eea
respectively. The spin, isospin and momentum assignments are indicated in
Fig.~\ref{fig_vtxrhopi} (for simplicity, we only consider neutral $\rho$ 
mesons characterized by the third isospin component of the
isovector $\vec\rho$ field). 
\begin{figure}
\begin{center}
\epsfig{figure=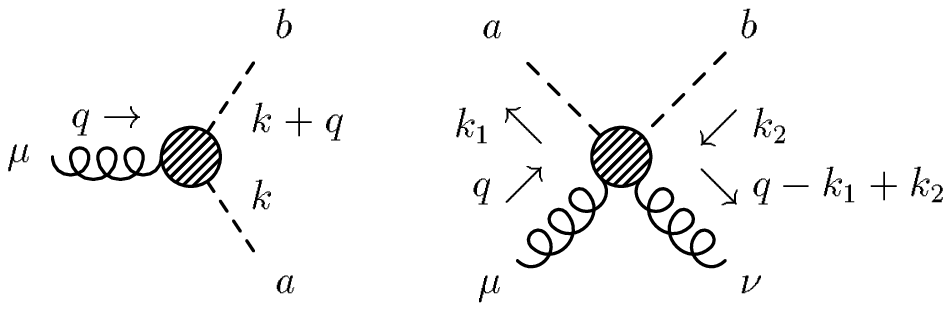,width=10cm}
\end{center}
\caption{$\rho\pi\pi$ (left) and $\rho\pi\pi$ (right) vertices involving
neutral $\rho$ mesons (curly lines); $\mu,\nu$ correspond to the 
$\rho$ meson polarizations, whereas $a,b$ are isospin indices of the pions  
(dashed lines); all other labels refer to the in-/outgoing four momenta.}
\label{fig_vtxrhopi}
\end{figure}
\begin{figure}[th]
\begin{center}
\epsfig{figure=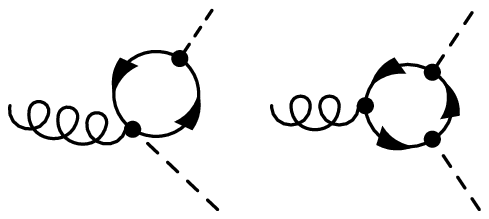,width=5cm}
\end{center}
\begin{center}
\epsfig{figure=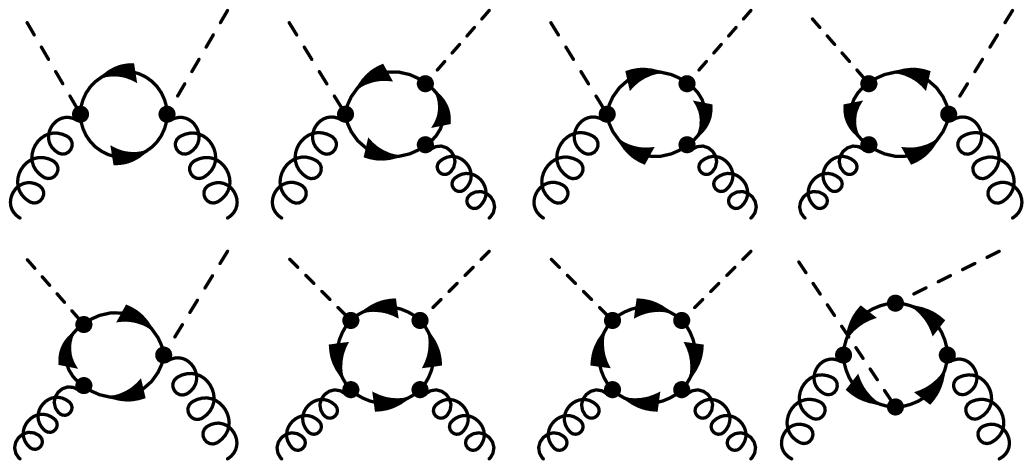,width=10cm}
\end{center}
\caption{In-medium corrections to the $\rho\pi\pi$ (upper panel)
and $\rho\rho\pi\pi$ vertex (middle and lower panel) when
including $NN^{-1}$ and $\Delta N^{-1}$ excitations in the intermediate
pion propagators of the $\rho$ selfenergy; curly lines: $\rho^0$'s,
dashed lines: $\pi$'s, solid lines: nucleons or deltas (when
forward-going) and nucleon holes (backward-going).}
\label{fig_vtxcorr}
\end{figure}
In free space, where 
\bea
\Gamma^{(3),0}_{\mu ab}(k,q) &=& g_{\rho\pi\pi} \ \epsilon_{3ab} \ 
(2k+q)_{\mu}
\nonumber \\
\Gamma^{(4),0}_{\mu\nu ab}(k_1,k_2,q) &=&
  2ig_{\rho\pi\pi}^2\ (\delta_{ab}-\delta_{3a}\delta_{3b}) \ g_{\mu\nu} \ ,  
\label{vtxfree}
\eea
Eqs.~(\ref{WT3}) and (\ref{WT4}) are trivially satisfied. In nuclear
matter, when including $NN^{-1}$ and $\Delta N^{-1}$ loops in the pion
propagation,  the required in-medium vertex corrections can be constructed 
by coupling the $\rho$ meson to the lines and vertices of the pion selfenergy
insertions in all possible ways, leading to the diagrams displayed 
in Fig.~\ref{fig_vtxcorr}. 
The final result for the $\rho$ meson selfenergy tensor in cold nuclear matter
can then be written as
\bea
\Sigma_{\rho\pi\pi}^{\mu\nu}(q) &=&
  i {1\over 2}\int{d^4 k\over {(2\pi)^4}} \ 
  iD_{\pi}(k) \ \Gamma_{ab}^{(3)\mu}(k,q) \ iD_{\pi}(k+q) \ 
  \Gamma_{ba}^{(3)\nu}(k+q,-q) 
\nonumber\\
&&+i{1\over 2}\int{d^4 k\over {(2\pi)^4}} \ iD_{\pi}(k) \ 
  \Gamma_{aa}^{(4)\mu\nu}(k,k,q) \ .   
\label{sgrppmunu}
\eea
The tensor can be decomposed into longitudinal and transverse components
in the usual way, see Eq.~(\ref{sgrhoLT}). 
In Fig.~\ref{fig_drhopicloud} the spin-averaged $\rho$ spectral 
function is shown for various densities (left panel) and 3-momenta
(right panel). 
\begin{figure}[!htb]
\begin{center}
\epsfig{figure=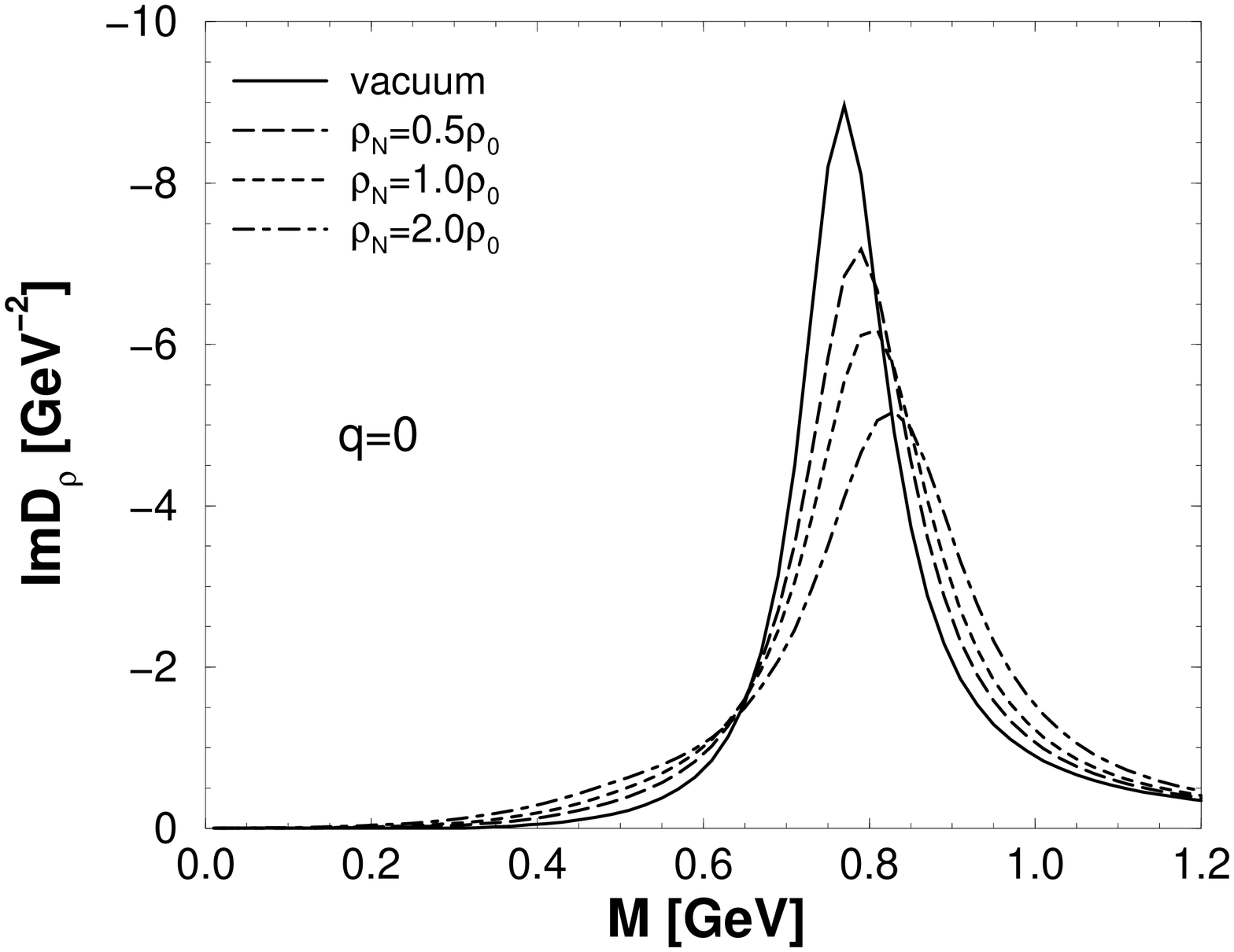,width=6cm,angle=-90}
\epsfig{figure=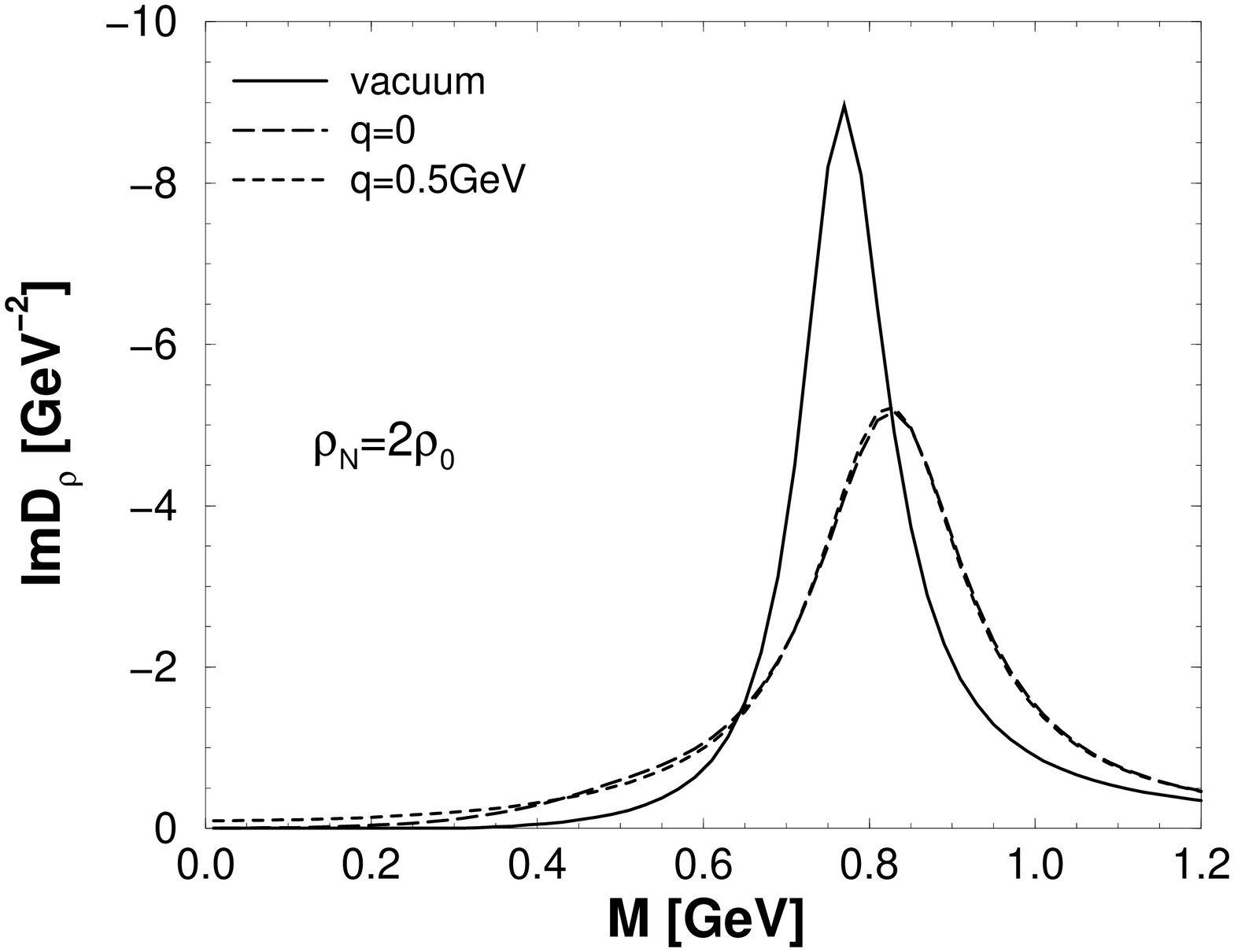,width=6cm,angle=-90}
\end{center}
\caption{Spin-averaged in-medium $\rho$ propagator when accounting
for $P$-wave  $NN^{-1}$ and $\Delta N^{-1}$ excitations in its
pion cloud; left panel: density dependence at fixed three momentum
$q=0$; right panel: three-momentum dependence at fixed nucleon density 
$\varrho_N=2\varrho_0$.}  
\label{fig_drhopicloud}
\end{figure}
As opposed to Fig.~\ref{fig_g2pic} the underlying $\pi NN$ and $\pi N\Delta$
form factor cutoffs (cf.~Eq.(\ref{ffpinn})), which essentially determine
the magnitude of the medium effects, have been fixed at 
$\Lambda_{\pi NN}=\Lambda_{\pi N\Delta}=300$~MeV (together with rather
small Migdal parameters of $g'_{NN}=0.6$, $g'_{N\Delta}=0.2$). These choices 
emerge as a consequence of model constraints imposed through 
the analysis of $\pi N\to \rho N$ scattering data (this will be discussed
in detail in Sect.~\ref{sec_hadscat} where we also elaborate on the
appearance of such 'unnaturally' soft form factors).    
In spite of the soft form factors, the $\rho$ meson spectral function still
exhibits a significant broadening due to the pion cloud modifications, 
being about 55~MeV at nuclear saturation density. However, this is 
substantially smaller as compared to earlier 
calculations~\cite{HeFN,ChSc,AKLQ,RCW,UBRW} where the 
$\pi\alpha$ form factors were used with the 
'Bonn value' (1.2~GeV)~\cite{MHE}
for the cutoff. On the other hand, the common feature shared by all previous 
calculations, \ie, an upward shift of the resonance peak persists.  
Furthermore, the relative smallness of the medium effects does
not allow for a strongly developed momentum dependence: the most
significant feature of finite momenta is an enhancement of the spectral
function for low invariant masses $M\le0.4$~GeV (see right
panel of Fig.~\ref{fig_drhopicloud}; for the chosen momentum
of $q=0.5$~GeV the effects are largest).

On more general grounds it is important to note that the models for
the in-medium $\rho$ propagator discussed above
contain a mixing of vector and axialvector correlators much in the same way
as for the purely thermal case discussed in Sect.~\ref{sec_vamix} 
and displayed in Fig.~\ref{fig_vamix}. This has been shown explicitly 
by Chanfray and collaborators~\cite{CDER} for the case
in which the $\rho$ meson couples to the pion-nucleon system (the conclusions 
remain valid when the $\Delta$-isobar is included in addition). 
To make the argument,
consider the longitudinal and transverse selfenergies of the $\rho$ meson.
According to (\ref{sgrhoLT}) these can be written as
\bea
\Sigma_\rho^L(q_0,\vec q)&=&
{q^2\over q_0^2}{q_iq_j\over \vec q^2}\Sigma^{ij}_\rho(q)
\equiv g_{\rho\pi\pi}^2{q^2\over q_0^2}{q_iq_j\over \vec q^2}V^{ij}(q)
\nonumber\\ 
\Sigma_\rho^T(q_0,\vec q)&=&{1\over 2}\biggl 
(\delta_{ij}-{q_iq_j\over \vec q^2} \biggr )\Sigma^{ij}_\rho(q)
\equiv g_{\rho\pi\pi}^2 {1\over 2}\biggl (\delta_{ij}-{q_iq_j\over \vec q^2}
\biggr )V^{ij}(q)
\label{spatialcor}
\eea
and (up to vertex form factors) define 
the spatial components of the vector correlator $V^{ij}$ ($i,j=1,2,3$). 
When evaluated by including nucleons and pions one obtains \cite{ChSc}
\begin{eqnarray}
 V_{ij}(q)& = &  i\int \frac{d^4k_1}{(2\pi)^4}\,
  \left[\big(1+\Pi^0(k_1)\big)k_{1i} -  \big(1+\Pi^0(k_2)\big)
  k_{2i}\right]   
\nonumber\\
 &  & \times D_\pi(k_1)D_\pi(k_2)\left[\big(1+\Pi^0(k_1)\big)k_{1j} - 
  \big(1+\Pi^0(k_2)\big) k_{2j}\right] 
\nonumber\\
 &  & + i\int\frac{d^4k_1}{(2\pi)^4}\, \left\{ 
\left[ \hat{k}_{1i}\hat{k}_{1j}\Pi^0(k_1) +
   (\delta_{ij}-\hat{k}_{1i}\hat{k}_{1j})\Pi^T(k_1)\right]
D_\pi(k_2) \right.  
\nonumber\\
 & & + \left. \left[\hat{k}_{2i}\hat{k}_{2j}\Pi^0(k_2) +
(\delta_{ij}-\hat{k}_{2i}\hat{k}_{2j})\Pi^T(k_2)\right] D_\pi(k_1)\right\} \ , 
\label{vcorr}
\end{eqnarray}
where $k_1$ and $k_2$ are the single-pion four-momenta 
(their sum is the incident momentum $q = k_1+k_2 $). 
In the above expression the dimensionless irreducible response 
function $\Pi^0$ is related to the pion selfenergy (including 
effects of short-range correlations via Migdal parameters) 
as $\Pi^0=\Sigma_\pi/{\vec k}^2$ (see Eq.~(\ref{Sigmapi})), while
$D_\pi(k)=1/[k_0^2-{\vec k}^2-m_\pi^2-{\vec k^2}\Pi^0(k)]$ 
denotes the fully dressed pion propagator with $NN^{-1}$ insertions, 
cf.~Eq.~(\ref{piprop}). Finally $\Pi^T$ denotes the fully iterated 
spin-transverse response function,   
where the nucleon bubble is iterated to all orders. 
Expression (\ref{vcorr}) coincides with Eq.~(\ref{sgrppmunu}). 
The corresponding axial correlator $A_{ij}$ can be defined in analogy
to (\ref{spatialcor}) and takes the form~\cite{CDER}   
\begin{eqnarray}
\frac{1}{f_\pi^2}A_{ij}(k) &= &  k_i k_j D_\pi(k) + 2k_i k_j\Pi^0(k)D_\pi(k) 
+ \hat{k}_i\hat{k}_j\Pi^L(k) + (\delta_{ij}-\hat{k}_i\hat{k}_j)\Pi^T(k)
   \nonumber   \\ 
      &= & k_i k_j\big(1 + \Pi^0(k)\big)^2 D_\pi(k) +\hat{k}_i\hat{k}_j\Pi^0(k)
      + (\delta_{ij}-\hat{k}_i\hat{k}_j)\Pi^T(k) \ ,  
      \label{acorr}
\end{eqnarray}
where also the fully iterated spin-longitudinal response function
$\Pi^L$ has been introduced. 
The pertinent diagrams are depicted in Fig.~\ref{fig_axcorr}.
\begin{figure}[!htb]
\begin{center}
\epsfig{figure=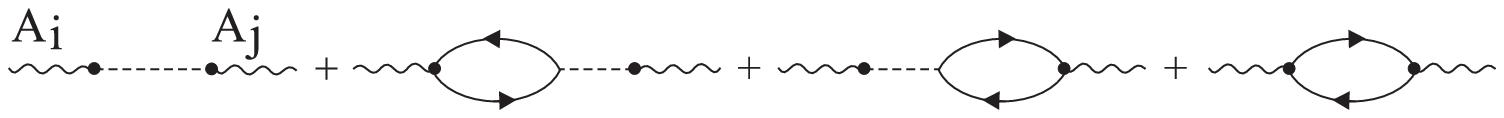,width=14cm}
\end{center}
\caption{The axial correlator in nuclear matter.}  
\label{fig_axcorr}
\end{figure}
The crucial observation of Chanfray \etal~ is now that $V_{ij}(q)$
can be expressed in terms of the axial correlator in a form 
\bea
V_{ij}(q) & = & i\int \frac{d^4k_1}{(2\pi)^4}\,
 \big[ \frac{1}{f_\pi^2}\big(A_{ij}(k_1)D_\pi(k_2)
  + A_{ij}(k_2)D_\pi(k_1)\big) 
\nonumber \\
&& - (1+\Pi^0(k_1)\big)\big(1+\Pi^0(k_2)\big)(k_{1i}k_{2j}+k_{2i}k_{1j})
  D_\pi(k_1)D_\pi(k_2)\big]    \ ,  
\label{mixing}
\eea
which displays the mixing effect
through the first term. This term arises from the vertex corrections displayed
in Fig.~\ref{fig_vtxrhopi}. Indeed, removing a (dressed) 
pion one is left with the axial correlator taken at the momentum of the 
other pion. In contrast to the thermal case the extra pion is, however, 
not provided by the heat bath but rather
by a nucleon from the medium. The second term in Eq.~(\ref{mixing}) does
not reduce to the product of the axial correlator with the pion
propagator. Its existence is due to the interaction of the photon with the
pion via the derivative term $\vec\pi\times\partial^\mu\vec\pi$ in the
interaction term of the VDM Lagrangian (\ref{Lintrhopi}). 
The extra term does not invalidate the basic mixing concept. In fact, 
removing a pion the remainder is still of axial nature.

It is very pleasing to see that partial restoration of chiral symmetry through
mixing of vector and axialvector correlators is also manifest in cold nuclear
matter. As pointed out in Ref.~\cite{ChSc,UBRW} the dominant mechanism for
shifting strength to lower energies in the $\rho$ spectral function
is provided by coupling to $\pi\Delta N^{-1}$-states through vertex
corrections. These are precisely of the type $A_{ij}D_\pi$ arising from the
last term of the axial correlator (\ref{acorr}) with a transverse 
$\Delta N^{-1}$-bubble.
Thus the manifestation of partial restoration of chiral symmetry is
in the broadening of the spectral function! Due to the lack of the 
$a_1$ meson as an explicit degree of freedom it is not clear at 
present whether the in-medium Weinberg sum rules (\ref{WSRm1}),
(\ref{WSRm2}) are fulfilled.

A somewhat different approach for evaluating vector meson spectral distributions
in cold nuclear matter has been pursued by Klingl \etal~\cite{KKW97}. They 
start from an the $SU(3)$ chiral Lagrangian for pseudoscalar mesons and
baryons~\cite{KSW}, based on pseudovector coupling as
\beq
{\cal L}_{\Phi B}=F \ \tr \left(\bar B \gamma_\mu\gamma_5 [u^\mu,B]\right)
+ D \ \tr \left(\bar B \gamma_\mu\gamma_5 \{u^\mu,B\}_+\right)  \ ,  
\eeq
where $\{.,.\}_+$ and $[.,.]$ denote anti-/commutators, respectively,
and the $SU(3)$ field matrices are given by 
\beq
B=\left(\begin{array}{ccc} 
\frac{\Lambda}{\sqrt{6}} + \frac{\Sigma^0}{\sqrt{2}} & \Sigma^+ & p \\
\Sigma^- & \frac{\Lambda}{\sqrt{6}}-\frac{\Sigma^0}{\sqrt{2}} & n \\
   \Xi^- & \Xi^0 & \frac{-2\Lambda}{\sqrt{6}} 
  \end{array} \right) \quad ,  \quad
\Phi=\left(\begin{array}{ccc}
\frac{\pi^0}{\sqrt{2}}+\frac{\eta}{\sqrt{6}} & \pi^+ & K^+ \\
\pi^- & \frac{-\pi^0}{\sqrt{2}}+\frac{\eta}{\sqrt{6}} & K^0 \\
   K^- & \bar K^0 & \frac{-2\eta}{\sqrt{6}} 
  \end{array} \right) \ . 
\eeq
The parameters $F\simeq 0.51$ and $D\simeq 0.75$ are chosen to 
comply with axial coupling constant $g_A=F+D=1.26$. 
The four vector
\beq
u^\mu=-\frac{1}{2f_\pi} \left( \partial^\mu \Phi 
-ie[{\cal Q},\Phi] A^\mu\right) 
\eeq 
contains the axialvector current of the pseudoscalar fields as well
as the minimal coupling term for the electromagnetic field $A^\mu$, 
${\cal Q}={\rm diag}(2/3, -1/3,-1/3)$ being the $SU(3)$ charge matrix. 
The vector meson-baryon interactions are then obtained from the 
minimal coupling scheme, \ie, replacing $e{\cal Q}A^\mu$ by $gV^\mu/2$
with
\beq
V^\mu=
{\rm diag}(\rho^\mu+\omega^\mu,-\rho^\mu+\omega^\mu,\sqrt{2}\phi^\mu) \ ,  
\eeq
and the relevant terms for $V$-$B$ interactions become
\bea
{\cal L}_{V\Phi B}&=&\frac{ig}{4f_\pi} \left\{ F \ \tr 
\left(\bar B \gamma_\mu\gamma_5 \left[ [V^\mu,\Phi],B\right]\right) 
+ D \ \tr \left(\bar B \gamma_\mu\gamma_5 \{[V^\mu,\Phi],B \}_+
\right)\right\}
\nonumber\\
{\cal L}_{VB}^{(1)} &=& \frac{g}{2}\left\{ 
\tr\left( \bar B \gamma_\mu [V^\mu,B]\right)
-\tr(\bar B\gamma_\mu B) \ \tr(V^\mu)\right\} 
\nonumber\\  
{\cal L}_{VN}^{(2)} &=& \frac{g\kappa_\rho}{4M_N} \bar N \vec\tau
\sigma_\mn N \partial^\mu \vec\rho^\nu +
\frac{g\kappa_\omega}{4M_N} \bar N 
\sigma_\mn N \partial^\mu \omega^\nu \ .  
\label{LVB}
\eea
The last part, ${\cal L}_{VN}^{(2)}$, has been added to include 
corrections due to anomalous $VN$ tensor couplings (using 
$\kappa_\rho=6$, $\kappa_\omega=0.1$, $\kappa_\phi=0$). Finally, 
the $VB$ and (axial) $\Phi B$ vertices are supplemented by phenomenological
monopole form factors
\beq
F_{VB}(k^2)=\frac{\Lambda_V^2-m_V^2}{\Lambda_V^2-k^2} \quad , \quad
F_A(k^2) = \frac{\Lambda_A^2}{\Lambda_A^2-k^2} \ ,  
\label{ffklingl}
\eeq
respectively, with  rather large cutoff parameters, $\Lambda_V=1.6$~GeV and 
$\Lambda_A=1$~GeV. 
 
The in-medium vector meson selfenergies are then constructed
from a low-density expansion as 
\beq
\Sigma_V(q_0,\vec q=0) \simeq \Sigma_V^\circ(q_0)-\varrho_N \ 
T_{VN}(q_0) \ , 
\eeq
which has been restricted to the case of vanishing three momentum
$\vec q=0$ where the longitudinal and transverse parts coincide.
The $V$-$N$ scattering amplitudes $T_{VN}$ are constructed from
the interaction vertices, Eqs.~(\ref{LVB}). The imaginary parts are 
evaluated from standard 
Cutkosky rules, and the real part is then obtained from a subtracted
dispersion relation,
\beq
{\rm Re} \ T_{VN}(q_0)=l_V + \int\limits_0^\infty \frac{d\omega^2}{\pi} 
\frac{{\rm Im} T_{VN}(\omega)}{(\omega^2-q_0^2)} \frac{q_0^2}{\omega^2} \ ,
\label{ReTVN}
\eeq
the subtraction constants being fixed by the Thompson limit ($q_0\to 0$) for 
Compton scattering of real photons. 
The set of diagrams
contributing to the $\rho N$ amplitude is depicted in Fig.~\ref{fig_rhoNamp}.
\begin{figure}[!tbh]
\vspace{3cm}
\begin{center}
\epsfig{figure=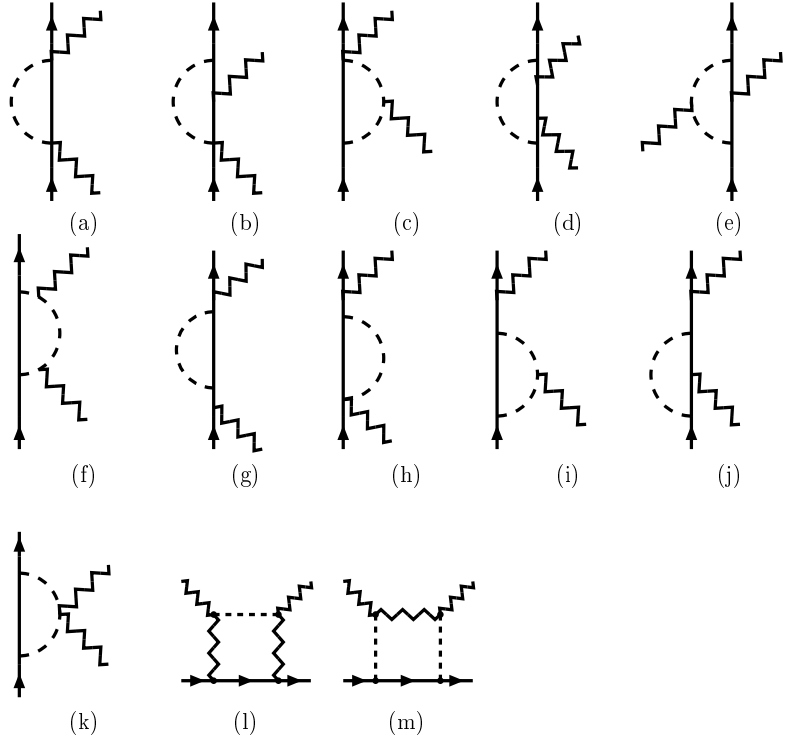,width=9cm,height=7cm}
\end{center}
\caption{Diagrams contributing to the $\rho$-$N$ scattering amplitude
in the approach of Klingl \etal~\protect\cite{KKW97}; dashed lines: pions,
wavy lines: $\rho$ mesons (in graphs (l) and (m) the internal wavy lines 
represent $\omega$ mesons), solid lines: nucleons or deltas (the latter
only for internal lines).}
\label{fig_rhoNamp}
\end{figure}
Except for the last two 'box' diagrams involving $\omega$ mesons, 
they are equivalent to the $\rho$ selfenergy contributions obtained by 
Urban \etal~\cite{UBRW} as encoded in the  various vertex corrections
displayed in Fig.~\ref{fig_vtxcorr}.  

Fig.~\ref{fig_rhoomphi} shows the resulting vector current correlators
(divided by the energy squared)
in the $\rho$, $\omega$ and $\phi$ meson channel for $\vec q=0$. 
\begin{figure}[!tbh]
\epsfig{figure=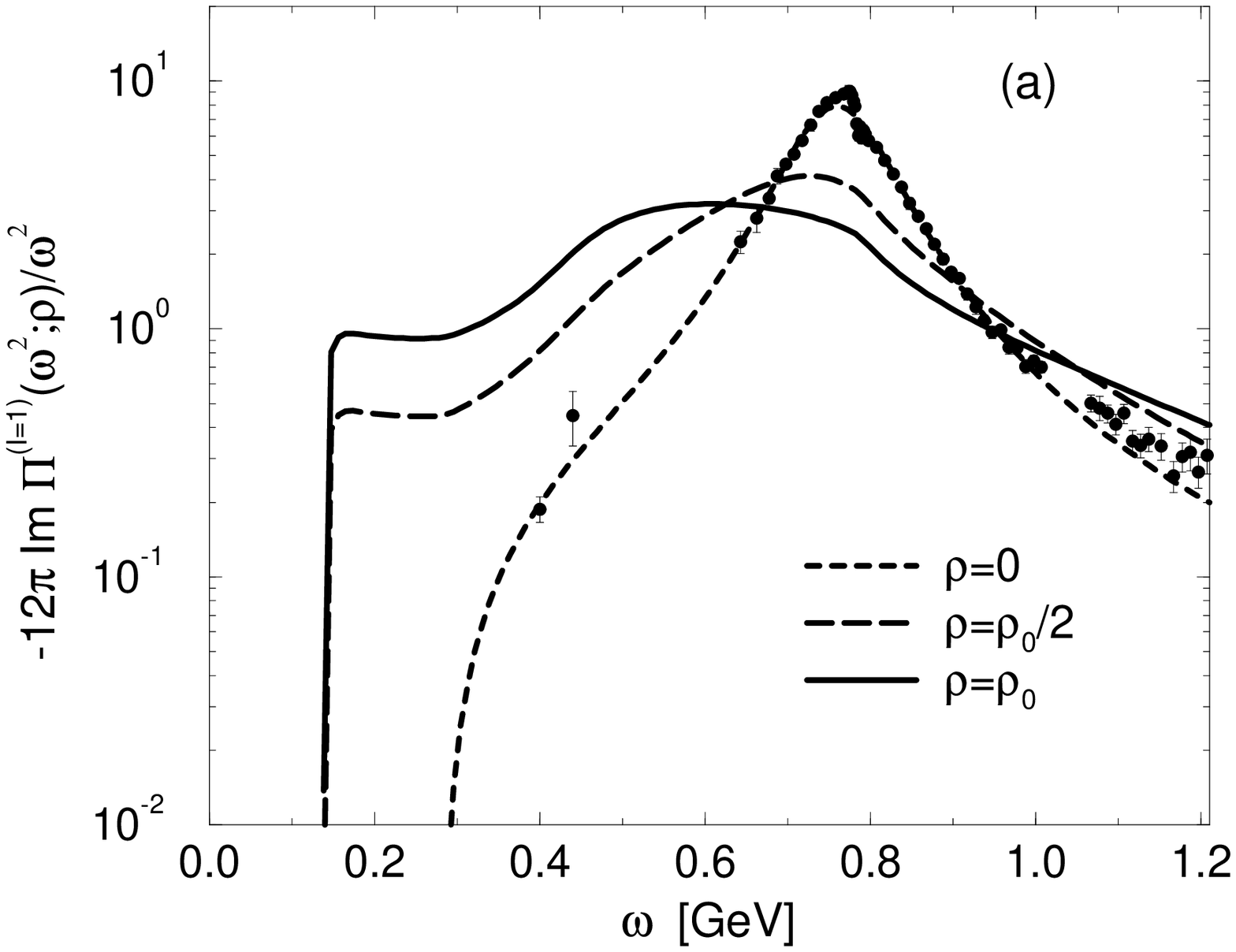,width=4.85cm}
\epsfig{figure=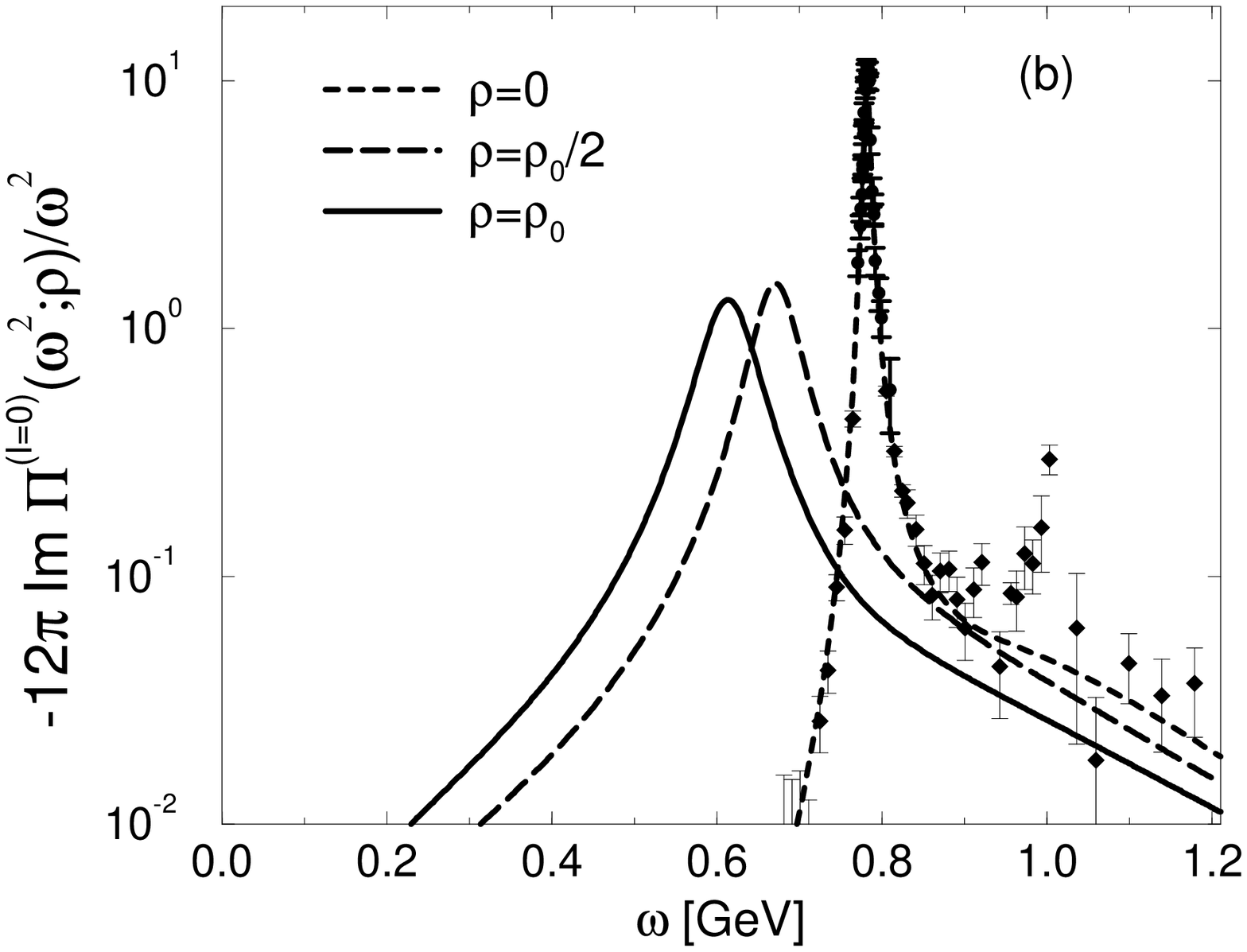,width=4.85cm}
\epsfig{figure=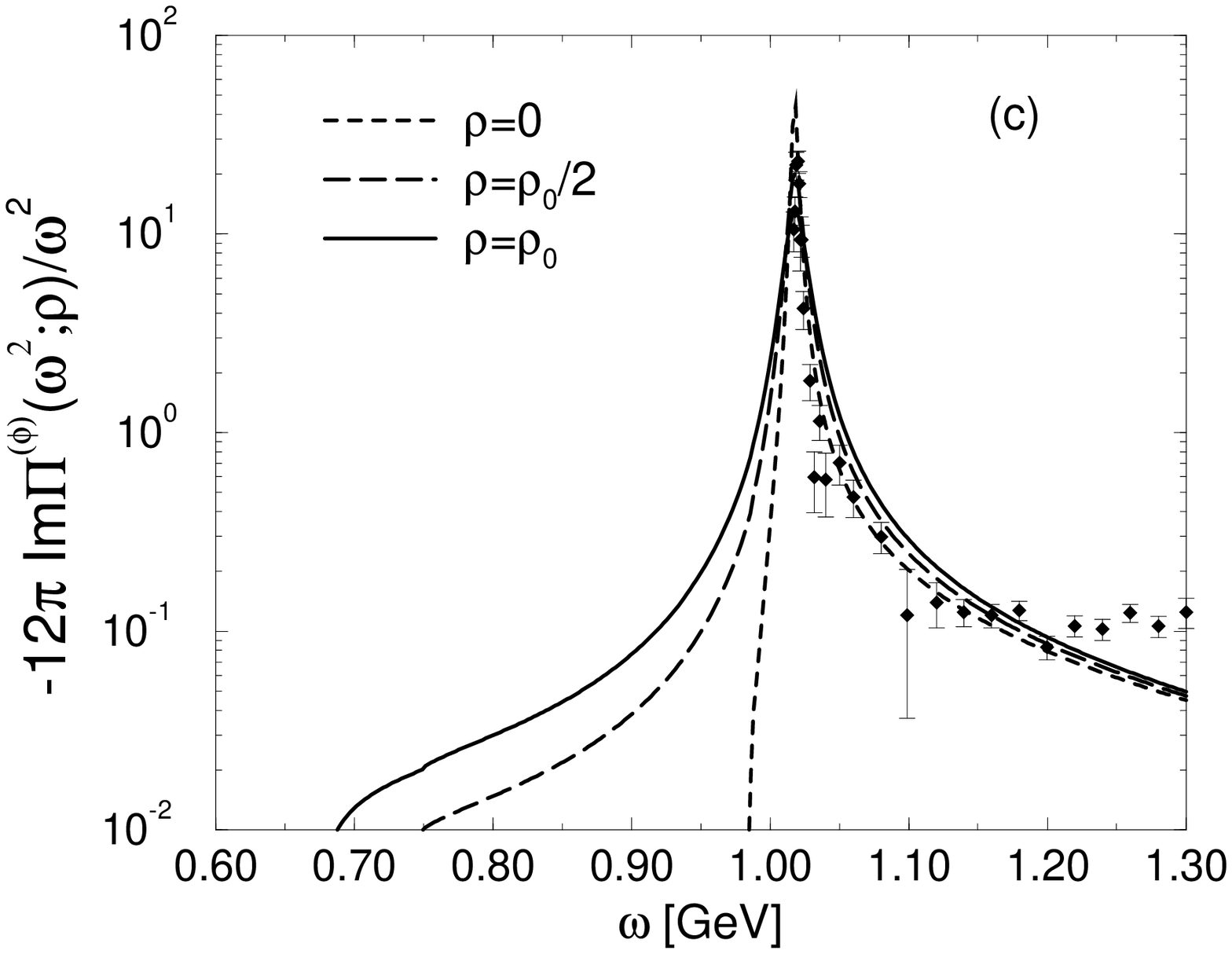,width=4.85cm}
\caption{Medium Modifications of the vector correlators in nuclear
matter within the approach of Klingl \etal~\protect\cite{KKW97};
panels (a), (b) and (c) correspond to the $\rho$, $\omega$ and
$\phi$ meson, respectively.}
\label{fig_rhoomphi}
\end{figure}
In the improved vector dominance scheme of Ref.~\cite{KKW97}, 
they are related to the vector meson selfenergies via
\beq 
 \Pi_V(q_0,\vec q=0) = \frac{1}{g_V^2}
\left( \Sigma_V(q_0,0)+[a_V q_0^4-\Sigma_V(q_0,0) q_0^2]
\ D_V(q_0,0) \right) 
\eeq
with constants $a_\rho=1.1$, $a_\omega=a_\phi=1$ (which give a good 
fit to the free $e^+e^-\to hadrons$ cross sections).  
As in the
other approaches discussed above, the $\rho$ (as well as the $\omega$) 
spectrum exhibit a strong
broadening with increasing nuclear density (note, however, that it is
quantitatively overestimated, as the hard form factors employed in this
calculation, Eqs.~(\ref{ffklingl}), entail a large overprediction of 
$\pi N \to \rho N$ and $\pi N \to \omega N$ scattering data).   
In addition, the in-medium $\omega$ meson mass is reduced, the 
$\rho$ meson mass 
being practically unchanged (as is revealed by inspection of the real part 
of the correlator (not shown); the apparent peak shift in the left
panel of Fig.~\ref{fig_rhoomphi} is due to an additional division
by $\omega^2$ in the plot).  
The modifications in the $\phi$ channel, which are due to modifications
in the kaon cloud,  are rather moderate.

\subsection{Direct $\rho$-Nucleon Resonances} 
\label{sec_rhoNres} 
Besides modifications of the pion cloud the (bare) $\rho$ meson may
couple directly to the surrounding nucleons. From meson-exchange
models such as the Bonn potential~\cite{MHE} one knows that, \eg, 
the $\rho NN$ (or $\rho N\Delta$) coupling constant can be quite large, 
even though the corresponding s-channel
process $\rho N \to N \to \rho N$ is
kinematically strongly disfavored. Such a kinematic suppression will
be much less pronounced with increasing energy
of the resonance in the intermediate state. 
There are indeed several baryonic resonances listed in the particle
data table~\cite{PDG96}  which exhibit a significant branching fraction 
into the $\rho N$ channel. 

Friman and Pirner~\cite{FrPi} first suggested to calculate a $\rho$ meson 
selfenergy from direct $\rho$-$N$ interactions in terms of
$N(1720)$ and $\Delta(1905)$ resonances. The latter two have  
large branching ratios of well above 50\% to $P$-wave $\rho$-$N$ states. 
Later on it was realized~\cite{PPLLM,RUBW} that some lower-mass 
resonances -- well below the naive $\rho N$-threshold of 
$m_N+m_\rho$ -- can have a strong coupling 
to predominantly $S$-wave $\rho$-nucleon states, most notably the $N(1520)$.
Its decay into $\rho N$ is only possible due to the finite width of the 
$\rho$ going into $\pi\pi$. However, 
the kinematically accessible fraction of the $\rho$ spectral function
in $N(1520)\to \rho N$ decays,  
\beq
F=\int\limits_{2m_\pi}^{M_{max}} \frac{MdM}{\pi} A_\rho(M) \  ,  
\label{F}
\eeq 
amounts to only about $F\simeq2$~\% (with 
$M_{max}=m_{N(1520)}-m_N\simeq 580$~MeV). Yet the 
experimental branching ratio for this decay is 
$\sim$~20\% out of the total width of $\Gamma_{N(1520)}^{tot}$=120~MeV. 
Another indication for the 
importance of the $N(1520)$ in its coupling to vector meson-nucleon states
is found in $\gamma N$ cross sections, where, next to the $\Delta(1232)$,
the $N(1520)$ represents the most prominent resonance structure. Besides the 
(not always well-known) hadronic branching ratios, the photoabsorption
cross sections are an important source of information to constrain 
the various $\rho NB$ couplings, as will be discussed in 
Sect.~\ref{sec_photoabs}.  

\begin{table}[!htb]
\bce
\begin{tabular}{c|ccccccc}
 B & $l_{\rho N}$ & $SI(\rho BN^{-1})$ & $\Gamma^0_{\rho N}$ [MeV] &
 $\Gamma^{0,fit}_{\rho N}$ [MeV] &
$\left(\frac{f_{\rho BN}^2}{4\pi}\right)$ & $\Lambda_{\rho BN}$ &
$\Gamma^{med}$ [MeV] \\
\hline
N(939)         & $P$ & 4    & --        & --   & 6.0  & 1500 & 0  \\
$\Delta$(1232) & $P$ & 16/9 & --        & --   & 16.2 &  700 &  25 \\
$N$(1440)      & $P$ &  4   & $<$28     & 0.5  & 1.1  & 600  & 200 \\
$N$(1520)      & $S$ &  8/3 & 24        & 23.5 & 6.8  & 600  & 300 \\
$\Delta$(1620) & $S$ &  8/3 & 24        & 36   & 1.5  & 700  & 200  \\
$\Delta$(1700) & $S$ & 16/9 & 128       & 111  & 2.5  & 1000 & 200  \\
$N$(1720)      & $P$ &  8/3 & 115       & 100  & 8.5  & 600  & 100  \\
$\Delta$(1905) & $P$ &  4/5 & $>$210    & 315  & 14.5 & 1200 &  50  \\
$N$(2000)      & $P$ &  6/5 & $\sim$300 & 75   & 1.0  & 1500 &  50 \\
\end{tabular}
\ece
\caption{\it Parameters of the $\rho BN$ vertices as obtained from the
interaction Lagrangians, Eqs.~(\protect\ref{LrhoNsw}) and
(\protect\ref{LrhoNpw}), when adjusted to photoabsorption spectra and
$\pi N \to \rho N$ scattering~\protect\cite{RUBW,Morio98};
table columns from left to
right: baryon resonance $B$, relative angular momentum in the
$\rho N$ decay as implicit in  the interaction Lagrangians,
spin-isospin factor (note that in its definition we have absorbed an
additional factor of $\frac{1}{2}$ as compared to table 2 in
Ref.~\protect\cite{RCW}), average value for the partial decay width
into $\rho N$ as extracted from Ref.~\protect\cite{PDG96} (including
all possible partial waves), partial decay width resulting from the fit
using the parameter values in the subsequent two columns,
and in-medium correction to the total
decay width.}
\label{tab_rhonb}
\end{table}
Appropriate $\rho NB$ interaction Lagrangians can be classified according
to the parity of the resonance $B$: negative/positive parity states are 
associated with  $S$-/$P$-waves in $\rho$-$N$, respectively, 
which are the dominant partial waves 
for moderate three-momenta. In the non-relativistic limit, gauge
invariant interaction vertices can be written down as
\bea
{\cal L}_{\rho BN}^{S-wave}  &=&  \frac{f_{\rho BN}}{m_\rho} \
\Psi^\dagger_{B} \ (q_0 \ {\vec s}\cdot \vec{\rho}_a -
\rho^0_a \ {\vec s}\cdot {\vec q}) \ t_a \ \Psi_N \ + \ {\rm h.c.} \
\label{LrhoNsw}
\\  
{\cal L}_{\rho BN}^{P-wave}  &=&  \frac{f_{\rho BN}}{m_\rho} \
\Psi^\dagger_{B} \ ({\vec s} \times {\vec q}) \cdot
 \vec{\rho}_a \ t_a \ \Psi_N \ + \  {\rm h.c.} \ . 
\label{LrhoNpw}
\eea  
The summation over $a$ is in isospin space with isospin 
matrices $\vec t= \vec\tau, \vec T$  depending on whether the resonance 
$B$ carries $I$=1/2 or 3/2, respectively. Analogously, the various 
vector/scalar products act in 
spin-momentum space with spin operators $\vec s=\vec\sigma, \vec S$
corresponding to $J$=1/2- or $J$=3/2-resonances (spin-5/2 resonances
such as $B=\Delta(1905)$ considered in Ref.~\cite{FrPi}  require a
tensor coupling of type $[R_{ij}q_i\rho_{j,a} T_a]$).  
From these interaction vertices one can derive in-medium
selfenergy tensors for $\rho$-induced $BN^{-1}$ excitations, which proceeds
in close analogy to the pionic case discussed above. Due to the spin-1
character of the $\rho$ meson one encounters both transverse and longitudinal 
components as   
\bea
\Sigma_{\rho\alpha}^{(0),T}(q_0,q) &=&
 -  \left(\frac{f_{\rho\alpha} \ F_{\rho\alpha}(q)}{m_\rho}
\right)^2 \ SI(\rho\alpha) \ Q^2 \ \phi_{\rho\alpha}(q_0,q) 
\\
\Sigma_{\rho\alpha}^{(0),L}(q_0,q) &=&
 -  \left(\frac{f_{\rho\alpha} \ F_{\rho\alpha}(q)}{m_\rho}
\right)^2 \ SI(\rho\alpha) \ M^2 \ \phi_{\alpha}(q_0,q)
\eea
with $Q^2=q^2,q_0^2$ for the transverse $P$- and $S$-wave contributions, 
respectively, whereas the longitudinal part appears only for the $S$-wave
interactions. The spin-isospin factors $SI$ for various resonances can 
be found in Table~\ref{tab_rhonb}, where also a typical set of coupling 
constants and cutoff parameters (entering the hadronic vertex form factor, 
taken to be of monopole form, cf.~Eq.~(\ref{ffpinn})) is quoted.
The Lindhard functions $\phi$ coincide with the pionic case, 
Eq.~(\ref{Lindhard1}). 

An important feature when calculating the corresponding
$\rho$ meson selfenergy and spectral function has been pointed out 
in Ref.~\cite{PPLLM}: as a result of low-energy strength 
appearing in the in-medium $\rho$ spectral function (due to broadening), 
the phase space for the in-medium $N(1520)\to\rho N$ decay  increases
substantially, \ie, the fraction $F$ defined in Eq.~(\ref{F}) becomes much 
larger than the 2\% in case of a free $\rho$ meson. This induces a strong 
density-dependent increase of the in-medium $N(1520)$ decay width, which
has to be reinserted into the expression for the $N(1520)$ width entering 
into the Lindhard function. This selfconsistency problem has  
been solved by numerical iteration in Ref.~\cite{PPLLM}, where besides 
the $N(1520)$ eight further $\rho N$ resonances have been included 
(essentially coinciding with the set given in Table~\ref{tab_rhonb}, 
although using a somewhat different parameter set: most notably, 
the cutoffs have been uniformly set to $\Lambda_{\rho BN}=1500$~MeV
with associated coupling constants to reproduce the $\rho N$ branching
ratios).  
The converged results for the  transverse and longitudinal parts of the 
$\rho$ spectral function, $A_\rho^T$ and $A_\rho^L$,  
at normal  nuclear matter density are shown in Fig.~\ref{fig_peters}. 
\begin{figure}[!tbh]
\epsfig{figure=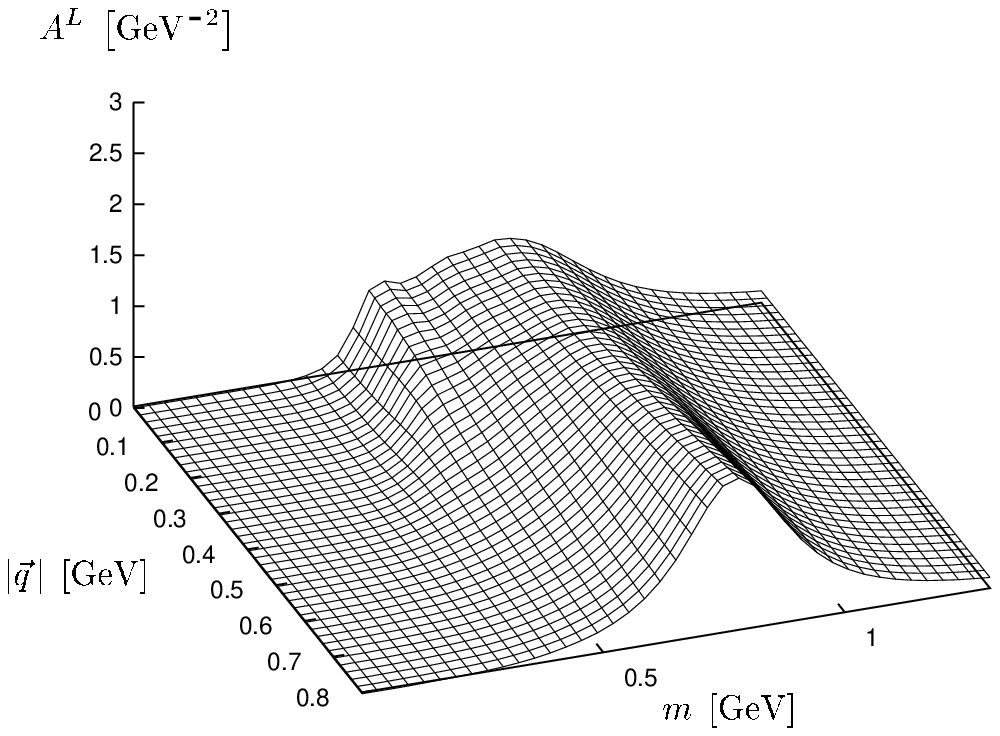,width=6.9cm}
\hspace{0.7cm}
\epsfig{figure=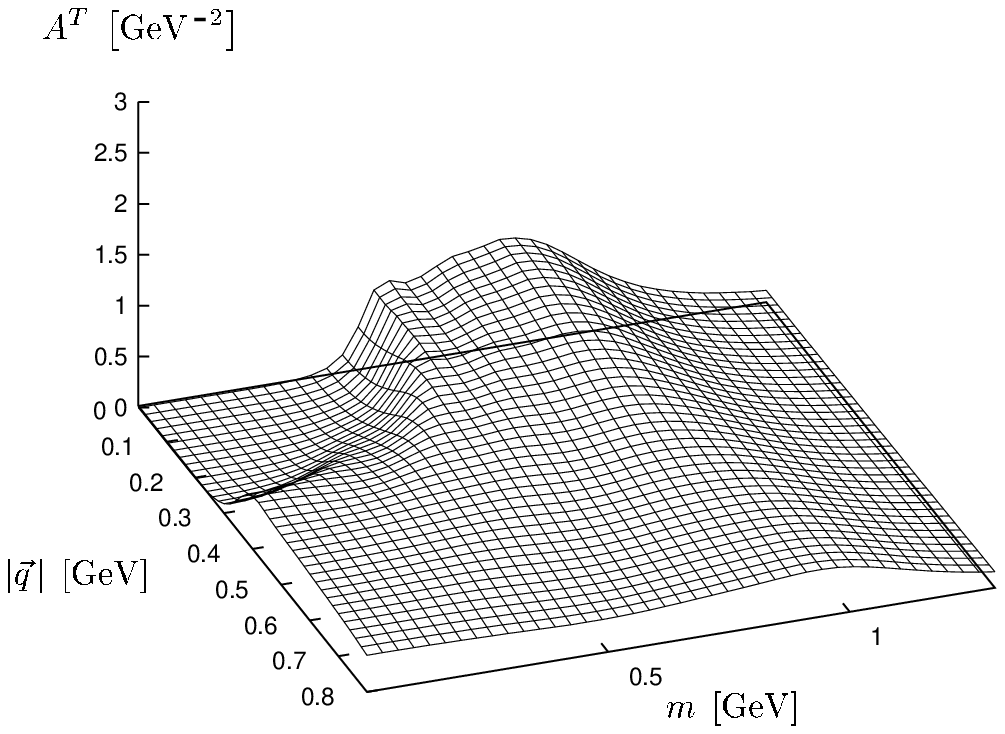,width=6.9cm}
\caption{Transverse and longitudinal parts of the $\rho$ spectral
function in cold nuclear matter when including direct $\rho$-induced
$BN^{-1}$ excitations according to Ref.~\protect\cite{PPLLM} (note that 
in this figure the spectral function has been defined as 
$A^{L,T}=\frac{-1}{\pi} {\rm Im} 
D_\rho^{L,T}$, which differs, \eg, from Fig.~\protect\ref{fig_drhopicloud} 
by a factor of $\frac{-1}{\pi}$. } 
\label{fig_peters}
\end{figure}
One observes a strong broadening of 200--300~MeV (on top of the free width)
with the only 
visible structure being a shoulder at $M\simeq 0.5$~GeV (at zero 
three-momentum) originating from the $N(1520)N^{-1}$ channel, which, 
however, is strongly washed out by a simultaneously emerging in-medium 
width of $\Gamma_{N(1520)}^{tot}\simeq 350$~MeV. Similar results 
arise in the calculations of Ref.~\cite{RUBW}. 

In Ref.~\cite{FLW98} a $\pi N$/$\rho N$/$\omega N$ coupled channel 
approach  has been pursued to asses the $\rho N$ scattering  amplitude. 
Using the standard $T$-$\varrho$ approximation to calculate 
in-medium selfenergies again reveals a strong coupling   
of both $\rho$ and $\omega$ to collective $N(1520)N^{-1}$ states with
roughly equal strength. In-medium corrections to the widths have not yet 
been accounted for. In Ref.~\cite{BLRRW} they have been shown 
to somewhat reduce the strength in the low-lying $N(1520)N^{-1}$ 
excitation in the $\rho$ spectral function (in addition to a strong
smearing).

\subsection{Dispersive Approaches at High Energies}
\label{sec_dispers}
For $\rho$ mesons of large energies, $q_0\gg 1$~GeV, the resonance 
descriptions discussed in the previous Sections should become less reliable.  
On the one hand, the couplings to resonances with masses beyond 2~GeV 
are not very well known. On the other hand, on general grounds, 
one should 
expect that a hadronic description  ceases to be the appropriate one. 
An alternative way to extract in-medium $\rho$ meson properties at large 
energies has been suggested in Ref.~\cite{EI97} using high-energy
$\gamma N$ cross sections in connection with vector dominance. 
Via the optical theorem
the imaginary part of the $\gamma N$ forward scattering amplitude 
can be related 
to the total cross section as
\beq
\sigma^{tot}_{\gamma N}(q_0)=-\frac{4\pi}{q} {\rm Im} T_{\gamma N}(q_0) \ ,  
\eeq
where $q_0=q$ denotes the incoming photon energy or laboratory momentum. 
A dispersion relation is applied to extract the real part
as
\beq
{\rm Re} T_{\gamma N}(q_0) =  \int\limits_0^\infty \frac{d\omega^2}{\pi}
\frac{{\rm Im} T_{\gamma N}(\omega)}{(\omega^2-q_0^2)} 
\frac{q_0^2}{\omega^2} \ ,   
\eeq
the subtraction point at zero energy being determined by the Thompson 
limit (cf. also eq.~(\ref{ReTVN})). VDM then provides the link to
the vector meson-nucleon scattering amplitude as
\beq
T_{\gamma N}=4\pi\alpha \left( \frac{1}{g_\rho^2} f_{\rho N} +
\frac{1}{g_\omega^2} f_{\omega N} + \frac{1}{g_\phi^2} f_{\phi N} 
\right) \ .  
\label{TgammaN}
\eeq
Both naive quark model arguments (where $T_{\rho N}\simeq T_{\omega N}$,
while $g_\omega^2/g_\rho^2\simeq 8$) and empirical information inferred 
from photoproduction data of $\rho$, $\omega$ and $\phi$
mesons (see, \eg, Ref.~\cite{Bauer78}) suggest that the second
and third term in eq.~(\ref{TgammaN}) are comparatively small. This 
allows the direct extraction of mass and width 
modifications for (on-shell) $\rho$ mesons in the low-density 
limit (neglecting Fermi motion) according to the expressions 
\bea
\Delta m_\rho(q_0) &\simeq& -2\pi\frac{\varrho_N}{m_\rho} {\rm Re} 
T_{\rho N}(q_0) 
\nonumber\\
\Delta \Gamma_\rho(q_0) &\simeq& -4\pi\frac{\varrho_N}{m_\rho} {\rm Im} 
T_{\rho N}(q_0) \ ,   
\eea  
which reflect a simple version of the well-known in-medium optical 
potentials.
In the applicable energy regime of $q_0\ge 2$~GeV, Eletsky and Ioffe
obtained $\Delta m_\rho\simeq (60-80)$~MeV and 
$\Delta \Gamma_\rho \simeq 300$~MeV for transversely polarized
$\rho$ mesons at nuclear saturation density.  
However, the thus obtained $\Delta \Gamma$ does not directly have 
the meaning of a resonance broadening as at high energies it includes 
contributions from both elastic and diffractive scattering processes. 

A similar analysis has been performed by Kondratyuk \etal~\cite{Kond98} 
based on photoproduction (rather than Compton scattering) data to extract 
$T_{\gamma N\to \rho N}$, and then using  
$T_{\rho N}=(e/g_\rho) T_{\gamma N\to \rho N}$ to obtain the $\rho N$
scattering amplitude. The results agree within $\sim$30\% with those of 
Ref.~\cite{EI97}. Moreover, using a resonance model for the low-energy
regime (based on similar $\rho N$ resonances as discussed in the previous
Section), they demonstrated that the $T-\varrho$ approximation 
does not lead to reliable predictions for mass shifts due to the 
importance of higher-order-in-density corrections as, \eg, 
induced through the $N^*$ broadening in matter. On the other hand, given 
the large broadening of the $\rho$ spectral 
function as found in previous Sections, the quasiparticle nature 
of the in-medium $\rho$ meson is lost
and its in-medium mass ceases to be a well-defined quantity (note that, 
in general, the calculation of an in-medium spectral function does not 
rely on the quasiparticle concept).

\subsection{Finite Temperature Effects in Baryonic Matter}
\label{sec_densT}
When applying hadronic models to calculate dilepton production in 
(ultra-) relativistic heavy-ion collisions at present lab-energies, ranging
from 1-200~AGeV, sizable temperatures and baryon densities
are encountered simultaneously. Whereas (in thermal equilibrium) the 
meson densities are exclusively determined by a given temperature $T$
(with an additional possibility of meson chemical potentials),
the composition of baryon matter at fixed density $\varrho_B$ 
changes appreciably with temperature. Apart from
the appearance of finite thermal meson abundances, the heating of 
a cold nuclear system induces  two additional features: 
\begin{itemize}
\item[(i)] the nucleon Fermi-distribution functions
experience a substantial smearing. Given a (kinetic) Fermi energy
of $\epsilon_N^F\simeq 40$~MeV at nuclear saturation
density, it is clear that even at BEVALAC/SIS energies of 1-2~AGeV, 
where the typical temperatures are in the 50-100~MeV range, thermal 
motion is quite significant as seen from the left panel in Fig.~\ref{fig_fermi}. 
\item[(ii)]  a certain fraction 
of the nucleons is thermally excited into baryonic resonances;
\eg, at a temperature of $T=170$~MeV and in chemical equilibrium 
the nucleon and $\Delta$ number densities are equal 
owing to the larger spin-isospin degeneracy factor of the $\Delta$
($g_\Delta=16$) as compared to nucleons ($g_N=4$) (right
panel of Fig.~\ref{fig_fermi}).  
\end{itemize}
\begin{figure}[!htb]
\begin{minipage}[t]{0mm}
{\makebox{\epsfig{file=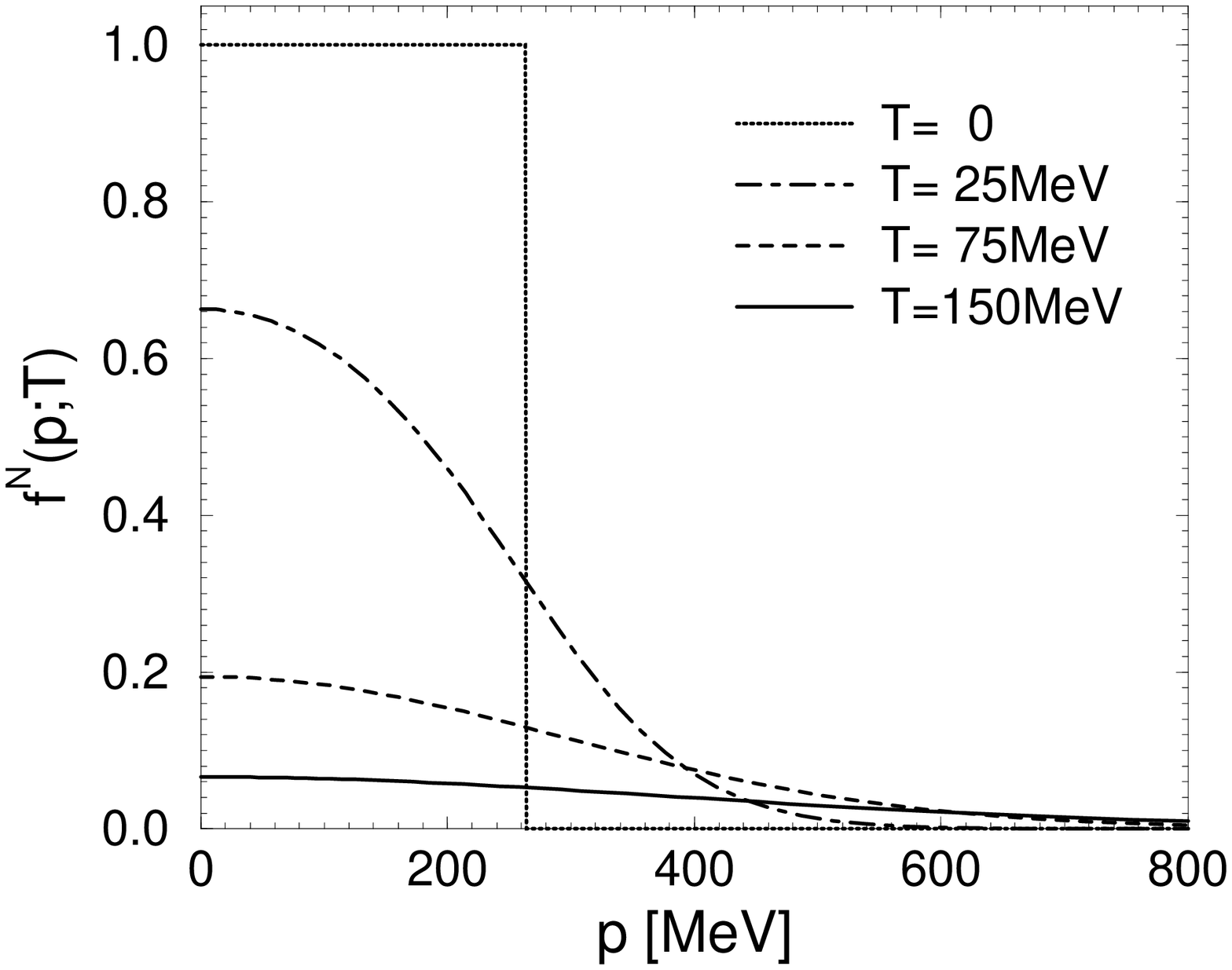,width=6.7cm}}}
\end{minipage}
\hspace{6.8cm}
\begin{minipage}[t]{0mm}
{\makebox{\epsfig{file=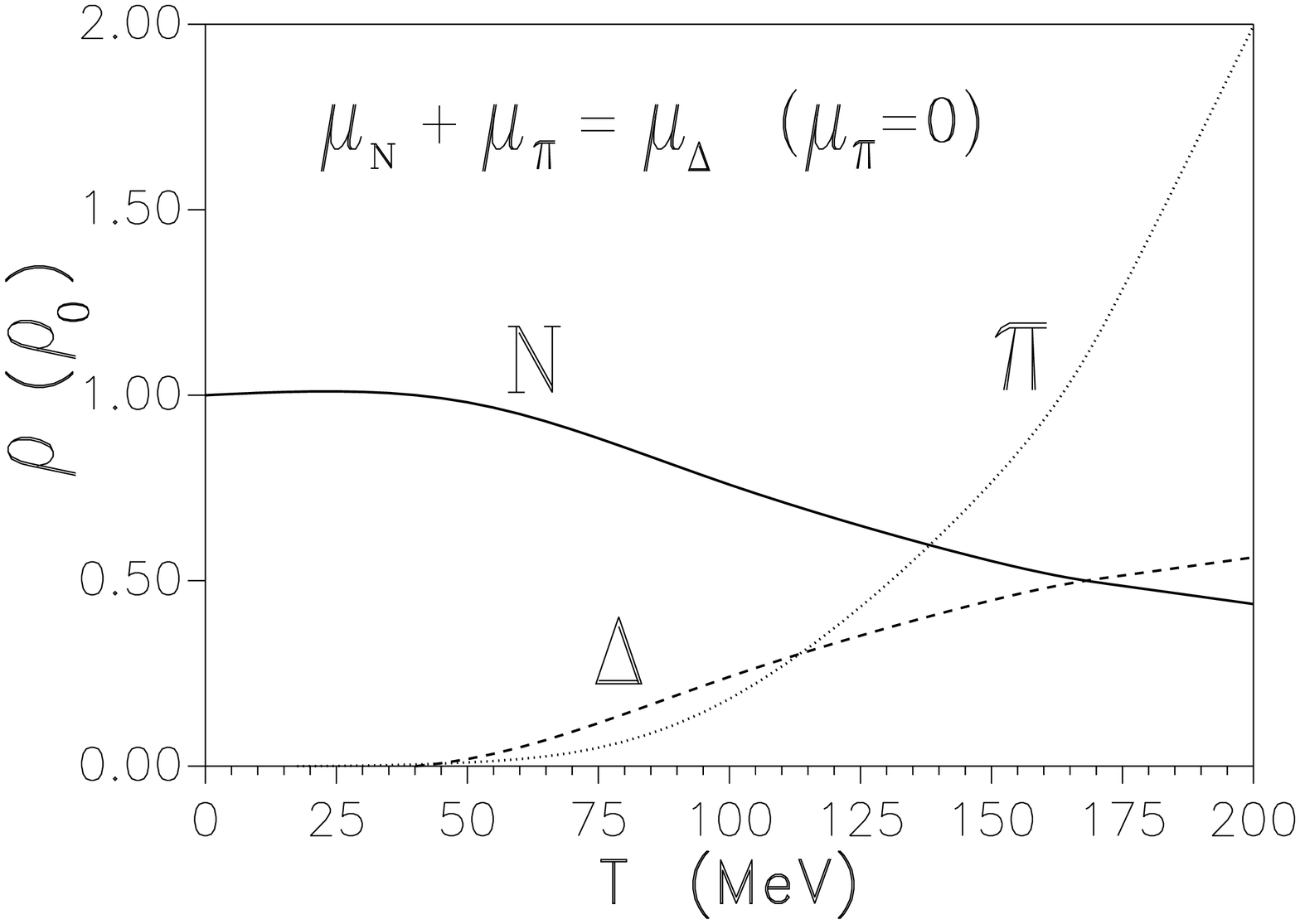,width=5.9cm,angle=90}}}
\end{minipage}
\caption{Finite-temperature effects in baryonic matter. 
Left panel: Fermi-distribution functions for nucleons at normal
nuclear matter density $\varrho_0=0.16$~fm$^{-3}$ at temperatures
$T$=0 (dotted curve), $T$=25~MeV (dashed-dotted  curve),
  $T$=75~MeV (dashed curve) and $T$=150~MeV (full curve).
The corresponding (relativistic) nucleon chemical potentials
are $\mu_N=975, 956, 832$ and 542~MeV, respectively. Note that the nuclear
(kinetic) Fermi energy at zero
temperature is about $\epsilon^F_N(p_F)\simeq 37$~MeV. 
Right panel: composition of a hot $\pi N\Delta$ gas as a function of 
temperature at a fixed baryon density of 
$\varrho_B=\varrho_N+\varrho_\Delta=0.16$~fm$^{-3}$.
For comparison, the thermal pion number density is also shown. Note that the 
nucleon density will be further depleted when including other
baryonic resonances.}
\label{fig_fermi}
\end{figure}

Realistic calculations of spectral functions in a hot and dense
meson/baryon mixture should include these effects. The first one
is readily incorporated by using finite-temperature Fermi distribution
functions in connection with thermal propagators; \eg, in the imaginary
time (Matsubara) formalism, the finite-$T$ Lindhard functions for an
excitation of a baryon resonance $B$ on a nucleon $N$  
takes the form
\beq
\phi_\alpha(q_0,q;\mu_B,T) =-\int \frac{p^2 dp} {(2\pi)^2}
\int\limits_{-1}^{+1}
 \ dx \ \sum_{\pm} \ \frac{f^N[E_N(p)]-f^B[E_B(\vec p +\vec k)]}
 {\pm q_0+E_N(p)-E_B(\vec p +\vec k) \pm \frac{i}{2}(\Gamma_B+\Gamma_N)}
\label{LindhardT}  
\eeq
which, in fact,  not only includes the direct $BN^{-1}$ bubble and its 
exchange term, but also the corresponding $NB^{-1}$ excitation 
(and exchange term) occurring on the finite (thermal) abundance of the 
resonance species $B$. Apparently, $\phi_\alpha$ has the correct
analytic (retarded) properties, \ie, $\phi_\alpha(q_0)=\phi_\alpha^*(-q_0)$.
       
Eq.~(\ref{LindhardT}) can be directly generalized to obtain baryonic
resonance excitations on thermally excited baryons. Of course, the 
corresponding coupling constants (and form-factor cutoff parameters) for
$\rho B_1 B_2$ vertices ($B_1,B_2\ne N$) are mostly unknown. 
In Ref.~\cite{BLRRW} 
it has been argued, however, that (in analogy to the 'Brink-Axel' hypothesis
for nuclear giant dipole resonances on excited states) the most 
important nucleonic excitation pattern, \ie, the $N(1520)N^{-1}$ should 
also be present on other baryonic resonances. From the particle 
data table one can indeed find some evidence for this conjecture: \eg, 
the $\Sigma(1670)$ (which is a well-established four-star resonance
with spin-isospin $IJ^P=1 \frac{3}{2}^-$), when interpreted  as a $\rho \Sigma$
(or $\rho\Lambda$) 'resonance',   very much resembles the quantum
numbers and excitation energy  ($\Delta E\simeq 500-700$~MeV) 
of the $\rho N\to N(1520)$ transition. In addition, 
the branching ratio  of $\Sigma(1670)$ decays into 'simple' final states 
such as $N\bar K, \Sigma \pi$ or $\Lambda\pi$ is substantially less than 100\% .
Similar excitations on non-strange baryonic resonances are even more
difficult to identify as the latter themselves decay strongly via
pion emission (\ie, the $B_1\to B_2 \rho$ decay is immediately followed 
by further $B_2\to N\pi$ and $\rho\to \pi\pi$ decays). Nevertheless, 
from pure quantum numbers it is tempting to associate, \eg, 
$\Delta(1930)\Delta^{-1}$ or $N(2080)N(1440)^{-1}$ excitations
with $S$-wave  'Rhosobar' states.

\begin{figure}[!htb]
\begin{center}
\epsfig{figure=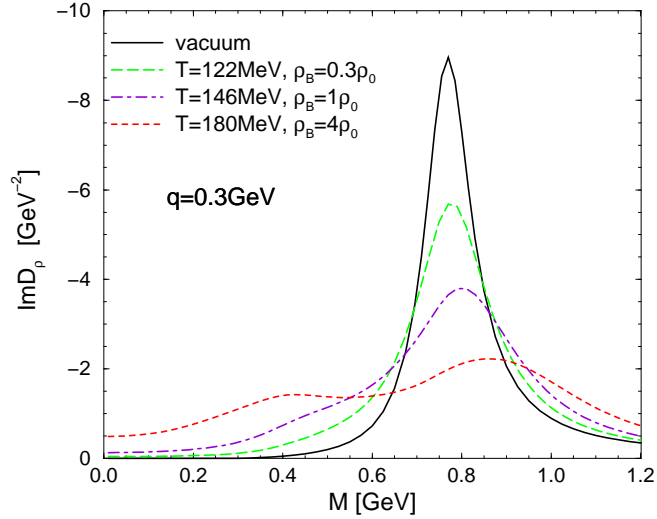,width=8cm,angle=-90}
\end{center}
\caption{Spin-averaged $\rho$ meson spectral function (at fixed 3-momentum
$q=0.3$~GeV)  in hot and dense hadronic matter
in the many-body approach of Refs.~\protect\cite{RW99} 
at various temperatures and total baryon densities corresponding
to a fixed baryon chemical potential of $\mu_B=0.408$~GeV and
vanishing meson chemical potentials.}
\label{fig_imdrho685}
\end{figure}
To end this Chapter we show in Fig.~\ref{fig_imdrho685} the final
result of a many-body calculation~\cite{RCW,Morio98,RG99,RW99} 
for the $\rho$ meson spectral function 
at finite temperatures and density including both thermal mesonic
(Sect.~\ref{sec_phenlagrang})  and baryonic resonances
(Sects.~\ref{sec_rhoNres} and \ref{sec_densT}) as well as 
pion cloud modifications (Sect.~\ref{sec_pioncloud}), also   
at both finite $T$ and $\mu_B$. One finds a very strong broadening 
of the $\rho$ in hot/dense matter, resulting in an almost entire
'melting' of the resonance structure at the highest temperatures
and densities. Possible consequences for experimentally measured 
dilepton spectra in heavy-ion collisions and theoretical 
interpretations will be discussed in detail in the following Chapter.

\chapter{Analysis of Dilepton Spectra: Constraints, Predictions and 
Implications}
\label{chap_dilepton}
As already mentioned in the Introduction, recent experimental analyses
of dilepton production in fixed target heavy-ion collisions at both 
relativistic (1-2~AGeV, BEVALAC) and ultrarelativistic  (158-200~AGeV, 
CERN-SpS) projectile energies 
have exhibited a strong enhancement of low-mass pairs as compared to 
expectations based on free decay processes of the various hadrons in the 
final state ('freezeout'). Once the hadronic composition of the freezeout 
state is known, 
this so-called 'hadronic cocktail' contribution to the observed spectra 
dilepton spectra can be quite reliably assessed without further assumptions, 
and has been shown to be in excellent agreement with the measurements performed 
under equivalent conditions in proton-induced collisions on various nuclei. 
This corroborated the naive expectation that the projectile protons essentially
traverse the target nuclei with the hadronization of the produced secondaries
mostly taking place outside the nucleus, thus leaving no traces of 
significant in-medium effects in the observed dilepton decays. 

As a first step further from the theoretical side, various 
authors have calculated the contributions from $\pi^+\pi^-$-annihilation 
occurring during the lifetime of the interacting hot fireball formed 
in the heavy-ion induced reactions. Even without invoking any in-medium 
modifications, this already requires some knowledge about the dynamics within
the fireball, in particular its pion abundance and momentum distributions. 
However, many different models for the URHIC-dynamics arrive at the 
same conclusion, namely that the experimentally observed enhancement at  
invariant masses $M_{ll}\simeq (0.2-0.6)$~GeV cannot be explained when
using free $\pi\pi$ annihilation (see also Fig.~\ref{fig_dlexth}). 

To learn about the type of medium modifications which eventually can account 
for the spectacular experimental results, it is most desirable to perform 
as 'parameter-free' calculations as possible. Although at the present stage, 
modeling of the collision dynamics unavoidably involves some uncertainty, 
the underlying microscopic processes for dilepton production can and 
should be determined imposing as much independent experimental
information as possible. 
This is particularly true for hadronic models involving medium 
effects, where the predictive power resides in the reliability of the 
density dependence. 
Such models typically comprise a large number of different processes,
some of which can be subjected to consistency checks.

The outline of this Chapter is as follows: in Sect.~\ref{sec_constraints}
 we discuss various constraints that the existing hadronic models have 
been (partially) exposed to. In Sect.~\ref{sec_dlrates}
the predictions for the dilepton production rates within different 
hadronic approaches and more exotic ones will be confronted. 
Photon rates are discussed in short in Sect.~\ref{sec_phrates}. 
Sect.~\ref{sec_spacetime} gives a brief (by no means complete) survey of the  
different ways to model heavy-ion collision dynamics with
focus on those that have been employed to calculate dilepton spectra. 
We then proceed to the analysis of low-mass dilepton spectra, starting 
in Sect.~\ref{sec_dls} from intermediate bombarding energies 
(1--2~AGeV) as performed at the BEVALAC (DLS collaboration) and 
to be remeasured 
in future precision measurements at SIS (HADES collaboration). 
The main part, presented in Sect.~\ref{sec_cern}, 
reviews the extensive theoretical 
efforts that have been made to date in studying the experimental data on 
low-mass dilepton production at the full CERN-SpS energies (158--200~AGeV), 
where most of our current understanding on the possible mechanisms 
involved is based on.  
Sect.~\ref{sec_phspectra} contains a much less comprehensive 
view at direct photon 
measurements, mainly to illustrate their potential for providing 
consistency checks on existing models for dilepton enhancement.     
Finally, in Sect.~\ref{sec_implications}, we attempt to give a critical 
assessment of the theoretical implications that have emerged so far.

\section{Constraints on Hadronic Dilepton Production}
\label{sec_constraints} 
As with most problems in nonperturbative QCD, the calculation of 
low-mass dilepton radiation from interacting hot and
dense hadronic matter has invariably to rely on effective approaches for 
the underlying 
production processes. To aim at quantitative predictions it is thus of 
essential importance to minimize the corresponding uncertainties 
in the calculations. In principle, two guidelines are at our disposal: 
(i) explicit implementation of (dynamical) symmetries shared with
QCD, (ii) model constraints imposed by independent
experimental information. Constraints of type (i) typically govern
the interaction dynamics, (approximate) chiral symmetry in our context, 
whereas (ii) usually serves to narrow the range of the parameters within 
effective models on a quantitative level. 
Ideally, both aspects should be satisfied; in practice, however, 
the complexity of the problem enforces compromises in one or the other 
way.  

In the previous two Chapters we have already elaborated on the chiral 
symmetry aspects of the various approaches in some detail, and how
models for the vector-vector correlator in free space can be constructed 
in accordance with empirical data. Here we will address the issue of 
imposing constraints of type (ii) on the in-medium behavior of the vector
mesons and their coupling to photons. Depending on the type of data,  
substantially different kinematical 
regimes may be probed, thus not only narrowing parameter choices
but also sensing dynamical (off-shell) properties of the model under 
consideration. The simplest constraints on coupling constants of
the interaction vertices one is interested in are provided by the 
partial decay widths of the corresponding resonances. These are usually
not very precise and, more importantly, do not contain information on 
energy-momentum dependencies or combined effects of several processes. 
More stringent constraints are thus obtained from scattering data. 
We will discuss both purely hadronic reactions and photoabsorption 
spectra -- which are more directly related to the dilepton regime -- 
where a wealth of high-precision data on both the nucleon and various nuclei
exists. The focus will again be on the $\rho$ meson which,
as repeatedly mentioned, is the  most important 
player in the game of low-mass dileptons from URHIC's.

\subsection{Decay Widths and Hadronic Scattering Data}
\label{sec_hadscat}
A standard approach to calculate modifications of a particle $a$ 
embedded in the medium is based on the two-body scattering amplitude 
$T_{ab}$, integrated over the momentum distributions of the matter 
particles $b$. 
In the simplest version this represents the standard $T$-$\varrho$
approximation and thus captures the linear-in-density effects. 
However, for the $\rho$ meson, which is a short-lived resonance 
(on strong interaction scales), the following complications arise:  
firstly, elastic $\rho$-hadron scattering amplitudes are 
not directly accessible from experiment. Secondly, the large (vacuum) 
width implies that in realistic calculations one cannot just use the 
physical pole mass, but has to account for its spectral mass 
distribution, $A_\rho(M)$. 
This, in turn, allows one to consistently incorporate  processes involving
off-shell $\rho$ mesons, in particular for masses $M\le m_\rho$ which 
are most relevant for low-mass dilepton spectra. 

As mentioned in Chap.~\ref{Vmodels},  a straightforward estimate of 
coupling constants for $\rho$-meson scattering on surrounding 
matter-hadrons, $h$, into a resonance $R$ can be obtained from the reverse
process, \ie, the decay $R\to\rho h$, cf.~Tab.\ref{tab_rhoPM}. 
\begin{table}[!htb]
\bce
\begin{tabular}{c|ccccc}
 $R$ & $I^GJ^P$ & $\Gamma_{tot}$ [MeV] & $\rho h$ Decay &
$\Gamma^0_{\rho h}$ [MeV] & $\Gamma^0_{\gamma h}$ [MeV] \\
\hline
$\qquad\omega(782)\qquad$ & $0^-1^-$ & 8.43 & $\rho\pi$ & $\sim 5$ & 0.72 \\
$h_1(1170)$   & $0^-1^+$ & $\sim 360$  & $\rho\pi$ & seen  &   ?  \\
$a_1(1260)$   & $1^-1^+$ & $\sim 400$  & $\rho\pi$ & dominant & 0.64 \\
$K_1(1270)$   & $\frac{1}{2}1^+$ & $\sim 90$ & $\rho K$  & $\sim 60$ &   ?  \\
$f_1(1285)$   & $0^+1^+$        & 25 & $\rho\rho$ & $\le$8   & 1.65  \\
$\pi(1300)$   & $1^-0^-$ & $\sim 400$ & $\rho\pi$ & seen & not seen \\
$a_2(1320)$   & $1^-2^+$ & 110        & $\rho\pi$ & 78   & 0.31   \\
\end{tabular}
\ece
\caption{\it Mesonic Resonances $R$ with masses $m_R\le 1300$~MeV
and substantial branching ratios
into final states involving either direct $\rho$'s (hadronic)
or $\rho$-like photons (radiative).}
\label{tab_rhoPM}
\end{table}
Based on the interaction 
Lagrangians given in Chap.~\ref{Vmodels}, one can derive the expression 
for the decay width $\Gamma_{R\to\rho h}$  and then adjust the coupling 
constant to reproduce the experimentally measured value.   

Let us first discuss purely mesonic interactions relevant for the
finite-temperature modifications of the $\rho$ meson~\cite{RG99}. 
For the axialvector meson resonances in $P\rho$ scattering the vertex
(\ref{LrhoPA}) leads to 
\bea
\Gamma_{A\to \rho P}(s) & = & \frac{G_{\rho PA}^2}{8\pi s} \  
\frac{IF (2I_\rho+1)}{(2I_A+1)(2J_A+1)} 
\int\limits_{2m_\pi}^{M_{max}} \frac{M dM}{\pi} \  A^\circ_\rho(M) \ q_{cm}
\nonumber\\
 & & \quad\quad \times \ [\frac{1}{2}(s-M^2-m_\pi^2)^2
+M^2 \omega_P(q_{cm})^2)] \ F_{\rho PA}(q_{cm})^2 \ . 
\label{gammaA}
\eea
From Eq.~(\ref{LrhoPV}) one obtains for vector mesons 
\bea
\Gamma_{V\to \rho P}(s) & = & \frac{G_{\rho PV}^2}{8\pi} \
\frac{IF (2I_\rho+1)}{(2I_V+1)(2J_V+1)}
\int\limits_{2m_\pi}^{M_{max}} \frac{M dM}{\pi} \  A^\circ_\rho(M)
\nonumber\\
 & & \qquad\qquad\quad \times \ 2 q_{cm}^3 \ F_{\rho PV}(q_{cm})^2 \ .  
\label{gammaV}
\eea 
Besides the coupling constant, $G$, the cutoff parameter $\Lambda$ 
entering the hadronic vertex form factors is a priori unknown. 
From principle reasoning it should
be in a sensible  range for hadronic processes, \ie, $\Lambda \le$~1-2~GeV. 
The hadronic widths are not very sensitive to the precise
value of $\Lambda$, \eg, variations in the above mentioned range entail
variations in $G^2$ of typically 10\% or less. However, as has been 
stressed in Ref.~\cite{RG99}, one can do better 
by simultaneously adjusting the radiative
decay widths of the resonances. 
Within the vector dominance model the
radiative decay widths follow from the hadronic ones by 
(i) taking the $M^2\to 0$ limit, \ie, substituting 
 $A_\rho^\circ(M)=2\pi \delta(M^2)$ for real photons, 
(ii) supplying the VDM coupling constant $(e/g)^2\simeq 0.052^2$ 
and (iii) omitting the $(2I_\rho+1)$ isospin degeneracy factor for 
the final state. This yields for both axialvector and vector resonances  
($R=A,V$)
\beq
\Gamma_{R\to \gamma P} = \frac{G_{\rho P R}^2}{8\pi} \
\left(\frac{e}{g}\right)^2 \ \frac{IF}{(2I_{R}+1)(2J_{R}+1)} \ 
2q_{cm}^3 \ F_{\rho PR}(q_{cm})^2 \ .
\label{gammaAV}
\eeq
Since the decay momentum, $q_{cm}$, acquires its maximum
value at the photon point it is clear that the latter is more sensitive
to the form-factor cutoff. With the dipole form factors of 
Eq.~(\ref{ffrhohR}) a universal $\Lambda_{\rho PR}=$~1~GeV yields quite 
satisfactory results for the decay widths of most resonances 
(cf.~Tabs.~\ref{tab_rhoPM}, \ref{tab_rhoPMfit}). 
\begin{table}[!htb]
\bce
\begin{tabular}{c|ccccc}
 $R$ & $IF(\rho hR)$ & $G_{\rho hR}$ [GeV$^{-1}$] & $\Lambda_{\rho hR}$ [MeV] &
$\Gamma^0_{\rho h}$ [MeV] & $\Gamma^0_{\gamma h}$ [MeV] \\
\hline
$\qquad\omega(782)\qquad$ & 1 & 25.8  & 1000 & 3.5 & 0.72 \\
$h_1(1170)$               & 1 & 11.37 & 1000 & 300 & 0.60 \\
$a_1(1260)$               & 2 & 13.27 & 1000 & 400 & 0.66 \\
$K_1(1270)$               & 2 & 9.42  & 1000 &  60 & 0.32 \\
$f_1(1285)$               & 1 & 35.7  &  800 &   3 & 1.67  \\
$\pi(1300)$               & 2 & 9.67  & 1000 & 300 &  0   \\
$a_2(1320)$               & 2 & 5.16  & 2000 &  80 & 0.24  \\
\end{tabular}
\ece
\caption{\it Results of a fit~\protect\cite{RG99}
to the decay properties of $\rho$-$h$
induced mesonic resonances $R$ with masses $m_R\le 1300$~MeV (the
$f_1(1285)$, $\pi(1300)$ and $a_2(1320)$ 
coupling constants are in units of GeV$^{-2}$). }
\label{tab_rhoPMfit}
\end{table}
If suitable data are available, an 
additional consistency check can be performed  by comparing
to the dilepton Dalitz decay spectra for $R\to Pl^+l^-$ as, \eg,  done in 
Ref.~\cite{KKW96} for the case of the $\omega$ meson. 

A similar procedure can be applied for baryonic resonances. With
the commonly employed non-relativistic $\rho NB$ Lagrangians given
in Sect.~\ref{sec_rhoNres}, the decay width into $\rho N$ states becomes
\bea
\Gamma_{B\to \rho N}(\sqrt{s}) & = & \frac{f_{\rho NB}^2}{4\pi m_\rho^2}
\ \frac{2m_N}{\sqrt{s}} \ \frac{(2I_\rho+1)}{(2J_B+1)(2I_B+1)} \
SI(\rho NB)
\nonumber\\
 & &  \quad  \int\limits_{2m_\pi}^{\sqrt{s}-m_N}
\frac{M dM}{\pi} A^\circ_\rho(M) \ q_{cm} \ F_{\rho NB}(q_{cm})^2 \ 
v_{\rho NB}(M) \ 
\label{gammaB}
\eea
with the vertex function $v_{\rho NB}(M)=(0.5M^2+q_0^2)$ or $q_{cm}^2$ 
for $S$- or $P$-wave resonances, respectively.  
However, the experimental values for $\Gamma_{B\to \rho N}$ are often 
beset with substantial uncertainties, especially for resonance
masses below the free $\rho N$ threshold. Moreover, the simple 
version of VDM seems to be less accurate in the baryonic sector. As
a consequence radiative decay widths, $\Gamma_{B\to \gamma N}$, can easily be 
overestimated~\cite{FrPi}. 
But unlike the case for the mesonic resonances,  much more 
quantitative constraints for the radiative couplings of baryonic 
excitations can be drawn from the analysis of photoabsorption spectra
on single nucleons as well as nuclei, to be discussed in 
the following Section in detail. Before we come to that 
let us elaborate here on some further purely hadronic reactions which
can provide valuable, comprehensive information on low-density 
nuclear effects in the vector correlator. 

As first pointed out by Friman~\cite{Fr98} the analysis of  
$\pi N \to \rho N $ scattering data is closely related to the
modifications of the $\rho$-propagator in nuclear matter. 
Diagrammatically, any cut through the in-medium $\rho$-selfenergy 
insertions represents a pertinent scattering process. In particular, 
all single cuts going through $NN^{-1}$ lines in the diagrams of 
Fig.~\ref{fig_vtxcorr} or in the dressed single-pion propagator 
constitute a contribution to the $\pi N \to \rho N$ reaction. Formally, 
this amounts to taking the imaginary part of the relevant 
contributions to the in-medium $\rho$-meson selfenergy.  
By using the optical theorem and detailed balance 
one finds for the isospin-averaged cross section   
\beq
\sigma_{\pi N\to \rho N} (s,M) = \frac{-3 q_{cm} m_N}{k_{cm}^2 \sqrt{s}} 
\lim_{\varrho_N\to 0} 
\frac{{\rm Im} \Sigma_\rho^\circ(M)-{\rm Im} \hat{\Sigma}_{\rho\pi N}(M)}
{\varrho_N} \ ,  
\eeq
where $\hat{\Sigma}_{\rho\pi N}$ denotes the in-medium selfenergy containing
only the diagrams with $\pi N N^{-1}$ cuts. The center-of-mass 
3-momenta $k_{cm}$ and $q_{cm}(M)$ belong to the incoming pion and 
outgoing $\rho$ meson
with fixed mass $M$, respectively. For comparison with experimental data, 
$\sigma_{\pi N\to \rho N} (s,M)$ has to be integrated over the free  
$\rho$-meson spectral mass distribution, \ie, 
\beq
\sigma_{\pi N\to \rho N} (s)= \int_{2m_\pi}^{\sqrt{s}-m_N} \frac{M dM}{\pi} 
\ A^\circ_\rho(M) \ \sigma_{\pi N\to \rho N} (s,M) \ .  
\label{Xpinrhon}
\eeq 
Two types of contributions to $\hat{\Sigma}_{\rho\pi N}$ arise: 
(i) from the medium modifications in the pion cloud of the $\rho$ meson, 
namely pion-induced $NN^{-1}$ excitations corresponding
to $t$-channel pion exchange in the $\pi N \to \rho N $ reaction, and 
(ii) from Rhosobar-type excitations through the $B\to \pi N$ partial decay 
widths corresponding to the $s$-channel 
processes $\pi N\to B\to \rho N$. 
The surprising result~\cite{Fr98} is that
the total cross section (\ref{Xpinrhon}) is very sensitive 
to the cutoff parameter $\Lambda_{\pi NN}$ in the $\pi NN$ form factor 
appearing in the pion cloud. Using the standard monopole form, 
$\Lambda_{\pi NN}$-values of slightly below 400 MeV already saturate 
the experimentally measured cross sections above the free $\rho N$ 
threshold. When additionally allowing
for the type (ii) $s$-channel baryon resonance contributions 
which are essential for the description of the photoabsorption 
spectra (see next Section), this number has to be further 
reduced to about $\Lambda_{\pi NN}\simeq 300$~MeV. Such on typical hadronic
scales 'unnaturally' small values are presumably related
to the lack of unitarity in the resulting Born-type scattering graphs for the
$\pi N\to \rho N$ process, as implicit  
in most models for the in-medium $\rho$-meson selfenergy. 
In fact, rather soft $\pi NN/\Delta$ form factors  
(with 300-500~MeV cutoffs) have been encountered in the literature, 
\eg,  in separable models of $\pi N$ scattering~\cite{Layson} (which  
resembles our model for the pion selfenergy) or pion 
photoproduction~\cite{KoMo}.  

Other ways to extract information on the $\rho$-meson properties 
in nuclear matter which actually go beyond the low-density limit
might be provided by two-pion production experiments on 
nuclei. Pion-induced experiments of the type $\pi +A\to A+\pi\pi$ 
have been performed
at TRIUMF for $\pi\pi$ invariant masses of up to 400~MeV. This mass regime 
is, however, dominated by $S$-wave states~\cite{CHAOS}. 
Alternatively, one could use proton-induced reactions as had been proposed
for  2.5-2.9~GeV energy beams at SATURNE~\cite{SATURNE}. At low scattering
angle one might be able to probe the in-medium $\rho$-meson spectral function 
close to the $q_0=q$ line around $q_0$=1~GeV, where $P$-wave Rhosobars
are expected to be important.
Very interesting results have also been reported by the TAGX 
collaboration~\cite{TAGX} for the reaction 
$^3{\rm He}(\gamma ,\pi^+ \pi^- )ppn$  
where, based on a partial wave analysis of the outgoing pions,
the data have been interpreted in terms of a large downward
shift of the in-medium $\rho$-meson mass, although no calculations
using medium-modified $\rho$-meson spectral functions are available yet. 
Another possibility might be provided by extracting phase information 
on electro-produced $\rho$-mesons in $A(e,e')$ reactions at 
TJNAF~\cite{Pree}. 

For the $\omega$ meson a similar analysis  has been 
performed in the $\pi N \to \omega N $ reaction~\cite{Fr98} which
also hints at a soft $\pi NN$ form factor. 
In Refs.~\cite{Fr98,FLW98} the effects of direct $\omega N$ scattering 
have been assessed. It seems that the $\omega$-$N(1520)N^{-1}$ excitation 
(the 'Omegasobar') plays an equally important
role as in the $\rho N$ interaction, leading to strong effects for the 
$\omega$-meson spectral function in nuclear matter.

A promising experiment to assess the in-medium $\omega$ properties
is planned at GSI~\cite{Metag}. In pion-induced reactions 
(via the elementary process $\pi N\to \omega N$),  
$\omega$ mesons can be produced in nuclei almost recoil free, thus
allowing sufficient time for decay in the nuclear environment. The 
invariant mass spectrum will then be measured through the 
dilepton channel. Another option is the proposed transfer
reaction $d+ A \to ^3$He~$ + \omega (A-1)$~\cite{Haya}.

\subsection{Photoabsorption Spectra}
\label{sec_photoabs}
A very important consistency check for any model of dilepton production 
can be inferred from photoabsorption spectra on both
proton and nuclei~\cite{SYZ2,KW96,RUBW,SZ99}. 
They represent the $M^2=0$ limit of the (time-like) dilepton
kinematics and are most relevant for heavy-ion energy regimes 
with sizable baryon densities (on the order of normal nuclear matter
density and above). These are clearly realized at BEVALAC/SIS energies but 
apparently also at the CERN-SpS where, as seen from rapidity 
spectra, significant baryonic stopping is exhibited. 
Whereas the absorption spectra on the proton 
provide low-density constraints, the various nucleus data constitute
a true finite-density test ground with the additional advantage over
hadronic probes that the incoming photon suffers little  
absorption thus probing the inner, most dense regions of nuclei. 
The proton photoabsorption cross sections have been used by several groups
to check their models for dilepton production.  

In the microscopic BUU transport model of Effenberger \etal~\cite{Effe1} 
the elementary $\gamma N$ cross sections have been calculated in 
terms of resonance contributions from $\Delta(1232)$, $N(1520)$, $N(1535)$
and $N(1680)$  as well as smooth background parameterizations of one- and
two-pion photoproduction amplitudes to reproduce the nucleon data. 
When moving to finite nuclei, medium effects have been accounted
for through modifications in the resonance widths (explicitely treated
in terms of collisional broadening and Pauli-blocking), collective
potentials in $\Delta$ and nucleon propagation as well as  
Fermi motion. Reasonable agreement with the experimental 
data is obtained. However, applications for dilepton production in 
nucleus-nucleus colisions are not available yet. 

In the 'master formula approach' for the vector correlator~\cite{SYZ1,SYZ2}
(cf.~Sect.~\ref{sec_chired}), 
$\gamma$ absorption spectra have been calculated in Refs.~\cite{SYZ2,SZ99}. 
Fig.~\ref{fig_gamsyz} shows the results for the nucleon (left panel)  and 
nuclei (right panel). 
\begin{figure}[!htb]
\vspace{0.8cm}
\epsfig{figure=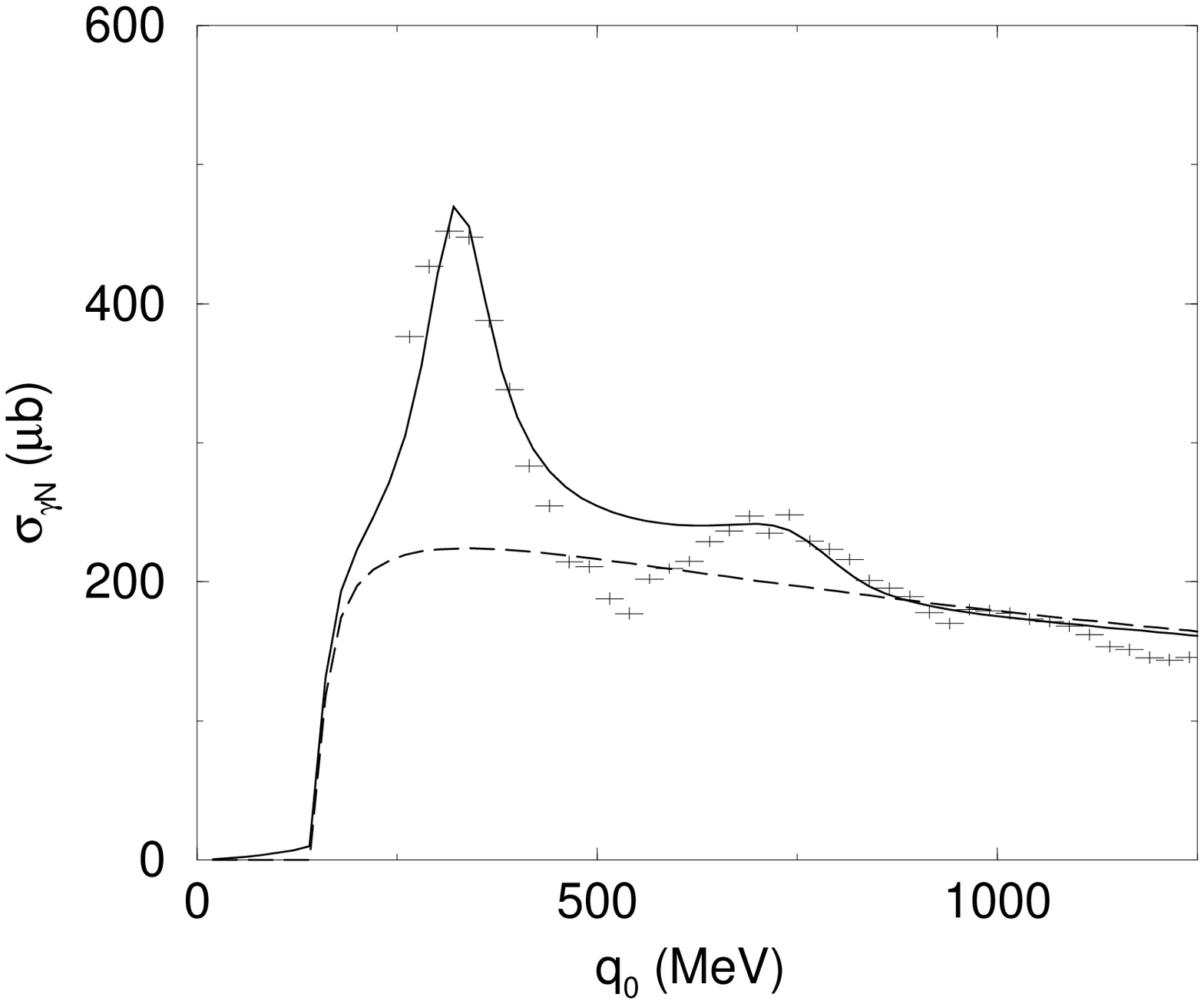,height=5cm,width=7cm}
\hspace{0.5cm}
\epsfig{figure=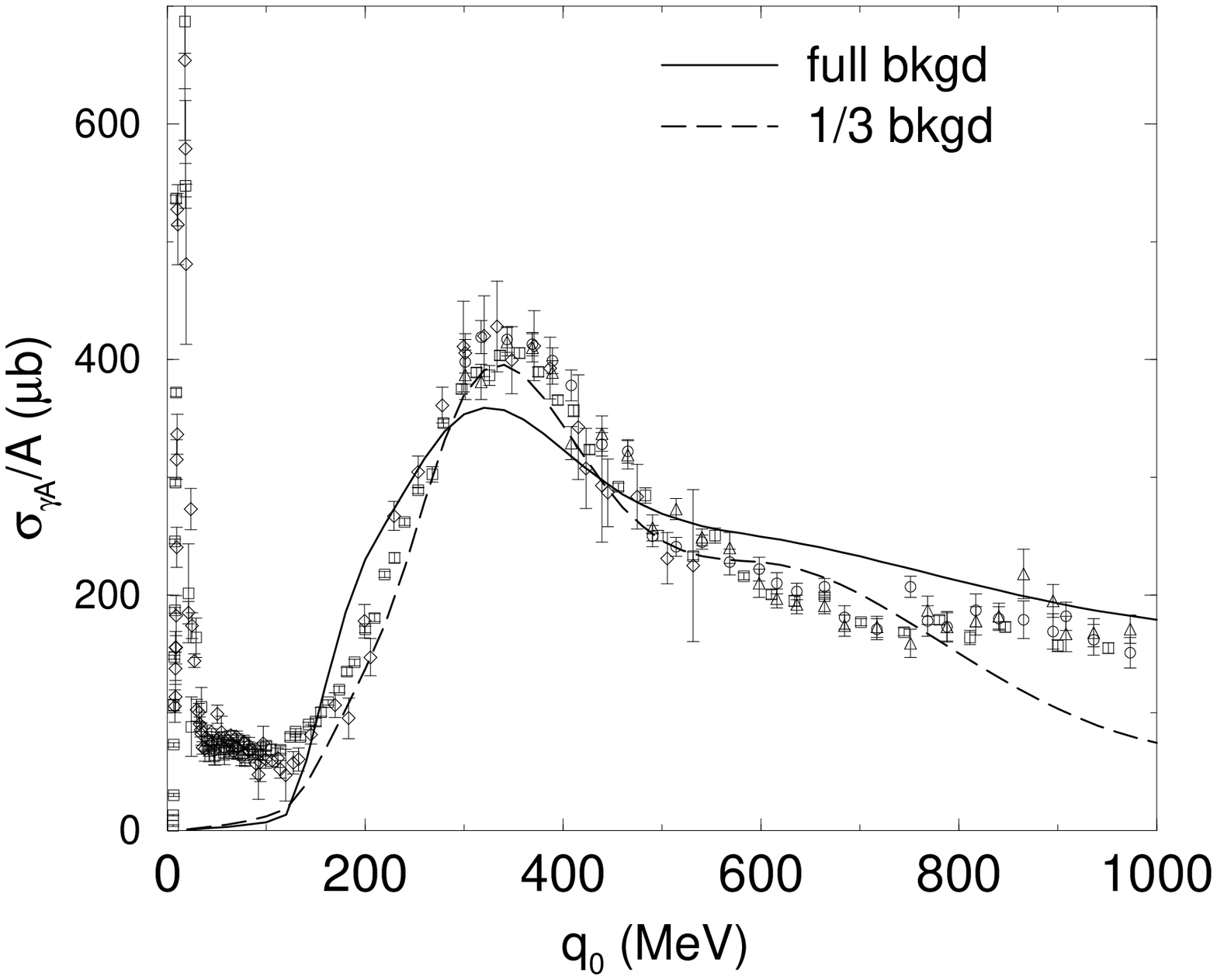,height=5cm,width=7cm}
\caption{Total photoabsorption cross section on the
nucleon (left panel) and on nuclei (right panel) within the chiral
reduction approach taken from Ref.~\protect\cite{SZ99}; the nucleon
and nuclei data are from Ref.~\protect\cite{PDG94} and
Refs.~\protect\cite{Saclay,Ahrens,Mainz,Frascati}, respectively.}
\label{fig_gamsyz}
\end{figure}
The dashed curve in the left panel represents a parameter-free prediction 
of the contribution from non-resonant $\pi N$ states 
(opening at the threshold energy of about $q_0=140$~MeV) as extracted from
$\pi N$ scattering data. It corresponds
to non-resonant one-pion photoproduction and  essentially saturates the 
data beyond the $\Delta$ resonance region. This is  
at variance with both the BUU calculations mentioned above~\cite{Effe1} 
as well as experimental phase analyses~\cite{Walk,PDG94,Krusch} which  
show that for incoming photon energies $q_0\ge 0.6$~GeV two-pion production 
processes start to prevail~\cite{Brag95}. The photoabsorption cross
sections on nuclei are calculated by averaging the nucleon cross section over
Fermi motion, 
\beq
\frac{\sigma_{\gamma A}}{A}
=\int \frac{d^3p}{4\pi p_F^3/3} \ \Theta(p_F-|\vec p|) \sigma_{\gamma N}(s) \ , 
\eeq
yielding a  reasonable fit to the experimental data.
The dashed curve in the right panel of 
Fig.~\ref{fig_gamsyz} was obtained with the $\pi N$ background
being artificially reduced  by a factor of three (compensated by larger 
$\Delta(1232)$ and $N(1520)$ contributions to still reproduce the nucleon 
data) which seems to improve the fit. Note here that a simple averaging
over Fermi motion seems to give a sufficient smearing of the 
$N(1520)$ resonance, in contrast to previous findings 
in the literature~\cite{AGL94,Kon94,Effe1,RUBW} 
where a strong in-medium resonance broadening was required.  

In the effective Lagarangian approaches where the in-medium vector meson
properties are calculated in terms of explicit interactions with 
surrounding matter particles, the resulting spectral functions can 
be related to total photoabsorption cross sections. In 
Refs.~\cite{KW96,KKW97} the electromagnetic current-current correlation
function has been expressed to lowest order in density 
in terms of the Compton tensor for forward $\gamma N$ scattering, 
$T_{\gamma N}^{\mu\nu}$, as
\beq
\Pi^{\mu\nu}_{\rm em}(q)=\Pi^{\circ\mu\nu}_{\rm em}(q)
+ g_N \int \frac{d^3p}{(2\pi)^3} \ \frac{M_N}{E_p^N} \ 
T_{\gamma N}^{\mu\nu}(q;p) \ \Theta(p_F-|\vec p|) \ .  
\eeq 
For real photons, only the transverse part of $T_{\gamma N}$
survives. Taking the low-density limit, $p_F\to 0$, and using the optical 
theorem, 
\beq
\sigma_{\gamma N}(q_0)=-\frac{4\pi\alpha}{q_0} \ 
{\rm Im} T^T_{\gamma N}(q_0,|\vec q|=q_0) \ ,  
\eeq
as well as VDM, 
\beq
{\rm Im} \Pi^{T}_{\rm em}(q_0,\vec q)
=\sum\limits_{V=\rho,\omega,\phi}
\frac{1}{g_V^2} \ {\rm Im} \Sigma_V^T(q_0,\vec q) \ ,  
\eeq
one obtains the desired relation between the isospin-averaged 
$\gamma$-nucleon cross section and the vector meson selfenergies:  
\beq
\sigma_{\gamma N}(q_0)= -\frac{4\pi\alpha}{q_0} 
\lim_{\varrho_N\to 0} \frac{1}{\varrho_N} \sum\limits_{V=\rho,\omega,\phi}
\frac{1}{g_V^2} \ {\rm Im} \Sigma_V^T(q_0,|\vec q|=q_0) 
\label{phcross0}
\eeq
(note that ${\rm Im} \Pi^\circ(M^2)$ vanishes below the two-pion threshold
$M=2m_\pi$). Unfortunately, no explicit results for photoabsorption spectra 
are available in the approach of Refs.~\cite{KW96,KKW97}.  

A slightly different way of deriving analogous relations has been pursued in 
Ref.~\cite{RUBW}. Here, the starting point is the total 
cross section of a photon per unit volume element $d^3x$ 
of cold nuclear matter, averaged over the incoming polarizations,  
\bea
\frac{d\sigma}{d^3x} &=& \frac{1}{2} \sum_\lambda 
\sum\limits_f \frac{1}{v_{in}} \frac{1}{2q_0}
\ |{\cal M}_{fi}|^2 \ (2\pi)^4 \ \delta^{(4)}(p_f-q)  
\nonumber\\
 &=& -\frac{4\pi\alpha}{q_0}  \ \frac{1}{2} \sum_\lambda 
\varepsilon_\mu(q,\lambda) \ \varepsilon_\nu(q,\lambda) 
\ {\rm Im}\Pi^{\mu\nu}_{\rm em}(q) \ , 
\label{phcross1}
\eea
where ${\cal M}_{fi}=e \ \langle f|j^{\rm em}_\mu(0)|0\rangle \ 
\varepsilon^\mu(q,\lambda)$ is the
transition matrix element of the electromagnetic current, taken between 
the initial nuclear ground state $|i\rangle = |0\rangle$ and 
final states $|f\rangle$ with $\varepsilon_\mu(q,\lambda)$ being the photon 
polarization vector. Neglecting small contributions from 
isoscalar vector mesons within the VDM, the electromagnetic correlator
can be saturated by the neutral $\rho$ meson using the field-current 
identity 
\beq
j^\mu_{\rm em}=(m_\rho^{(0)})^2/g_\rho \ \rho^\mu_3 \ .  
\label{fci} 
\eeq
Using the completenes relation for photon polarization vectors, 
\beq
\sum_\lambda \varepsilon_\mu(q,\lambda) \ \varepsilon_\nu(q,\lambda)
=-g_{\mu\nu} \ ,  
\eeq
the total photoabsorption cross section, normalized to the number 
of nucleons $A$, takes the form  
\bea
\frac{\sigma_{\gamma A}^{\rm abs}}{A} 
&=& \frac{1}{\varrho_N} \ \frac{d\sigma}{d^3x}
\nonumber\\ 
 &=& -\frac{4\pi\alpha}{q_0} \
\frac{(m_\rho^{(0)})^4}{g_\rho^2} \frac {1}{\varrho_N} \
{\rm Im}D_\rho^T(q_0,|\vec q|=q_0) \ 
\label{phcross2}
\eea
with the transverse in-medium $\rho$-meson propagator  
\begin{equation}
{\rm Im}D_\rho^T=\frac{{\rm Im} \Sigma^T_\rho}{|M^2-(m_\rho^{(0)})^2
-\Sigma^T_\rho|^2} \ 
\end{equation}
(for $M^2=0$ the longitudinal part vanishes identically).   
Note that Eq.~(\ref{phcross2}) does not involve any low-density
approximations. However, for $\varrho_N\to 0$ and $A=1$ (corresponding
to the absorption process on a single nucleon), one has
\beq
\lim_{\varrho_N\to 0}{\rm Im} D_\rho^T(q_0,\vec q)= 
\lim_{\varrho_N\to 0}{\rm Im}\Sigma^T_\rho(q_0,\vec q)
/(m_\rho^{(0)})^4 \ , 
\eeq
thus readily recovering Eq.~(\ref{phcross0}). 
In the model of Ref.~\cite{RUBW}
the $\rho$-meson selfenergy in nuclear matter receives  
two contributions, 
\begin{equation}
\Sigma_\rho^{T} = \Sigma_{\rho\pi\pi}^{T} + \Sigma_{\rho N}^{T} \ , 
\label{Siglt}
\end{equation}
representing  the renormalization of the pion cloud through
$\pi NN^{-1}$ and $\pi\Delta N^{-1}$ excitations as well as 
direct $\rho BN^{-1}$ interactions, respectively. However,  
as has been noted long ago, the most simple version of the VDM
(\ref{fci}) typically results in an overestimation of the
$B\to N\gamma$ branching fractions when using the hadronic coupling
constants deduced from the $B\to N\rho$ partial widths. One
can correct for this by employing an improved version of the VDM~\cite{KLZ},
which allows to adjust the $BN\gamma$ coupling $\mu_B$
(the transition magnetic moment) at the photon
point independently~\cite{FrPi}. It amounts to replacing the combination
$(m_\rho^{(0)})^4\;{\rm Im}D_\rho^T(q_0,\vec q)$
entering Eq.~(\ref{phcross2}) by the following 'transition form factor':
\begin{eqnarray}
{\cal F}^T(q_0,\vec q) & = & -{\rm Im}\Sigma^T_{\rho\pi\pi} |d_\rho-1|^2
-{\rm Im}\Sigma^T_{\rho N} |d_\rho-r_B|^2
\nonumber\\
d_\rho(q_0,\vec q) & = &
\frac{ M^2-\Sigma^T_{\rho\pi\pi}- r_B \Sigma^T_{\rho N} }
{ M^2-(m_\rho^{(0)})^2-\Sigma^T_{\rho\pi\pi}-\Sigma^T_{\rho N} } \ , 
\label{KLZ}
\end{eqnarray}
where
\begin{equation}
r_B=\frac{\mu_B}{\frac{f_{\rho BN}}{m_\rho} \
\frac{e}{g}}
\end{equation}
denotes the ratio of the photon coupling to its value in the naive VDM.
In principle, each resonance state $B$ can be assigned a separate
value for $r_B$ but, as will be seen below, reasonable fits to the
photoabsorption spectra can be achieved with a single value 
making use of some latitude in the hadronic
couplings $f_{\rho BN}$ and form-factor cutoff parameters $\Lambda_\rho$  
within the experimental uncertainties of the
partial widths (\ref{gammaB}). 
The final expression to be used for the photoabsorption calculations then
reads
\begin{equation}
\frac{\sigma_{\gamma A}^{\rm abs}}{A}=\frac{4\pi\alpha}{g^2 q_0} \
 \frac {1}{\varrho_N} \ {\cal F}^T(q_0,|\vec q|=q_0) \ .
\label{cross3}
\end{equation}

Let us first discuss the $\gamma N$ spectra, in particular 
the role of background contributions (which in this context 
can be regarded as 'meson exchange' processes), encoded in the 
low-density limit of $\Sigma_{\rho\pi\pi}$. With the coupling constants of 
the $\pi NN$- and $\pi\Delta N$-vertices fixed at their standard values 
(cf. Tab.~\ref{tab_pind}), the strength of the background is 
controlled by the cutoff $\Lambda_{\pi NN}$ in the 
phenomenological (hadronic) vertex form factors. In early  applications
to dilepton spectra~\cite{HeFN,ChSc,AKLQ,CRW,RCW,KKW97} the values 
were chosen around (1--1.2)~GeV in reminiscence to the Bonn 
potential~\cite{MHE}.
However, with the analysis of photoabsorption spectra~\cite{RUBW} 
it became clear that much lower values
are required, at most of around 600~MeV. Shortly thereafter, Friman pointed
out~\cite{Fr98} that this is still too large to be  
compatible with available $\pi N\to \rho N$ scattering data as discussed 
in the previous Section, enforcing even smaller values~\cite{Morio98}, 
thereby further suppressing the 'background' contribution in the 
photoabsorption spectra. Using the value of 
$\Lambda_\pi\simeq 300$~MeV deduced in Sect.~\ref{sec_hadscat} 
actually improves the description of 
the $\gamma N$ spectra in the 'dip region' between the $\Delta(1232)$
and the $N(1520)$ resonances~\cite{Morio98} as compared to the 
results obtained with $\Lambda_{\pi NN}=550$~MeV in Ref.~\cite{RUBW}, 
see left panel of Fig.~\ref{fig_gam69}. 
\begin{figure}[t]
\epsfig{figure=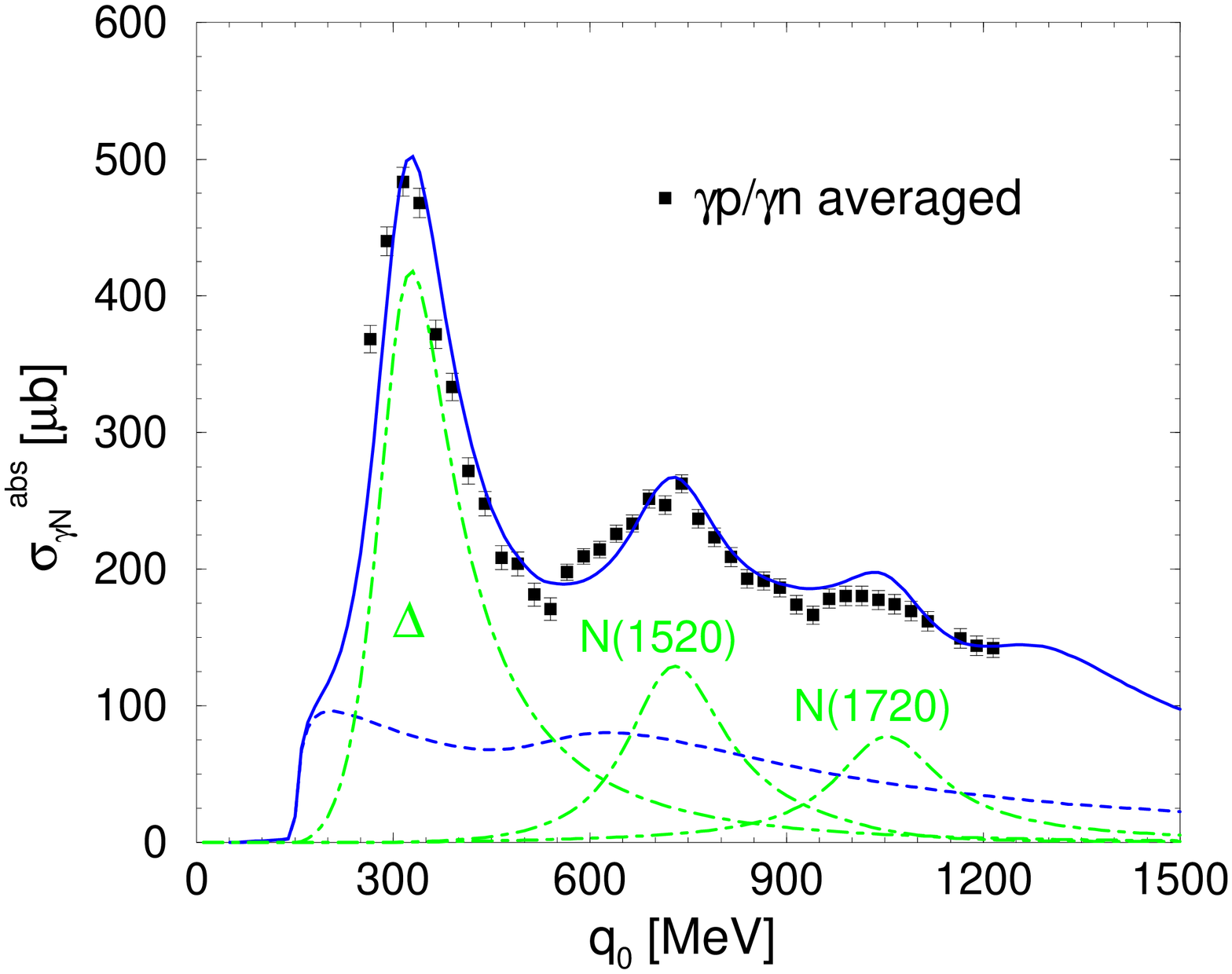,width=6cm,angle=-90}
\epsfig{figure=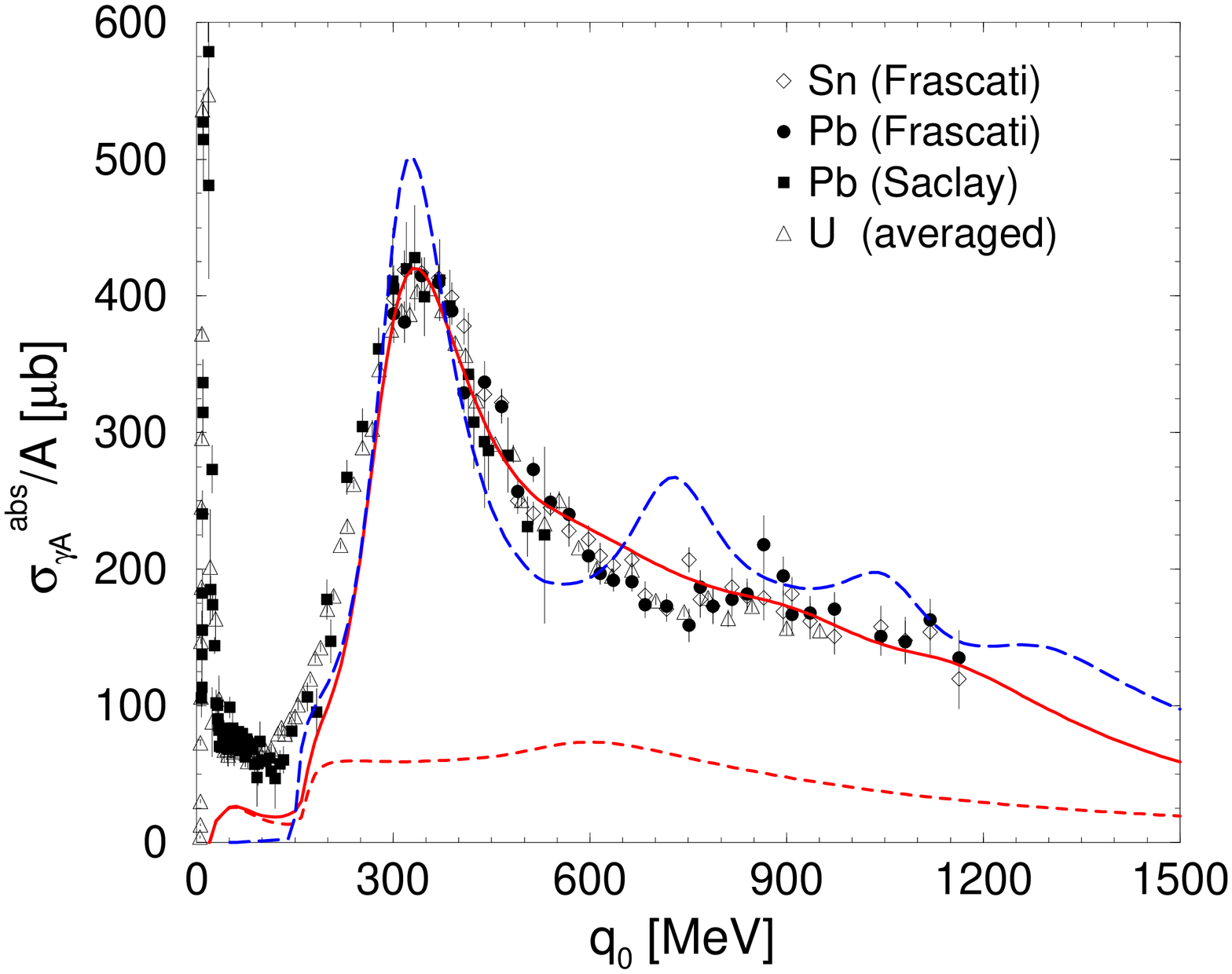,width=6cm,angle=-90}
\caption{Total photoabsorption cross section on the nucleon (left panel) 
and on nuclei (right panel) as obtained in the $\rho$ spectral function 
approach~\protect\cite{RUBW,RW99}.
Left panel: full result of
the fit using the parameters of Tab.~\protect\ref{tab_rhonb} (solid line),
$\pi\pi$ 'background' (dashed line) as well as
the three dominant $\rho N$ resonances $\Delta$(1232), $N$(1520)
and $N$(1720) (dashed-dotted lines); the data are averaged over 
proton~\protect\cite{Arm72p} and neutron~\protect\cite{Arm72n} 
measurements. Right panel: full result (solid line) and 
non-resonant background contributions (short-dashed line) for
$\varrho_N=0.8\varrho_0$, as well as  the lowest-order-density result from 
the nucleon fit (long-dashed line); the data are taken from 
Refs.~\protect\cite{Saclay,Ahrens,Mainz,Frascati}.}
\label{fig_gam69}
\end{figure}
Moreover, the non-resonant contribution
in the $\Delta$ region now amounts to about 70~$\mu b$, which 
coincides with what has been extracted from experimental phase 
analyses~\cite{Walk,PDG94,Krusch}. On the other hand, 
the parameter-free assessment 
of this background component in the 'master formula approach'~\cite{SYZ2} 
differs by approximately a factor of 3 across the entire photon energy 
range under consideration
(a very similar result is found when using the Bonn-value 
of 1.2~GeV deduced from $\pi N$ data in the framework of 
Ref.~\cite{UBRW}). Given the magnitude 
of the background, the resonance contributions encoded in $\Sigma_{\rho B}$
are readily adjusted to obtain a good fit to the $\gamma N$ data. 
The employed resonances and vertex 
parameters are summarized in Tab.~\ref{tab_rhonb}. One should note that 
an unambiguous determination of coupling constants and form-factor cutoffs
paramaters is not possible from the total absorption cross section alone. 
To further disentangle them 
more exclusive reaction channels (\eg, one- and two-pion
photoproduction) need to be analyzed.
Nevertheless, for the 
purpose of predicting reliable dilepton production rates the constraints
from the total absorption cross sections give reasonable 
confidence. 

For absorption spectra on nuclei,
one experimentally observes an almost independent scaling with the
mass number $A$ of different nuclei (cf.~right panel of 
Fig.~\ref{fig_gam69}).
This suggests that both surface and nuclear structure effects play a minor 
role as might be expected since the incoming photon predominantly 
probes the interior of the nuclei. Therefore it appears justified  
to perform the calculations  for the idealized situation of 
infinite nuclear matter at an average
density, which has been taken as $\bar{\varrho}_N$=0.8$\varrho_0$ 
(in fact, the results for the normalized cross section, Eq.~(\ref{cross3}),
depend only weakly on density within reasonable limits of  
$0.6\le \varrho/\varrho_0 \le 1$). 
As compared to the free nucleon two additional features appear in the nuclear
medium: short-range correlation effects in the resummation of
the particle-hole bubbles and in-medium
corrections to the resonance widths. Due to the rather soft
form factors involved, the $P$-wave pion-induced excitations turn out to 
favor rather small Landau-Migdal parameters of
$g'_{NN}$=0.6 and $g'_{\alpha\beta}$=0.2 for all other transitions
including $P$-wave Rhosobars. 
The $S$-wave $\rho BN^{-1}$ bubbles show only marginal evidence
for short-range correlations with a slight tendency towards larger 
values (the results in the left panel of Fig.~\ref{fig_gam69} have 
been obtained with $g'_{S-wave}=0.6$).  
However, as already mentioned above, the observed disappearance of the
$N$(1520) resonance in the nuclear medium requires a large in-medium
increase of its width. Such a behavior has indeed been found in a 
selfconsistent microscopic 
treatment of the $\rho$-meson spectral function and the $N$(1520) width
in nuclear matter~\cite{PPLLM}, which is based on a very similar framework as
employed here. The actual value used for
$\Gamma_{N(1520)}^{med}$ in Fig.\ref{fig_gam69} 
is in accordance with Ref.~\cite{PPLLM}.   
On the other hand, the net in-medium
correction to the $\Delta$(1232) width is quite small. This
reflects the fact that a moderate in-medium
broadening is largely compensated by Pauli blocking effects on the
decay nucleon. The sensitivity of the results with respect to the in-medium
widths of the higher lying resonances is comparatively small. 
It is also noteworthy that below the pion
threshold some strength appears. This is nothing but the
well-known 'quasi deuteron' tail above the giant dipole resonance, arising
from pion-exchange currents. These are naturally included in the 
spectral function framework through higher orders in density in  
the pion cloud modifications.
Also note that a linear-density approximation to Eq.~(\ref{KLZ}), which is
equivalent to the $\gamma N$ result, does not properly reproduce 
the $\gamma A$ data.

\section{Dilepton Rates in Hot and Dense Matter}  
\label{sec_dlrates}
Having discussed the approaches that have been constructed to 
compute the vector correlator in hot and dense matter as well as the 
corresponding efforts
and philosophies to constrain the underlying assumptions, we now
turn our attention to the results for the  dilepton production rates. 
To be able to draw any conclusions from the eventual analysis
of dilepton spectra in heavy-ion collisions it is essential
to assess  the differences and similarities in the 
model predictions on an equal footing, \ie, without the complications
arising from modeling the space-time history of the collisions or 
experimental acceptance cuts. 
In practically all microscopic approaches one calculates 
an eight-fold differential rate per unit four momentum
and four volume. To facilitate the comparison, it has become common 
to focus on the 3-momentum integrated rates 
\beq
\frac{dR_{l^+l^-}}{dM^2}(M) = \int 
\frac{d^3q}{2q_0} \  \frac{dR_{l^+l^-}}{d^4q}(q_0,\vec q)
\label{dRdM2} 
\eeq
at fixed temperature $T$ and baryon density $\varrho_B$ 
(or, equivalently, baryon chemical potential $\mu_B$). In the following
we will first address the class of more 'conventional', hadronic calculations 
based on essentially known interactions,  and then put these into context 
with results from scenarios associated with 
'new' physics such as  QGP or DCC formation.

\subsection{Comparison of Hadronic Approaches}                           
\label{sec_hadrates}
As elaborated in Sect.~\ref{sec_DPR+VM} the general expression for the 
dilepton production rate in a hadronic medium of given temperature $T$ 
and baryon chemical potential $\mu_B$ can be written as
\begin{equation}
\frac{dR_{l^+l^-}}{d^4q}=L_{\mu\nu}(q) W^{\mu\nu}(q) \ ,  
\label{dRd4q} 
\end{equation}
where $L_{\mu\nu}$ and $W_{\mu\nu}$ denote the leptonic and hadronic tensor,
respectively, cf.~Eqs.~(\ref{Lmunu}) and (\ref{Wmunu}). The latter 
can also be written as  
\begin{equation} 
W^{\mu\nu}(q;T)=\sum\limits_i \frac{e^{E_i/T}}{\cal Z} 
\sum\limits_f \langle i| j^\mu(0) |f\rangle \langle f| j^\nu(0)|i\rangle
 (2\pi)^4 \delta^{(4)}(q+p_f-p_i) \ ,  
\label{Wmunu2}
\end{equation} 
which is straightforwardly related to 
the retarded current-current correlation function according to 
\begin{equation} 
-2 \ {\rm Im} \Pi^{\mu\nu}_{\rm em}(q)=(e^{q_0/T}-1) \ W^{\mu\nu}(q) \ . 
\label{ImPImunu}
\end{equation} 
Inserting Eqs.~(\ref{Lmunu}) and (\ref{ImPImunu}) into (\ref{dRd4q}),  
and exploiting gauge invariance, $q_\mu \Pi^{\mu\nu}_{\rm em}=0$, 
one obtains the general result 
\begin{equation} 
\frac{dR_{l^+l^-}}{d^4q}=-\frac{\alpha^2}{\pi^3 M^2} \ f^B(q_0;T) \
{\rm Im} \Pi_{\rm em}(q_0,\vec q) 
\label{dRd4q_2}
\end{equation}
with the thermal Bose occupation factor $f^B(q_0;T)=(e^{q_0/T}-1)^{-1}$
and the spin-averaged correlator 
\begin{equation}
{\rm Im} \Pi_{\rm em}(q_0,\vec q)=\frac{1}{3} 
\left[ {\rm Im}\Pi^L_{\rm em}(q_0,\vec q) + 
2 {\rm Im}\Pi^T_{\rm em}(q_0,\vec q) \right] \ , 
\end{equation} 
given in terms of its standard decomposition into longitudinal and
transverse projections.

\subsubsection{Effects of a Hot Meson Gas}

Let us start by considering a hot meson gas without any baryons. 
The most obvious source of dilepton radiation from such a system 
is the free $\pi^+\pi^-\to l^+l^-$ annihilation  process with  
no further medium effects included. This process is reliably described 
within the simple VDM framework, so that very little model uncertainty
is involved once the experimental data for the pion electromagnetic
form factor in the time-like region are properly accounted for. Therefore we  
will use this process as a standard baseline for comparing medium 
effects within  various approaches. In VDM (\ie, invoking 
the field-current identity 
(\ref{fci})) $\pi\pi$ annihilation proceeds via the formation of an 
intermediate 
$\rho$ meson. The corresponding hadronic tensor is then saturated 
by the $\rho$-meson propagator, 
\begin{equation}
W^{\mu\nu}(q)=-2 \ f^\rho(q_0;T) \ \frac{(m_\rho^{(0)})^4}
{g_{\rho\pi\pi}^2} \ {\rm Im}D_\rho^{\mu\nu}(q) \ .
\end{equation} 
Thus the dilepton production rate for $\pi^+\pi^-\to\rho\to\gamma^*\to e^+e^-$ 
becomes 
\begin{eqnarray}
\frac{dR_{\pi\pi\to ee}}{d^4q}(q_0,\vec q) &=& 
-\frac{\alpha^2 (m_\rho^{(0)})^4}{\pi^3 g_{\rho\pi\pi}^2} \
\frac{f^\rho(q_0;T)}{M^2} \ {\rm Im} D_\rho(q_0,\vec q) 
\nonumber\\
            &=&  -\frac{\alpha^2}{\pi^3 g_{\rho\pi\pi}^2} \ 
\frac{f^\rho(q_0;T)}{M^2} \
 {\rm Im} \Sigma_{\rho\pi\pi}^\circ(M) \ |F_\pi^\circ(M)|^2 \ , 
\label{pipirate}
\end{eqnarray}
where the second equality implies the use of the free $\rho$-meson 
propagator. As is well-known a thermodynamically equivalent way 
to describe the {\em same} 
contribution is to consider the decay of thermal $\rho$ mesons
in the system. Accounting for both $\pi\pi$ annihilation 
and $\rho$ decays clearly constitutes double counting. 

In-medium modifications to the free $\pi\pi$ annihilation process
in a hot meson 
gas have been studied by several groups, as we have already discussed 
in some detail in Sect.~\ref{sec_Vmodtemp}.  
At a fixed temperature of $T=150$~MeV the three-momentum integrated 
rates (as given by Eq.~(\ref{dRdM2})) are confronted 
in Fig.~\ref{fig_mesrates} for some of the different approaches available 
in the literature. One of the selected examples is of model-independent 
nature~\cite{SYZ1} putting the 
emphasis on chiral symmetry aspects of the interactions, whereas the 
other two ~\cite{GaLi,RG99} represent the more 
phenomenological approaches based on effective Lagrangians.  
\begin{figure}[!htb]
\bce
\epsfig{figure=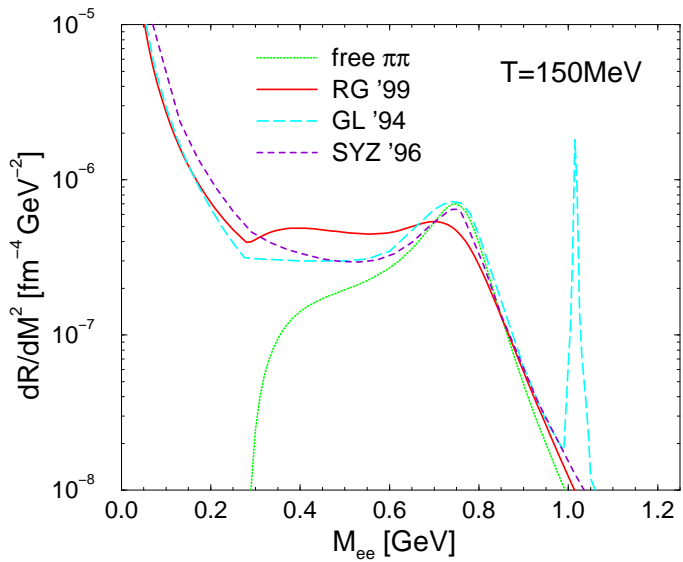,height=10cm,angle=-90}
\ece
\caption{Three-momentum integrated dilepton rates in a hot meson gas
at $T=150$~MeV in the hadron-based approaches of
Gale/Lichard~\protect\cite{GaLi} (long-dashed line),
Steele \etal~\protect\cite{SYZ1} (short-dashed line) and
Rapp \etal~\protect\cite{RCW,RG99} (full line). The dotted line represents
free $\pi\pi$ annihilation.}
\label{fig_mesrates}
\end{figure}


The chiral reduction formalism employed by Steele \etal~(which combines 
chiral Ward identities with experimental data on spectral functions 
in the vector and axialvector channels, dominated by the $\rho$ and 
$a_1(1260)$ mesons, respectively, cf.~Sect.~\ref{sec_chired})
is shown as the short-dashed line in Fig.~\ref{fig_mesrates}.
The appearance of the $a_1$ (predominantly formed in resonant
$\rho\pi$ scattering) in the electromagnetic rates 
represents the mixing effect of the vector and axialvector correlators
as dictated by chiral symmetry. 
Most of the enhancement over the free $\pi\pi$ rate for invariant masses below
$M_{ee}\simeq 0.6$~GeV can indeed be assigned to the (Dalitz-decay) 
tails of the $a_1(1260)$.  
On the other hand, note that there is practically 
no depletion of the $\rho$-meson peak around $M_{ee}\simeq m_\rho$. This is 
a consequence of the density expansion inherent in this virial-type 
expansion 
which evades any kind of diagrammatic resummations. Such resummations 
naturally occur in the propagator formalism:  
even if the (in-medium) selfenergy correction 
is evaluated to lowest order, the solution of the 
corresponding Dyson equation automatically generates iterations to 
all orders which typically leads to a downward shift of strength as we 
will see below.

In Ref.~\cite{GaLi} a large number of mesonic tree level scattering 
amplitudes involving $e^+e^-$ final states has been computed within  
the VDM, employing phenomenological Lagrangians compatible with
gauge invariance and inspired by the chiral properties of 
low-energy QCD~\cite{Meis88}.  The corresponding coupling constants 
have been determined along the lines discussed in Sect.~\ref{sec_hadscat} 
using the experimental branching ratios for radiative
decays, etc.. 
Within standard relativistic kinetic theory the dilepton 
production rate has then been obtained by suitable momentum integrations 
over the squared matrix elements including appropriate thermal occupation 
factors (for a given process, the kinetic theory expressions can be 
shown to be equivalent to standard finite-temperature field theoretic 
results to leading order in temperature).  
The long-dashed curve in Fig.~\ref{fig_mesrates} represents 
the final result of this analysis, also exhibiting 
substantial excess  over the free $\pi\pi$ rate for low masses 
$M_{ee}\le0.6$~GeV, which is  
predominantly generated by the radiative decay of the 
omega, $\omega\to \pi^0 e^+e^-$. Contributions from $a_1(1260)$
 mesons have not been included. The incoherent summation of the various
processes does again not induce any depletion of the peak, leading 
to a  close overall resemblance with the rate from the chiral 
reduction formalism~\cite{SYZ1} (short-dashed curve). 

In Ref.~\cite{RCW} the imaginary time (Matsubara) formalism has been 
employed to calculate the $\rho$ propagator in hot hadronic matter 
accounting for the full off-shell dynamics of the considered interactions. 
As far as meson gas effects are concerned, only resonant 
$\rho\pi\to a_1(1260)$ and $\rho K/\bar K \to K_1(1270)/\bar K_1(1270)$
contributions were included at the time, yielding  very 
similar results as have been 
obtained before in an (on-shell) kinetic theory 
framework~\cite{Ha95} (which is mainly 
due to the fact that both the $a_1$ and $K_1$ resonances are situated above 
the free $\rho\pi$ and $\rho K$ thresholds, respectively, resulting 
in little off-shell sensitivity). A more comprehensive analysis of the 
finite temperature $\rho$-meson selfenergy was performed in 
Ref.~\cite{RG99} along the same lines as in Ref.~\cite{RCW}.  The most 
notable of the additionally included mesonic resonances turned out 
to be the $\omega(782)$ meson accessible through interactions of  
off-shell $\rho$ mesons (with mass $M\simeq500$~MeV, 
cf.~Fig.~\ref{fig_sgrhoT}) with thermal pions.  
The corresponding
dilepton production process is $\rho\pi\to\omega\to \pi ee$, \ie,
the radiative Dalitz decay of thermal $\omega$ mesons. It indeed quantitatively
coincides with the equivalent contribution calculated in Ref.~\cite{GaLi}. 
In addition, the most simple temperature effect in the pion cloud
of the $\rho$ meson has been included, consisting of  
a Bose-Einstein enhancement in the in-medium
$\rho\to \pi\pi$ decay width.  Schematically written it modifies the 
imaginary part of the $\rho\pi\pi$ selfenergy as 
\beq
{\rm Im} \Sigma_{\rho\pi\pi}^\circ(M) \to {\rm Im} \Sigma_{\rho\pi\pi}(M;T)=
{\rm Im} \Sigma_{\rho\pi\pi}^\circ(M) \left[1+2 f^\pi(M/2)\right] \ , 
\eeq
cf.~also ~Eq.~(\ref{GpipiT}). 
The full result of Ref.~\cite{RG99} is shown by the solid  line in 
Fig.~\ref{fig_mesrates}: in the low-mass region it  
exceeds the results of both  
the chiral reduction approach~\cite{SYZ1} and of the incoherent 
summation of decay/scattering processes~\cite{GaLi}. 
The discrepancy to the latter can be traced back to the $a_1$-meson 
contribution and, more importantly, to the Bose enhancement in the 
$\rho\to \pi\pi$ decay width~\cite{RG99}. 
Also note that the $\rho$-meson spectral function calculation induces
a $\sim$~40\% suppression of the signal in the vicinity of the free 
$\rho$ mass. This is a characteristic feature of many-body type
approaches and is not present in density-expansion schemes as implicit
in the calculations of Gale/Lichard and  Steele \etal.   It can be easily
understood as follows~\cite{RG99}: the $\rho$
spectral function (which governs the dilepton rate in the many-body 
framework) can be schematically written in terms of the selfenergy as
\begin{equation}
{\rm Im} D_\rho=
\frac{{\rm Im} \Sigma_\rho}{|M^2-m_\rho^2|^2+|{\rm Im}\Sigma_\rho|^2} \ ,
\label{schematic}
\end{equation}
where we have absorbed the real part of the selfenergy in the (physical)
$\rho$ mass $m_\rho$. In the low-mass region, where
$m_\rho\gg M$ and $m_\rho^2\gg|{\rm Im} \Sigma_\rho|$, the denominator
is dominated by $m_\rho$ so that
\begin{equation}
{\rm Im} D_\rho(M \ll m_\rho)
\propto \frac{{\rm Im} \Sigma_\rho}{m_\rho^4}  \ .
\end{equation}
Since ${\rm Im} \Sigma_\rho$  basically encodes a summation of scattering
amplitudes times (pion-) density, one immediately recognizes the close
analogy to kinetic theory or low-density expansions. On
the other hand, in the vicinity of the $\rho$-peak, where
$M\simeq m_\rho$, the denominator
in Eq.~(\ref{schematic}) is dominated by ${\rm Im}\Sigma_\rho$ so that
\begin{equation}
{\rm Im} D_\rho(M\simeq m_\rho)\propto \frac{1}{{\rm Im} \Sigma_\rho} \ ,
\end{equation}
demonstrating that the consequence of an increase in density
is a suppression of the maximum, which cannot be straightforwardly casted
in a low-density expansion.

\subsubsection{Effects of Finite Baryon Density}
The situation becomes more involved when comparing dilepton production
rates in the presence of baryons. Most of the investigations so far  
have been  restricted to the case of   
 nucleons at zero temperature, which is particularly obvious for model  
constraints inferred from $\rho N$ or $\gamma N / \gamma A$ scattering 
data as discussed in Sect.~\ref{sec_constraints}. The first complication 
arises from the fact that in any finite temperature system with a
net baryon density, some fraction of the nucleons will 
be thermally excited into  
baryonic resonances and therefore, in principle, should not 
be included in medium modifications generated by nucleons. For instance, at 
a temperature of $T=150$~MeV and total baryon density $\varrho_B=\varrho_0$ 
(which, when accounting for all baryonic resonances 
with masses $m_B\le 1.7$~GeV as well as the lowest-lying $\Lambda$
and $\Sigma$ hyperons, translates into a common baryon chemical potential 
of $\mu_B\simeq 385$~MeV) only about one third of the  
baryons are actually nucleons.  On the other 
hand, also excited resonances will have nonzero cross sections  
with pions or $\rho$ mesons which are, however, usually somewhat smaller 
in  magnitude. The second subtlety consists of a substantial smearing 
of the zero temperature nucleon distribution functions 
(Fig.~\ref{fig_fermi}) which might further suppress any 
nucleon-driven medium effects in high-energy heavy-ion collisions. 
Nevertheless, meson-nucleon interactions are typically much 
stronger than meson-meson ones, such that even at full CERN-SpS energies
(158--200~AGeV),
where the final pion-to-nucleon ratio is about 5:1, baryons have 
a substantial impact on pion and $\rho$-meson properties as has been 
demonstrated by several authors.  

In one class of models, medium effects in dilepton production rates 
are again studied by focusing on the role of the $\rho$-meson spectral 
function~\cite{HeFN,ChSc,RCW,FrPi}. 
As a representative we choose the most 
recent version of Refs.~\cite{RUBW,Morio98,RG99,RW99}, where both effects 
of finite temperature and finite density have been incorporated. In the
baryonic sector the naive VDM, which works well for the description of 
purely mesonic processes (see previous Section), is improved by the 
Kroll-Lee-Zumino coupling~\cite{KLZ} in the  
transverse part to optimally reproduce photoabsorption data 
(cf.~Sect.~\ref{sec_photoabs}).  
The dilepton rate is then given by 
\beq
\frac{dR_{\pi\pi\to ee}}{d^4q}(q_0,\vec q) =
\frac{\alpha^2} {\pi^3 g_{\rho\pi\pi}^2} \
\frac{f^\rho(q_0;T)}{M^2} \ \frac{1}{3} \left[ {\cal F}^L(q_0,\vec q)+
2 {\cal F}^T(q_0,\vec q) \right]  
\label{dRd4q_klz}
\eeq
with the transverse transition form factor ${\cal F}^T$ from Eq.~(\ref{KLZ}), 
whereas the longitudinal part, not being constrained 
by photon data,  is obtained in  the 
naive VDM, 
\beq
{\cal F}^L(q_0,\vec q)=-(m_\rho^{(0)})^4 \  
{\rm Im} D_\rho^L(q_0,\vec q) \ .  
\eeq

Within model-independent approaches, a simultaneous assessment of 
finite temperature and finite density effects has been performed 
by Steele \etal~\cite{SYZ2,SZ99} using the chiral reduction formalism. 
Here the in-medium dilepton rates are based on the in-medium vector
correlator as outlined in Sect.~\ref{sec_chired}.

The thermal rate results employing the in-medium $\rho$ spectral function 
and the chiral reduction formalism are confronted in Fig.~\ref{fig_Brates}. 
\begin{figure}[!htb]
\bce
\epsfig{figure=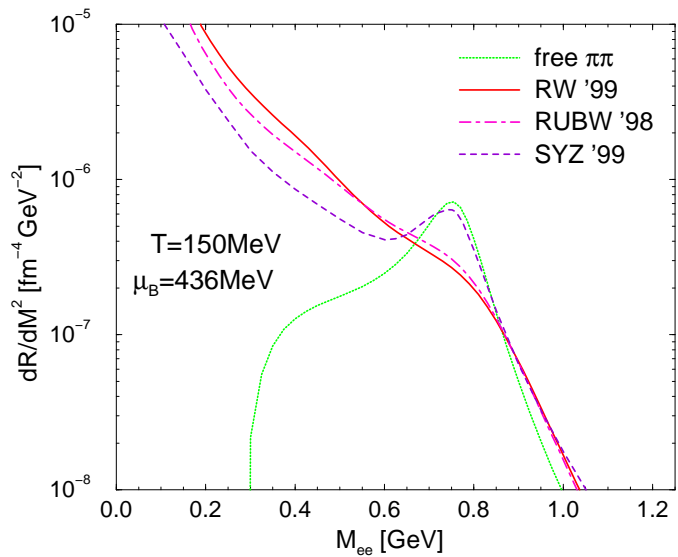,height=10cm,angle=-90}
\ece
\caption{Three-momentum integrated dilepton rates in a hot meson-nucleon
gas  at $T=150$~MeV and a nucleon density $\rho_N=0.5\rho_0$ (corresponding
to a nucleon chemical potential $\mu_N=436$~MeV) in the hadronic
 approaches of Steele \etal~\protect\cite{SYZ2} (dashed line) and
Rapp \etal~\protect\cite{RUBW,RG99} (dashed-dotted line); the solid line
corresponds to the full results of Ref.~\protect\cite{RW99},  
\ie,  when additionally
including scattering contributions off  thermally excited baryonic
resonances at a common baryon chemical potential  
$\mu_B=\mu_N$ as described in Sect.~\ref{sec_densT}.}
\label{fig_Brates}
\end{figure}
We should stress again 
that both approaches have been thoroughly constrained in both their
finite density and finite temperature behavior (cf.~Sects.~\ref{sec_hadscat}
and \ref{sec_photoabs}). However, some differences emerge in the dilepton 
regime. At low invariant masses $M_{ee}\le 0.6$~GeV
the rates qualitatively agree in that a strong additional enhancement due
to the presence of nucleons is observed. Quantitatively, the spectral
function results give up to a factor of $\sim$~2 more $e^+e^-$ yield around
$M_{ee}\simeq 0.4$~GeV -- exactly the region where the $N(1520)$ 
contribution figures in most importantly. This discrepancy indeed originates
from the different assignment of 'background' and resonance contributions 
in the photoabsorption spectra: in the chiral reduction formalism, the 
$\gamma N$ cross sections are dominated by the non-resonant 'background'
(see left panel of Fig.~\ref{fig_gamsyz}), 
whereas the most recent $\rho$ spectral function calculations attribute
the major strength to direct $\rho N$ resonances, most relevant 
the $N(1520)$ (see left panel of Fig.~\ref{fig_gam69}). 
The important point is now that when moving from the
photon point to the time-like dilepton regime (at small three-momentum), 
the resonance contributions are much more enhanced than the more or less
structureless background, which is essentially a kinematic 
effect\footnote{Similar conclusions have been reached in Ref.~\cite{SZ99};
it has been shown there that when 'artificially' reducing the $\pi N$ 
background obtained from the chiral reduction formalism in the 
photoabsorption spectra 
by a factor of 3 and assigning the missing strength to the $N(1520)$, the 
resulting prediction for the dilepton rate in the $M=0.4$~GeV region 
increases by a factor of 2--3.}. As we have pointed out in 
Sect.~\ref{sec_photoabs}, these deviations can be resolved by analyzing 
more exclusive channels in the photoabsorption data.    
A more severe, {\em qualitative} difference in the rate predictions 
of Fig.~\ref{fig_Brates} again shows up around the 
free  $\rho/\omega$ mass. Whereas the virial-type density expansion 
of the chiral reduction formalism leaves the dilepton yield essentially 
unchanged,  the spectral function result exhibits a strong  
reduction of the resonance peak due to a large in-medium broadening 
of the $\rho$ meson (see the discussion at the end of the previous
Section). Thus,  
contrary to collective effects, which are typically driven by strong 
resummation effects in the {\it real} part of the selfenergy, higher
order effects in the {\it imaginary} part of the vector meson 
propagators seem to play an important role. Finally we remark that 
the additional inclusion of excitations on thermally excited resonances
of type $B_1 B_2^{-1}$ (cf.~Sect.~\ref{sec_densT})~\cite{RW99}
 further reinforces
the broadening and low-mass enhancement by up to 20\%, see solid curve
in Fig.~\ref{fig_Brates}.  \\ 

To summarize this Section about the various hadronic dilepton rate 
calculations one may conclude that quite different approaches pursued 
in the literature so far have reached a reasonable consensus
in the pure mesonic sector. The corresponding enhancement over 
the 'standard candle' of free $\pi\pi$ annihilation below the free 
$\rho/\omega$ masses  amounts to a factor of $\sim 3-5$ at typical temperatures
around $T=150$~MeV. At finite baryon density, due to the stronger nature 
of meson-baryon interactions, a substantially stronger   
impact on in-medium dilepton rates has been found, entailing more pronounced 
discrepancies between various models, differing {\em quantitatively} 
in the low-mass region (by factors of 2--3), and,  more importantly, 
{\em qualitatively} as far as the fate of the vector meson resonance 
peaks is concerned.

\subsection{Beyond Conventional Scenarios for Dilepton Enhancement}
\label{sec_beycon}
 
In this Section we will present some more speculative mechanisms  
for dilepton production which conceptually deviate from the rather well 
established hadronic reactions discussed in the previous Section. 
In particular, we will address radiation from quark-antiquark annihilation,  
disoriented chiral condensates and a dropping $\rho$-meson mass as
implied by BR scaling.

\subsubsection{Quark-Gluon Plasma}
At sufficiently high invariant masses as well as temperatures
and densities asymptotic freedom of the quark interactions 
implies that the rate can be described by perturbation theory.
In lowest order of the strong coupling constant, $O(\alpha_S^0)$, 
it is determined by the free $q \bar q \to \gamma^* \to e^+e^-$ annihilation 
process through a convolution over anti-/quark three-momenta 
$p_{\bar q}$, $p_q$  according to 
\begin{equation} 
\frac{dR_{q\bar q \to ee}}{d^4q} = \int \frac{d^3p_q}{(2\pi)^3} 
\frac{d^3p_{\bar q}}{(2\pi)^3} \ \sum\limits_q
v_{q\bar q} \  \sigma_{q\bar q \to ee}(M) 
\ f^q(p_q^0) \ f^{\bar q}(p_{\bar q}^0) \ 
\delta^{(4)}(q-p_q-p_{\bar q}) \ ,  
\label{qqrate}
\end{equation} 
where the total color-averaged 
$q\bar q \to ee$ cross section for each flavor $q=u,d,s,\dots$ 
is given by 
\begin{equation} 
\sigma_{q\bar q \to ee}(M)=e_q^2 \  \frac{4\pi\alpha^2}{9M^2} 
\left( 1-\frac{4m_q^2}{M^2}\right)^{-\frac{1}{2}} \ 
\left( 1+\frac{2m_q^2}{M^2}\right) \  
\end{equation} 
with current quark masses $m_q$ and fractional quark charges 
$e_q=-\frac{1}{3},+\frac{2}{3}$. In 
Eq.~(\ref{qqrate})
\begin{equation} 
v_{q\bar q}=\frac{M\sqrt{M^2-4m_q^2}}{2\omega_{p_q}\omega_{p_{\bar q}}}
\end{equation} 
denotes the relative velocity between quark and antiquark, and 
\begin{eqnarray}
f^q(p_q) &=& \frac{N_s N_c}{\exp[u\cdot p_q -\mu_q] +1}
\nonumber\\
f^{\bar q}(p_{\bar q}) &=& 
\frac{N_s N_c}{\exp[u\cdot p_{\bar q} +\mu_q] +1}
\end{eqnarray}
their Fermi distribution functions including spin-color degeneracy 
factors as well as the quark chemical potential $\mu_q$.
The fluid velocity of the plasma relative to the thermal frame is denoted
by $u^\mu$ . 
For a plasma at rest $u^\mu =(1,0,0,0)$ 
and in the limit of vanishing quark masses  
the momentum integrations can be performed analytically
yielding~\cite{CFR,SchBl} 
\begin{equation} 
\frac{dR_{q\bar q \to ee}}{d^4q} = \frac{\alpha^2}{4\pi^4} 
\frac{T}{q} f^B(q_0;T) \sum\limits_q e_q^2 \ \ln
\frac{\left(x_-+\exp[-(q_0+\mu_q)/T]\right) \left( x_++\exp[-\mu_q/T]\right)}
{\left(x_++\exp[-(q_0+\mu_q)/T]\right) \left( x_-+\exp[-\mu_q/T]\right)} 
\label{qqrate2} 
\end{equation} 
with $x_\pm=\exp[-(q_0\pm q)/2T]$.    
For $\mu_q=0$   
Eq.~(\ref{qqrate2}) simplifies to  
\beq 
\frac{dR_{q\bar q \to ee}}{d^4q} = \frac{\alpha^2}{4\pi^4} f^B(q_0;T) 
\left( \sum\limits_q e_q^2 \right)   
\left( 1+ \frac{2T}{q} \ln \left[\frac{1+x_+}{1+x_-}\right] \right) \ . 
\label{qqrate0} 
\eeq
Note that, apart from the overall Bose factor $f^B$, this result carries 
a temperature-dependent correction factor as compared 
to the widely used $O(T^0)$ approximation. The additional 
$\ln$-term gives in fact a negative contribution which becomes significant 
for invariant masses below $M_{ee}\simeq 1$~GeV, cf.~Fig.~\ref{fig_qqrates}. 
Another noteworthy feature is that with increasing 
net quark density ($\mu_q>0$) the perturbative emission rate from a QGP
decreases slightly at low masses 
due to the mismatch between the quark and antiquark Fermi spheres.  
\begin{figure}[!htb]
\begin{center}
\epsfig{figure=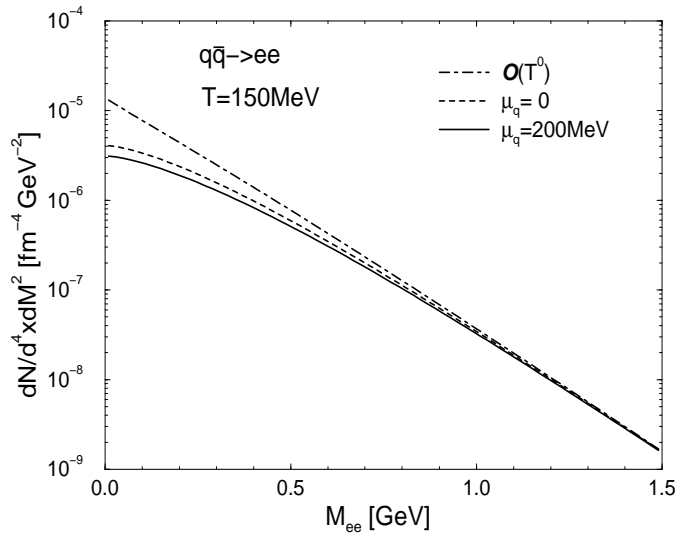,height=10cm,width=8cm,angle=-90}
\end{center}
\caption{Three-momentum integrated perturbative $q\bar q\to e^+e^-$ 
annihilation rates at temperature
$T$=150~MeV: for $\mu_q=0$ (dashed line) and $\mu_q=200$~MeV (solid line)
using Eq.~(\protect\ref{qqrate2}), and at $\mu_q=0$ using 
Eq.~(\ref{qqrate0}) to order $O(T^0)$ (dashed-dotted line); 
all curves have been  obtained for three massless quark flavors.} 
\label{fig_qqrates}
\end{figure}

Perturbative QCD corrections to the $q\bar q$ rate have been studied, \eg, in 
Refs.~\cite{BPY,ChWo,MWCW} and turned out to be appreciable, especially when 
extrapolated to invariant masses well below  1~GeV where the results 
cease  to be reliable. 
Furthermore, since in the plasma phase at moderate 
temperatures, $T\simeq (1-2) T_c$, the strong coupling constant is still 
of order 1, nonperturbative effects might not be small either   
in this regime. 
In particular, the gluon condensate is non-vanishing as discussed in 
Sect.~\ref{sec_lattice}.  
In Ref.~\cite{LWZH}, \eg, the somewhat speculative existence 
of an (euclidean) $A_4$ condensate of type 
$\langle \frac{\alpha_s}{\pi}A_4^2 \rangle$ has been shown to generate a 
strong enhancement in the rate at low invariant masses. 
More conservative approaches include instanton-induced 
interactions~\cite{SS98} (known to be of prime importance for the 
nonperturbative QCD vacuum structure and the
low-lying hadronic spectrum) which, in somewhat modified appearance 
at finite temperature/density
  (\ie, as so-called instanton-antiinstanton ($I$-$A$) 
molecules) might still 
prevail at  moderate plasma conditions~\cite{SSV95,VS97,RSSV99};  
in fact, $I$-$A$ molecule induced $q\bar q$ interactions, as opposed to 
single instantons, lead to nonzero contributions in the vector channel,
which, however, seem to be quite small~\cite{VS97}. 

A generic feature of dilepton production rates from the plasma phase 
are the so-called van-Hove singularities~\cite{BPY,Weld99,PT99}, \ie, 
(possibly)  sharp structures as a function of the dilepton energy. They 
originate from a softening of the
quark dispersion relation in the medium (the so-called
'plasmino' branch~\cite{BPY}) and  
are typically located at rather small energies below $\sim$~0.5~GeV.
On the other hand, finite imaginary parts in the
quark selfenergy as well as smearing effects when accounting for finite 
three-momenta of the virtual photon~\cite{Wong92}
will damp these peaks. Also, 
soft Bremsstrahlung-type processes involving 
gluons such as $q\bar q\to g\gamma^*$ or $q g\to \gamma^*q$ may easily
overshine the van-Hove structures at low $M$~\cite{BPY}, even after 
including Landau-Pomeranchuk type effects~\cite{CGR93}, \ie, destructive
interferences in the coherent emission.

\subsubsection{Disoriented Chiral Condensate}
Another possibility of increased (soft) dilepton radiation related 
to the chiral 
phase transition has been associated with so-called Disoriented Chiral 
Condensates (DCC's), 
which have been suggested to develop in the cooling process of 
high-energy heavy-ion collisions~\cite{BKT,RaWi}. Assuming the formation of 
a chirally restored plasma phase in the early stages of the collision,  
a sufficiently rapid transition into the chirally
broken phase might not select  
the standard ground state characterized by a single scalar (sigma)
condensate $\langle \sigma \rangle = \langle \bar qq\rangle\ne 0$, 
but rather 'jump' into a chirally
rotated (metastable) state carrying a nonvanishing pseudoscalar (\ie,  
pion-) condensate $\langle \vec \pi \rangle = \langle \bar q \gamma_5 
\vec \tau q\rangle \ne \vec 0$. In particular, this state  
carries nonzero net isospin, which provides the basis for detecting it in  
event-by-event analyses of heavy-ion collisions   
through anomalous fluctuations in, \eg, the number of $\pi^0$'s.  
Since the pion condensates inherent in the DCC
constitute an enhanced source of predominantly soft pions, it 
has soon been realized that within this coherent state a  
copious annihilation into soft dilepton pairs might occur, similar to the 
standard $\pi\pi$ annihilation, only at much lower invariant masses. 
In fact, the dominant yield from such coherent radiation will 
be concentrated well below the two-pion threshold~\cite{HuWa}. Therefore, 
such a signal will be very difficult to discriminate 
in a heavy-ion collision due to the notoriously  large background from 
$\pi^0\to \gamma e^+e^-$ Dalitz decays after freezeout.   
However, as suggested in 
Ref.~\cite{KKRW}, non-coherent pions from the surrounding heat bath 
may annihilate on the coherent state, forming dilepton pairs of typical
invariant masses in a rather narrow window around $M\simeq 2m_\pi$, 
thus avoiding the $\pi^0$ Dalitz-decay region.
Starting from the  standard linear $\sigma$-model Lagrangian, 
\begin{equation}
{\cal L}_{lsm}=\frac{1}{2} \partial_\mu \phi \partial^\mu \phi 
-\frac{1}{4} \lambda (\phi^2-v^2)^2 + H\sigma \  ,  
\end{equation}
the coupling to the electromagnetic current is realized through the 
third component of the isovector current as 
\bea
j^{\rm em}_\mu(x) &=& \left[ \vec \pi \times \partial_\mu \vec \pi \right]_{3} 
\nonumber\\
   &=& \frac{i}{2} \left[ \pi^\dagger(x) 
\stackrel{\leftrightarrow}{\partial}_\mu \pi(x) \pi(x) 
\stackrel{\leftrightarrow}{\partial}_\mu \pi^\dagger(x) \right] \ ,  
\eea
where the pion fields in charge basis are related to the isospin fields as  
\bea
\pi(x)&=&\frac{1}{\sqrt 2} \left[\pi_1(x) -i\pi_2(x) \right]
\nonumber\\
\pi^\dagger(x)&=&\frac{1}{\sqrt 2} \left[\pi_1(x) +i\pi_2(x) \right] \ .  
\eea
The electromagnetic current-current correlator entering the dilepton rate 
can then, employing a mean-field treatment, be expressed by pion two-point 
functions as
\bea 
W_{\mu\nu}(x,y) &=& \langle \pi^\dagger(x) \pi(x)\rangle  \ 
\langle \partial_\mu \pi(x) \partial_\nu \pi^\dagger(y)\rangle
+\langle \partial_\mu \pi^\dagger(x)\partial_\nu  \pi(x)\rangle \ \langle \pi(x)
\pi^\dagger(y)\rangle
\nonumber\\ 
 && -\langle \partial_\mu \pi(x) \pi^\dagger(x)\rangle \langle \pi^\dagger(x)
\partial_\nu \pi(y)\rangle
-\langle \partial\pi^\dagger(x) \pi(x)\rangle \langle \pi(x)
\partial_\nu \pi^\dagger(y)\rangle \ .  
\eea
Further following Ref.~\cite{KKRW}, the charged pion fields are expanded in 
creation and annihilation operators 
($a_k^\dagger, a_k$ for positive and $b_k^\dagger, 
b_k$ for negative pions) with accompanying mode functions $f_k$ as 
\beq
\pi(\vec x ,t)= \int \frac{d^3k}{(2\pi)^3} \ e^{i \vec k \cdot \vec x}  
\ \left[ f_k(t) a_k + f_k^*(t) b_{-k}^\dagger \right] \ .   
\eeq
The time evolution of the pion fields, and thus of the dilepton production 
rate, is then determined by specifying the mode functions together with 
appropriate initial conditions for the $\sigma$ mean-field. Results have 
been obtained  for both a purely thermal scenario (pion gas) and  
a DCC ('quench') scenario at  equivalent initial energy density. 
A strong enhancement (factors of $\sim$~10--50)
of the DCC based rates over the thermal ones, Eq.~(\ref{pipirate}), 
was found at invariant masses around 2$m_\pi$, but 
restricted to rather low total three-momenta $|\vec q|\le 300-500$~MeV.
The quoted ranges roughly reflect the uncertainty in the underlying 
approximations as estimated from a second calculation using  classical 
equation-of-motion techniques to describe the time evolution of pion and 
$\sigma$ fields. 

\subsubsection{'Dropping' Rho-Meson Mass}
The most prominent approach that has been successfully applied 
to explain the low-mass dilepton enhancement in the CERN-SpS experiments
in connection with the chiral restoration transition is based on the 
BR scaling conjecture~\cite{BR91} for effective chiral Lagrangians. 
In the dilepton context, the most relevant feature is the 
decrease of the $\rho$-meson mass at finite temperature and density which,
reinforced through enhanced thermal occupation factors at lower masses, 
leads to a strong excess of $e^+e^-$ pairs below the free $\rho$ mass 
through the  $\pi\pi$-annihilation channel. As long as no collisional 
broadening in the $\rho$ width is included, the $e^+e^-$ yield 
is sharply centered around  
the corresponding in-medium mass $m_\rho^*$. 
Unlike in the DCC scenario, where the enhancement is localized at  
invariant masses around $2m_\pi$, $m_\rho^*$ will 'sweep' over 
the entire low-mass region in the course of a heavy-ion collision 
due to the continuous cooling and dilution of the system from the 
chiral restoration point towards freezeout. More realistically, 
also in the dropping mass scenario the $\rho$-meson spectral function will 
undergo a substantial broadening in the hot and dense hadronic 
medium. In particular, if the decrease in the $\rho$ mass is (partially)  
identified with resonant $S$-wave $\rho$-$N(1520)N^{-1}$ interactions
(as discussed in detail in Ref.~\cite{BLRRW}), 
then large in-medium widths  of the N(1520) resonance (as inferred 
from nuclear photoabsorption data)  inevitably  induce a large 
width of the low-lying $N(1520)N^{-1}$ state. 
Since such an increase in width affects both the denominator
{\it and} the numerator of ${\rm Im} D_\rho$, entering the dilepton rate 
(\ref{pipirate}), the broadening does in essence not reduce the total
number of produced pairs.

\subsubsection{Off Equilibrium Pion Gas}
Another class of  non-conventional scenarios with potential impact  
on dilepton radiation is associated with off-equilibrium situations.
Deviations from thermal equilibrium are usually addressed within kinetic 
theory or transport-type approaches, and the preferred method to assess 
dilepton yields under these conditions  should be numerical simulations. 
Fortunately, in the
case of heavy-ion collisions, thermalization of the hadronic system 
seems to require only a few rescatterings, so that local thermal 
equilibrium (in the comoving frame of collective expansion) is 
established on time scales which are much  shorter than the 
typical lifetime of the hadronic fireball~\cite{Gavin,HSD}. 
A fast approach to thermal equilibrium is further corroborated 
in scenarios with initially deconfined matter as demonstrated,  
\eg, in QCD inspired event generators such as HIJING~\cite{WaGy} or parton 
cascades~\cite{Geiger}. Complete chemical equilibration, however, is by far 
less certain; in an expanding pion gas, \eg, the empirical absence of 
pion number changing processes in low-energy $\pi\pi$ interactions 
(\ie, for invariant masses below $M_{\pi\pi}\simeq 1~$GeV) 
together with 
the assumption of isentropic expansion entails the build-up of 
a finite pion chemical potential $\mu_\pi>0$ towards lower 
temperatures~\cite{Bebie}.    
Another possibility is that early hadron formation processes do not
transform into a chemically equilibrated initial hadron gas but rather 
follow, \eg, string dynamics~\cite{LUND}.  

The field-theoretical implementation of a finte $\mu_\pi$ into the 
thermal dilepton 
production rate from $\pi\pi$ annihilation has been studied 
by Baier \etal~\cite{BDR971,BDR972} in the real-time formalism of 
finite temperature field theory. For practical purposes they 
worked in an approximation which amounts to introducing the chemical 
potential through Bose distribution functions as 
\beq
\tilde f^B(k_0;\mu_\pi,T)=\left\{ \begin{array}{cc} 
f^B(|k_0|;\mu_\pi,T) & , k_0>0  \\
 -[1+f^B(|k_0|;\mu_\pi,T)] & , k_0<0 \end{array} \right.  
\eeq
with the standard finite-$\mu_\pi$ Bose factor
$f^B(|k_0|;\mu_\pi,T)=[e^{(|k_0|-\mu_\pi)/T}-1]^{-1}$.  
In the Boltzmann approximation and for situations not too far 
off equilibrium it was shown that the production rate in the 
standard one loop approximation (\ie, Eq.~(\ref{pipirate}) with 
$\Sigma_{\rho\pi\pi}$ given by the free $\pi\pi$ bubble) 
simply picks up an overall enhancement factor according to 
\beq
\frac{dR_{\pi\pi\to ee}}{d^4q}(q_0,\vec q;\mu_\pi)=
(1+2 \delta \lambda) \ 
\frac{dR_{\pi\pi\to ee}}{d^4q}(q_0,\vec q;\mu_\pi=0)
\eeq
with $\delta \lambda =\lambda -1$ and the 'fugacity' 
$\lambda=e^{\mu_\pi /T}$ (to lowest order in $\mu_\pi/T$
one recognizes $(1+2 \delta \lambda)\simeq e^{2\mu_\pi /T}$).

The situation becomes more complicated if one includes higher thermal 
loop corrections in the pion propagators of the intermediate 
two-pion states. Without going into further details, we mention 
that the use of   off-equilibrium distribution functions 
then implies the appearance of terms involving products of 
retarded and advanced propagators, schematically given as
\bea
D_\pi^R(k) \ D_\pi^A(k) &=& \frac{1}{k^2-m_\pi^2+ i sgn(k_0) \epsilon}
 \ \frac{1}{k^2-m_\pi^2- i sgn(k_0) \epsilon} 
\nonumber\\
 &=& \frac{1}{(k^2-m_\pi^2)^2 - \epsilon^2} \ , 
\eea
which apparently exhibit ill-defined poles, the so-called 
'pinch-singularities'. They have to be regularized, \eg, by accounting
for thermal pion selfenergies $\Sigma_\pi$ with finite imaginary parts
${\rm Im} \Sigma_\pi \ne 0$, as was 
done in Refs.~\cite{BDR971,BDR972}. As a somewhat surprising result, 
to leading order in $1/{\rm Im}\Sigma_\pi$ the pinch term actually 
generates an overall reduction (enhancement) of the dilepton production 
rate for positive (negative) $\mu_\pi$.

\subsection{Quark-Hadron Duality}
\label{sec_duality}
A key question that is at the heart of the entire heavy-ion physics
program concerns discriminating signatures for the possible occurrence
of a chiral restoration phase transition.  
In our context of low-mass dilepton observables we have already 
qualitatively eluded 
to some of them in the preceeding Section, most notably DCC formation 
(characterized by a strong enhancement in the two-pion threshold
region at low three-momenta) and the dropping $\rho$-meson mass (characterized
by a complete extinction of the yield around the $\rho$ peak, 
accompanied by a strong enhancement for lower invariant masses). 
However, there is also the possibility that in the vicinity 
of the phase transition the dilepton radiation
from the hot and dense fireball does {\em not} depend on whether it
is in the chirally broken confined phase or the chirally restored 
QGP phase. In other words,  in a certain temperature and 
density window around the transition region the dilepton rate calculations  
using either  hadronic or quark-gluon degrees of freedom merge together, \ie,
the two descriptions become {\em dual}. Although this would imply    
that there is no unique dilepton signature which could distinguish the 
two phases, it is a highly non-trivial scenario.   
In the low-mass region, say for invariant masses $M_{ll}\le 1.5$~GeV, 
it requires that the hadronic side, which at  
low densities/temperatures is dominated by rather narrow resonances
$\rho, \omega, \phi, \rho'$, etc., develops into a supposedly more or less
structureless quark-gluon world (although the latter might still
involve nonperturbative interactions).   
To illustrate that there are indeed indications for such a 
scenario~\cite{RW99,Rqm99}, 
we compare in Fig.~\ref{fig_qqvpp} the lowest-order perturbative 
QCD $q\bar q$ annihilation rates with the (most recent) full hadronic 
in-medium spectral function calculations
at identical temperatures and equivalent baryon-/quark-chemical potential.  
\begin{figure}[!htb]
\epsfig{figure=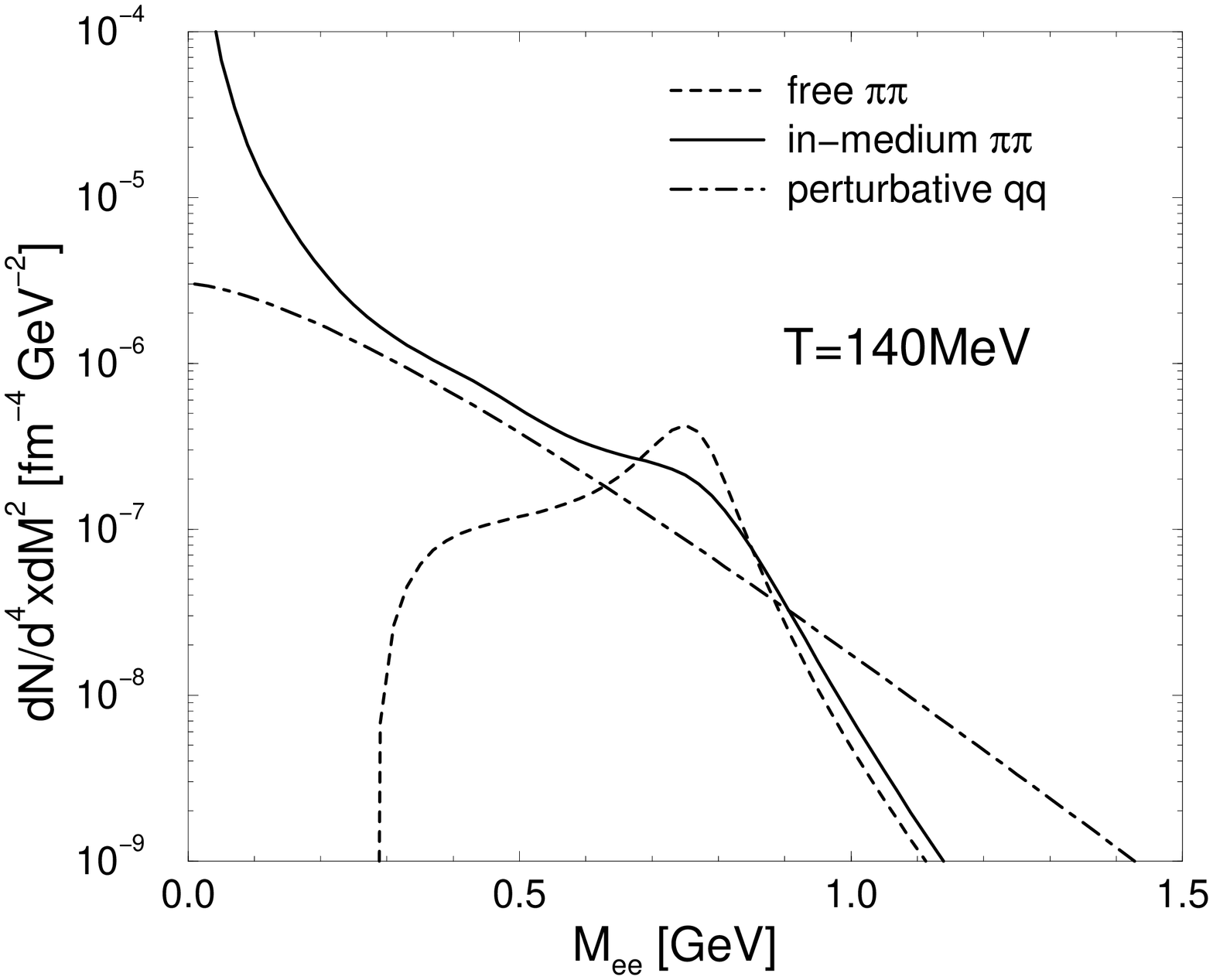,width=6.2cm,angle=-90}
\epsfig{figure=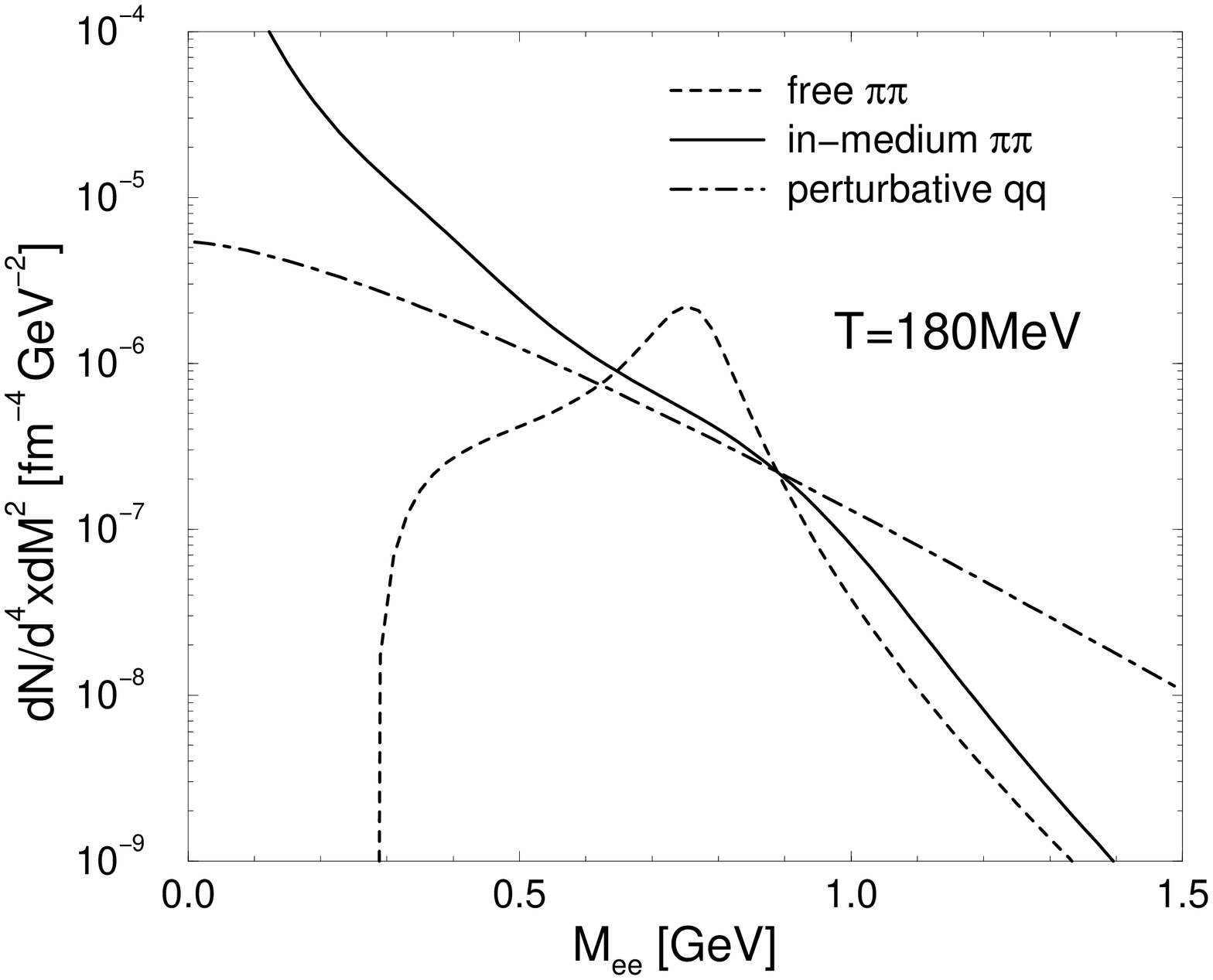,width=6.2cm,angle=-90}
\caption{Three-momentum integrated dilepton production rates at a baryon
chemical potential $\mu_B=408$~MeV and temperatures
$T=140$~MeV (left panel) and $T=180$~MeV from  free $\pi\pi$ annihilation
(dashed curves), in-medium $\pi\pi$ annihilation within the most recent 
hadronic many-body approach of Refs.~\protect\cite{RCW,RUBW,RG99,RW99}, 
and from lowest order QCD $q\bar q$ annihilation (dashed-dotted curves);
the latter two are calculated at equivalent baryon/quark chemical
potentials of $\mu_q=\mu_B/3=136$~MeV.}
\label{fig_qqvpp}
\end{figure}
Already at moderate temperature ($T=140$~MeV) and density 
($\rho_B=0.75\rho_0$)  
they are not  very different from each other (left panel),  
especially when contrasted with the result for free $\pi\pi$ 
annihilation. Nevertheless, the in-medium $\pi\pi$ curve  
still exhibits a clear trace of the $\rho$ peak. The latter is completely
'melted' at $T=180$~MeV, $\rho_B=4\rho_0$ leading to remarkably close
agreement with the $q\bar q$ result in the invariant mass range
of $M_{ee}\simeq (0.5-1.0)$~GeV  (one should note that the 
deviations between the partonic and the in-medium hadronic results towards
low $M_{ee}$ might be reduced once soft 'Bremsstrahlung'-type
graphs are accounted for in the QGP environment. On the other hand, 
as discussed in Sect.~\ref{sec_beycon}, plasmino modes  
can lead to additional nontrivial structures in the dilepton production 
rate~\cite{BPY,Weld99,PT99} below $M_{ee}\simeq 0.5$~GeV. 
The disagreement beyond 1~GeV is mainly caused by missing states 
involving more than two pions (such as the $\rho'(1450)$) in the 
hadronic description of the free vector correlator, most notably
four-pion type $\pi a_1$ annihilation graphs, see also below). 
From an experimental point of view this means that, even if no  
{\it distinct} signatures for the appearance of new phases are extractable 
from low-mass dilepton production, the absolute yields and spectral shape 
are very different from free $\pi\pi$ annihilation and contain rather 
specific information on properties of 
strongly interacting matter in the vicinity of the phase boundary.  
We will reiterate this point within a detailed analysis of low-mass
dilepton observables below. 

The duality arguments can be made more rigorous starting from the   
intermediate-mass region, 1.5~GeV$ \lsim M_{ll} \lsim $3~GeV.  
Conceptually the situation there is more transparent: firstly, 
one might expect that at these energies the  
$q\bar q$ annihilation process is already  rather well described 
by a perturbative treatment. In fact, there is strong empirical 
support for this expectation from the well-known 
$e^+e^- \to hadrons$ cross section, which can be accounted for by the 
perturbative result for $e^+e^-\to q\bar q$ within a 30\% accuracy
in the above mentioned range, \ie,  
\beq
\sigma(e^+e^-\to hadrons) \simeq \frac{4\pi\alpha^2}{3M^2} \ R^{\rm pert}  
\eeq
with the famous 
$\sigma(e^+e^-\to hadrons)/\sigma(e^+e^-\to \mu^+\mu^-)$-ratio  
\beq
R^{\rm pert}=N_c \sum\limits_q (e_q)^2 \ .  
\label{Rpert}
\eeq
Secondly, since at $\sim$~2~GeV one is probing space-time distances of 
the order of 0.1~fm, possible corrections from the surrounding heat bath 
should also be small. 
This gives  some confidence that the perturbative expression 
for the dilepton emission rate from a 
quark-gluon plasma (\ref{qqrate}) is a reasonable 
approximation at intermediate masses.  
The challenge then 
is to match this result to a hadron-based calculation at temperatures
in the vicinity of $T_c$. 

In Ref.~\cite{Leon} this type of duality has been {\em enforced} in 
the spirit of a Hagedorn-type hadronic mass spectrum, 
where the complicated structure of overlapping, interacting hadronic
resonances is encoded in some simple spectral density $\xi_h(M)$. 
Using VDM for the dilepton decays of the vector mesons, 
\beq
\Gamma_{V\to ll}(M)= \frac{1}{g^2(M)} \frac{4\alpha^2 M}{3} \ ,  
\eeq
the corresponding three-momentum integrated dilepton production rate at 
temperature $T$ in Boltz\-mann approximation, 
\beq
\frac{dR_{V\to ll}}{dM^2} =\xi_V(M) \ \frac{\alpha^2 M^2 T}{6\pi g^2(M)}   
\ K_1(M/T) \  
\eeq
($K_1$: modified Bessel function),  
is then determined by the subspectrum of vector mesons, $\xi_V(M)$, 
and the corresponding VDM coupling $1/g(M)$. 
Further assuming that the $e^+e^- \to hadrons$ cross section is
saturated by vector mesons,  
\bea
\sigma(e^+e^- \to V) &=& \frac{(2\pi)^3\alpha^2}{g^2(M) M} \xi_V(M) 
\nonumber \\
 &\simeq& \sigma(e^+e^-\to h)  
\eea
one can trade the dependence on $g(M), \xi_V(M)$ for  the experimental 
cross section ratio as~\cite{Leon}
\beq
\frac{dR_{V\to ll}}{dM^2}=\frac{\alpha^2 M T }{6\pi^3} \ K_1(M/T) \ 
  R^{\rm exp}(M) \ ,  
\label{dualrate}
\eeq
where $R^{\rm exp}(M)$ now also accounts for (moderate) deviations from the 
lowest-order QCD result, Eq.~(\ref{Rpert}). 
A similar procedure has been pursued by Huang~\cite{Huang} who 
additionally included lowest-order temperature effects through 
the vector-axialvector mixing phenomenon
induced by soft pion contributions from the heat bath 
(cf.~Sect.~\ref{sec_vamix}), leading to
\beq
\frac{dR_{ll}}{dM^2} = 
\frac{4\alpha^2 M T }{2\pi} \ K_1(M/T) \
\left(\rho^\circ_{\rm em}(M)-(\epsilon-\frac{\epsilon^2}{2})
\left[ \rho^\circ_V(M)-\rho^\circ_A(M)\right] \right)  
\label{dRdM2huang}
\eeq
with $\epsilon=T^2/6f_\pi^2$ (in the chiral limit and for two flavors). 
The first term involving the full vacuum electromagnetic correlator 
coincides with (\ref{dualrate}), whereas the free vector and 
axialvector correlators 
are responsible for the soft pion corrections. Huang also extracted  
them from data, as delineated in Sect.~\ref{sec_vamix}, see 
Eqs.~(\ref{rhoV0}) and (\ref{rhoA0}). 
It turns out that the finite-temperature corrections are 
marginal for dilepton invariant masses 1.5~GeV $\le M_{ee} \le 2.5$~GeV 
and, moreover, that the 'empirical' results using Eq.~(\ref{dualrate}) 
closely follow the perturbative results (\ref{Rpert}),   
cf.~the open circles and the dotted line in Fig.~\ref{fig_dRdM2int}.  
\begin{figure}[!htb]
\vspace{-1cm}
\begin{center}
\epsfig{figure=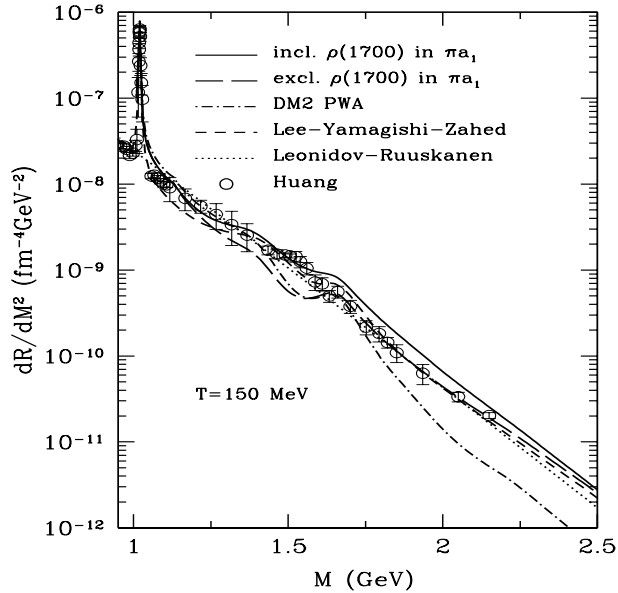,height=9cm,width=11cm}
\end{center}
\vspace{-1cm}
\caption{Three-momentum integrated dilepton production rates in the 
intermediate-mass region at a temperature $T=150$~MeV, as obtained in 
various approaches; solid, long-dashed and dashed-dotted curve: 
hadronic calculations of 
Ref.~\protect\cite{LiGa} with different ways of estimating the $\pi a_1$ 
contribution; short-dashed curve: chiral reduction 
approach~\protect\cite{LYZ}; dotted curve:  using the perturbative
value $R^{pert}=2$ in Eq.~(\protect\ref{dualrate})~\protect\cite{Leon};
open circles: using empirical spectral densities including
lowest order temperature corrections according to 
Eq.~(\protect\ref{dRdM2huang})~\protect\cite{Huang}. The plot is taken 
from Ref.~\protect\cite{LiGa}.}
\label{fig_dRdM2int}
\end{figure}
This is not surprising recalling that the 'duality threshold' in vacuum
is situated at $M\simeq 1.5$~GeV. 
On the other hand, it was pointed out in Ref.~\cite{Leon} 
that, when comparing the 'empirically' inferred rate (\ref{dualrate}) 
to the phenomenological hadronic rate calculations 
in terms of binary collisions of Ref.~\cite{GaLi}, the latter  
fall short by a factor of 2--3. However, in the hadronic treatment of 
Ref.~\cite{GaLi} contributions from 
$a_1(1260)$ mesons (in particular $\pi a_1 \to ll$), 
which are most relevant for the intermediate-mass region, had not been 
included at the time. This has been improved later 
on~\cite{Song94,Kim96,LiGa} and shown to resolve the 
afore mentioned discrepancy 
(Fig.~\ref{fig_dRdM2int}). The yield from $\pi a_1$ processes was indeed 
found to dominate other sources such as $\pi\omega, \pi\rho, \pi\pi$ 
or $K\bar K$ channels for invariant masses 
1.2~GeV$\le M_{ll} \le 2.2$~GeV. In particular, it results in good agreement
with both the perturbative $q\bar q$ and the 'empirical' calculations 
including the lowest order $V$-$A$ mixing effect~\cite{Huang}, 
Eq.~(\ref{dRdM2huang}), down to invariant masses of about 1~GeV, 
cf.~solid line, dotted line and open points, respectively, 
in Fig.~\ref{fig_dRdM2int}. In fact, 
the $\pi a_1\to \rho'\to e^+e^-$ contributions exactly correspond to 
the mixing effect since thermal pions colliding with an $a_1$ meson
'move' strength from the axialvector to the vector channel! 
Also shown in Fig.~\ref{fig_dRdM2int} are the results obtained within 
the chiral reduction formalism~\cite{LYZ}; here the free correlators
have been determined from a set of 12 $\rho, \omega, \phi, a_1$ and 
$K_1$ resonances below $M=2$~GeV, and by a parameterization of  
$R^{\rm exp}(M)$ above. Single-meson final-state corrections have then been 
inferred via chiral Ward identities. Again, the agreement with the other 
approaches is quite satisfactory.

As we will see in Sect.~\ref{sec_specint}, the intermediate-mass 
emission rates as deduced above provide a good description of experimental
$\mu^+\mu^-$ spectra measured by the HELIOS-3 collaboration~\cite{helios3} at 
full CERN-SpS energies (200~AGeV), which show an excess of a 
factor of 2--3 in S+W 
as compared to p+W collisions. Similar observations have been made by the 
NA50/NA38 collaboration~\cite{na38,na50int}, 
and the most natural explanation seems to be thermal 
radiation.   This further supports the theoretical 
arguments for quark-hadron 'duality' in intermediate-mass 
dilepton production. Since the intrinsic energy scales in this
region are already suggestive for a perturbative treatment within 
the partonic picture, the agreement with the phenomenological hadronic 
calculations above 1.5~GeV may be considered as the more intriguing part 
(it will still be worthwhile 
to understand the origin of the 30--40\% deviations of $R_{exp}(M)$
from the lowest-order perturbative $q \bar q$ results). As we have argued 
already in Sect.~\ref{sec_vamix}, medium effects start to become visible
below 1.5~GeV, where the lowest-order in temperature $V$-$A$ mixing
establishes a three-fold degeneracy between vector and axialvector 
correlators on the one hand (dictated by chiral symmetry), and the 
perturbatively calculated $q\bar q$ rates on the other hand, 
reaching down to about 1~GeV.

\section{Photon Production Rates}
\label{sec_phrates}
In our analysis of electromagnetic observables in heavy-ion reactions
we will also address spectra of single (real) photons.  This is motivated
by the observation that  
every process capable of creating a dilepton pair can, in
principle, lead to the radiation of a (real) photon: the
latter simply constitutes the $M^2\to 0$ limit of the virtual
(time-like) photon which occurs as an intermediate state in each
dilepton-producing reaction. This intimate relation has already been
extensively exploited for imposing model constraints on in-medium effects
through photoabsorption data on the nucleon and nuclei,
cf.~Sect.~\ref{sec_photoabs}. In Sect.~\ref{sec_phspectra} 
the reverse process, \ie, photon production in heavy-ion 
reactions, will be used as an additional consistency check.  

Theoretical calculations for photon production rates have been performed
for both the hadronic and the plasma phases. In the latter, one has mainly
focused on lowest-order QCD calculations for two-body reactions of the 
annihilation type,  $q\bar q\to g \gamma$,  as well as the Compton type,
$q(\bar q) g  \to q(\bar q) \gamma$.
The first analysis of hadronic photon rates has been performed in  
Ref.~\cite{KLS91} for a heat bath of the lightest, non-strange pseudoscalar
($\pi$, $\eta$) and vector mesons ($\rho$, $\omega$). The most important
reactions have been attributed to the analogues of the QGP processes,
$\pi\pi\to\rho\gamma$ and $\pi\rho\to\pi\gamma$, as well as the vector meson 
decay channels $\omega\to\pi\gamma$ and $\rho\to\pi\pi\gamma$.
The conclusion of Ref.~\cite{KLS91} was summarized as  'the  
hadron gas shines as brightly as the quark-gluon plasma', \ie, the
photon production rates of both phases closely coincided over a large
range of photon energies $E_\gamma \ge 0.5$~GeV.  
Later on the importance of the  $a_1(1260)$ meson -- not included in 
Ref.~\cite{KLS91} -- as an
intermediate state in $\pi\rho\to a_1\to \pi\gamma$
has been realized, especially for photon energies above 0.5~GeV.
The impact of nucleons on photon production has been explicitely
discussed, \eg, in Ref.~\cite{SYZ2}. In fact, analogous information 
can be readily extracted from any dilepton rate
calculation involving baryonic processes~\cite{RCW,FrPi,KKW97}
by extrapolating it to the photon point. To be specific, 
the differential photon production rate can be obtained from 
the dilepton expression (\ref{dRd4q}) by simply replacing 
the lepton tensor by the photon tensor
\beq
P_{\mu\nu}=4\pi\alpha \int \frac{d^3p}{(2\pi)^3 2p_0} 
\sum_\lambda {\varepsilon_\mu}^*(\lambda,p) \ \varepsilon_\nu(\lambda,p) 
\ \delta^{(4)}(p-q) 
\eeq
such that 
\beq
\frac{dR_\gamma}{d^4q}=P_{\mu\nu} W^{\mu\nu} \ ,   
\eeq
which can be simplified to 
\beq
q_0 \frac{dR_\gamma}{d^3q}=\frac{\alpha}{4\pi^2} \ W(q) 
\eeq
with $q_0=|\vec q|$ and $W(q)\equiv g_{\mu\nu} W^{\mu\nu}$.
In the (improved) VDM, \eg, the photon rate becomes
\beq
q_0 \frac{dR_\gamma}{d^3q}=\frac{\alpha}{\pi^2 g_{\rho}^2}
 \ f^\rho(q_0) \ {\cal F}^T (q_0,|\vec q|=q_0) 
\eeq
with the transverse transition form factor from Eq.~(\ref{KLZ}).

\section{Space-Time Evolution of Heavy-Ion Collisions}
\label{sec_spacetime}
Dilepton spectra as measured in (ultra-) relativistic heavy-ion 
collision experiments might be, at least conceptually, split up into 
two components: the first one arises from the phase where the 
system is characterized by strong interactions among its constituents, 
as a result of which a certain amount of photons (real and virtual) 
is radiated. Once the hadronic
system has reached a degree of diluteness where the short-range strong
interactions are no longer effective (the so-called hadronic freezeout
stage), all unstable resonances decay according to their vacuum lifetimes, 
with some probability into radiative channels. Therefore, this second 
component (which has become known as the 'hadronic cocktail' contribution
to the dilepton spectra) does not contain any information on in-medium
properties of the parent particles. However, it can be reasonably well  
reconstructed once the hadronic abundances at freezeout are known 
(this may not be that evident and we will come back to this issue  
in Sect.~\ref{sec_cocktail}). On the other hand,  
given one's favorite dilepton production rate in hot and dense matter,  
the calculation of the first component  
requires the knowledge of the space-time history of the colliding
and expanding nuclear system. This is one of the major objectives 
in relativistic heavy-ion physics by itself. In the following we will 
discuss three different approaches to simulate the reaction dynamics 
in their  application to evaluate dilepton spectra.


\subsection{Hydrodynamical Approach}
\label{sec_hydro} 
The hydrodynamic description of heavy-ion collisions is based on the 
assumption that the strong interactions in the matter are able 
to maintain local thermal equilibrium throughout 
the expansion of the nuclear system until some breakup stage (the 
freezeout). Thus, each fluid cell in its rest frame is characterized 
by standard thermodynamic variables such as  pressure, temperature 
and (energy-) density. For this reason the hydrodynamic framework 
is the most natural one for the implementation of equilibrium
dilepton rates, as the latter are formulated in exactly the same 
variables. It has a long tradition in its application to high-energy 
reactions involving  high-multiplicity hadronic final states, 
starting from hadron-hadron~\cite{Land53,Shur74} 
(or even $e^+e^-$~\cite{CF75}) 
collisions to the more modern field of relativistic nucleus-nucleus 
reactions~\cite{Gers86,Heinz92,Kat93,HuSh95,Prak97}. The basic 
equations are the conservation of energy and momentum, which 
can be expressed in a Lorentz-covariant form as
\beq
\partial_\mu T^{\mu\nu}(x)=0 
\eeq
through the energy-momentum tensor $T^{\mu\nu}(x)$. For an ideal fluid, 
\ie, neglecting any viscosity, the latter is given by
\beq
T^{\mu\nu}(x)=\left[ \epsilon(x)+p(x)\right] u^\mu(x) u^\nu(x)
-p(x) g^{\mu\nu} 
\eeq
with local energy-density $\epsilon(x)$, pressure $p(x)$ and fluid 4-velocity
$u^\mu(x)$. Additionally conserved currents, such as  the  baryon 
number current, $j_B^\mu= \varrho_B u^\mu$, or the strangeness current 
are enforced by pertinent continuity equations,  
\beq
\partial_\mu j_B^\mu=0 \ , \quad  {\rm etc.} \ . 
\eeq
Let us also mention here that in the later stages of heavy-ion collisions
the pion number might be effectively conserved due to the empirical 
absence of pion-number changing processes at small $cm$ energies
(see also Sect.~\ref{sec_beycon}). In this case, one has a 
further continuity equation, 
\beq
\partial_\mu j_\pi^\mu=0 \ ,  
\eeq
which induces a nonzero pion chemical potential. 

The basic ingredient governing the hydrodynamic evolution of the 
system is the equation of state (EoS), \ie,  the dependence of the 
pressure on energy- and baryon-density, $p=p(\epsilon,\varrho_B)$. 
For heavy-ion collisions at (ultra-) relativistic energies, the 
early stages can be barely considered as proceeding under any kind  
of equilibrium conditions. In the hydrodynamic description, one therefore
has to assume a so-called formation time $\tau_0$ 
(typically around 1-2~fm/c)
together with some initial conditions for the energy density. 
Given a specific EoS, these have to be determined by requiring a reasonable 
fit to the finally observed hadronic spectra, with some additional 
freedom of how and at which temperature the freezeout occurs. 
The differential equations for the evolution are usually solved numerically 
on a space-time grid using finite-differencing methods. 

Many interesting issues can be addressed within the hydrodynamic 
framework, as, \eg, the interrelation between collective
flow and freezeout temperature~\cite{Heinz92} or 
properties of the EoS in connection with potential phase transitions
in strongly interacting matter~\cite{Risch95,HuSh95,Schlei}. 
Here, we restrict ourselves to the evaluation of dilepton spectra. 
Given a thermal production rate, $dR/d^4q$, the total spectrum 
from in-medium radiation is straightforwardly
obtained as a sum over all timesteps in the evolution and over all 
fluid cells of the grid  of individual temperature and density above the 
freezeout value. The contribution  of a single 
cell is given by 
\beq
\left(\frac{dN_{ll}}{d^4q}\right)_{cell}=
\left(\frac{dN_{ll}}{d^4qd^4x}(q_0,\vec q ; T_{cell}, \mu_{B,cell} )\right)
 \ V_{cell} \ \Delta t \ , 
\eeq
where the time-step width $\Delta t$ and cell 3-volume $V_{cell}$ are 
to be taken in the local rest frame of the cell. As there is no preferred 
direction in this frame, the virtual photon of invariant mass 
$M=(q_0^2-\vec q^2)^{1/2}$ and three-momentum 
$\vec q$ can be assumed to decay isotropically into dileptons. The  
resulting two lepton tracks then have to be boosted
to the lab-system according to the local fluid velocity $u^\mu$, where
possible experimental acceptance cuts can be readily applied.  
However, in some practical applications, this procedure might be too time 
consuming. An approximate but more efficient way is to first integrate 
over the (possibly complicated) three-momentum dependencies to obtain 
$dR/dM^2$  and then regenerate
rapidity and transverse momentum distributions according 
to~\cite{SPEG}
\beq
\frac{dR}{dM^2dyq_tdq_t} 
=\frac{1}{2M T K_1(M/T)} \ e^{-E/T} \ \frac{dR}{dM^2}(M,T)
\eeq
with $E=q_\mu u^\mu$. This procedure has been employed, \eg, in the 
hydrodynamic models of Refs.~\cite{Prak97,HuSh97}, 
where dilepton spectra at the fulll CERN-SpS energies have been analyzed.  
In both works the hydrodynamic equations have been solved 
locally in (2+1) dimensions assuming cylindrical symmetry of the 
collision system (which implies a restriction to central collisions). 

A further simplification can be made in the limit of ultrarelativistic
collision energies, as suggested by Bjorken~\cite{Bj83}. In this case 
the longitudinal expansion is dominant and 
boost invariance can be assumed so that the longitudinal 
velocity scales with the distance from the central region as $v_L=z/t$. 
Neglecting transverse expansion, the 4-volume element simply becomes 
\beq
d^4x=d^2r_t dz dt = \pi R_t^2 dy \tau d\tau  
\eeq
with the proper time $\tau =(t^2-z^2)^{1/2}$ and $cms$ rapidity 
\beq
y=\frac{1}{2} \ln \frac{t+z}{t-z} \ .  
\eeq
The transverse extension  $R_t$ is typically taken close to the overlap 
radius of the colliding nuclei. 
From local entropy conservation,
\beq
\frac{\partial (s u_\mu)}{\partial x_\mu}=0   \ ,  
\eeq
one finds the entropy-density to behave as $s(\tau)=s(\tau_0) \tau_0/\tau$.  
Moreover, for isentropic expansion one can relate the final-state
hadron multiplicity to the formation time $\tau_0$, initial entropy-density
and transverse size as~\cite{Hwa85}
\beq
\frac{dN_h}{dy}\simeq \pi R_t^2 \ \tau_0 \ s(\tau_0) \ . 
\eeq
If one further assumes an ideal gas EoS, the temperature scales as 
\beq
T(\tau)=T_0 \left(\frac{\tau}{\tau_0}\right)^{-1/3} \ . 
\eeq
The dilepton rates are then easily integrated over the time history 
of the hot nuclear system, as has been done, \eg, in 
Refs.~\cite{shur80,CFR,Dom91,KPPS95} or, specifically for CERN-SpS conditions  
at 158--200~AGeV in  Refs.~\cite{SSG96,Ha96,BDR971,SchBl}. 

The contribution to the dilepton spectrum from hadron decays after
freezeout proceeds along similar lines as in transport calculations 
(see the following Section).

The discussion of the actual results for the final dilepton spectra
is deferred to Sect.~\ref{sec_cern} where it will be put into  
perspective in comparison with  other dynamical approaches for the 
heavy-ion reaction dynamics.

\subsection{Transport Simulations}
\label{sec_transport}
The transport-theoretical approach has been extensively used in the past 
in various facets to describe heavy-ion reaction dynamics over a broad 
range of  collision energies. Among these are the  
Boltzmann-Uehling-Uhlenbeck (BUU) approach~\cite{Bert88} and its relativistic
extensions (RBUU)~\cite{Ko88,Blatt88}, Quantum Molecular Dynamics~\cite{Aich93}
and its relativistic versions RQMD~\cite{SSG89}, UrQMD~\cite{Bass98}, or the 
Hadron-String Dynamics (HSD)~\cite{HSD}. 

In the relativistic treatments
the evolution dynamics of the two colliding nuclei are governed by a
coupled set of (covariant) 
transport equations for the phase-space distributions $f_h(x,p)$ 
of hadron $h$, 
\begin{eqnarray}  
\lefteqn{\left\{ \left( \Pi_{\mu}-\Pi_{\nu}\partial_{\mu}^p U_{h}^{\nu}
-M_{h}^*\partial^p_{\mu} U_{h}^{S} \right)\partial_x^{\mu}
+ \left( \Pi_{\nu} \partial^x_{\mu} U^{\nu}_{h}+
M^*_{h} \partial^x_{\mu}U^{S}_{h}\right) \partial^{\mu}_p
\right\} f_{h}(x,p) } 
\nonumber \\
 & & = \sum_{h_2 h_3 h_4\ldots} \int d2 d3 d4 \ldots
 [G^{\dagger}G]_{12\to 34\ldots}
\delta^{(4)}(\Pi +\Pi_2-\Pi_3-\Pi_4 \ldots )  \nonumber\\
& & \times \left\{ f_{h_3}(x,p_3)f_{h_4}(x,p_4)\bar{f}_{h}(x,p)
\bar{f}_{h_2}(x,p_2)\right. 
-\left. f_{h}(x,p)f_{h_2}(x,p_2)\bar{f}_{h_3}(x,p_3)
\bar{f}_{h_4}(x,p_4) \right\} \ldots \ .
\nonumber\\
\label{transport} 
\end{eqnarray}
The {\it lhs} describes the motion of particle 1 under consideration in 
momentum-dependent relativistic mean fields $U_{h}^{S}(x,p)$ 
and $U_{h}^{\mu}(x,p)$, which 
correspond to the real part of the scalar and vector hadron selfenergies, 
respectively.  The {\it rhs} represents the collision term for the 
process  $1+2\to 3+4+\ldots$  involving 
momentum integrations for incoming particle 2 as well as all outgoing 
particles $3,4,\dots$. The associated `transition rate'   
$W\equiv [G^+G]_{12\to 34\ldots} \delta^{(4)}(\Pi+\Pi_2-\Pi_3-\Pi_4\ldots )$
is given
in terms of the relativistic G-matrix (\ie, the in-medium 
scattering amplitude). For on-shell two-body
scattering the transition rate can be expressed through the differential 
cross section as
\beq
W= s \ \frac{d\sigma}{d\Omega}(s,\theta) \ \delta^{(4)}(\Pi+\Pi_2-\Pi_3-\Pi_4)
\eeq
with $cms$ energy $s=(\Pi+\Pi_2)^2$ and scattering angle $\theta$. 
The hadron quasiparticle properties in Eq.~(\ref{transport}) are defined 
via the mass-shell constraint~\cite{Weber1}, characterized by
$\delta (\Pi_{\mu}\Pi^{\mu}-M_{h}^{*2})$, with effective masses
and momenta given by
\begin{eqnarray} 
M_{h}^* (x,p)&=&M_h + U_h^{{S}}(x,p) \nonumber \\
\Pi^{\mu} (x,p)&=&p^{\mu}-U^{\mu}_h (x,p)\ \ ,
\end{eqnarray}
while the phase-space factors
\begin{equation}
\bar{f}_{h} (x,p)=1 \pm f_{{h}} (x,p)
\end{equation}
account for Pauli blocking or Bose enhancement,
depending on the type of hadron in the final and initial
state. The ellipsis in Eq.~(\ref{transport}) indicate further contributions
to the collision term with more than two hadrons in the final/initial
channels. The transport approach (\ref{transport}) is fully specified by
in-medium potentials $U_{h}^{S}(x,p)$ and $U_{h}^{\mu}(x,p)$ 
$(\mu =0,1,2,3)$, which
determine the mean-field propagation of the hadrons, and by the
transition rates $G^\dagger G\,\delta^{(4)} (\ldots)$ in the collision
term that describe the scattering and hadron production/absorption
rates. Clearly, these quantities should be in accordance with empirical 
information as much as possible. Therefore, a  frequently used model for 
the underlying microscopic mean-field
potentials is the $\sigma-\omega$ (Walecka) model~\cite{SeWa} (also
known as 'Quantum Hadrodynamics' or QHD) which accounts for the ground-state
properties of nuclear matter as well as proton-nucleus scattering, 
 once momentum-dependent corrections are properly included. To make
a more direct link to the underlying quark structure, which should become
relevant at full CERN-SpS energies, the ideas of the Walecka model have been
extended to couple the mean fields to the constituent quarks within the
hadrons~\cite{LKB}, without distorting the nuclear matter properties. 
In the HSD approach, where the (soft) hadronic dynamics are also based on 
(chiral) quark dynamics along the lines of NJL-models, extra care has been
taken to correctly handle the 'hard' processes (relevant in the collision
term). This has been achieved by employing the LUND string fragmentation 
model~\cite{LUND}, which correctly describes inelastic hadronic reactions 
over a wide energy regime. The  HSD model has been successfully applied 
to heavy-ion reactions ranging from SIS (1~AGeV) to CERN-SpS (200~AGeV) 
energies, see Ref.~\cite{HSD} for a comprehensive presentation.

In the transport framework, dilepton observables can be calculated 
with relative ease by  incorporating the relevant process in the 
collision term using the corresponding cross sections. 
In the case of $\pi\pi$ annihilation, which in VDM proceeds through 
$\rho$-meson formation, one has 
\bea
\sigma_{\pi^+\pi^-\to e^+e^-}(M) &=& \sigma_{\pi^+\pi^-\to \rho^0}(M)
\ \Gamma_{\rho^0\to e^+e^-}(M)/\Gamma_{\rho^0}^{tot} 
\nonumber\\
&=&\frac{4\pi\alpha^2}{3} \ \frac{p_\pi}{M^3} \ |F_\pi(M)|^2  \  
\label{xsppee}
\eea
with the electromagnetic form factor $F_\pi$ and the pion decay 
momentum $p_\pi=(M^2/4-m_\pi^2)^{1/2}$. 
Since electromagnetic processes are suppressed by a factor $\alpha=1/137$ 
(for dilepton production by even a factor of $\alpha^2$, \ie, four orders of 
magnitude), the feedback on the heavy-ion reaction dynamics can be neglected. 
To account for decays of $\rho$ mesons which are not produced in the
$\pi\pi$ channel but, \eg, in $\pi N$ or $NN$ collisions, one integrates 
their (time-dependent) abundance in the fireball over its
lifetime $t_{fo}$ according to~\cite{LKB,BC97}
\beq
\frac{dN_{\rho\to ee}}{dM} = \int\limits_0^{t_{fo}} \ dt \
\frac{dN_{\rho^0}(t)}{dM} \ \Gamma_{\rho^0\to ee}(M) \ , 
\label{rhoee} 
\eeq
where $dN_{\rho^0}(t)/dM$ is the number of $\rho$ mesons per invariant 
mass bin $dM$ at time $t$. In fact, the same treatment can be applied 
to $\rho$ mesons produced in the $\pi\pi$ channel, 
but then the latter should no longer be accounted for through 
Eq.~(\ref{xsppee}) to avoid double counting. Without any
further medium modifications 
both variants have been shown to be equivalent~\cite{LK95}.  
For $\omega$ 
and $\phi$ mesons relations analogous to Eq.~(\ref{rhoee}) hold. 
Note, however, that apart from the absolute abundances of the vector 
mesons (which should be similar for $\rho^0$ and $\omega$ and somewhat 
suppressed for the $\phi$ due to its higher mass) the key quantity which
determines the dilepton yield in Eq.~(\ref{rhoee}) are the {\em absolute} 
values of the dilepton decay widths. The (on-shell) numbers are  
\bea
\Gamma(\rho^0\to ee) &=& 6.77\pm 0.32~{\rm keV}  \nonumber\\ 
\Gamma(\omega\to ee) &=& 0.60\pm 0.02~{\rm keV}  \nonumber\\
\Gamma(\phi\to ee)   &=& 1.37\pm 0.05~{\rm keV}  \ ,
\label{emwidth}
\eea
clearly indicating the prevailing character of the $\rho$ meson
for radiation originating from the interacting hadronic system.  
Three-body decays into dilepton channels are evaluated in a 
similar fashion, \eg, for $a_1(1260)\to \pi ee$: 
\beq
\frac{dN_{a_1\to\pi ee}}{dM}= \int\limits_0^{t_{fo}} \ dt \ N_{a_1}(t) \ 
\frac{d\Gamma_{a_1\to\pi ee}}{dM} 
\eeq
with $N_{a_1}(t)$ being the number of $a_1$ mesons present at time $t$.
The differential Dalitz-decay width is given via the
radiative decay width $\Gamma_{a_1\to\pi\gamma}=0.64$~MeV as~\cite{BC97} 
\beq
\frac{d\Gamma_{a_1\to\pi ee}}{dM}=\frac{2\alpha}{3\pi} \ 
\frac{\Gamma_{a_1\to\pi\gamma}}{M} 
\frac{\left[(m_{a_1}^2+M^2-m_\pi^2)^2-4m_{a_1}^2 M^2)\right]^{3/2}}
{(M_{a_1}^2-m_\pi^2)^3}
\eeq
(this expression is reminiscent of Eq.~(\ref{gammaA}) after stripping off 
the hadronic and VDM form factors from the latter). 

Due to the above argument, also transport analyses have primarily  
investigated in-medium effects in dilepton production by  
focusing on modifications of the $\rho$ meson (or $\pi\pi$ annihilation). 
The most straightforward effect to incorporate is to simply change
the in-medium $\rho$ mass (as motivated by theoretical predictions  
such as BR scaling discussed in Sect.~\ref{sec_drop}). It can be 
accommodated by appropriate 
mean-fields in the transport equations, together with replacing 
the free masses $m_\rho$ by $m_\rho^*$ in
Eqs.~(\ref{xsppee}) or (\ref{rhoee}), as has been done, \eg, in 
Refs.~\cite{CEK95,LKB,CEK96,LKBS,BC97,BK98}.  
On the other hand, medium modifications which go beyond 
the mean-field treatment of on-shell quasiparticles are not easily 
implemented. This is obvious for the case in which the in-medium widths 
of the  propagated hadrons become so large that they lose their 
quasiparticle nature. Although there is no satisfactory 
solution to this problem  yet (see Ref.~\cite{CJ99,Knoll98} for recent 
progress), some attempts have been made to account 
for the broadening effect of the $\rho$ in a transport framework.  
In Ref.~\cite{Koch96} the cross section approach based on 
Eq.~(\ref{xsppee}) has been employed by simply multiplying 
$\sigma_{\pi^+\pi^-\to e^+e^-}(M)$ with a temperature-dependent 
'in-medium correction factor', which is determined 
as the ratio 
\beq
R_{\rm med}(M,T)\equiv \frac{\left[dR/dM^2d^3q\right]^{med}_{q=0} }
{\left[dR/dM^2d^3q\right]^{free}_{q=0} }
\label{Rmed}
\eeq
of the (equilibrium) dilepton production rate for free $\pi\pi$ 
annihilation over the in-medium one~\cite{SK96}. To assign a  temperature for 
a given $\pi\pi$ event necessary to evaluate $R_{\rm med}$, the local 
(invariant) pion density in the transport has been used assuming 
local thermal equilibrium. A possible three-momentum dependence
(at fixed invariant mass $M$) as well as dispersion corrections
to the explicit pion propagation in the transport have been neglected
(the impact of a modified pion dispersion relation on the overall
transport dynamics has been shown to be small in Ref.~\cite{KB93}). 
Along similar lines Refs.~\cite{CBRW,BCRW} proceeded in taking
the $\pi\pi$ annihilation cross section as
\beq
\sigma_{\pi^+\pi^-\to e^+e^-}(M) = - {16\pi^2\alpha^2\over g_{\rho\pi\pi}^2} \
 {1\over p_\pi^2 M^2} \ (m_\rho^{(0)})^4 \ {\rm Im} D_\rho (q_0, \vec q)
\label{xsppeemed}
\eeq
with the full in-medium spectral function based on the many-body calculations
of Refs.~\cite{RCW,RUBW,UBRW}, including the non-trivial three-momentum
dependence (for the vacuum spectral function Eq.~(\ref{xsppee}) is
recovered). Here the temperature has been deduced from the slope 
parameter associated with the pion-momentum distributions in the local
rest frame ('comoving' frame). Both calculations are in principle plagued 
by a singularity towards the two-pion threshold $M=2m_\pi$, caused  
by the inherent inconsistency of how the pions are treated in the
transport and in the dilepton rate. 
However, in practice this failure does not seem to entail severe 
disturbances except for very close to threshold. In Refs.~\cite{CBRW,BCRW}
the additional decays of $\rho$ mesons produced in meson-baryon and 
baryon-baryon interactions have been accounted for by using
a medium modified mass distribution according to 
\beq
{dN_{\rho}\over dM}= -{2M\over \pi} \ {\rm Im} D_\rho (q_0,\vec q) \ ,
\label{dNrhodM}
\eeq
together with a constant branching ratio of 
$\Gamma_{\rho\to ee}/\Gamma_\rho^{tot}=4.5\cdot 10^{-5}$ for the dilepton
channel. This leads to a quite different spectral shape as
compared to the $\pi\pi$ component included via Eq.~(\ref{xsppee}), 
and might not be realistic (see also the criticism in Ref.~\cite{Knoll98}). 
It has been improved in due course  
by introducing an additional phase space factor $(M/m_\rho)^2$ into
the electromagnetic branching ratio. 

To avoid double counting when using medium-modified dilepton rates, it is 
important to omit explicit production channels that are included 
in the medium effects~\cite{CBRW}; \eg, if the $\rho$ selfenergy in the 
in-medium propagator contains $\pi \rho \to a_1$ contributions, 
the explicit $a_1\to \pi e^+e^-$ decays have to be switched off. 
This holds for any process that can be generated by cutting any
 selfenergy diagram of the many-body spectral function. 

\subsection{Thermal Fireball Expansion}
\label{sec_fireball}
A much simplified attempt to capture the basic features of a heavy-ion
reaction relevant for dilepton production is represented by 
fireball models, some of which are reminiscent to the 
Bjorken hydrodynamical description. One class of approaches~\cite{RCW} is 
based on temperature evolutions parameterized in accordance
with microspcopic transport simulations~\cite{LKBS}, \eg, 
\beq
T(t)=(T^i-T^\infty) \ e^{-t/\tau} + T^\infty \
\label{cooling}
\eeq
with the initial temperature of the hadronic phase,
$T^i$, a time constant $\tau$  and an 'asymptotic' temperature
$T^\infty$. The baryon density is determined
by the number of participants supplememted with an
(isotropic) volume expansion, which can be approximately described by
an 'average'  baryon chemical potential, \eg,  $\mu_B\simeq 350-400$~MeV
for lab-energies of 158--200~AGeV. The total dilepton yield from $\pi\pi$ 
annihilation is then  normalized to the transport results for the 
case when no-medium effects are included. The pertinent normalization factor
turns out to be $N_0\simeq 2-3$ and can be understood as an overpoulation
of the pion phase space due to a finite chemical
potential. In fact, for typical SpS freezeout temperatures of
$T_{fo}\simeq 120$~MeV, the simplistic fireball model 
results in a total pion number which falls short by about 50\%
as compared to the experimentally observed pion-to-baryon ratio
of 5:1 if no pion chemical potential is involved.
This can be corrected for by introducing a finite value
of $\mu_\pi\simeq 50$~MeV,  amounting to a pion  fugacity
(in Boltzmann approximation) of $z_\pi=e^{\mu_\pi/T_{fo}}\simeq 1.5$.
On the level of dilepton production from $\pi\pi$
annihilation~\cite{PKoch,KKP94} this
results in an enhancement factor of $N_0\simeq z_\rho= z_\pi^2\simeq 2.3$. 

A more microscopic understanding of the emergence  of finite (meson) 
chemical potentials can be gained by noticing that the strong interactions
of, \eg,  pions -- $\pi\pi$ scattering or $\pi N$ interactions dominated
by baryonic resonances -- are essentially elastic, \ie, pion-number
conserving. Maintaining the assumption of local thermal equilibrium 
pion-number and entropy conservation then enforces the build-up 
of a pion chemical potential 
in the expansion and cooling process within the hadronic phase of
a heavy-ion collision, as has been first pointed out in Ref.~\cite{Bebie}.     
Such a scenario fits in fact nicely into the picture of recent
hadro-chemical analysis~\cite{PSWX,pbm96,pbm98}, 
where a large body of data for finally observed particle abundances at 
SpS energies (158~AGeV) could be accomodated by a universal 
temperature and baryon chemical potential of 
$(T,\mu_N)_{ch}\simeq (175,270)$~MeV, characterizing the 
{\it chemical} freezeout of the system (the same method successfully describes
the AGS data as well, cf.~Fig.~\ref{fig_freeze}). From here
on the particle composition in terms of ({\it w.r.t.}~to strong interactions)
stable particles does no longer change, although the hadronic systen still
interacts via elastic collisions sustaining thermal equilibrium until the 
{\it thermal} freezeout is reached. The evolution proceeds along a trajectory
in the $T-\mu_N$ plane which can be determined by imposing entropy and 
baryon-number conservation within, 
\eg, a hadronic resonance gas equation of state. 
The induced  pion ($\mu_\pi(T)$) and kaon 
($\mu_K(T)\simeq \mu_{\bar K}(T)$) chemical potentials along 
this trajectory increase approximately linearly, reaching typical values of 
$\mu_\pi^{fo}\simeq 60-80$~MeV and $\mu_{K,\bar K}^{fo}\simeq 100-130$~MeV
at thermal freezeout, being located around 
$(T,\mu_N)_{fo}\simeq (115\pm 10,430\pm 30)$~MeV.
Other chemical potentials associated with strong interactions are  
kept in relative chemical
equilibrium, \eg, $\mu_\Delta=\mu_N+\mu_\pi$ or $\mu_\rho=2\mu_\pi$ according
to elastic reactions $\pi N\to \Delta \to \pi N$ or
$\pi\pi\to\rho\to\pi\pi$.   

Finally one needs to introduce a time scale to obtain the volume expansion.
For SpS energies the latter is realistically approximated 
by a cylindrical geometry as
\beq
V_{FC}^{(2)}(t)= 2 \ (z_0+v_z t +\frac{1}{2} a_z t^2) \ \pi \
(r_0+\frac{1}{2} a_\perp t^2)^2 \ , 
\eeq
where two firecylinders expanding
in the $\pm z$ direction have been employed to allow for a sufficient
spread in the particle rapidity distributions. Guided by hydrodynamical 
simulations~\cite{HuSh97}
the primordial longitudinal motion for Pb(158~AGeV)+Au reactions
is taken to be $v_{z}=0.5c$, and the longitudinal and transverse
acceleration are fixed to give final velocities $v_{z}(t_{fo})\simeq 0.75c$,
$v_\perp(t_{fo})\simeq 0.55c$ as borne out from experiment~\cite{Stach}
 (this, in turn,
requires fireball lifetimes of about $t_{fo}=10-12$~fm/c and implies
transverse expansion by 3-4~fm, consistent with HBT analyses~\cite{hbt}).
The parameter $r_0$ denotes the initial nuclear overlap radius,
\eg, $r_0=4.6$~fm for collisions with impact parameter $b=5$~fm and
$N_B\simeq 260$ participant baryons. The parameter $z_0$ is equivalent to
a formation time and fixes the starting point of the trajectory
in the $(T,\mu_N)$ plane.
Estimates for the initial baryon densities can be taken, \eg, from
RQMD calculations which for  CERN-SpS energies typically lie around 
$\varrho_B^i\simeq 2-4 \varrho_0$~\cite{Sorge96}, corresponding 
to, \eg, $(T,\varrho_B)_{ini}$=(190~MeV,2.55$\varrho_0$) on the above
specified trajectory.  
\begin{figure}[!htb]
\vspace{-0.5cm}
\begin{center} 
\epsfig{figure=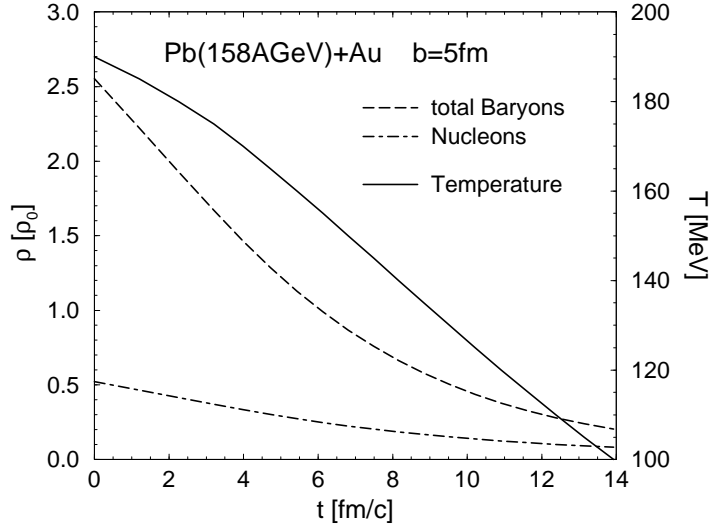,width=8cm,angle=-90}
\end{center} 
\caption{Time evolution of temperature (full line, right scale), 
total baryon density (long-dashed line, left scale) and nucleon density 
(dashed-dotted line, left scale) 
as typical for central Pb(158A~GeV)+Au collisions at impact parameter
$b$=5~fm with fixed entropy per baryon and assuming effective
pion- and kaon-number conservation.}  
\label{fig_Tevo}
\end{figure}

Dilepton spectra from in-medium $\pi\pi$ annihilation (or, equivalently, 
$\rho$ decays) are now straightforwardly calculated by integrating
the thermal rate, Eq.~(\ref{dRd4q_klz}). Using $q_0 dq_0= M dM$, one has  
\bea
\frac{dN_{\pi\pi\to ee}}{dM d\eta} &=&
\int\limits_{0}^{t_{fo}}
dt \ V_{FC}(t) \int d^3q \ \frac{M}{q_0} \ 
\frac{dR_{\pi\pi\to ee}}{d^4q}(q_0,\vec q;\mu_B,\mu_\pi,T) \
Acc(M,\vec q) \
\nonumber\\
&=&  \frac{\alpha^2}{\pi^3 g_{\pi\pi\rho}^2 M} \
\int\limits_0^{t_{fo}} dt \ V_{FC}(t)
\int \frac{d^3q}{q_0} \ f^\rho(q_0;\mu_\rho,T) \
{\cal F}(M,\vec q; \mu_B,\mu_\pi,T) \ Acc(M,\vec q) \ ,  
\nonumber\\
\label{dlspecfb}
\eea
where the function $Acc(M,\vec q)$ accounts for the 
experimental acceptance cuts specific to the detector characteristics
(\eg, in the CERES experiment each electron/positron track is required 
to have transverse momentum $p_T>0.2$~GeV, to fall in the (pseudo-)
rapidity interval $2.1< \eta < 2.65$, and to have a pair  
opening angle $\Theta_{ee}>35$~mrad). The meson chemical potentials 
have throughout been introduced in Boltzmann approximation, \ie, 
$f^\pi(\omega;\mu_\pi,T)\equiv f^\pi(\omega;T) \ e^{\mu_\pi/T}$, etc..  

The fireball models certainly oversimplify the dynamics
present in more realistic descriptions of relativistic heavy-ion 
collisions.  However, their relative simplicity enables more transparent 
comparisons between various underlying models for the
microscopic rates~\cite{CRW,RCW,SYZ2,LYZ,GKP99,RW99}  at the level of 
experimentally observed data,  
potentially discriminating different temperature and  
density dependencies of bare rates once integrated over a 
common 'cooling-curve'.

\section{Dilepton Spectra at BEVALAC/SIS Energies}
\label{sec_dls}
The first measurements of dilepton invariant mass spectra in proton and 
heavy-ion induced reactions at bombarding energies in the 1-5~AGeV
range have been performed by the DLS collaboration for p+Be collisions at 
1, 2.1 and 4.9~AGeV,  
for Ca+Ca at 1 and 2~AGeV and for 
Nb+Nb at 1.05~AGeV~\cite{DLS1}. Although the $cms$ energy available in a 
primary nucleon-nucleon collision at 1~GeV laboratory energy only suffices to 
produce dilepton pairs of invariant masses up to $M\simeq0.45$~GeV, 
significant yields have been observed beyond this naive kinematical limit
in the collisions involving heavy ions. 
Therefore, these first generation DLS data have been interpreted as the 
first evidence for the $\pi\pi\to e^+e^-$ annihilation channel, predominantly 
populating the invariant mass region $M_{ee}\ge 0.4$~GeV (up to 
about 1~GeV).  Theoretical calculations based on  BUU transport 
models~\cite{Xiong90,Wolf} have confirmed this conjecture. Also, 
for lower invariant masses, $M\le0.4$~GeV, the major contributions 
to the dilepton spectra were identified as Dalitz decays of $\Delta$'s
and $\eta$'s as well as proton-neutron Bremsstrahlung processes, resulting
in  a fair agreement with the first generation DLS 
data~\cite{Xiong90,Wolf}. The limited statistics of the latter 
did not allow for any further conclusions. 

From more recent
publications of the DLS collaboration~\cite{DLS2} it turns out that
new measurements of dilepton yields in $p$+$p$, $p$+d as well as  in 
1~AGeV d+C, He+Ca, C+C and Ca+Ca  collisions have been substantially
revised in comparison to the previous data
set~\cite{DLS1} due to improvements of the DLS detector and of 
the data analysis, correcting for dead-time losses. 
The $p$+$p$ data for 1--5~GeV incident energies are reasonably 
well described by standard Dalitz and vector meson decay sources 
(for recent transport calculations see Ref.~\cite{Ernst} using 
the UrQMD and Ref.~\cite{BCEM99} using the HSD approach; both analyses
essentially agree up to slight uncertainties in some meson production 
channels; also the role of $N(1520)\to N e^+e^-$ decays, not included
in Ref.~\cite{Ernst}, was found to be significant in Ref.~\cite{BCEM99}
especially for the higher energies). This gives confidence in the 
elementary production channels when carrying the calculations to the
more complicated systems. Nevertheless, significant      
discrepancies start to build up already for light projectiles such 
as in d+Ca and He+Ca systems~\cite{Ernst}. 
In the 1~AGeV C+C and Ca+Ca reactions the 
experimental reanalysis~\cite{DLS2} resulted in corrections of up
to  factors of 6--7 as compared to the first generation data.  
Consequently, the afore mentioned theoretical approaches now appreciably 
underestimate the dilepton yield in the invariant mass 
region $0.15~{\rm GeV} \le M_{ee} \le 0.65~{\rm GeV}$. This is apparent,  
\eg,  from the left panels of Fig.~\ref{fig_dlshsd}, where a rather 
complete calculation 
of the various dilepton sources including (free) $\pi\pi$ annihilation 
employing the HSD model is displayed~\cite{BCRW}.    
\begin{figure}
\vspace{-1cm}
\epsfig{figure=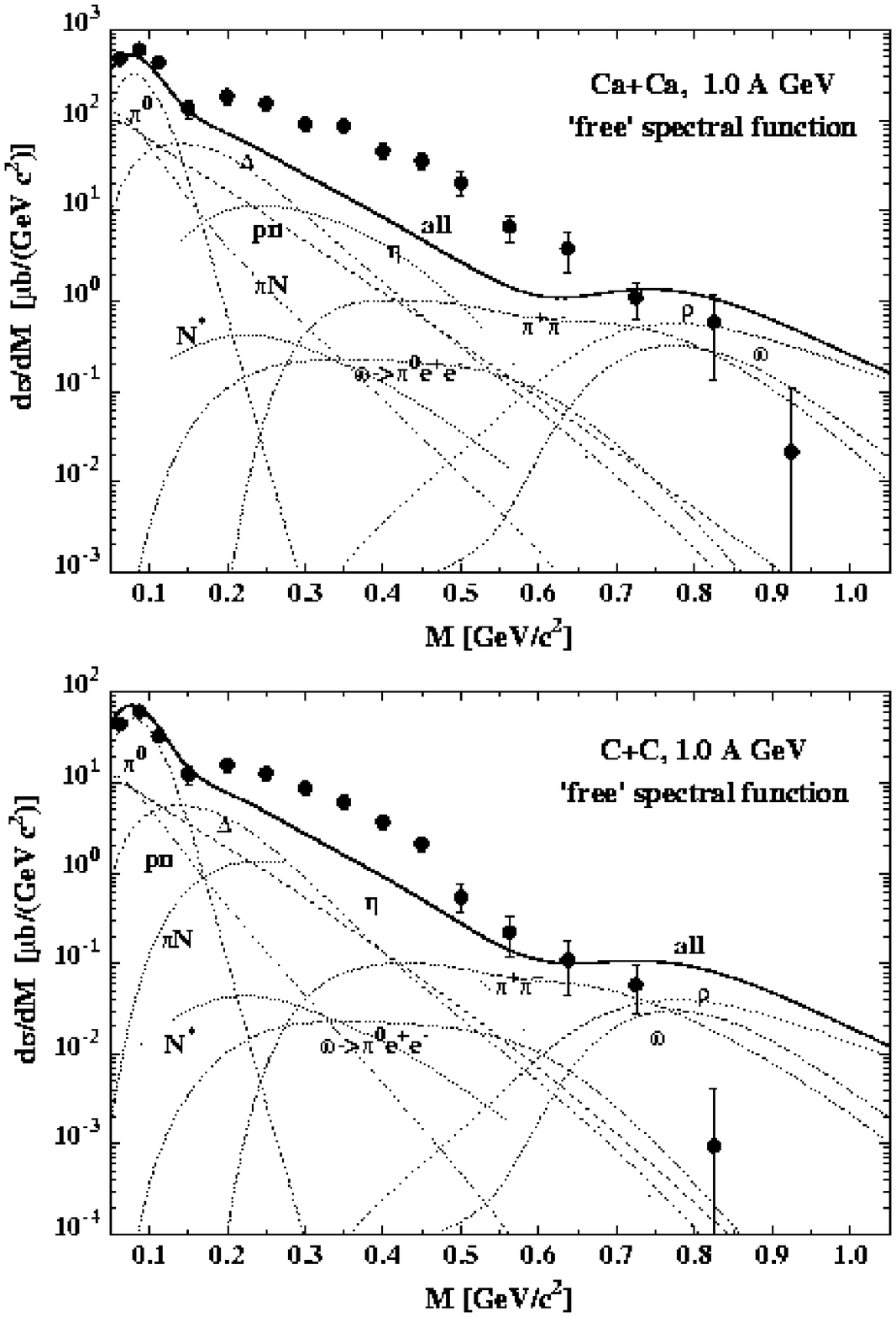,height=11cm,width=7.7cm}
\hspace{-0.6cm}
\epsfig{figure=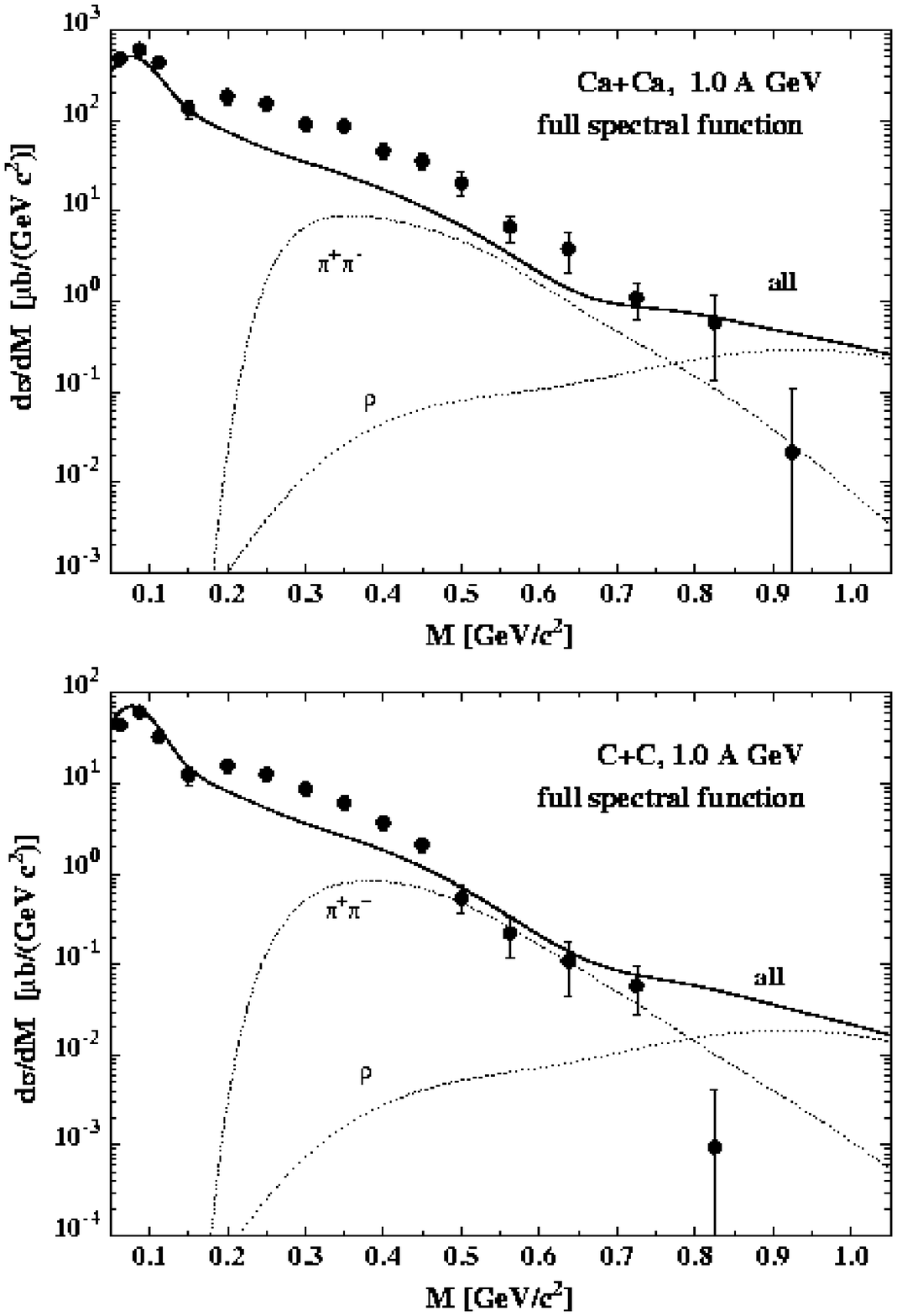,height=11cm,width=7.7cm}
\vspace{-0.6cm}
\caption{Dilepton spectra as measured in 
Ca+Ca (upper panels) and C+C (lower panels) reactions at 1.0~AGeV 
projectile energies. The experimental data from the DLS
collaboration~\protect\cite{DLS2} are compared to  
HSD transport calculations, using either a 'free' $\rho$ spectral 
function (left panels) or the in-medium one  
from Refs.~\protect\cite{RCW,RUBW} (right panels) for 
both $\pi\pi$ annihilation and direct decays of $\rho$ mesons produced 
in baryonic collisions.  
The DLS acceptance filter (version 4.1) as well as a mass resolution 
of $\Delta M/M=10\%$ are included.
The thick solid lines represent the total results. 
The thin lines in the left panels indicate the individual 
contributions from the different production channels, \ie,  
starting from low $M$: Dalitz decays $\pi^0 \to \gamma e^+ e^-$ (dashed
line), $\eta \to \gamma e^+ e^-$ (dotted line), $\Delta \to N e^+ e^-$
(dashed line), $\omega \to \pi^0 e^+ e^-$ (dot-dashed line), $N^* \to N
e^+ e^-$ (dotted line), proton-neutron bremsstrahlung (dot-dashed
line), $\pi N$ bremsstrahlung (dot-dot-dashed line); for $M \approx $
0.8 GeV:  $\omega \to e^+e^-$ (dot-dashed line), $\rho^0 \to e^+e^-$
(dashed line), $\pi^+ \pi^- \to \rho \to e^+e^-$ (dot-dashed line).
The plots are taken from Ref.~\protect\cite{BCRW}.}
\label{fig_dlshsd}
\end{figure}
In particular, this calculation reproduces well the 
total yields and transverse momentum spectra of $\pi$'s and $\eta$'s  
as measured for the same collision systems and energies by the 
TAPS collaboration~\cite{TAPS1}. This imposes stringent constraints 
on the dilepton yields from $\pi^0$ and $\eta$ Dalitz decays. Note 
that the former result in a satisfactory description of the mass region 
$M_{ee}\le 0.15$~GeV.
On the other hand, it was noted in Ref.~\cite{DLS2} that, assuming an 
isotropic dilepton emission from a thermal source, the new DLS data 
could be accounted for by (arbitrarily) increasing the $\eta$ 
contribution by a large factor close to 10. This observation could be
reconciled with the TAPS data imposing a dramatic anisotropy in the 
$\eta$ production, since the TAPS data are taken mainly at mid-rapidity, 
whereas the DLS data are at forward rapidities. Although the HSD calculations
show some enhancement of the $\eta$ distributions~\cite{BCRW} at 
forward angles, this would at maximum allow for a 20\% enhancement 
of the total $\eta$ yields~\cite{TAPS1} which is nowhere close 
to providing  an excess of the $\eta$ signal as required to reproduce 
the DLS data. 
 
The question thus arises whether in-medium effects can resolve 
the discrepancy. 
As an example of the conditions probed at BEVALAC energies, 
Fig.~\ref{fig_dlsevo} shows a typical time dependence  
of the baryon density for inclusive  Ca+Ca collisions at 1~AGeV  
bombarding energy as extracted  from the transport model of
Ref.~\cite{ZG94}. 
\begin{figure}[!htb]
\centerline{\epsfig{file=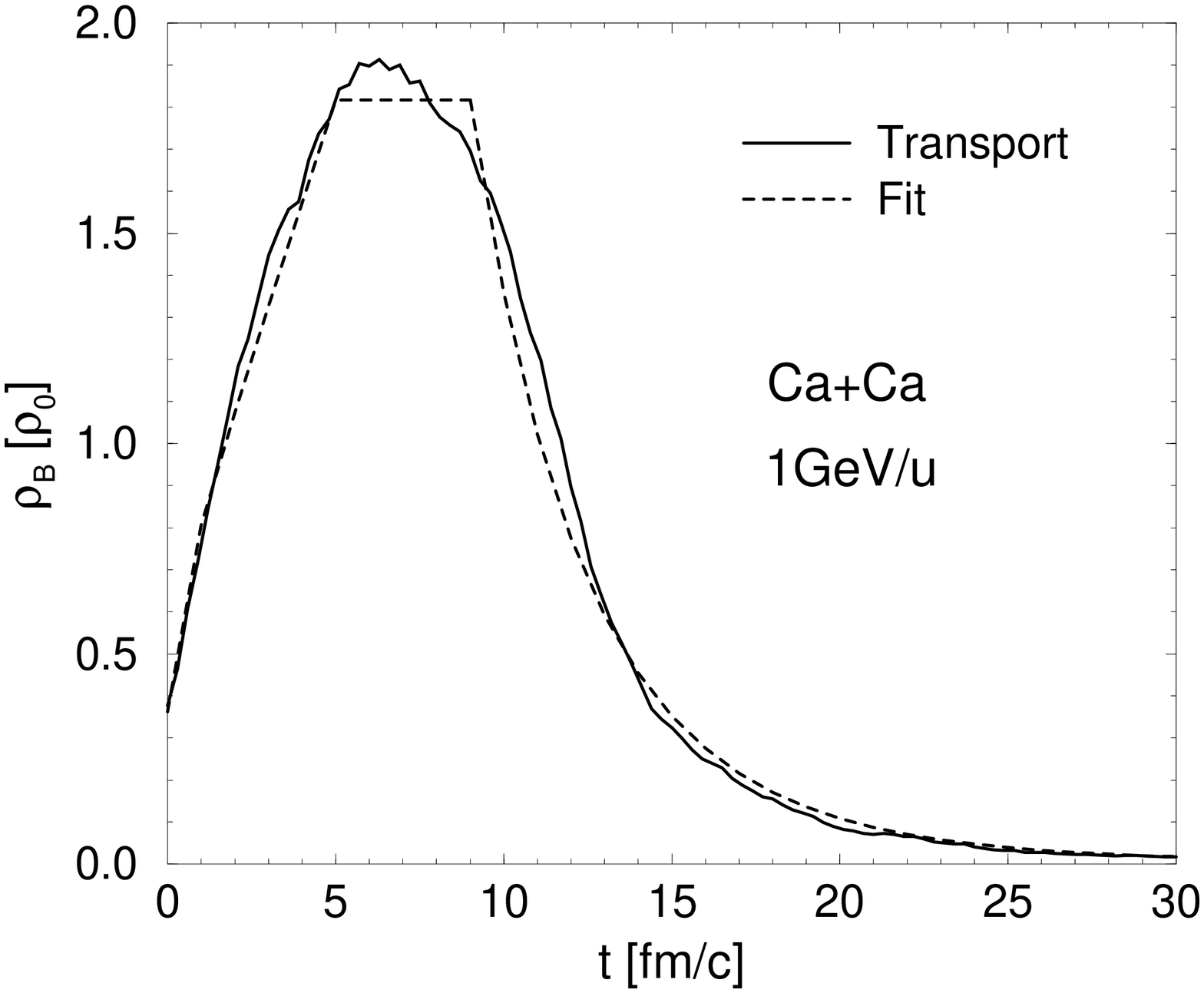,width=8cm,angle=-90}}
\caption{Evolution of average baryon
density as a function of time in
inclusive 1~AGeV Ca+Ca collisions as extracted from the transport model
of Ref.~\protect\cite{ZG94} (solid curve); the dashed curve is 
a simple parameterization
thereof with a maximal temperature of $T_{max}=100$~MeV in the high density
phase and a time-independent baryon 
chemical potential of $\mu_B=0.76$~GeV.}
\label{fig_dlsevo}
\end{figure}
The evolution roughly proceeds in 3 stages:
in the first 5~fm/c the nuclei penetrate each other to form
a high-density phase at about $2\varrho_0$ which then quite rapidly
dilutes towards freezeout. The accompanying temperatures at the highest
densities are around $T\simeq 80-100$~MeV and therefore one should
expect the system to stay in the hadronic phase throughout.
Since the pion densities are rather small ($n_\pi=0.03$~fm$^{-3}$
at $T$=100~MeV), the dominant medium effects should be driven
by nucleons and baryonic resonances in the system (note that at $T$=100~MeV
already $\sim$25\% of the baryons are thermally excited into 
$\Delta$'s).

In Ref.~\cite{BCRW}, the impact of many-body effects~\cite{RCW,RUBW} 
on the $\rho$-meson spectral function, generated through $\pi N$ and $\rho N$
interactions (also including finite-temperature effects, which are, however,  
much smaller), has been investigated. 
The results are shown in the two right panels of Fig.~\ref{fig_dlshsd}. 
Apparently, the full results still underestimate the second 
generation DLS data by a factor of 2--3. Very similar results are 
obtained~\cite{BCRW} when employing  
the finite-density/zero-temperature $\rho$-meson spectral function of
Ref.~\cite{PPLLM}, which is based on a selfconsistent calculation 
of resonant $\rho N$ interactions. 
Since present transport approaches cannot fully account for the off-shell 
dynamics of the pions (especially close to and below the two-pion
threshold), additional   
calculations using the thermal fireball along the lines of 
Sect.~\ref{sec_fireball}  have been performed. Integrating the thermal
dilepton emission rates over the density and temperature profile of 
Fig.~\ref{fig_dlsevo} leads to very similar results as obtained 
in the HSD calculation. 

Alternative theoretical attempts were made by including dropping vector meson 
masses in transport calculations. Ref.~\cite{BK98} focused on the 
role of the $N(1520)$-resonance, which exhibits a strong coupling to 
the $\rho N / \gamma N$ channel, see Sects.~\ref{sec_rhoNres} 
and \ref{sec_photoabs}. Rather than including it via the 
in-medium $\rho$-meson spectral function as was done
in Ref.~\cite{BCRW}, the Dalitz decays $N(1520) \to N e^+e^-$ were evaluated 
explicitely, confirming its relative importance for 
the low-mass dilepton spectra.
However, once an additional reduction of the $\rho$ mass is introduced
(using, \eg,  the QCD sum rule results, $m_V^*=m_V(1-C\varrho/\varrho_0)$ 
with $C\simeq 0.18$), the hadronic decay width of the $N(1520)$ strongly
increases due to the opening of phase space in the $\rho N$ decay, 
which results in a net reduction of the dilepton yield from $N(1520)$ 
decays (note that the in-medium broadening is included in the many-body
calculations of Refs.~\cite{RUBW,PPLLM}). Although the direct 
$\rho$ decay contributions are enhanced 
by a factor of about 3 with a dropping $\rho$ mass, the total spectra
still underestimate the DLS data by a factor of 3--4 for invariant
masses $0.15~{\rm GeV} \le M \le 0.5~{\rm GeV}$. Similar conclusions 
have been reached in Ref.~\cite{Ernst} where the use of dropping 
vector meson masses has been found to give a small net increase of     
the spectra around $M\simeq 0.4$~GeV, together with the reduction
around the free $\rho/\omega$ peak. 

Concluding this Section we emphasize that there is currently 
no theoretical explanation of the second generation DLS data for 
dilepton production in 1--2~AGeV heavy-ion collisions, with the  
various model predictions falling short by large factors of 2--3 above
the $\pi^0$ Dalitz region and below $M\simeq 0.5$~GeV.  
The upcoming high-precision dilepton measurements with the HADES detector
at GSI will be crucial to shed new light on this situation.

\section{Dilepton Spectra at CERN-SpS Energies}
\label{sec_cern}
The dilepton program at the CERN-SpS started in 1990--1992 with the $^{32}$S 
beam at 200~AGeV, followed by a 450~GeV proton run in 1993 and 
$^{208}$Pb nuclei accelerated to 158~AGeV in 1995, 1996 and 1998. Data 
have been taken by three collaborations: CERES/NA45 for $e^+e^-$ pairs 
using $^{197}$Au targets (as well as $^9$Be in the proton 
run)~\cite{ceres95,ceres96,ceres98}, HELIOS-3~\cite{helios3} for $\mu^+\mu^-$ 
pairs using $^{184}$W targets, and NA38/NA50~\cite{na38,na50psi,na50int} 
also for $\mu^+\mu^-$ pairs using $^{32}$S, Cu and  $^{238}$U as well as 
$^{208}$Pb targets for the lead runs. The major challenge in these
experiments is the large background due to both charged hadrons and, 
more severely,  
combinatorial misidentification of pairs in the same event, \ie, the 
pairing of $l^+$ and $l^-$ tracks which did not arise from the decay
of the {\em same} virtual photon (or other {\em correlated} physical 
processes, as, \eg, the so-called 'open charm' contributions, where a 
pairwise production of $D\bar D$ mesons is followed by subsequent
weak decays $D\to l^+ X$ and $\bar D\to l^- X$). In the CERES 
experiment, \eg, charged hadrons are suppressed with Cerenkov
detectors, whereas the combinatorial background to 
the 'physical' $e^+e^-$ signal is typically 
determined through pairing of like-sign pairs, \ie, $e^+e^+$ and 
$e^-e^-$.     

As a result of the different ways in handling these problems,
the kinematical regions covered by the three experiments are 
quite distinct, see Fig.~\ref{fig_expcuts}. 
Before one can identify non-trivial signals from the highly
complicated measurements in nucleus-nucleus collisions,
\begin{figure}[!htb]
\bce
\epsfig{figure=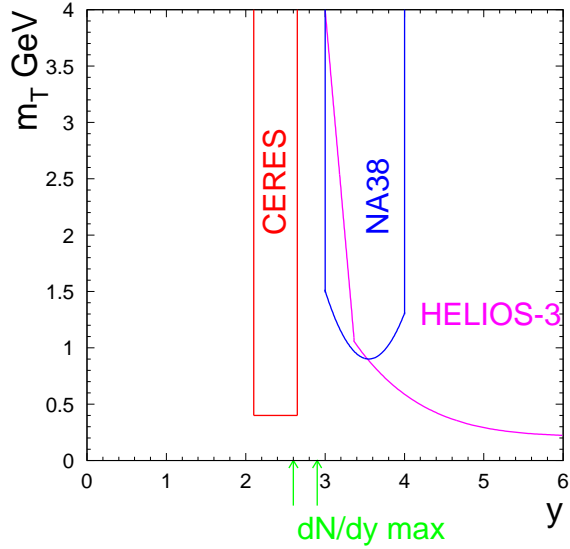,width=8cm}
\ece
\caption{Kinematical regions covered by the three collaborations 
that have measured dilepton spectra at full CERN-SpS energies of
158--200~AGeV. The plot is taken from Ref.~\protect\cite{ceres96d}.} 
\label{fig_expcuts}
\end{figure}
one has to have good control over the more simple systems first.  
The proton-induced reactions, supposedly governed by the mere
free decays of the produced hadrons involving no significant 
rescattering, thus serve as an important aid in understanding 
the detector systematics. The free hadronic decay contributions have 
become known as the ´hadronic cocktail´ and will be discussed in 
the next Section.     
Since at full CERN-SpS energies the pion-to-baryon ratio observed in the final
state of heavy-ion induced reactions is about 5:1 (with little dependence 
on the impact parameter $b$), the 
dominant in-medium source is expected to stem from $\pi\pi$ annihilation.
The assessment of this contribution without any medium-modifications  
will be addressed in sect.~\ref{sec_freepipi}. 
In Sects.~\ref{sec_mspec} and \ref{sec_ptspec} we proceed to the analysis 
of the experimental data in the available kinematic 
projections, \ie, invariant mass and transverse pair-momentum, 
with emphasis on in-medium effects that have been proposed.  
In Sect.~\ref{timeinfo} we discuss  how the various parts 
of the in-medium signals relate to the time of
emission  within different theoretical models. In particular, the  
issue of quark-gluon/hadron duality, raised in Sect.~\ref{sec_duality},  
will be reiterated for both the low- and intermediate-mass region.

\subsection{Decays after Freezeout: Hadronic Cocktail versus 
Experiment}   
\label{sec_cocktail} 
From the experimental side, a systematic study of the cocktail 
contributions has been performed by the CERES/NA45 
collaboration. It has been shown~\cite{ceres95} that the 
$e^+e^-$ invariant mass spectra in $p$+Be and $p$+Au, normalized to the 
number of charged particles observed in the same rapidity window, 
can be well reproduced in terms of known hadron decays using 
particle production multiplicities from $p$+$p$ data, cf.~Fig.~\ref{fig_dlpa}
in the Introduction.    
The low-mass end of the spectrum, $M_{ee}\le 0.15$~GeV, is completely 
saturated  by $\pi^0$ Dalitz decays, whereas for 
$0.15~{\rm GeV} \le M_{ee} \le 0.6$~GeV $\eta$ and $\omega$ Dalitz decays
are prevailing. Beyond $M_{ee}= 0.6$~GeV up to about 1.5~GeV the direct
decays of the light vector mesons, $\rho, \omega, \phi \to e^+e^-$,
are the dominant sources, substantially smeared due to the finite 
mass resolution of about 10\% in the CERES detector (which, in fact, stems
from the finite momentum resolution of the individual lepton tracks).  
Here one should note that, for an equal number of produced $\rho^0$'s and 
$\omega$'s, the dilepton yield from the latter is by almost a factor 
of 2 larger than the former, since in free space the probability 
$P_{V\to ee}$ for decaying into the dilepton channel is determined by the 
relative branching ratio of electromagnetic over total decay width, \ie,
\beq
P_{V\to ee} = \frac{\Gamma_{V\to ee}}{\Gamma_V^{tot}} = 
\left\{ \begin{array}{cc}
 0.0045\% \ , &  V=\rho  \\
 0.0071\% \ , &  V=\omega  \\
 0.0311\% \ , &  V=\phi  
  \end{array} \right. \  .  
\eeq 
This is a quite different characteristics as compared to the 
signal from an interacting (thermalized) system, cf. Eq.~(\ref{rhoee})
and the subsequent remarks.      
The CERES assessment of the cocktail in proton-induced reactions has 
been confirmed by microscopic transport calculations~\cite{CEK95,LKBS}. The 
latter give equivalent results for the dimuon data of HELIOS-3~\cite{helios3}
taken in $p$+W reactions. Thus the measured dilepton spectra in 
proton-induced reactions at CERN-SpS energies can be well understood 
by the final-state hadron decays in a consistent way.  

\begin{figure}[!htb]
\epsfig{figure=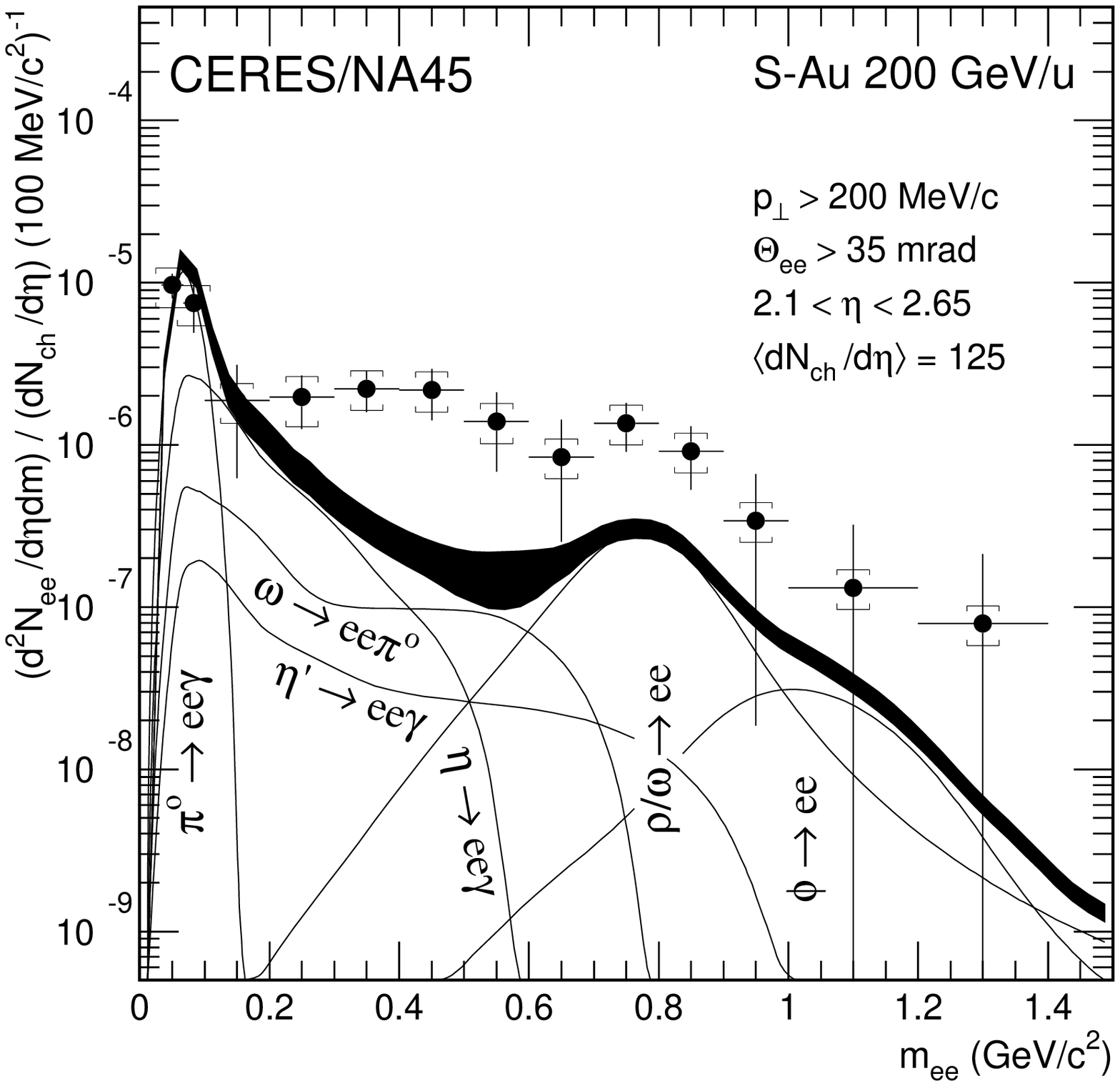,width=7cm}
\epsfig{figure=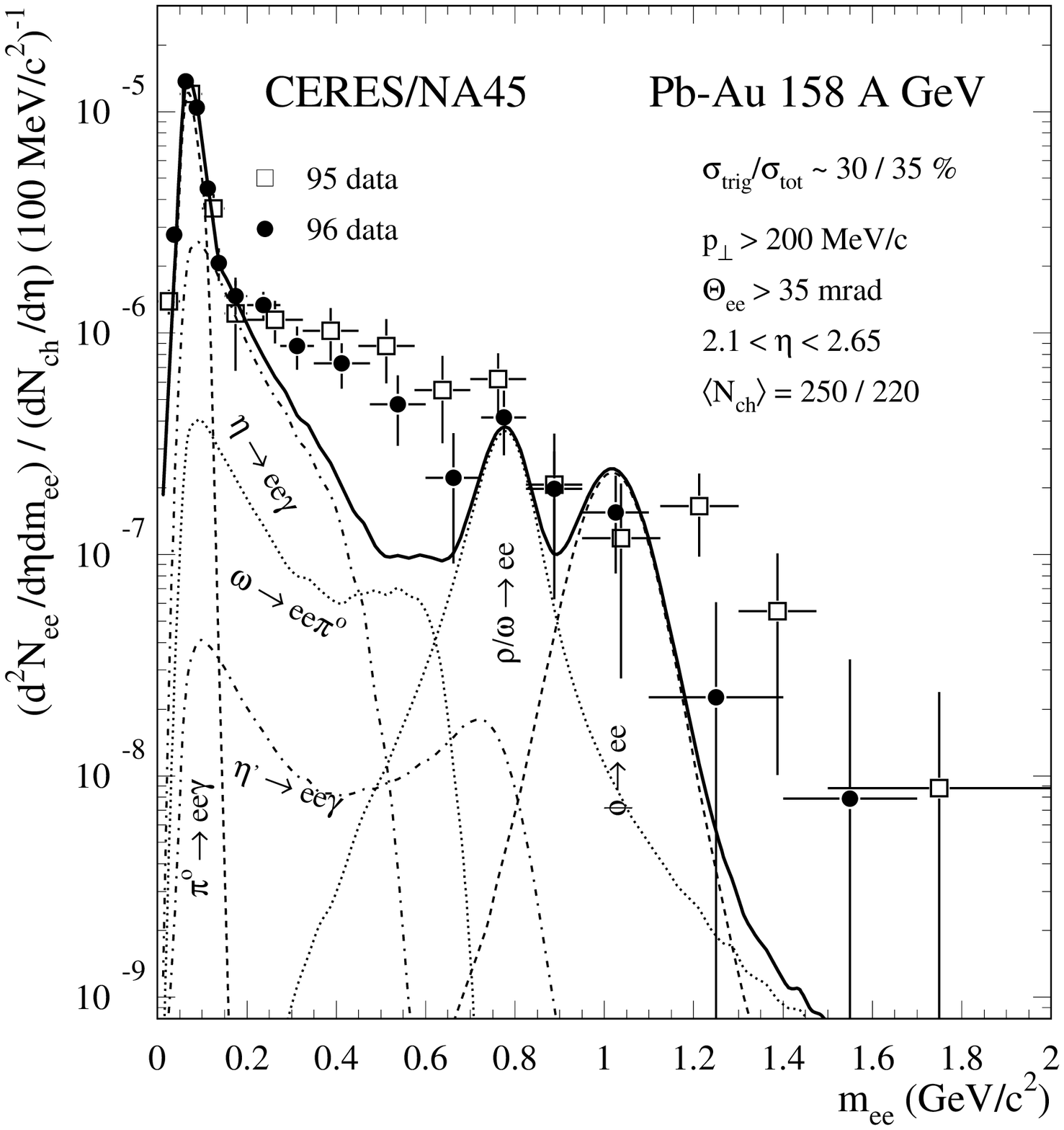,width=7.5cm,height=8cm}
\caption{Dilepton spectra from heavy-ion collisions as 
measured by the CERES/NA45 collaboration. Left panel:
central 200~AGeV S+Au collisions, contrasted with the hadronic
cocktail contributions as extrapolated from hadron multiplicities in 
$p$+$p$ data. The plot is taken from Ref.~\protect\cite{ceres95}.
Right panel: 35\% central 158~AGeV Pb+Au collisions (open squares: '95 data, 
full circles: '96 data),  compared to a hadronic cocktail inferred from 
a thermal model with $T=175$~MeV and $\mu_B=270$~MeV~\protect\cite{pbm98},
which reproduces the measured hadronic multiplicities in Pb+Pb  
collisions at identical projectile energy.} 
\label{fig_ceres93-96}
\end{figure}
The situation changes drastically when moving to heavy-ion projectiles.
In the first measurements at the CERN-SpS, which were performed with a
200~AGeV $^{32}$S beam, the CERES collaboration found a total enhancement
factor of measured pairs over the expectation based on the cocktail of
$5.0 \pm 0.7 ({\rm stat}) \pm 2.0 ({\rm syst})$ (integrated over the
invariant mass range 0.2--1.5~GeV)~\cite{ceres95} in
central collisions with $^{197}$Au targets. The
enhancement is in fact most pronounced around $M_{ee}\simeq 0.45$~GeV,
reaching a factor of 10 (Fig.~\ref{fig_ceres93-96}).
Whereas the CERES data are taken close 
to midrapidity ($2.1\le \eta \le 2.65$),
the HELIOS-3 experiment~\cite{helios3} covered more  forward 
rapidities $3.7\le \eta \le 5.2$. Here the enhancement, when comparing
to transport calculations~\cite{CEK96,LKBS}, is less developed but still 
significant (although the HELIOS-3 collaboration did not quote 
any systematic errors). 

The '95 and '96  runs with 158~AGeV $^{208}$Pb projectiles in essence  
confirmed the sulfur results, cf.~right panel of Fig.~\ref{fig_ceres93-96}. 
Here, the final state hadron decay contributions have been evaluated 
in an alternative way as recently developed by the CERES collaboration.  
It is based on hadron abundances from the thermal model of Ref.~\cite{pbm98},
where the (chemical) freezeout conditions for temperature and
baryon chemical potential have been deduced from an optimal fit to
a large body of hadronic observables 
at SpS and AGS energies (the such obtained cocktail agrees with 
sources scaled from $p$+$p$ collisions within 20--30\%, with 
the only exception of the $\phi$ meson -- related to strangeness 
enhancement --, which we will not address here). 
The resulting enhancement of the '96 data over the 'thermal-model' cocktail 
in the 30\% most central Pb+Au collisions then amounts to  
$2.6 \pm 0.5 ({\rm stat.}) \pm 0.9 ({\rm syst.})$~\cite{ceres98} 
in the invariant mass range 0.25--0.7~GeV.  
In terms of the experimental analysis, the '96 data
set is the best understood with the highest statistics (a factor of 5 larger
than in '95). It is consistent with both the '95 sample and the 
'93 sulfur results within two standard deviations (note that the S+Au
data are based on a higher centrality selection).  
Nevertheless, the
net signal in the '96 data seems to lie systematically below
the '95 data, this trend being more accentuated towards higher 
masses. In particular, at the free vector meson masses 
the '96 data are basically accounted for by the hadronic cocktail.   
If this feature will be confirmed in future measurements, it
has severe consequences for the theoretical interpretation 
of the spectra. A rather precise determination of the $\omega$ 
contribution, which is the dominant cocktail component at the 
$\rho/\omega$ mass, will be most important to draw firm conclusions. 
Unfortunately, the $\omega$ yield in heavy-ion reactions is not 
very well determined  so far. 
In hydrodynamical calculations as,  \eg,  
presented in Ref.~\cite{Pasi99}, the $\omega$-meson yield might be
substantially smaller than in the CERES cocktail if its final abundance
is determined by a simultaneous thermal {\em and} chemical 
freezeout temperature as low as 
$T_{fo}=120$~MeV. The most promising way to resolve this issue will be  
provided by an improved mass resolution in the dilepton 
measurements. If the latter can be reduced to about 2\%, the $\omega$
peak will clearly stick out, thus putting valuable 
constraints on other (in-medium) sources in its vicinity.  An upgrade of the 
CERES experiment using an additional time projection chamber (TPC) is 
expected to achieve the required resolution. An accordingly small 
mass binning will critically depend on a sufficiently large data statistics.  

Once the direct $\omega\to e^+e^-$ decays are known, also the 
Dalitz-decay contributions $\omega\to \pi^0 e^+e^-$ are fixed. The latter 
constitute an important part of the cocktail in the mass region where 
the experimental excess of  dilepton pairs is the largest, \ie, 
0.3~GeV~$\le M_{ee} \le$~0.6~GeV. The other important hadronic
decay in this region is $\eta \to \gamma e^+e^-$ and, to a lesser 
extent, $\eta'\to \gamma e^+e^-$. Therefore, an enhanced production 
of $\eta, \eta'$ mesons in heavy-ion as compared to proton-induced 
reactions could significantly alter the cocktail composition. 
Mechanisms for such a behavior have indeed been proposed in 
connection with the (partial) restoration of the $U_A(1)$ symmetry 
(see also Sect.~\ref{sec_sym-ano})
in high density/temperature matter, reducing the $\eta$ and $\eta'$ masses
and thus increasing their (final) abundances~\cite{eta+}.       
However, an enhanced $\eta, \eta'$ production also entails an increase 
of the direct photon yield from the two-photon decay modes
$\eta,\eta'\to \gamma\gamma$. In Ref.~\cite{drees96} upper limits on 
inclusive photon measurements in heavy-ion collisions at the SpS have 
been converted to a maximally allowed $\eta$ production. Assuming that  
the upper limit of the photon signal is entirely saturated by extra
$\eta$ decays, it has been shown that the $\eta$ yield in $^{32}$S
induced reactions cannot be enhanced by more than a factor of 1.5 
as compared to the $p$+$p$ case. Similar arguments have been drawn 
to limit the $\eta'$ enhancement to a factor of 2.5. As a result, 
using the upper bounds on $\eta$ and $\eta'$ numbers, the 
CERES cocktail in 200~AGeV S+Au collisions (displayed in the left panel of 
Fig.~\ref{fig_ceres93-96}) is increased by at most 
40\% which is far from accounting for the observed excess in
the 0.3--0.6~GeV region (this situation is reminiscent to 
BEVALAC/SIS energies, where an increased $\eta$ yield as the 
source for the dilepton enhancement found by DLS in $A$-$A$ collisions 
has also been ruled out by means of the TAPS two-photon data, see
Sect.~\ref{sec_dls}).   
The failure of the hadronic cocktail to describe the low-mass dilepton 
spectra in  nucleus-nucleus collisions at the SpS thus inevitably points 
towards radiation originating from processes occurring during 
the interaction phase of the collisions, which will be discussed
in the following sections.

\subsection{Free $\pi^+\pi^-$ Annihilation in the Hadronic Fireball}  
\label{sec_freepipi} 
One of the strongest evidences for a 'non-trivial' source of dilepton  
pairs in heavy-ion reactions is illustrated in Fig.~\ref{fig_multidep}.
The total $e^+e^-$ pair yield, normalized to the number of charged 
particles in the final state, exhibits a clear increase with 
multiplicity, indicating two- (or more) body annihilations.   
\begin{figure}[!htb]
\centerline{\epsfig{figure=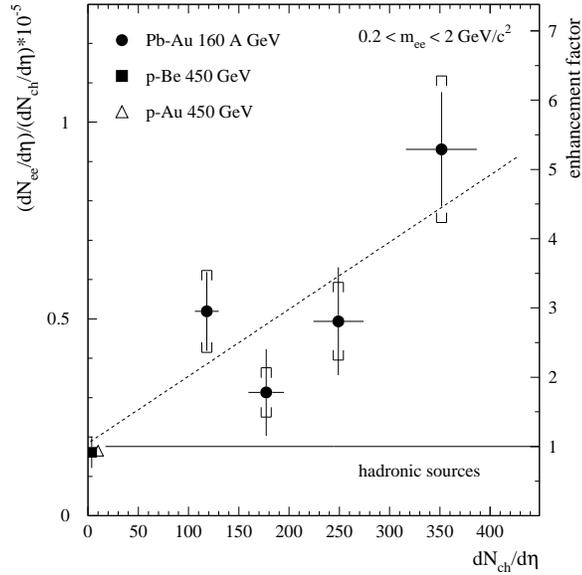,width=8cm}}
\caption{Total dilepton pair yield, normalized to the number
of observed charged particles in the corresponding rapidity 
interval,  and resulting  enhancement factor over the 
hadronic cocktail (right vertical scale) as a function of 
charged particle multiplicity as found in different collision 
systems; the horizontal solid line indicates the expectation 
from final-state hadron decays whereas the dashed line is a linear
interpolation of  the data which implies a quadratic dependence on the 
number of charged particles as expected from an additional two-body
annihilation source. The plot is taken from 
Ref.~\protect\cite{ceres96}.}
\label{fig_multidep}
\end{figure}
Since the most abundant particles at SpS energies are pions, the 
obvious candidate for this behavior is the 
$\pi^+\pi^-\to \rho^0\to e^+e^-$ process. Many authors have 
calculated its contribution in various 
approaches to model the heavy-ion reaction dynamics,  
leading to rather good agreement with each other as we have already eluded 
to in the Introduction, see Fig.~\ref{fig_dlexth}
for the case of $^{32}$S-induced reactions. 
All calculations share the common feature that, although the total 
yield is appreciably increased, the shape of the spectra strongly 
deviates from the data in that one finds too much yield around the 
free $\rho$ mass and too little below, which is a trivial consequence
of the free pion electromagnetic form factor peaking at the $\rho$
resonance.  

At this point it is useful to notice an important difference between 
transport and hydrodynamical approaches. In the latter the total yield is 
typically by a factor of at least 2 smaller than in the former ones if 
no chemical potentials for pions are involved as is the case, \eg, for the
(dotted) Hung-Shuryak curve~\cite{HuSh97}
 in the left panel of Fig.~\ref{fig_dlexth} (see also  
Ref.~\cite{Prak97}). Although in both schemes  
the final hadronic spectra are usually equally well accounted for, 
hydrodynamic calculations involve smaller (average) pion densities $n_\pi$ 
(due to the restriction to $\mu_\pi=0$) and hence larger fireball 
volumes $V_{FB}$ to obtain an identical final number of pions.  
This means that, for a given total  
number of pions, $N_\pi=n_\pi\times V$, larger average densities in 
the transport simulations  
lead to a larger dilepton signal from $\pi\pi$ annihilation, since
the latter is basically proportional to the pion density {\em squared},
$N_{\pi\pi\to ee} \propto n_\pi^2 \times V$. On the other hand, 
in the Bjorken-type hydrodynamical calculations of Baier \etal~\cite{BDR971}
where a pion chemical potential
of $\mu_\pi=100$~MeV has been employed, the dilepton yield is even
slightly larger than in most of the transport results, 
cf.~solid line in the left panel of Fig.~\ref{fig_dlexth}.
This is understandable, as the 'average' squared pion density in 
this calculation is increased over the $\mu_\pi=0$ case by roughly the 
squared fugacity $(\exp[\mu_\pi/T])^2\simeq \exp[2\times 100/150]\simeq 4$.    
Another quantity which  governs the amount of dileptons
radiated from the hadronic fireball is its total lifetime. For Pb+Au
collisions at the full SpS energy (158~AGeV) the latter is around 
10-15~fm/c. However, as pointed out by Shuryak and Hung~\cite{HuSh95},
the upcoming low-energy run at 40~AGeV may lead to initial conditions 
that are close to the so-called 'softest' point in the EoS of the 
quark-hadron transition, associated with  a very small initial pressure.
In this case, the system expands very slowly entailing a 
much increased fireball lifetime which would have to leave
its trace in the total dilepton yield.      
  
The incompatibility of free $\pi\pi$ annihilation persists  
in the Pb+Au data.   
Although the excess signal in the more recent CERES measurements 
is somewhat reduced as compared to the early sulfur runs 
the inclusion of free $\pi\pi$ annihilation in theoretical models 
can still not resolve the discrepancy with 
the data. This statement is corroborated by the trend that in the 
Pb+Au system the 
cocktail is close to saturating the data in the $\rho/\omega$ region, 
where the free $\pi^+\pi^-\to\rho^0\to e^+e^-$ process has its 
maximal contribution! Thus one is seemingly led to the following 
two alternatives: \\ 
(i) $\pi\pi$ annihilation is not an important ingredient 
    in the dilepton spectra, but rather some very different processes 
    with flat characteristics as a function of invariant mass, 
    \eg,  $q\bar q$ annihilation. However, this is not easy to 
    imagine for the SpS conditions, where one expects the excited nuclear
    system to spend the major part of its space-time history in a hadronic
    phase with a large pion component; \\  
(ii) $\pi\pi$ annihilation {\em is} the dominant process. In this case
     drastic medium modifications are inevitable to fill in the 0.3--0.6~GeV
     mass region without giving too much yield around the free 
     $\rho$-meson mass.\\  
Both possibilities will be considered in the forthcoming Sections.

\subsection{Medium Effects I: Invariant Mass Spectra}
\label{sec_mspec}
Most of the in-medium effects proposed so far have drawn  
their attention to  the pion-pion annihilation channel. They can be roughly 
divided into the following two categories:  
\begin{itemize}
\item[(I)] a temperature- and density-dependent reduction ('dropping') 
    of the $\rho$ meson mass, 
    $m_\rho^*$, according to BR scaling or the Hatsuda-Lee QCD sum 
    rule calculations, usually applied without invoking  any changes 
    in the pion propagation. This then entails a reduction of the 
    $\rho$-meson width due to the shrinking pion phase space at smaller 
    $m_\rho^*$  as well as a sharp threshold -- at twice the free pion 
    mass $2m_\pi$ -- for the onset of the enhancement in the invariant
    mass dilepton spectra; 
\item[(II)] a modification of both $\pi$ and $\rho$ properties due to 
    phenomenologically inferred interactions with the surrounding 
    hadrons in the hot and dense gas which, depending on the language used, 
    are encoded in the $\rho$-meson spectral function (\eg,  
    Refs.~\cite{RCW,FrPi,PPLLM}), in vector current correlation 
    functions (\eg, in Refs.~\cite{SYZ2,KKW97}), in the pion electromagnetic 
    form factor (Ref.~\cite{SK96}), etc.. Also the rate 
    calculations for individual processes as, \eg,  performed in 
    Refs.~\cite{GaLi,Ha96}, should be assigned to this category, as
    was discussed in Sect.~\ref{sec_hadrates}. 
\end{itemize}

\subsubsection{200~AGeV Sulfur Beam Runs}
For dilepton spectra in  the 200~AGeV $^{32}$S-induced reactions 
the consequences of a dropping $\rho$-meson mass 
have been explored in Refs.~\cite{CEK95,LKB,CEK96,LKBS}.  
All these analyses find good agreement with the experimental data
(Fig.~\ref{fig_lkbs}).  
\begin{figure}[!htb]
\bce
\epsfig{figure=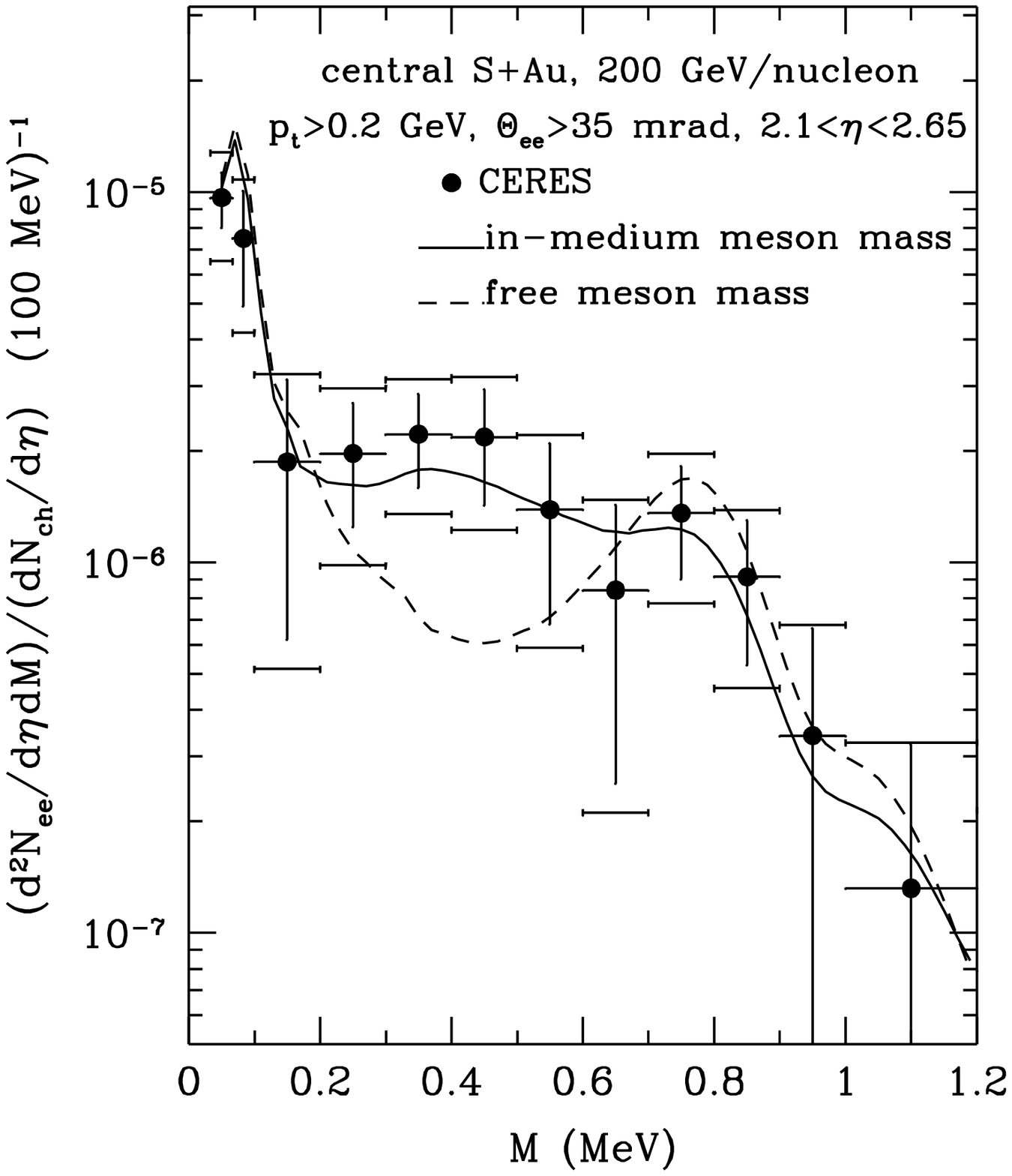,width=7.4cm}
\hspace{-0.7cm}
\epsfig{figure=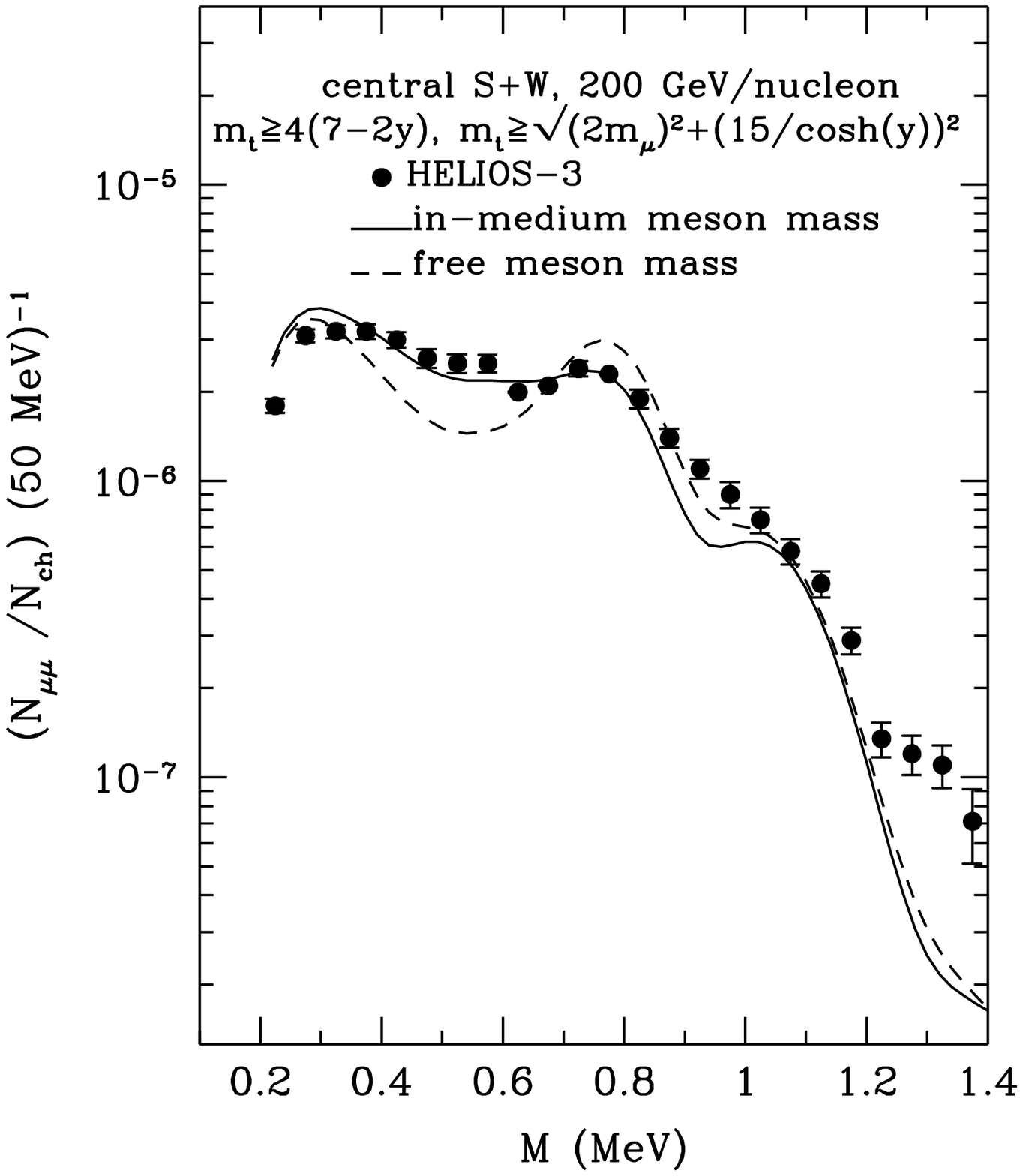,width=7.4cm}
\ece
\caption{Comparison of dilepton data from central 200~AGeV sulfur-induced
reactions on heavy nuclei with transport calculations employing 
a dropping $\rho$ mass (full curves) as opposed to a free $\rho$ mass  
(dashed curves)~\protect\cite{LKBS};  
left panel: CERES dielectron spectra on Au targets; right panel:
HELIOS-3 dimuon spectra on W targets.}
\label{fig_lkbs}
\end{figure}
The mechanism is clear: in the early  phase, characterized by hadronic 
initial conditions of $T^{i}\simeq 170$~MeV and 
$\varrho_{B}^{i}\simeq 2.5\varrho_0$, 
the in-medium mass $m_\rho^*$ is close to the two-pion 
threshold. As the hot fireball expands thereby diluting and cooling, 
$m_\rho^*$ starts to rise and sweeps across the low-mass 
region thus filling the dilepton continuum between 
0.3 and 0.6~GeV (at freezeout, the $\rho$-meson has regained about 
80\% of its vacuum mass). At the same time, the $\rho$ (or $\pi\pi$) 
contribution around the free $\rho$ peak  is strongly reduced, 
which is also in line with the 
experimental data. However, between the $\omega$ and the $\phi$ mass
the situation is less clear, mainly due to limited experimental 
mass resolution and statistics (the former amounting to $\sim$8-11\% 
in this mass region for the '92-'95 CERES data).    
The medium modifications of the $\omega$ meson itself  
have only  little impact on the dilepton spectra, although in BR scaling the  
$\omega$ mass is subjected to the same reduction as the $\rho$ mass. 
The reason is simply that the $\omega\to e^+e^-$ decays mostly 
occur after the hadronic freezeout where medium effects 
are absent. The final  number of $\omega$ mesons is roughly equal
to the case where no dropping masses are assumed. This is so because  
in Refs.~\cite{LKB,LKBS} rather large pion chemical potentials 
$\mu_\pi\simeq 100$~MeV are present in the initial conditions when 
using the free masses to correctly reproduce the observed number 
of final pions. On the other hand, when using in-medium masses, 
much smaller $\mu_\pi$ are required to obtain about the same final 
number of pions and $\rho/\omega$ mesons. Within the Walecka-type
mean-field potentials employed in the transport equations of 
Refs.~\cite{LKB,LKBS} the scalar field has been assumed to act 
on the constituent $u$-/$d$- quark content of the hadrons only, thus 
leaving the mass of the $\phi$ meson, which is an almost pure 
$s\bar s$ state, unchanged. An important point to note is that 
the baryons (rather than pions which govern the finite-temperature effects) 
in the hadronic fireball are the key component in 
generating the large (attractive) scalar fields which are at the origin  
of the dilepton enhancement. 

Dropping meson masses have also been implemented in hydrodynamical
simulations~\cite{BDR971,HuSh97}.  Although the total dilepton signal in the
latter is typically smaller than in the transport frameworks 
(if $\mu_\pi\equiv 0$, see previous Section), 
they also give reasonable agreement with the S+Au data~\cite{BDR971,HuSh97}    

\begin{figure}[!htb]
\begin{minipage}[t]{0mm}
{\makebox{\epsfig{file=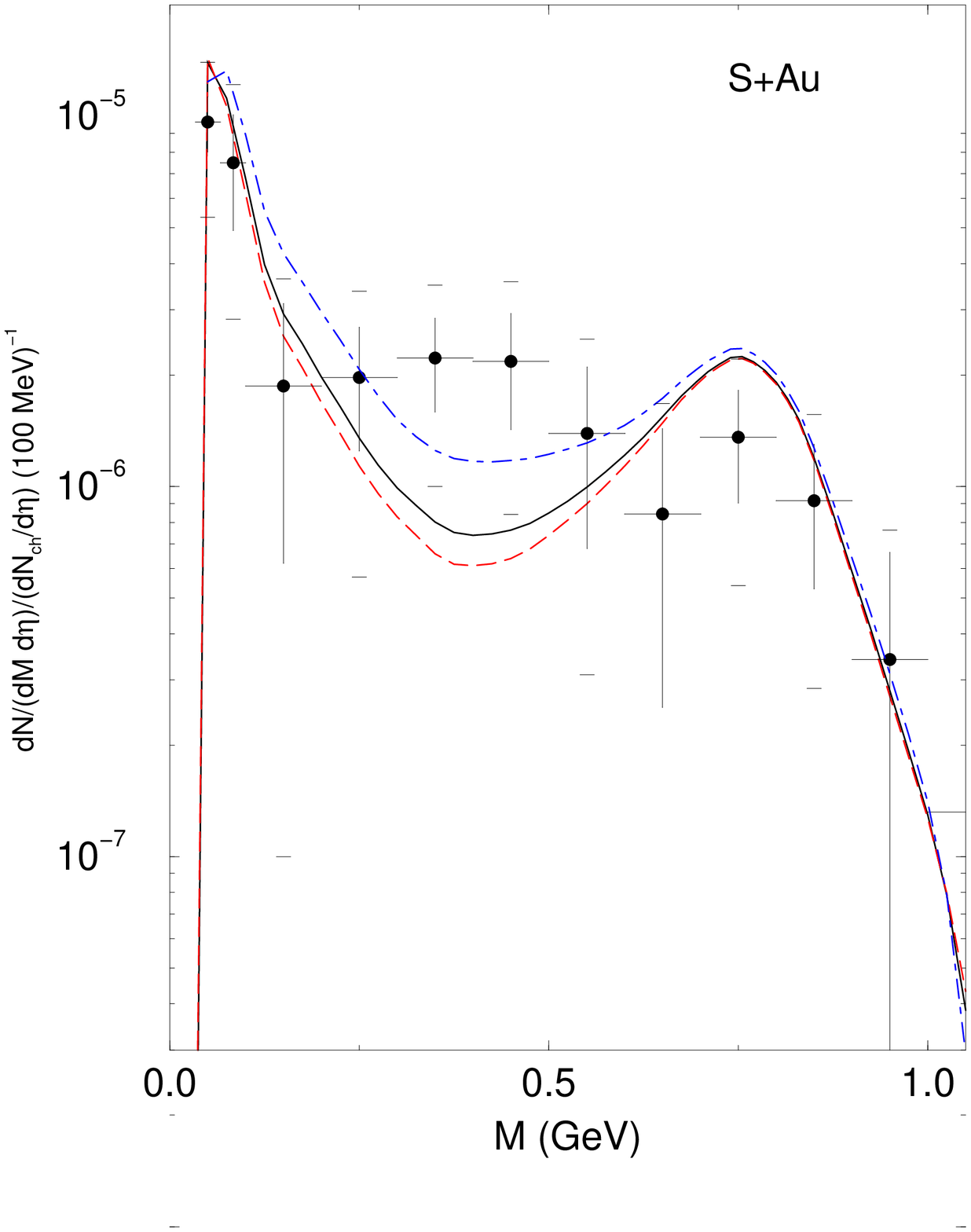,width=7.3cm,height=6cm}}}
\end{minipage}
\hspace{7.3cm}
\begin{minipage}[t]{0mm}
{\makebox{\epsfig{file=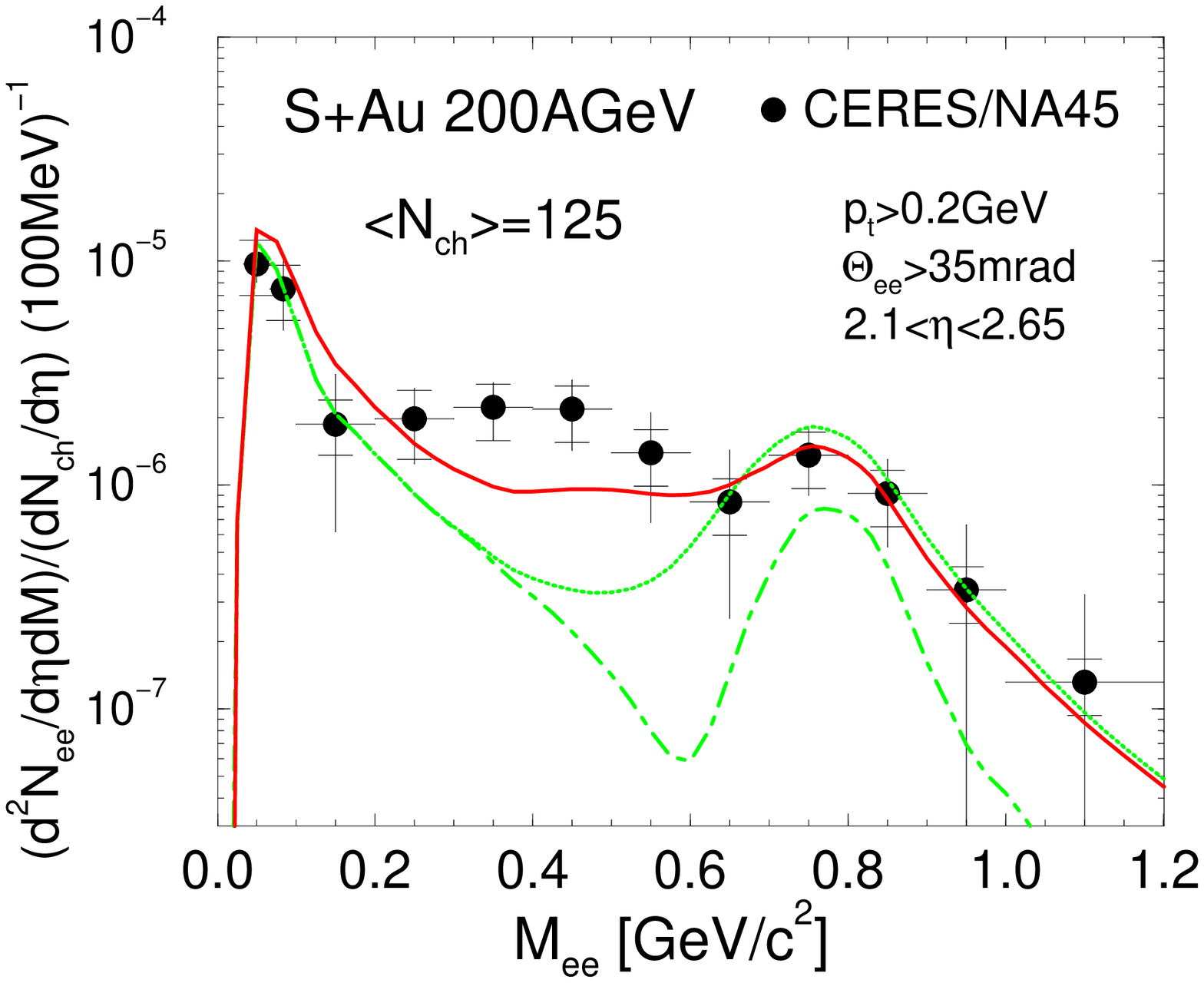,width=7.6cm}}}
\end{minipage}
\caption{Comparison of CERES data from central 200~AGeV S+Au
reactions with thermal fireball calculations including in-medium effects
according to the chiral reduction formalism~\protect\cite{SYZ2} (left panel;
all curves include the hadronic cocktail as given by  transport
results~\protect\cite{LKB}; in addition, in-medium radiation is accounted
for in the following ways: pure pion gas (dashed curve),  
pion-nucleon gas using $\varrho_N^i=0.7\varrho_0$ 
corresponding to the realistic case of $\mu_B=0.39$~GeV (full curve), 
and pion-nucleon gas  using the full baryon density 
$\varrho_N^i\equiv\varrho_B^i=2.5\varrho_0$ (dashed-dotted curve)),  
and within the many-body approach for the $\rho$ spectral 
function~\protect\cite{RCW,UBRW,Morio98,RG99} (right panel; dashed-dotted 
curve: hadronic cocktail; dotted curve:
cocktail plus $\pi\pi$ annihilation using the 
free $\rho$ spectral function; full curve: cocktail plus 
$\pi\pi$ annihilation using the in-medium $\rho$ spectral function).}   
\label{fig_sauconv}
\end{figure}
As representatives for the analyses of the CERES S+Au data that
are based on more 'conventional' scenarios  we have chosen the
calculations within the chiral reduction approach~\cite{SYZ2} and
the many-body approach for the $\rho$-meson spectral function~\cite{RCW}
in its recent version (including constraints from nuclear
photoabsorption~\cite{RUBW} and
$\pi N\to \rho N$ data~\cite{Morio98}, Rhosobar excitations on thermally
excited baryon resonances  as well as a more complete assessment
of the mesonic contributions~\cite{RG99}).
The results are confronted in Fig.~\ref{fig_sauconv} with experiment.
To facilitate the direct comparison both spectra have been 
computed in the thermal fireball 
expansion of Ref.~\cite{RCW} (with  $T^i=170$~MeV, $T^\infty=110$~MeV, 
$\tau=8$~fm/c, $t_{fo}=10$~fm/c in Eq.~(\ref{cooling}),  
$N_B^{part}=110$, a constant isotropic expansion velocity $v=0.4$c and 
$\mu_B=0.39$~GeV 
which translates into initial/freezeout baryon densities of 
$\varrho_B^i=2.49\varrho_0$~/~$\varrho_B^{fo}=0.32\varrho_0$ and a freezeout 
temperature of $T^{fo}=127$~MeV; in addition an overall
normalization factor $N_0$=3 has been introduced in reminiscence to 
transport results~\cite{LKB}, corresponding to an 'average' pion chemical 
potential of $\sim$~80~MeV).   
Neither of the two 'conventional' approaches gives as good agreement 
with the S+Au data as the dropping $\rho$ mass scenarios (this 
is even more pronounced for other attempts~\cite{SSG96,Ha96,Koch96,BDR971}), 
although the experimental uncertainty is not small.   
In the chiral reduction formalism (left panel in Fig.~\ref{fig_sauconv}) 
the incoherent summation 
of individual rate contributions in a low-density expansion (for both
pions and nucleons) generates some  enhancement over the results
based on free $\pi\pi$ annihilation (dotted curve in the right panel), 
but does not lead to any depletion of the free $\rho$ peak. Consequently, 
the {\em shape} of the theoretical curves does not match the experimental
data very well. This is qualitatively different in the  many-body approach
(right panel in Fig.~\ref{fig_sauconv}). The strong broadening of the 
$\rho$-meson spectral function 
yields a factor of $\sim$~2 more enhancement below $M_{ee}\simeq 0.6$~GeV 
together with some reduction in the $\rho/\omega$ region, 
which makes it somewhat more compatible  with the data. 
A possible caveat might 
be given by the fact that the (experimental) systematic errors 
presumably  have little (or at least a very smooth) dependence
on invariant mass. This could mean that an agreement of the theoretical
curves in the 0.3--0.5~GeV region entails a disagreement around     
$M_{ee}\simeq 0.2$~GeV and vice versa. We will come back to this point
further below.  Another noteworthy feature is that, although the 
in-medium spectral function is larger than the free one at the high 
mass end for $M_{ee}\ge 0.9$~GeV (cf.~Fig.~\ref{fig_imdrho685}), this feature  
does not show up in  the dilepton spectrum in the right panel of
Fig.~\ref{fig_sauconv}, which can  be traced back to the mass 
resolution of the CERES detector in the '92 setup 
($\delta M/M\simeq 11\%$ around $M\simeq 1$~GeV).   

The spectral function approach has also been employed 
using a more realistic description of the heavy-ion reaction 
dynamics within the HSD transport simulations~\cite{CBRW}. 
Fig.~\ref{fig_swhsd} shows the results for the HELIOS-3 data in 
200~AGeV S+W collisions using the free (upper panel) and 
in-medium (lower panel) $\rho$ spectral function, the latter based
on the model of Refs.~\cite{RUBW,UBRW}.  
\begin{figure}[!htb]
\centerline{\epsfig{figure=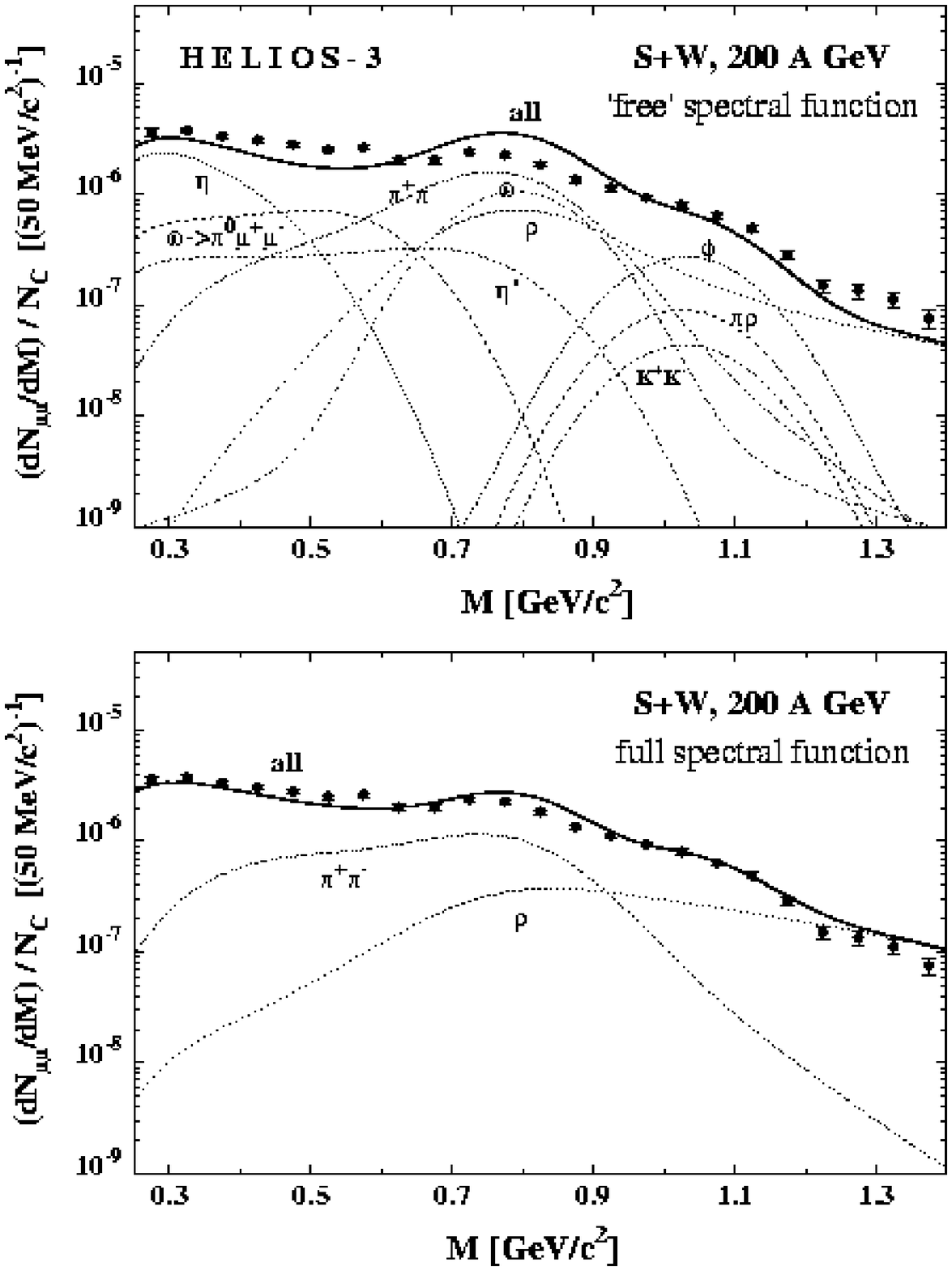,height=10cm,width=8cm}}
\caption{Comparison of HELIOS-3 data from central 200~AGeV S+W
reactions with HSD transport calculations~\protect\cite{CBRW} employing 
free (upper panel) and in-medium (lower panel) $\rho$ spectral
functions from Refs.~\protect\cite{RCW,RUBW};
note that the high-mass end of the  contribution labeled 
by '$\rho$' (which stem from meson-baryon and baryon-baryon collisions) 
is probably somewhat overestimated, cf.~the remarks following
Eq.~(\protect\ref{dNrhodM}).}  
\label{fig_swhsd}
\end{figure}
Again, the broadening of the $\rho$ spectral function
significantly improves the agreement with experiment.

\subsubsection{158~AGeV Lead Beam Runs}
Again, let us first address analyses that involve 
dropping  meson masses. Fig.~\ref{fig_pbauhsd} shows
results of the HSD transport approach~\cite{BC97}, which once 
more demonstrate that a reduced $\rho$-meson mass is very well in line 
with the experimentally observed low-mass dilepton enhancement at 
full CERN-SpS energies, most notably around the 0.5~GeV region. 
\begin{figure}[!htb]
\centerline{\epsfig{figure=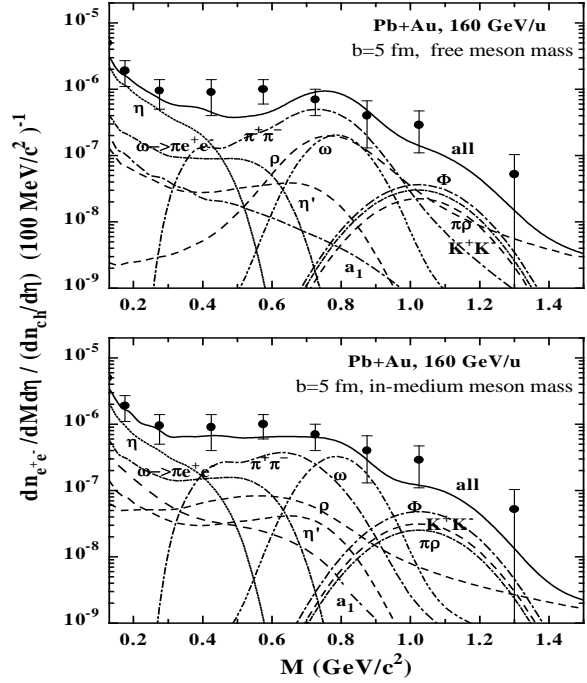,height=11cm,width=9cm}}
\caption{Comparison of the (preliminary) '95 CERES 
data~\protect\cite{ceres96a} from 35\% central 158~AGeV Pb+Au  
reactions with HSD transport calculations~\protect\cite{BC97} employing
free (upper panel) and  dropping (lower panel) meson masses.}  
\label{fig_pbauhsd}
\end{figure}
Similar conclusions have been  drawn in the transport calculations of  
Ref.~\cite{LKB98} for the 8\% most central sample of the '95 CERES
data, as well as in the hydrodynamical description of Ref.~\cite{HuSh97}.   

Proceeding to category II ('conventional' medium modifications),  
we display in Fig.~\ref{fig_pbauconv} a 
comparison of transport and fireball calculations. 
\begin{figure}[!htb]
\epsfig{figure=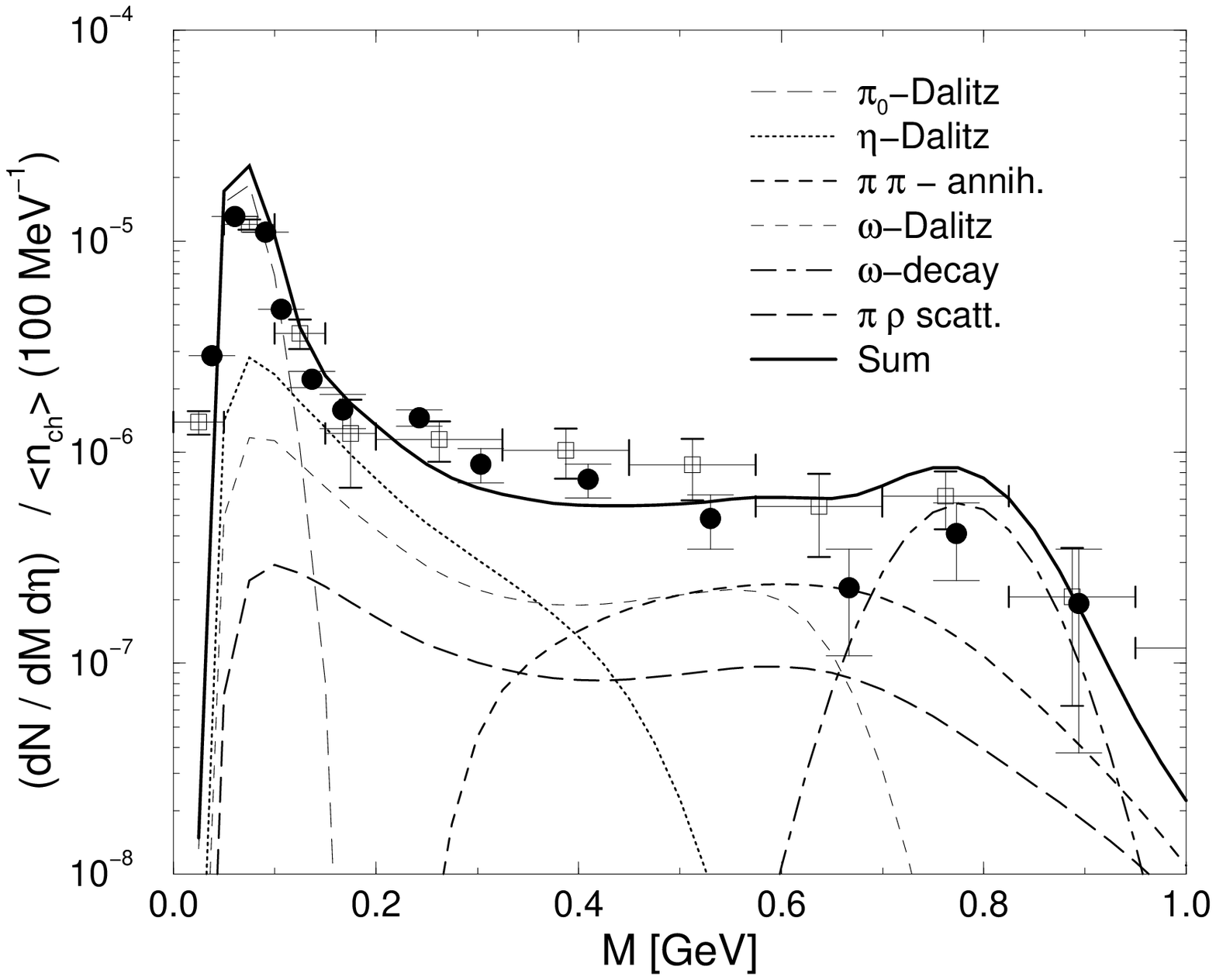,width=6.6cm}
\hspace{0.3cm}
\epsfig{figure=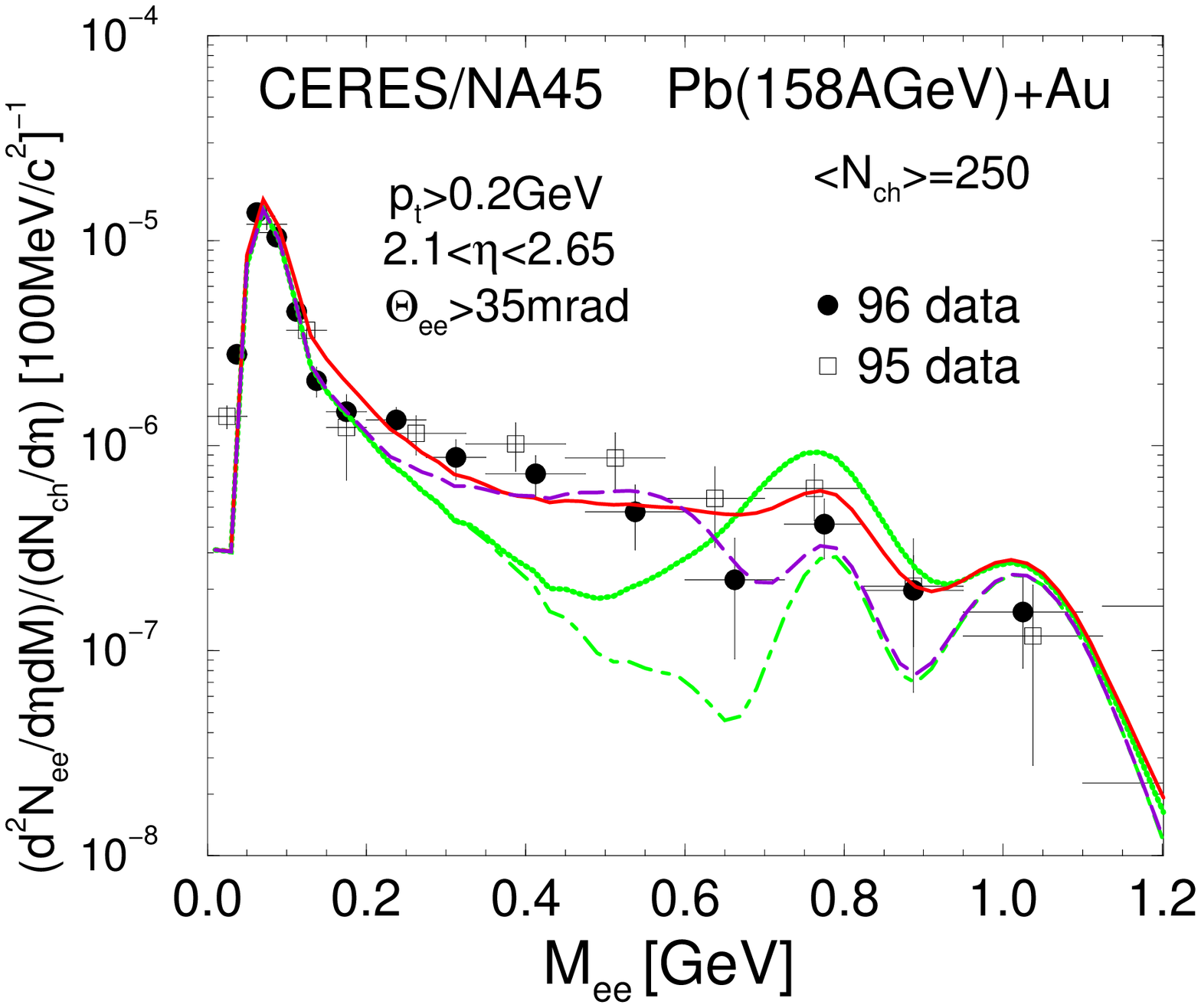,width=7.7cm}
\caption{Comparison of the CERES data from 35\%/30\% central 158~AGeV Pb+Au
collisions  with BUU transport calculations~\protect\cite{Koch96,Koch99}
including finite-temperature effects in the pion electromagnetic form
factor and through explcit $\pi\rho$ scattering (left panel). Right panel:
fireball calculations~\protect\cite{RCW} using
thermal production rates from in-medium $\pi\pi$ annihilation in
three scenarios;  dashed-dotted line: CERES cocktail (no in-medium
contribution, no $\rho$ decays); dotted line: cocktail plus
free $\pi\pi$ annihilation; solid line: cocktail plus
$\pi\pi$ annihilation employing the in-medium $\rho$ spectral
function~\protect\cite{RUBW,RG99,RW99}; long-dashed line: cocktail plus
$\pi\pi$ annihilation using a dropping $\rho$ mass.}
\label{fig_pbauconv}
\end{figure}
The left panel shows  
recent BUU transport calculations~\cite{Koch99}
along the lines of Ref.~\cite{Koch96} where the in-medium effects entirely
reside in finite temperature  effects in the $\pi\pi$ annihilation channel
and $\pi\rho$-type contributions (dominated by the $a_1(1260)$). 
Due to collisional broadening introduced in the (denominator of) the 
pion electromagnetic form factor the significance of the in-medium $\pi\pi$ 
channel is rather moderate; the processes inducing the broadening (\ie, 
$\pi\rho$ collisions) are treated  explicitely for dilepton
production to approximately restore unitarity in the transport framework.  
Within 1.5 standard deviations, all data points for 30\% 
central Pb+Au are reproduced; in the low-mass region, 
$M_{ee}\simeq 0.3-0.6$~GeV, this is largely achieved through a strong 
contribution of the $\omega\to \pi^0 e^+e^-$ and $\eta\to \gamma  e^+e^-$
Dalitz decays, each about a factor of 2 larger than in 
the CERES cocktail. As we have mentioned earlier, 
especially the $\omega$ contribution is as of now  not very well 
under control (and has been introduced through suitable initial conditions 
in the transport); on the other hand, one also realizes from 
the left panel of Fig.~\ref{fig_pbauconv} that at the free 
$\rho/\omega$ mass the 
direct decays $\omega\to e^+e^-$ tend to overestimate the experimental 
data, \ie, the freezeout $\omega$ abundance has been pushed to its limit. 
In contrast, the  in-medium spectral function 
approach~\cite{CRW,RCW,FrPi,KKW97,PPLLM} assigns the major part of 
the low-mass enhancement to the (modified) $\pi\pi$ channel including the 
effects of baryons. The right panel of  
Fig.~\ref{fig_pbauconv} shows the pertinent results~\cite{RG99,RW99} 
employing a thermal fireball model including the experimentally
determined hadro-chemical freezeout as well as the subsequent build-up 
of finite pion chemical potentials (cf.~Sect.~\ref{sec_fireball}) 
in 30\% central Pb+Au (the time evolution is specified by initial/freezeout 
conditions $(T,\varrho_B)_{ini}$=(190~MeV,2.55~$\varrho_0$),    
$(T,\varrho_B)_{fo}$=(115~MeV,0.33~$\varrho_0$) with  
$N_B^{part}=260$).   
The hadronic cocktail part has been taken from the  
CERES collaboration~\cite{cktl} based on an identical chemical freezeout 
but with the $\rho$-meson contribution removed, 
since the latter is accounted for by  the in-medium $\pi\pi$ annihilation 
at the freezeout stage of the fireball.        
Also included in Fig.~\ref{fig_pbauconv} is the result obtained with a  
dropping $\rho$ mass based on the same fireball 
evolution. The density and temperature dependence of $m_\rho^*$ has been 
assumed to resemble QCD sum rule estimates, 
\beq
m_\rho^*= m_\rho \ (1-C \ \varrho_B/\varrho_0)
\ \left(1-(T/T_c^\chi)^2\right)^\alpha
\label{mrhostar}
\eeq
with $C=0.15$, $T_c^\chi=200$~MeV and $\alpha=0.3$.  
Given the experimental uncertainties both the dropping $\rho$ mass
and the in-medium broadening give reasonable account for the dilepton
enhancement in the 0.3--0.6~GeV region. Substantial differences set in 
beyond, where the down-shifted $\rho$ mass does no longer
contribute, as opposed to the broadening scenario. At the $\rho/\omega$
peak, the more recent data seem to favor the former, but between the 
$\omega$ and $\phi$, the in-medium spectral function might do better, 
providing sufficient yield. Once again we see that an improved mass
resolution of the measurements, separating $\omega$ and $\phi$
cocktail ingredients more distinctly, is crucial to reach definite 
conclusions on these issues.  
 
Hydrodynamical calculations employing in-medium rates of category II
have been performed recently in Ref.~\cite{Pasi99}. In particular, 
the differences between the many-body~\cite{RCW,RUBW} and the 
chiral reduction approach~\cite{SYZ2} and their consequences for 
dilepton production in central Pb+Au have been 
explored in some detail. On the level of the bare rates the two
approaches agree reasonably well in the pion gas sector, but differ 
by factors of $\sim$~2 already at finite (nucleon) densities as low as 
$0.5 \varrho_0$, cf.~Sect.~\ref{sec_hadrates} and 
Fig.~\ref{fig_Brates}\footnote{One should note that the 
$\rho$ spectral function on which the calculations in Ref.~\cite{Pasi99}
are based, is not the most recent (more complete) one, although it includes 
constraints from photoabsorption spectra according to Ref.~\cite{RUBW}.}. 
This directly translates into a similar discrepancy in the $\pi\pi$ 
induced dilepton signal, with the many-body rates resulting in the 
larger yield (left panel of Fig.~\ref{fig_pbauhydro}), once more 
demonstrating that baryons play an important role at full SpS energies
of 158--200~AGeV. 
\begin{figure}[!htb]
\bce
\epsfig{figure=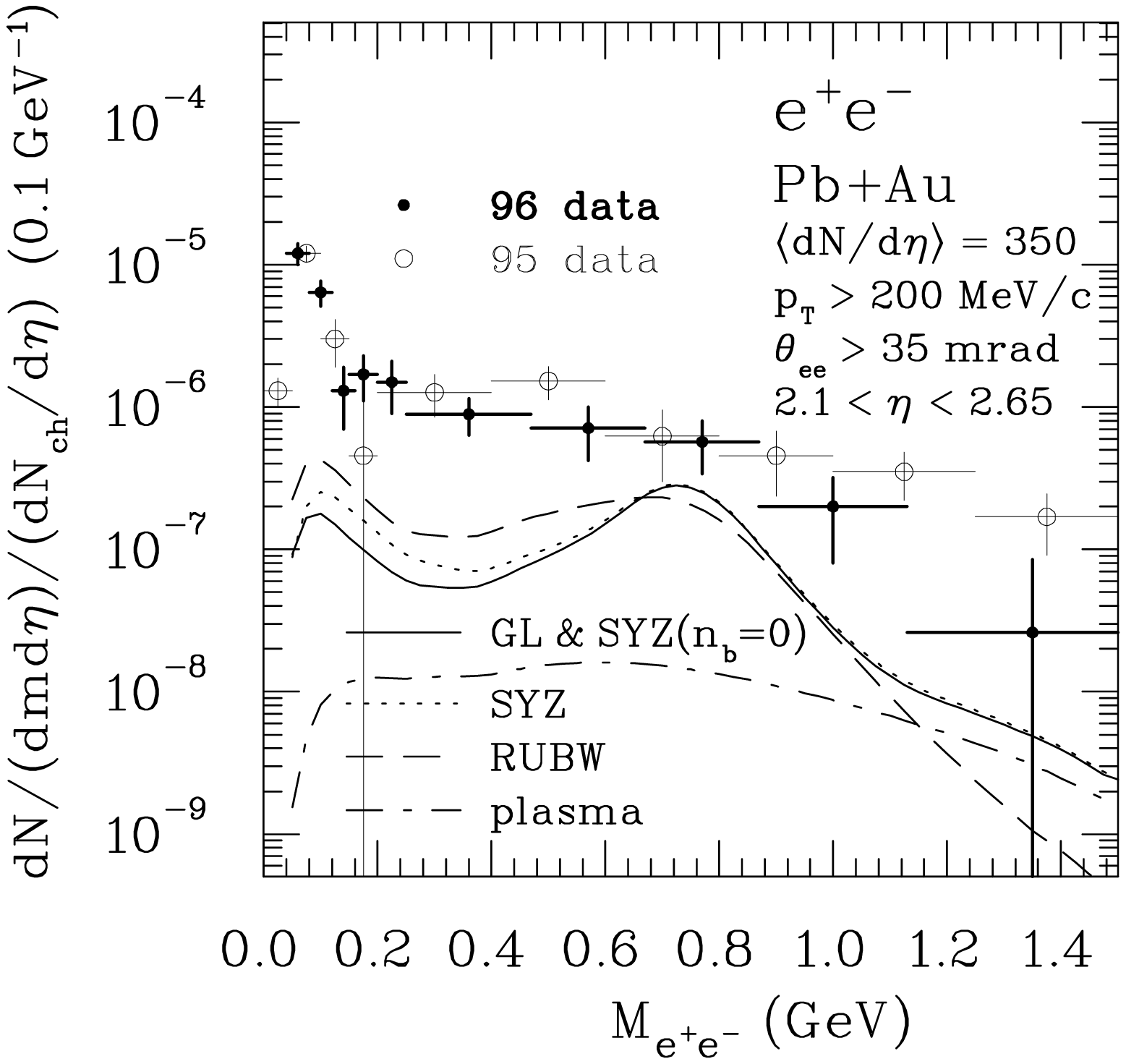,width=6.8cm}
\hspace{0.8cm}
\epsfig{figure=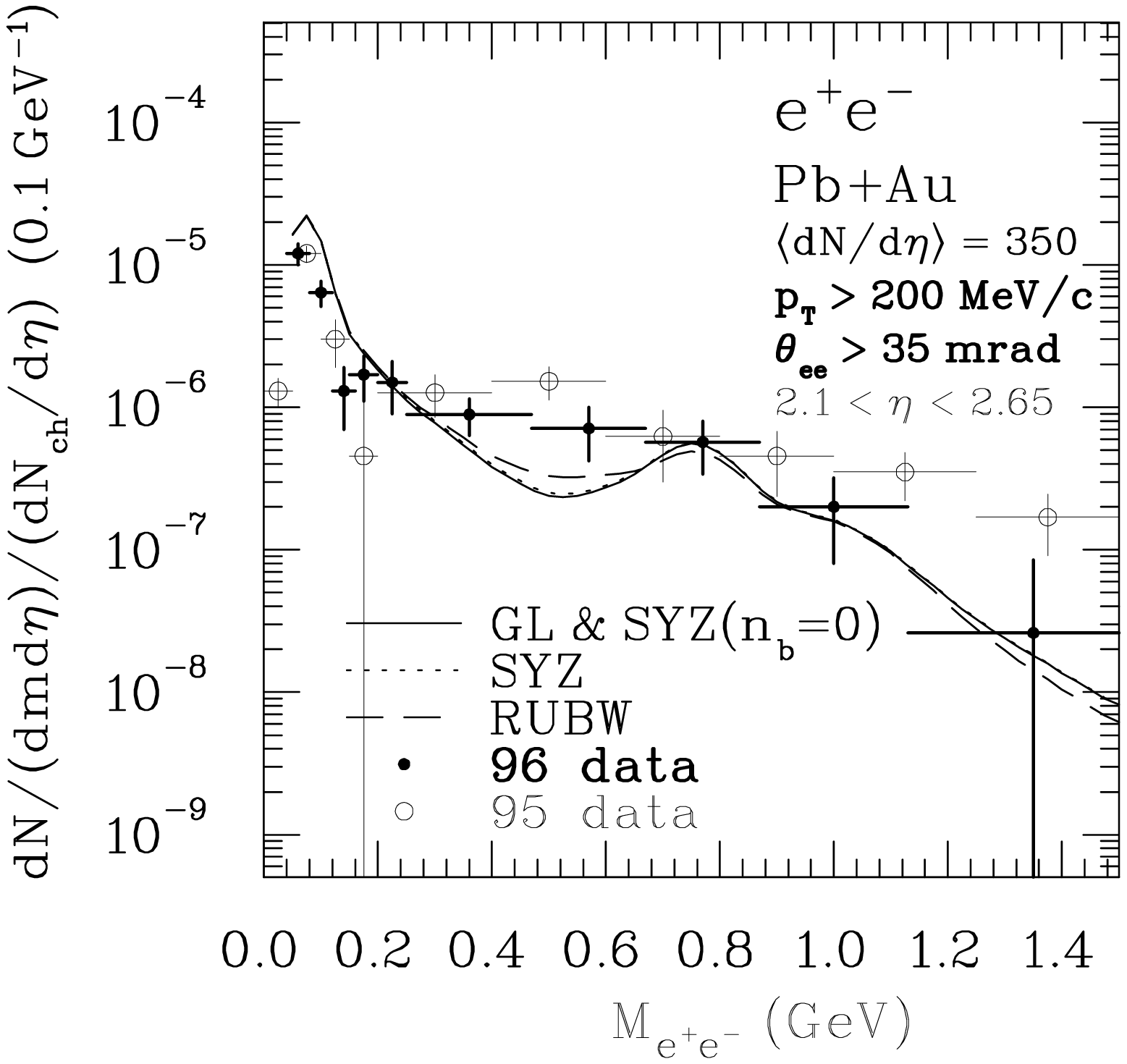,width=6.8cm}
\ece
\caption{Comparison of the '95 and '96 CERES data from 8\% central 158~AGeV
Pb+Au with hydrodynamical calculations~\protect\cite{Pasi99};
left panel: dielectron yields from the interaction phase using hadronic
rates without baryonic effects~\protect\cite{GaLi,SYZ1}
(solid line) and including
baryonic effects according to  the chiral reduction~\protect\cite{SYZ2}
(dotted line) or the many-body framework~\protect\cite{RUBW} (long-dashed
line), and from
the QGP phase using perturbative $q\bar q$ annihilation (dashed-dotted line);
right panel: final spectra including hadron decays after freezeout (line
identification as in left panel).}
\label{fig_pbauhydro}
\end{figure}
These findings are very
reminiscent to the naive fireball calculations in central S+Au displayed
in Fig.~\ref{fig_sauconv}. Note that the absolute magnitude of 
the $\pi\pi$-induced signal in Fig.~\ref{fig_pbauhydro} is appreciably  
smaller than in transport calculations, owing to the vanishing pion
chemical potential implicit in the hydrodynamical framework. The final results
are therefore not quite able to account for the CERES data in the 
$M_{ee}\simeq 0.5$~GeV region.  At the same time, as a consequence of
the rather small thermal $\omega$ meson abundance at freezeout, the 
signal from direct $\omega\to e^+e^-$ decays amounts to only about 
50\% of the one in the CERES cocktail; thus, there is no issue of 
overpredicting the $\rho/\omega$ region (right panel of 
Fig.~\ref{fig_pbauhydro}). 

One of the most attractive features of hydrodynamic simulations is 
their capability of incorporating phase transitions in the time evolution in 
a well-defined way via the equation of state. For SpS energies at
158--200~AGeV, however, 
the general findings are~\cite{SSG96,HuSh97,Prak97} that the   
dilepton signal from a possibly formed quark-gluon plasma, 
as estimated by employing perturbative $q\bar q\to ee$ annihilation rates,  
is  down by about an order of magnitude as compared to the $\pi\pi$ 
channel for invariant masses below 1~GeV (dashed-dotted curve 
in the left panel of Fig.~\ref{fig_pbauhydro}).  
Recalling that the thermal $q\bar q$ production rates are
not very different from the in-medium hadronic ones (cf.~Fig~\ref{fig_qqvpp} 
in Sect.~\ref{sec_duality}), one has to conclude that at SpS energies 
the space-time volume occupied by the QGP phase is rather small
(this might not be the case for the mixed phase).  

The CERES collaboration has also analyzed their data with respect to  
centrality dependence of the invariant mass spectra, \ie, 
dividing them in four distinct event classes with average charged 
multiplicities $\langle N_{ch}\rangle =$ 150, 210, 270 and 350.  
Although one finds a clear increase of the enhancement with 
$\langle N_{ch}\rangle$, especially the low-multiplicity events
do not allow for more quantitative statements. No systematic
theoretical analyses are available yet.  

\subsection{Medium Effects II: Transverse Momentum Dependencies}
\label{sec_ptspec}
An additional observable to help discriminate different mechanisms that 
lead to a similar enhancement in the invariant mass spectra is the 
dilepton transverse momentum $q_t$, \ie,  
the total momentum of the dilepton pair perpendicular to the 
beam axis of the colliding nuclei.   
From a theoretical point of view this possibility is provided by the 
fact that the specification of a preferred reference frame -- that is, 
the thermal frame, in which the matter as a whole is at rest -- 
breaks Lorentz invariance of space-time.  
It implies that the in-medium propagators of the vector mesons  
(or, equivalently, their spectral functions) separately depend   
on energy $q_0$ and three-momentum modulus $|\vec q|$ (or on invariant mass 
$M=(q_0^2-{\vec q}^2)^{1/2}$ and three-momentum). Moreover, their 
polarization states are no longer isotropic, 
but split up into two completely independent modes, most conveniently
described in terms of longitudinal and transverse components,  
see Sect.~\ref{sec_vamix}, Eqs.~(\ref{Vcorr}), (\ref{Plt}).   
A different behavior of the latter might induce anisotropies 
in the dilepton yield which
are, however, extremely difficult to measure. No such attempt 
has been made to date. On the other hand, transverse momentum spectra
in three (four) adjacent invariant mass bins have been extracted by the CERES 
collaboration in the '95 ('96) lead runs. Since in the CERES experiment
the full kinematic information on the individual lepton tracks is 
recorded, their $q_t$-spectra are subject to the same 
statistical and systematic errors as the invariant mass spectra.     
 
In the previous Section we have seen that both the dropping $\rho$ mass
and the in-medium spectral function scenarios can reproduce
the invariant mass spectra at full CERN-SpS energies reasonably well.  
Both approaches have also been employed to calculate transverse
momentum spectra~\cite{Morio98,BLRRW} 
(Figs.~\ref{fig_pbauqtBR}, \ref{fig_pbauqtRW}). 
\begin{figure}[!htb]
\centerline{\epsfig{figure=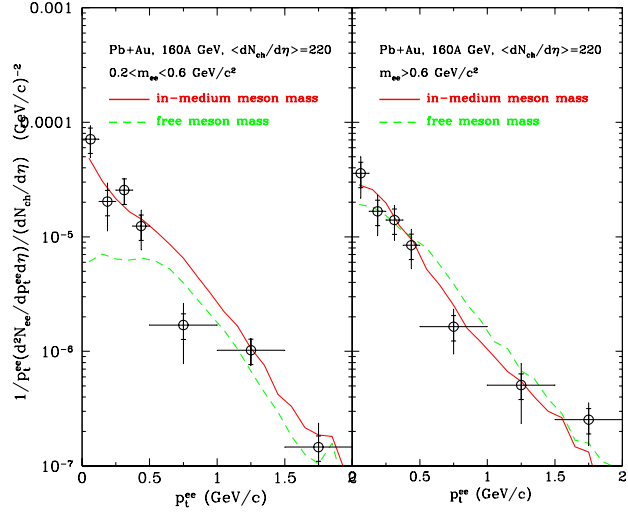,width=8cm,angle=-90}}
\caption{Comparison of the CERES transverse momentum
spectra in two invariant mass bins from 30\% central 158~AGeV
Pb+Au~\protect\cite{ceres96} with dropping $\rho$ mass 
calculations in a transport model~\protect\cite{BLRRW};
dashed and full curves: using free and in-medium masses, respectively.}
\label{fig_pbauqtBR}
\end{figure}
Naively one would expect that a mere reduction of the $\rho$ mass 
does not entail any distinct traces in the $q_t$-dependence; 
this is, however, not true due to a subtle interplay with the thermal
occupation factors $f^\rho(q_0)$, which depend on {\em energy}. Thus, 
for a small $\rho$ mass $m_\rho^*$, the three-momentum dependence of 
$q_0=[(m_\rho^*)^2+{\vec q}^2]^{1/2}$ is more pronounced, leading to a 
relative enhancement of $\rho$ mesons of small three-momentum. This 
is nicely reflected by the 
left panel of Fig.~\ref{fig_pbauqtBR}, where the enhancement of the 
dropping $\rho$ mass curve in the $0.2 {~\rm GeV} < M < 0.6$~GeV 
invariant mass bin  is predominantly concentrated 
at  transverse momenta $q_t\lsim 0.6$~GeV, in good agreement with
the '95 CERES data.  More complicated three-momentum dependencies 
may arise in the spectral function
approach. This was first pointed out in Ref.~\cite{FrPi}, where substantial 
effects with increasing three-momentum were predicted on the basis of  
 a strong $\rho N$ $P$-wave coupling to the $N(1720)$ and $\Delta(1905)$ 
resonances. The accompanying hadronic vertex form factors, 
\beq
F_{\rho BN}(q)= \Lambda_{\rho BN}^2/(\Lambda_{\rho BN}^2+{\vec q}^2) \ , 
\eeq
which govern the suppression of large three-momenta of the $\rho$, were 
used with rather hard cutoff parameters of $\Lambda_{\rho BN}=1.5$~GeV. 
However, in a subsequent analysis of photoabsorption spectra~\cite{RUBW} 
it turned out that such values are not compatible with $\gamma p$ and
$\gamma A$ data, requiring much softer $\Lambda_{\rho BN}\simeq 0.6$~GeV. 
These constraints have been extracted before any data on dilepton 
$q_t$-spectra were available. A spectral function calculation of the 
latter~\cite{RW99}, including these constraints, is shown in 
Fig.~\ref{fig_pbauqtRW}. 
\begin{figure}[!htb]
\centerline{\epsfig{figure=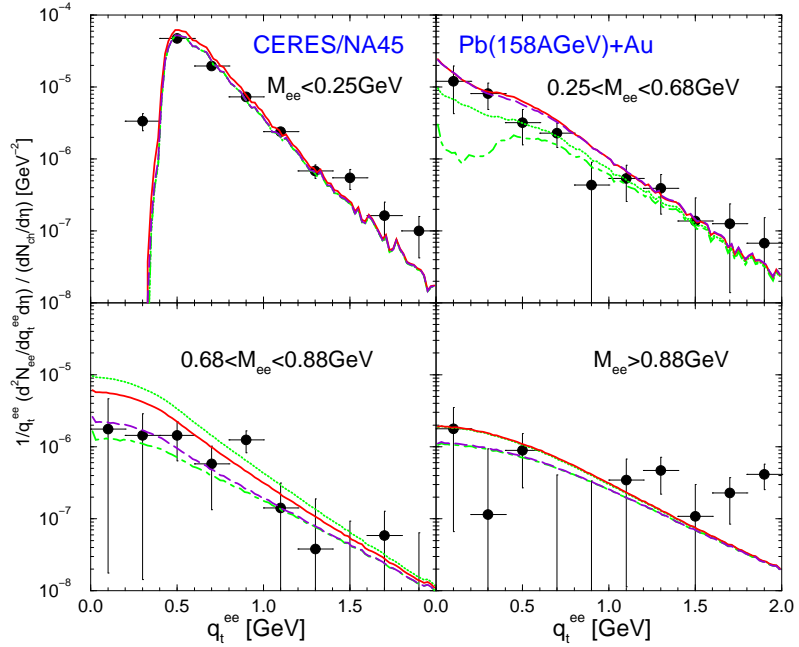,width=9cm,angle=-90}}
\caption{Comparison of the CERES transverse momentum
spectra in four invariant mass bins 
from 30\% central 158~AGeV Pb+Au~\protect\cite{ceres98} 
using a thermal fireball model including finite $\mu_\pi$~\protect\cite{RW99};
dashed-dotted curves: CERES cocktail; dotted curves: cocktail + free
$\pi\pi$ annihilation, dashed curves: cocktail + $\pi\pi$ annihilation 
with a dropping $\rho$ mass, full curves: cocktail + $\pi\pi$ annihilation 
using the in-medium $\rho$ spectral function.}
\label{fig_pbauqtRW}
\end{figure}
Similar to the dropping $\rho$ mass 
results,  its basic features agree with the data.    
 
In another projection of the data, the CERES collaboration generated
invariant mass spectra for two distinct regions of transverse pair 
momentum, \ie, $q_t<0.5$~GeV and $q_t>0.5$~GeV~\cite{ceres98}. 
\begin{figure}[!htb]
\centerline{\epsfig{figure=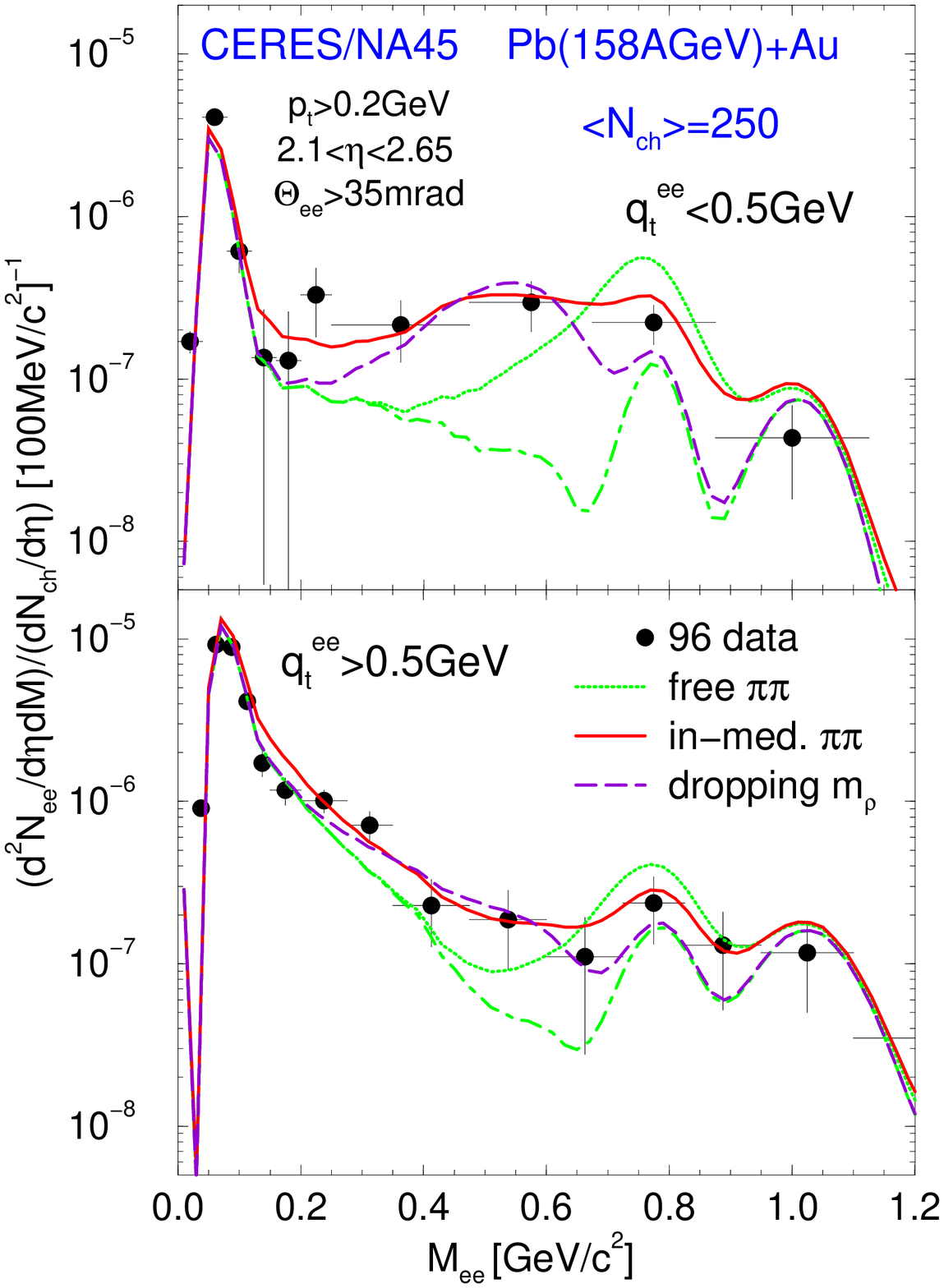,width=7.5cm}}
\caption{Comparison of 30\% central 158~AGeV Pb+Au CERES
invariant mass spectra in two
transverse momentum bins~\protect\cite{ceres98},
$q_t < 0.5$ (upper panel) and $q_t > 0.5$
(lower panel) with thermal fireball calculations~\protect\cite{RW99} 
(including finite $\mu_\pi$) employing
a dropping $\rho$ mass (dashed curves) and an in-medium spectral function
(full curves); the dashed-dotted curve is the CERES cocktail (without
the $\rho\to e^+e^-$ contribution), added
to the respective $\pi\pi$ annihilation yields from the fireball.}
\label{fig_pbauqt500}
\end{figure}
Again, one clearly observes 
that the major part of the low-mass enhancement is concentrated in 
the low-momentum bin, whereas the high-momentum bin is essentially 
consistent with the cocktail (Fig.~\ref{fig_pbauqt500}). 
This is just opposite to the rate calculations based on $P$-wave 
$\rho N$ scattering performed in Ref.~\cite{FrPi}, which confirms   
the necessity for rather soft form factors as predicted on the basis 
of photoabsorption data. 
The theoretical calculations shown in Fig.~\ref{fig_pbauqt500} contrast
once more the results of a dropping $\rho$ mass and the in-medium 
broadened spectral function. 
At the present status of the data, both explanations are viable.

\subsection{Time Dependence of In-Medium Signals}
\label{timeinfo}

The great hope that has been associated with dilepton observables 
as penetrating probes is to learn about the innermost zones
of high-density and high-temperature matter formed in the early stages 
of nuclear collisions. Thus, after our detailed study of various 
models in their application to experimental low-mass dilepton 
spectra we would like to address the question as to what extent signals 
from the highest excitation phases can be disentangled, \ie, how certain
features in the spectra might be related to the time 
(or temperature/density) of emission. 
Unfortunately, the answers are  beset with strong model 
dependencies, even if the 'background' from the hadronic cocktail 
were accurately known (as we will for simplicity pretend in the following). 
The dropping $\rho$ mass scenario implies an  obvious  
correlation  between invariant mass and emission time for the 
in-medium signal: using as a rough guideline the fireball evolution for
30\% central Pb+Au collisions~\cite{RW99} (cf.~Fig.~\ref{fig_Tevo}) 
together with a temperature- 
and density-dependent mass given by the QCD sum rule-type relation 
(\ref{mrhostar}), the time instances  
$t$=1, 6 and 11~fm/c correspond to masses $m_\rho^*(t)$=275, 465 and
650~MeV, respectively, which directly reflect the populated 
dilepton invariant mass regions.   
The situation is less straightforward when an in-medium spectral 
function is employed. 
\begin{figure}[!htb]
\vspace{-0.5cm}
\bce
\epsfig{figure=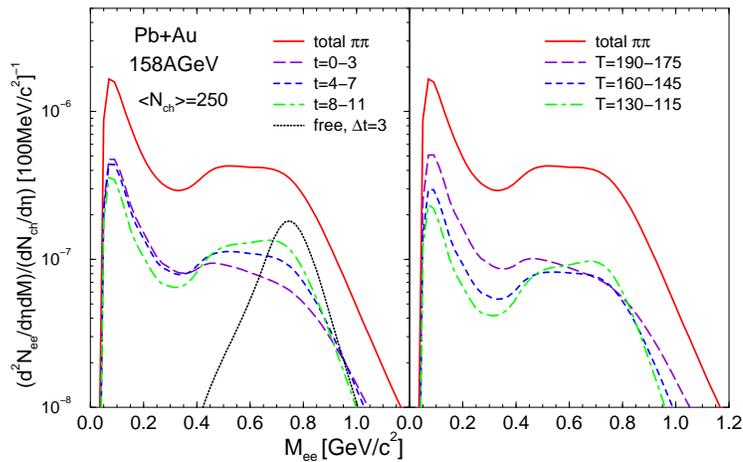,height=10cm,angle=-90}
\ece
\vspace{-0.7cm}
\caption{Decomposition of the $\pi\pi\to ee$ signal
employing the in-medium $\rho$ spectral function from
Refs.~\protect\cite{RUBW,RG99,RW99}
in 30\% central Pb+Au (including the acceptance of the CERES detector);
the solid line is the total yield from a thermal
fireball of lifetime $t_{fo}$=11.5~fm/c. In the left panel, the long-dashed,
short-dashed
and dashed-dotted curves represent the contributions from the time
intervals $t$=0--3~fm/c, $t$=4--7~fm/c and $t$=8--11~fm/c,
respectively, and the dotted curve arises for an emission via free
$\pi\pi$ annihilation over a duration of $\Delta t=3$~fm/c being
almost  independent on the evolution stage. In the right panel
the yields arising during temperature intervals $T$=175--190~MeV (long-dashed
curve), $T$=145--160~MeV (short-dashed curve) and $T$=115--140~MeV
(dashed-dotted curve) are displayed.}
\label{fig_timedeco}
\end{figure}
The left panel of Fig.~\ref{fig_timedeco} shows a (partial)
decomposition of the total in-medium signal in three equidistant 
time slabs. 
Within a few percent the integrated yield from each
of the three time intervals is essentially equal which is due
to a trade-off between increasing volume and decreasing temperature
during the expansion. 
Although lower masses are preferably populated at earlier stages, 
the time dependence of the spectral shape is rather smooth especially 
when comparing to the dropping mass scenario. 
The most prominent feature associated with early emission times 
is a strong depletion of the free $\rho$ peak around $M\simeq 0.75$~GeV. 
The total final spectrum (solid line in Fig.~\ref{fig_timedeco}) 
in fact closely resembles the
contribution from intermediate times (4--7~fm/c, short-dashed curve)
multiplied by a factor of 4. This means that the time-integrated  
in-medium signal actually probes
a hadronic resonance gas at an average temperature and density of about
$T\simeq 150-160$~MeV and $\rho_B\simeq \rho_0$, not very far
from the expected phase boundary to the quark-gluon plasma.
Also note that a typical emission spectrum from free $\pi\pi$ annihilation
is quite different from the in-medium pattern even close to freezeout. 

At low masses a more pronounced differentiation emerges when one
divides the emission contributions  into temperature (density) 
slices, cf.~right panel of Fig.~\ref{fig_timedeco}.  
The strongest fingerprint of a high temperature/density  phase 
seems to be around the free two-pion threshold, 
$M_{ee}\simeq 0.3$~GeV. The difference to the time decomposition
arises since the system spends somewhat longer time (3~fm/c) in the high
temperature interval than in the two lower temperature bins (about 2fm/c).
This effect originates from a (slight) softening of the equation of state 
as borne out of hydrodynamic simulations~\cite{HuSh95}. It is expected 
to be much more pronounced in the low-energy (40~AGeV) run at the SpS.

%


\subsection{Intermediate-Mass Spectra}
\label{sec_specint}
In this Section we would like to investigate in how far
medium effects that have been invoked to explain the low-mass 
enhancement are relevant for/consistent with
  the intermediate-mass regime (IMR). 
In Sect.~\ref{sec_duality} it has already been eluded to the 
conjecture that in vacuum, starting
from invariant masses of about $M_{ee}\simeq 1.5$~GeV, one enters
the 'dual' regime, \ie, the thermal dilepton emission rate 
can be reasonably well accounted for without further medium effects
by either\\
(i) perturbative $q\bar q\to e^+e^-$ annihilation, based on the 
observation  that the reverse process accounts for the total inclusive 
cross section for $e^+e^-\to hadrons$ within $\sim$~30\%, or \\
(ii) binary hadronic collisions, once an appropriate set of meson 
states is included, being similarly constrained by 
$e^+e^-\to hadrons$ cross sections in the corresponding 
(exclusive) channels.\\ 
Then the following questions have to be asked:\\
(1) Are such 'dual' rates compatible with 
experimental spectra?\\
(2) How do medium effects, which are crucial at low masses,
influence the intermediate-mass region? \\
For the quark-gluon description it has been argued that medium effects
should play a minor role as the small distance annihilation
of (nearly) massless quarks and antiquarks inhibits large corrections 
from the surrounding heat bath. 
However, this is much less obvious within the hadronic picture
since the interacting mesons (such as in the dominant 
$\pi a_1$ channel) carry substantial rest masses which already 
make up a large fraction of the $cm$ energy so that the 
annihilation reactions involve fairly  small momentum
transfers.  

At the CERN-SpS, intermediate-mass dilepton spectra have been 
measured by the HELIOS-3 and NA38/NA50 experiments. In analogy to 
the low-mass case one can divide the spectra into  
a (physical) background  part and an in-medium  signal radiated from the 
interaction phase of the fireball. If one again defines the background  
as the contributions arising in $p$+$p$ collisions, 
the higher masses probed necessitate a somewhat different composition 
that now mainly stems from hard processes 
occurring in the primordial stage of the collision. Most notably
these are Drell-Yan annihilation as well as open-charm decays 
(\ie, an $l^+l^-$ pair originating from the separate decay
of an associatedly produced pair of $D$ and $\bar D$ mesons) 
which are negligible in the low-mass region~\cite{pbm97}, but start to 
dominate over the final-state meson decays for $M_{ll}\gsim 1.5$~GeV. 
Contrary to the final-state meson decays, the initial hard processes  
are assumed  to scale with the number of primary nucleon-nucleon
collisions to provide their contribution in $p$+$A$ and $A$+$A$ reactions  
(for the open charm, \eg, this has been verified  
for $p$+$A$ collisions in Ref.~\cite{pbm97}). 
Fig.~\ref{fig_pwswint} shows a comparison of the  
the HELIOS-3 $\mu^+\mu^-$ data from $p$+W and S+W collisions with  
the various background sources~\cite{drees98,LiGa}.
\begin{figure}[t]
\vspace{-1cm}
\begin{center}
\epsfig{figure=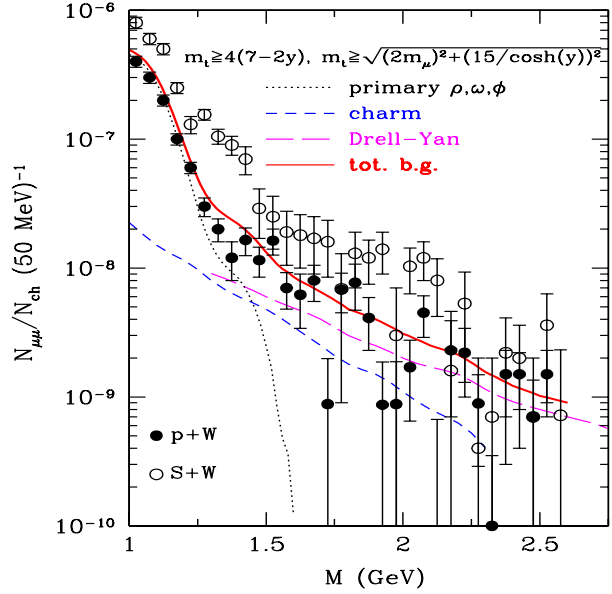,height=9cm,width=11cm}
\end{center} 
\vspace{-1cm}
\caption{Dimuon invariant mass spectra from the HELIOS-3
collaboration~\protect\cite{helios3} taken in 450~GeV
proton-induced and 200~AGeV sulfur-induced reactions on tungsten 
targets (full and open circles, respectively), compared to 
the expected background yields from light vector mesons (dotted
curve), Drell-Yan processes (long-dashed curve), open charm
decays (short-dashed curve) and their sum (solid curve). The plot
is taken from Ref.~\protect\cite{LiGa}.} 
\label{fig_pwswint}
\end{figure}    
Whereas the total background
reproduces the $p$+W spectra quite satisfactorily, the S+W data
are underestimated by a factor of 2--3 throughout the entire 
mass range from 1--2.5~GeV (equivalent observations have been reported 
from the NA38/NA50 collaboration~\cite{na50int}). 
 
Li and Gale have performed transport calculations including 
the radiation from the fireball using the mesonic production
rates from binary collisions~\cite{LiGa}. They find
good agreement with the HELIOS-3 data in the intermediate-mass region
if no further medium effects are included (dotted line in 
Fig~\ref{fig_swdrop}). In particular, this implies that both perturbative
$q\bar q$ rates and the lowest-order in temperature mixing effect in the
axial-/vector correlator are compatible with the data (as follows from the 
rate comparison discussed in Sect.~\ref{sec_duality}).  
\begin{figure}[t]
\vspace{-1cm}
\begin{center}
\epsfig{figure=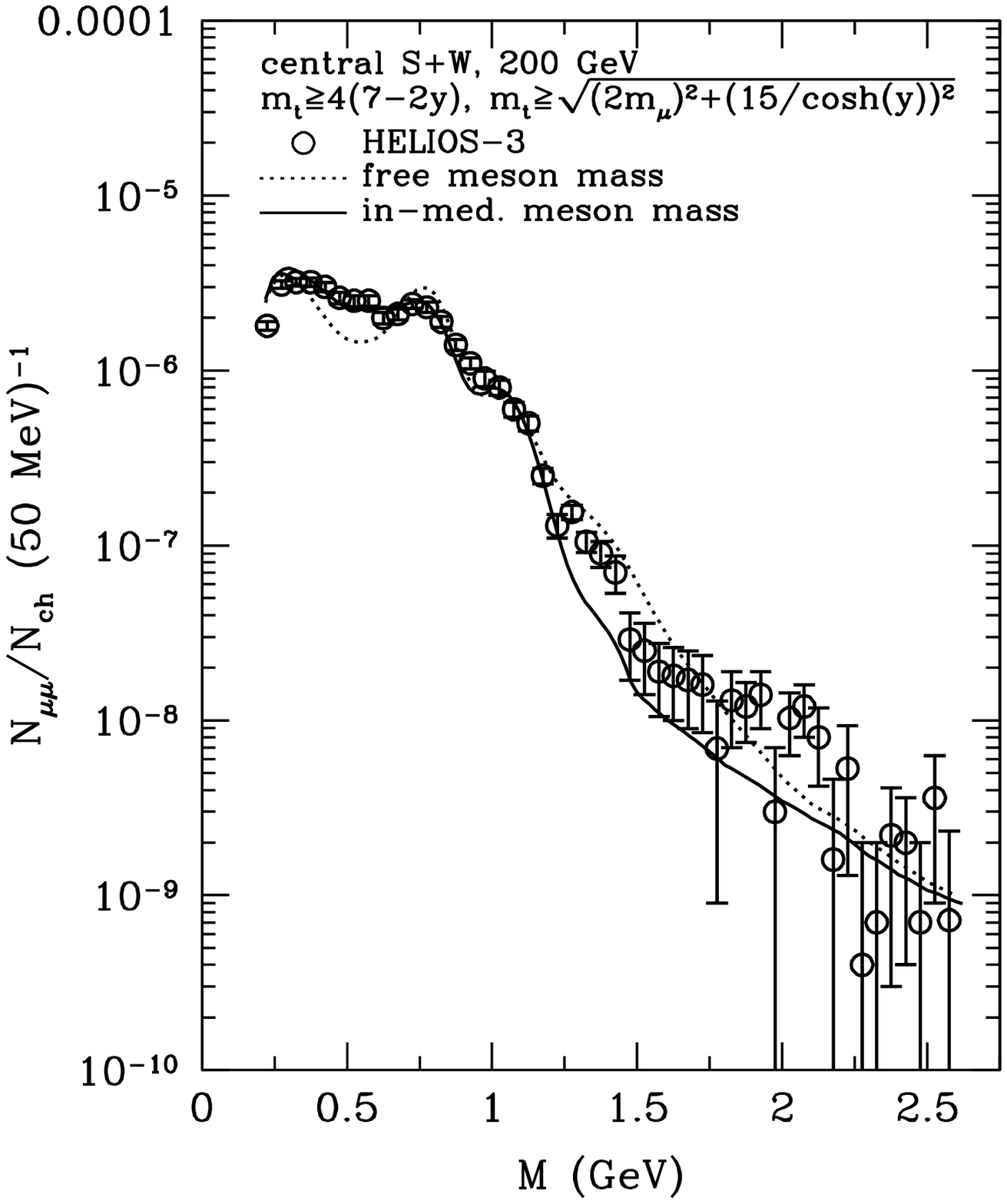,height=9cm,width=11cm}
\end{center}
\vspace{-1cm}
\caption{Dimuon invariant mass spectra as measured by the HELIOS-3
collaboration~\protect\cite{helios3} in 200~AGeV S+W reactions,
compared to transport calculations using a purely hadronic description
for the dilepton radiation from the fireball; dotted line: employing
free hadron masses; full line: dropping mass scenario. The plot
is taken from Ref.~\protect\cite{LiGa}.}
\label{fig_swdrop}
\end{figure}
On the other hand, the low-mass end of the spectrum (below $\sim$~1~GeV) 
cannot be explained in terms of binary collisions without invoking
any further medium effects. Li and Gale therefore employed  
 the dropping mass scenario extrapolated to include the higher  
mass (non-strange) vector resonances as well.   
The data are then nicely reproduced from the two-muon threshold 
up to about 1.2~GeV, but seem to be underestimated 
beyond (possibly also between the $\omega$ and $\phi$ mass). 

Using the emission rates from the chiral reduction formalism, 
together with an expanding thermal fireball model~\cite{RCW}, 
Lee \etal~\cite{LYZ} have obtained similar results to the Li-Gale 
calculations with free meson masses.

As of now there are no intermediate-mass dilepton
calculations available using in-medium many-body spectral 
functions. Here the question is whether a strong broadening 
of, \eg, the $\rho$ resonance could lead to an {\em over-}estimation 
of the data in the 1--1.5~GeV mass region. However, from a theoretical 
point of view, if the broadening scenario indeed 
approaches chiral restoration by merging into the perturbative
plateau-value for the vector and axialvector correlators (as we have 
argued in Sects.~\ref{sec_vamix},~\ref{sec_duality}), 
the HELIOS-3 data will be reproduced,
see the above remarks. Also note that  the in-medium $\rho$-meson 
spectral function as used for the low-mass region 
actually exhibits only a rather moderate enhancement (factor of $\sim$~2) 
over the vacuum one for invariant masses somewhat above the free
$\rho$ peak (Fig.~\ref{fig_imdrho685}). 
Above 1~GeV further contributions need to be included for a more complete
description of the vector correlator, such as $\pi a_1\to\rho'$ 
selfenergies corresponding to four-pion like processes. 
This will closely resemble the kinetic theory treatment 
in terms of the various binary scattering processes, as performed 
in Ref.~\cite{LiGa}. 
Since coherence effects in the many-body treatment are expected 
to be small especially towards higher energies 
(already in the low-mass region they were found to be quite moderate),  
there should be little discrepancies to the (incoherent) kinetic 
approach, even on a quantitative level.  
Beyond $M=1.5$~GeV the 'dual' (hadronic or partonic) 
production rates are characterized by an essentially flat spectral shape. 
Here the main 
issue therefore is whether the space-time description used for the 
calculations in the low-mass region will lead to 
a total yield that reproduces   
the experimentally observed enhancement. Another possible source of this
excess has been pointed out by the NA50 collaboration: they 
showed~\cite{na50int} that the excess can be accounted for by 
introducing an anomalously increased production of open 
charm mesons by a factor of $\sim$~3. However, there are no theoretical 
indications for a suitable mechanism of this kind (the final answer
will be provided by a direct measurement of produced $D$ mesons at the
CERN-SpS). Lin and Wang~\cite{LW98}
have addressed the possibility of $D$ meson rescattering to enrich the 
dilepton yield in the NA50 acceptance; however, this does not constitute 
more than a 20\% effect.    
On the other hand, in the recent analysis of Ref.~\cite{GKP99} it has been
shown that the use of the dual lowest-order $q\bar q$ annihilation rate
throughout  the entire mass range from 0--3~GeV folded over a schematic 
fireball evolution~\cite{RCW}  leads to very similar yields in the IMR as 
obtained with a factor of 3 open charm enhancement; at the same time 
the (low-mass) CERES data are approximately accounted for  
(see also dashed curve in Fig.~\ref{fig_pbauqq}). 
Similar conclusions are reached in Ref.~\cite{RS99}: 
the NA50 enhancement
between 1.5 and 3~GeV can be explained with the dual $q\bar q$ rate 
employing the same fireball model~\cite{RW99} (being consistent
with hadro-chemical analysis of CERN-SpS data including finite pion
chemical potentials towards thermal freezeout) 
that leads to a satisfactory description of the CERES data once 
medium effects in the $\rho$ spectral function are incorporated.

\section{Direct Photon Spectra}
\label{sec_phspectra}
As we have stressed in Sect.~\ref{sec_phrates}, real photons and dileptons  
can be considered as two kinematical realizations of otherwise 
identical electromagnetic production mechanisms.   
For heavy-ion reactions it follows  
that any model which claims success in describing the observed dilepton 
spectra must also be consistent with direct photon spectra. The notion 
'direct' has been introduced to refer to only those photons which 
are originating
from the interaction phase of the fireball, \ie, unlike the dilepton 
case, the contributions from hadron decays after freezeout are 
considered as a background that ought to be subtracted to obtain the 
final spectra. This, in fact, represents the main experimental 
difficulty, since around 90\% or more of the 
photons produced in heavy-ion  collisions at full SpS energies stem from the 
$\pi^0\to \gamma\gamma$ and $\eta\to \gamma\gamma$ decays. As a consequence
direct photon observables are about two orders of magnitude less sensitive
to any in-medium signals than dileptons~\cite{ceres96b}. 
Also note that the observed single-photon energies pick up the 
laboratory three-momentum of the 
decaying hadron and hence are not restricted by the hadron rest mass 
-- as opposed to 
the invariant masses of dileptons from the Dalitz decays, where 
$M_{ll}\le m_{\pi}$, etc.. The major 
systematic error then arises from the uncertainty in the  
$\pi^0$ and $\eta$ abundances. The contributions from other mesons
are usually estimated from the so-called '$m_T$ scaling', \ie,  
a $\sim$$\exp(-\beta m_T)$ dependence of the transverse mass spectra
with a universal slope parameter $\beta$.     
Photon measurements at the CERN-SpS have been performed  by  
 the HELIOS-2~\cite{helios2}, CERES~\cite{ceresph} and 
WA80/WA98~\cite{wa80,wa98} collaborations. 
In central S+Au,  WA80 found a photon excess of 
$5\% \pm 5.8\% ({\rm syst}) \pm 0.8 \%({\rm stat})$ consistent
with CERES and HELIOS-2 results which carry somewhat larger errors. 
From these
measurements they extracted an upper limit for direct photons at 
the 90\% confidence level (see below), which has been used to 
test theoretical models. More recently, a preliminary direct
photon spectrum  for central Pb+Pb has been published~\cite{wa98}.  
 
In hydrodynamical simulations~\cite{SiSi94,frank95}
it has been claimed that the upper limits set by WA80
are not compatible with purely hadronic scenarios. However, these 
conclusions 
have been drawn using a very limited number of degrees of 
freedom in the hadronic gas phase ($\pi, \eta, \omega, \rho$).  
For equal initial energy densities this leads to  
much larger initial temperatures in the hadronic phase than 
in a quark-gluon plasma, entailing much higher photon yields.  
Purely hadronic models with lower initial temperatures cannot 
be ruled out by this reasoning.
In fact, the thermal production rates of photons  
from a QGP are presumably not very different from  those of  
a hadron gas at the same temperature, as has been first noted in 
Ref.~\cite{KLS91}. 
Hence, similar to what has been found for dileptons, one should expect 
that at CERN-SpS energies the
rather small space-time volumes occupied by a possibly formed QGP
do not generate substantial photon signals as compared to the hadronic
phase. 
This has been explicitly demonstrated in 
the hydrodynamic calculations of Refs.~\cite{SiSi94,Prak97}.
Thus one is led to focus on  the multiple possibilities for hadronic
photon production (Sect.~\ref{sec_phrates}).   

\begin{figure}[!htb]
\bce
\epsfig{figure=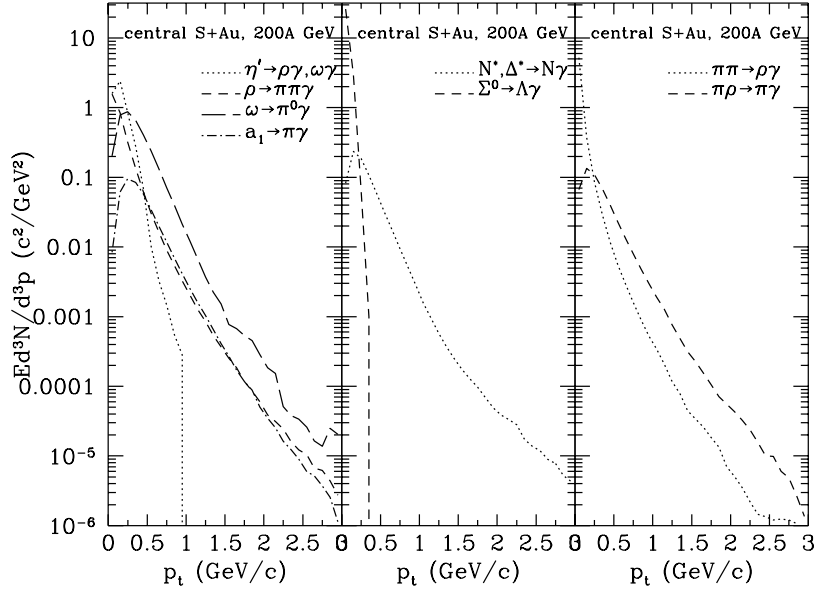,width=9.5cm,angle=-90}
\ece
\vspace{-0.6cm}
\caption{Thermal single-photon spectra 
from meson decays (left panel), baryon decays (middle panel) 
and two-body reactions (right panel) in transport calculations 
for central 200~AGeV S+Au collisions~\protect\cite{LB98}.} 
\label{fig_phspecli1}
\end{figure}
Fig.~\ref{fig_phspecli1} shows the photon transverse momentum
spectra from various sources
in central S+Au collisions at 200~AGeV, evaluated in the transport
framework~\cite{LB98}.
A kinematical cut in pseudorapidity of $2.1 < \eta < 2.9 $ is applied 
to comply with the WA80 experiment. For transverse momenta 
$q_t\ge 0.5$~GeV the dominant processes are radiative
$\omega$ and $a_1$ decays, the latter significantly exceeding the 
nonresonant $\pi\rho\to \pi\gamma$ reactions. Note that the baryonic
decays seem to have little relevance here. 
Fig.~\ref{fig_phspecli2} shows that the 
incoherent sum of all contributions (short-dashed curve) 
respects the experimental upper limits of WA80. 
\begin{figure}[!htb]
\centerline{\epsfig{figure=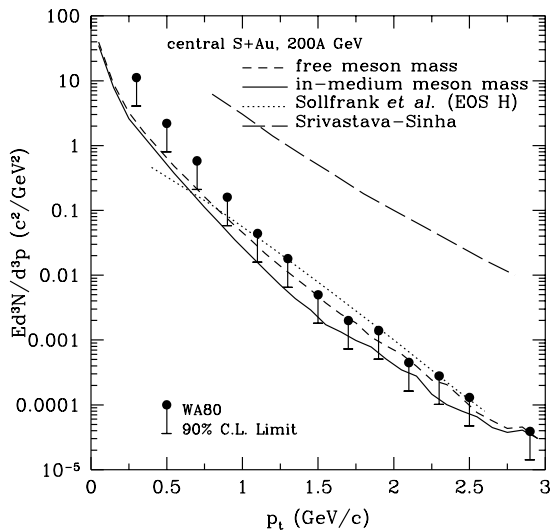,width=9cm}}
\caption{Total direct photon spectra
in central 200~AGeV S+Au collisions compared to the upper limits
of the WA80 collaboration~\protect\cite{wa80}; full and short-dashed
curves: transport calculations of Ref.~\protect\cite{LB98} with
and without dropping masses, respectively; long-dashed curve: Bjorken-type
hydro-calculations with a purely hadronic equation of state including
$\pi,\eta,\rho$ and $\omega$ mesons only~\protect\cite{SiSi94};
dotted curve: 2+1 dimensional hydro-calculations employing a purely
hadronic equation of state with a larger set of mesons as well as
baryons~\protect\cite{Prak97}; both hydro-approaches have used the photon
production rates of Ref.~\protect\cite{KLS91}. The compilation is taken
from Ref.~\protect\cite{LB98}.}
\label{fig_phspecli2}
\end{figure}
Furthermore, applying the dropping-mass 
scenario within the same transport approach does not induce major
changes in the final spectrum  
(full curve in Fig.~\ref{fig_phspecli2}), 
which can be traced back to compensating 
mechanisms:  for the radiative decays of the vector mesons 
$\rho$, $\omega$  and $a_1$, their increased abundance (due to the smaller
masses) is balanced by a reduced phase space for the decay products (see
also Ref.~\cite{sarkar98}, where similar observations have been made
at the relevant temperatures of about $T\simeq 160$~MeV).  
Analogous  features have been found for other variants of the
dropping $\rho$-meson mass scenario when implemented in the Hidden 
Local Symmetry 
approach for $\pi$-$\rho$-$a_1$ dynamics~\cite{halasz98}. 
Also shown in Fig.~\ref{fig_phspecli2} are the hydrodynamical results
of Ref.~\cite{SiSi94} (long-dashed curve) which, as mentioned above,  
strongly overshoot the WA80 bounds due to a high initial temperature 
in a purely hadronic description with rather few degrees of freedom. 
On the other hand, using a larger set of
hadronic states (including the lowest-lying pseudoscalar and vector meson 
nonet as well as baryon octet and decuplet), the hydrodynamical calculations of 
Ref.~\cite{Prak97} are essentially compatible with the data and also
not very different from the transport calculations.  

Concerning the role of baryons,  
both the chiral reduction formalism~\cite{SYZ2} and 
the $\rho$-meson spectral function approach~\cite{RCW,RUBW,Morio98,RG99} 
have reached conclusions which are at some 
variance with the relative assignments of the 
incoherent decomposition given in Fig.~\ref{fig_phspecli1}. 
Using the simple thermal fireball expansion for central S+Au collisions
(where the baryon-density evolution is taken from the transport simulations)  
yields the photon $q_t$-spectra displayed in Fig.~\ref{fig_phsyz-rw}; 
for each of the two approaches, they are based on exactly the same 
ingredients as the corresponding dilepton spectra of Fig.~\ref{fig_sauconv}. 
\begin{figure}[!htb]
\epsfig{figure=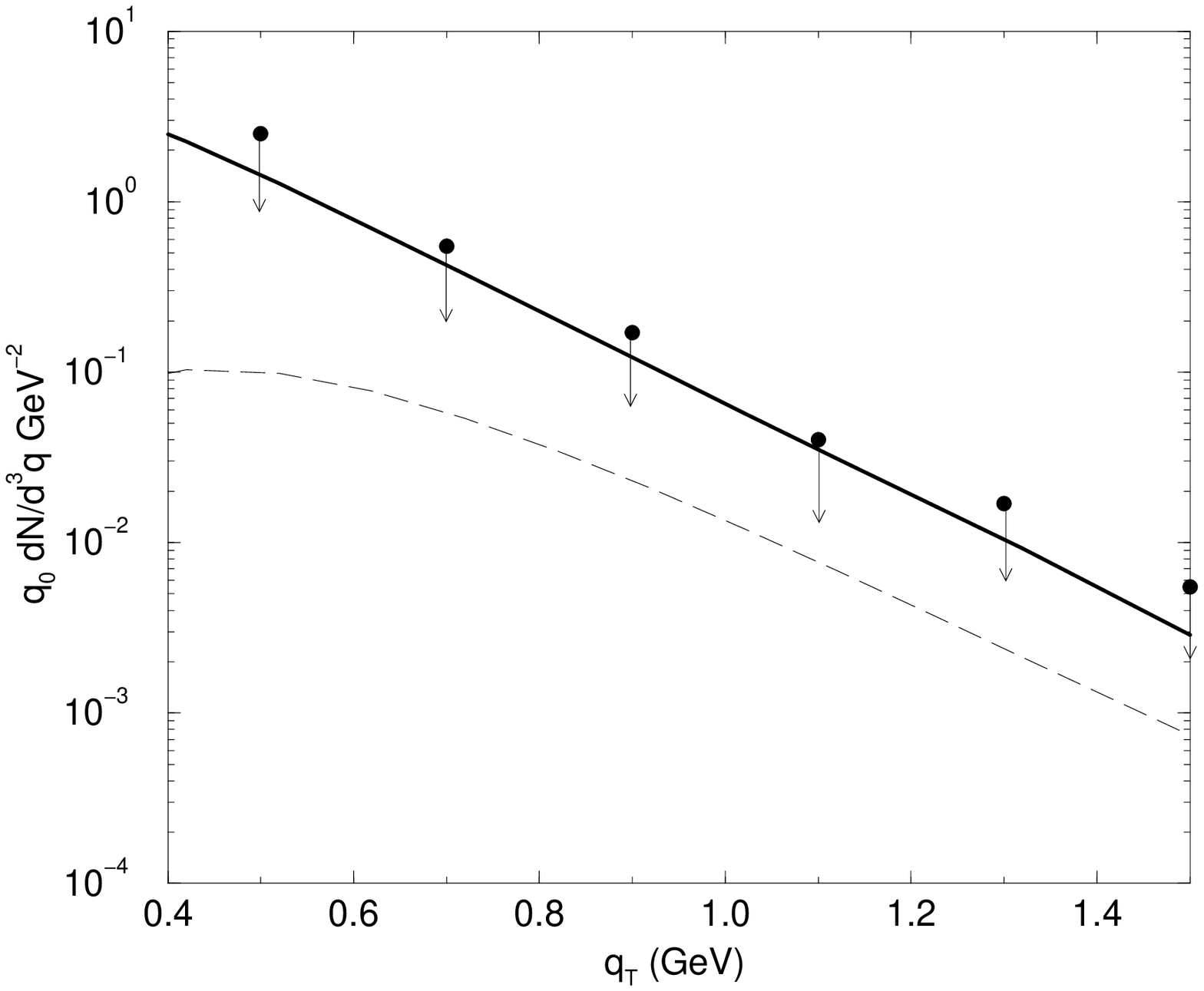,width=7cm,height=6.3cm}
\epsfig{figure=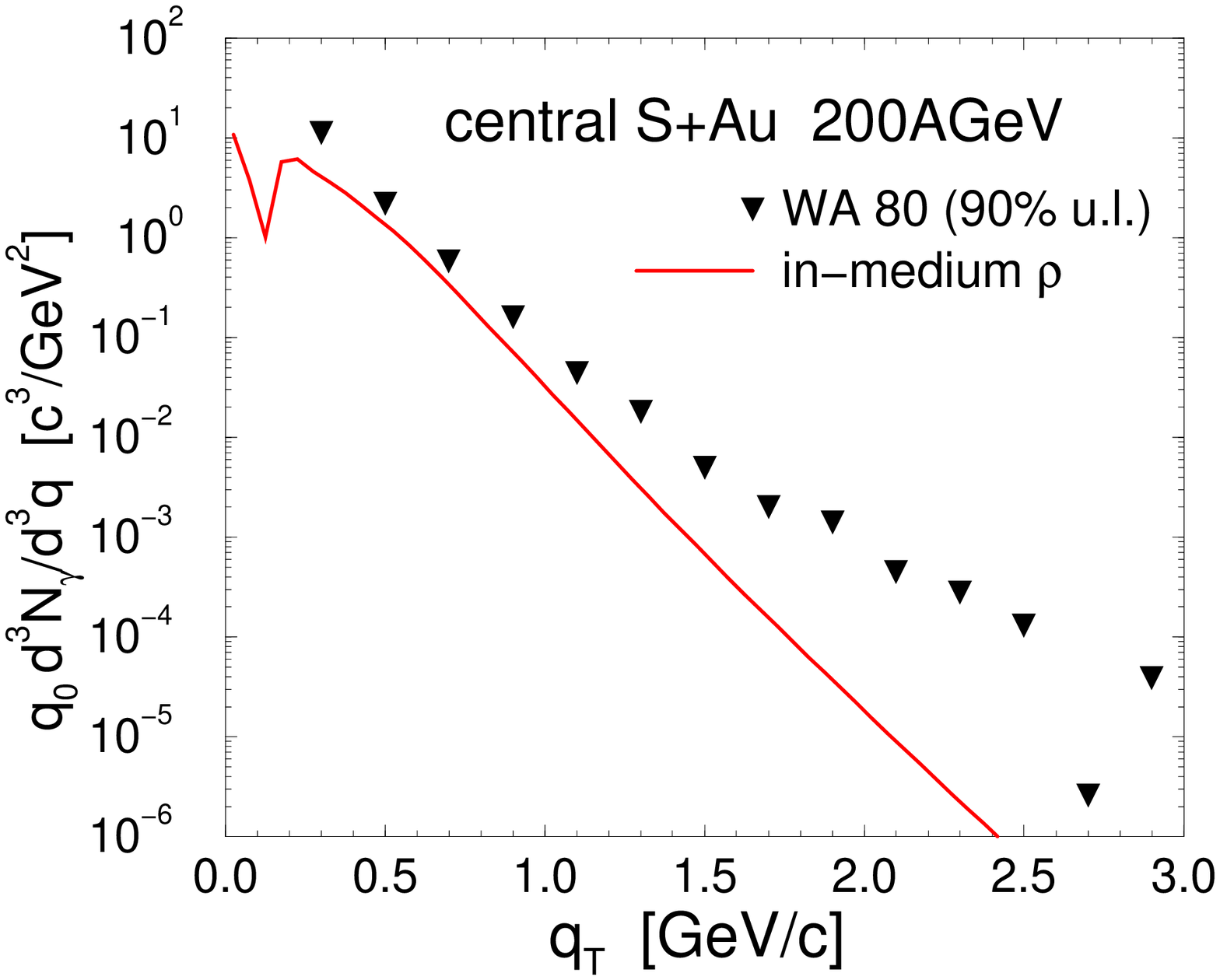,width=7.3cm}
\caption{Direct photon spectra, compared to WA80 upper 
limits~\protect\cite{wa80}, 
in central 200~AGeV S+Au collisions using a simple fireball evolution 
with thermal rates from the chiral reduction 
formalism~\protect\cite{SYZ2} (left panel; solid and dashed curve are 
obtained with and without the baryonic contributions, respectively) 
and the in-medium $\rho$ spectral 
function~\protect\cite{RCW,RUBW,Morio98,RG99} (right panel).}
\label{fig_phsyz-rw}
\end{figure}
In the chiral reduction formalism the baryonic contributions are 
exclusively associated with the finite nucleon density in the fireball. 
According to the decomposition inferred from photoabsorption
data (Fig.~\ref{fig_gamsyz}) the effects can be attributed to  
the $\Delta(1232)$ for $q_t\le0.4$~GeV and to the $\pi N$-'background' 
above. 
In the photon production spectra for central S+Au reactions they induce an
enhancement over the mesonic contributions by a sizable  
factor (left panel of Fig.~\ref{fig_phsyz-rw}). Very similar
results follow from the $\rho$-meson spectral function approach, which
also reproduces well the nuclear photoabsorption spectra, albeit
with a somewhat different decomposition, \ie, a smaller $\pi N$ 
'background' together with much larger contributions from direct 
$\rho N$ resonances. 
This difference is not relevant  for the direct photon spectra. 
On the other hand, recalling the comparison to the CERES dilepton
spectra in central S+Au, it has been found that the $\rho$-meson spectral 
function approach {\em does} lead to a larger low-mass enhancement. This
can be readily gleaned from the rate comparison exhibited in 
Fig.~\ref{fig_Brates}:  compared to the Steele \etal-rates, 
the Rapp \etal-rates are by a factor 2--3 larger around $M=0.5$~GeV 
(but agree with the former towards the photon point).


\section{Theoretical Implications}  
\label{sec_implications}
After the preceeding rather detailed discussion of the various 
efforts made in exploring low-mass dilepton spectra in (ultra-)
relativistic heavy-ion collisions (with additional impact from photon
and intermediate-mass dilepton spectra),  we have to face the question 
in which respects it has advanced our understanding of strongly interacting 
hot and dense matter. 
For that purpose let us try to critically review and compare the successes 
and failures as well as the interrelations of the different theoretical 
attempts that have been pursued to describe the various experiments.   

Clearly, the BR scaling conjecture has been very successful
in its application to low-mass dilepton data at CERN-SpS energies of
158--200~AGeV. 
In its original form it predicts a very 
specific realization of chiral symmetry restoration, namely that all 
masses of the light (non-Goldstone) hadrons merge to zero. 
Most of the underlying arguments in its favor rely on mean-field
type approximations, related to the decrease of the chiral quark 
condensate (via constituent quark masses) or to the presence of strong 
scalar fields in hot/dense hadronic matter. 
More recently connections have been drawn to link a reduction 
in the $\rho$ and $\omega$ masses in nuclear matter to strong
collective excitations in $\rho$-$N$ and $\omega$-$N$ interactions, 
most notably through $N(1520)N^{-1}$ states. The ramifications of this 
identification are still under debate. 

On the other hand, the consequences of a strong in-medium broadening of 
the $\rho$-meson spectral function, as predicted on the basis of 
phenomenologically 
well-established hadronic interactions combined with standard 
many-body techniques,  also seem to reproduce the
SpS low-mass dilepton data fairly well. 
If this scenario holds true close to the phase transition, it implies 
that the chiral partner of the $\rho$, the $a_1(1260)$, becomes 
as broad and structureless: chiral symmetry restoration manifests itself
through a merging of both vector and axialvector
correlators into a flat continuum (such a behavior has been conjectured
to signal deconfinement in Ref.~\cite{Dom89}). 
If, in addition, the height of the continuum corresponds to the 
perturbative $q\bar q$ plateau value characterized by the famous 
cross section ratio  
$R=\sigma(e^+e^-\to hadrons) / \sigma(e^+e^-\to \mu^+\mu^-)= 5/3$ 
(=2, when including strange quarks),  
nonperturbative effects would be marked as no longer relevant which, 
after all, constitutes the very essence of the quark-gluon plasma.  
This provokes the following theoretical exercise:    
we simply replace the in-medium hadronic rates in the expanding fireball
model (\ref{dlspecfb}) by  the perturbative $q\bar q$ annihilation 
rates, (\ref{qqrate2}), using the same time evolution of 
temperature and density.  
\begin{figure}[!htb]
\vspace{-0.7cm}
\centerline{\epsfig{figure=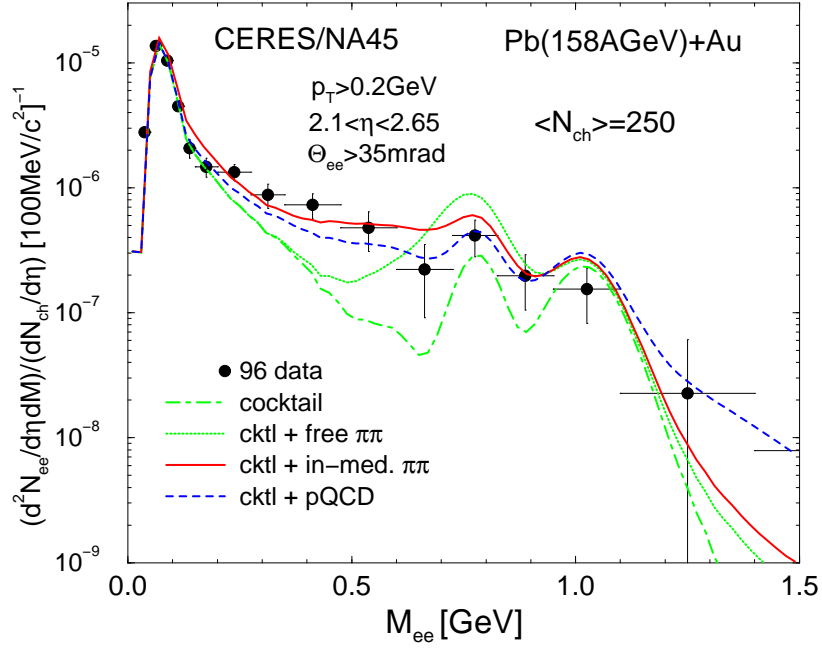,width=10cm,angle=-90}}
\caption{CERES dilepton spectra in 30\% central Pb+Au 
collisions~\protect\cite{ceres98} compared to a thermal
fireball calculation using lowest order $q\bar q\to ee$ annihilation 
rates only (dashed line); also shown is the full in-medium $\rho$ 
spectral function calculation (solid line).
Both fireball yields are supplemented with the CERES cocktail.}
\label{fig_pbauqq}
\end{figure}
The result shown in Fig.~\ref{fig_pbauqq} indicates that the global
use of the pQCD rate looks quite reasonable but cannot fully account for 
the low-mass enhancement
observed in the CERES data (unless one invokes nonperturbative
effects, see, \eg, Ref.~\cite{LWZH}). More importantly, one should note 
that,  whereas the hadronic medium effects are more sensitive  to finite 
baryon densities, the lowest-order $q\bar q$-rates exhibit a negligible 
dependence on a finite quark chemical potential. This obvious 
'duality-mismatch' clearly deserves further studies. 

Nevertheless, let us return to the hadronic approaches and first  
address the question what the discriminating features for different 
fates of the $\rho$ meson could be, \ie, 'dropping mass' versus 
'melting resonance'. Both lead to enhanced dilepton yields below
the free $\rho$ mass, but behave very differently above. Whereas
the in-medium broadening implies an {\em enhancement} over free $\pi\pi$
annihilation between, say, the $\omega$ and the $\phi$ mass, the 
dropping mass scenario predicts a {\em depletion} (the possible impact
from higher states, such as the $\rho'(1450)$, entering this region  
should be small, since
(a) it is not obvious that a large fraction of their mass is due  
to chiral symmetry breaking, and  
(b) they have little significance in the pion electromagnetic form 
factor). As stressed above,  
an improved mass resolution of the currently available dilepton 
measurements, together with sufficient statistics, 
should be able to settle this issue. 
The mass region just above the $\rho/\omega$ resonances is also  
supposed to be accessible by the PHENIX experiment  
at RHIC, although plagued by a tremendous  
combinatorial background which may limit any sensitivity to 
resonance structures with sufficiently narrow widths. 
Also, both the broadening and the dropping mass are predominantly
driven by finite baryon density effects, which are presumably small
at RHIC.  In this respect the upcoming low-energy 
run at 40~AGeV at the CERN-SpS will be ideally suited, probing
even lower temperatures and higher baryon densities than with  the
158~AGeV beams. 

The lack of 'dramatic' effects in the $\rho$ properties 
in a thermal gas of mesons, as found by many 
authors~\cite{Ha95,RCW,EK99,Gao99,RG99}, seems, however, somewhat puzzling.
Phenomenologically this can be traced back to the
weaker  meson-meson interactions (especially with 'Goldstone-protected' 
pions), as compared to the meson-nucleon case. 
Thus, when carrying the calculations to higher densities it seems
plausible that the $\rho$ and the $a_1$ could 'degenerate through
broadening'. On the other hand, even at temperatures as high as 
$T=200$~MeV, quite moderate in-medium corrections to the $\rho$ width
have been deduced, reaching at most $\Gamma_{med}\simeq 200$~MeV.  
Since this leaves a well-defined resonance structure, it appears 
unclear how the $\rho$ and $a_1$ spectral distributions merge
at this point. These features might either point at  some shortcomings
in the phenomenological approaches or indicate that the lowest-order 
in temperature vector-axialvector
mixing is the prevailing effect in pure {\em meson} matter.   
Also note that in most model calculations the chiral condensate
exhibits a strong, linear decrease with nuclear density (leading
to an appreciable reduction already at normal nuclear density), 
whereas its temperature dependence is much less pronounced until
close to the critical temperature $T_c^\chi$. 
Again, dilepton measurements at RHIC will hopefully shed more
light on the behavior of $\rho$, $\omega$ and $\phi$  mesons in 
a high-temperature and low-baryon-density environment. 

Next, we compare the hadronic many-body calculations to the 
chiral reduction formalism put forward in its application 
to electromagnetic emission rates in  Refs.~\cite{SYZ1,SYZ2}. 
As far as low-mass dilepton production at the CERN-SpS is concerned, 
the latter seems to give a factor of 2--3 less enhancement in the 
relevant region around $M_{ll}\simeq 0.4$~GeV than 
the results obtained with in-medium $\rho$ spectral 
functions~\cite{RCW,RUBW,Morio98,RG99,RW99}, 
see Figs.~\ref{fig_sauconv},~\ref{fig_pbauhydro}. 
Although both approaches have been constrained
by photoabsorption spectra on nucleons and nuclei 
the differences in the dilepton excess do mainly emerge from finite 
density effects. This is {\em not} due to a deviation at large
densities (\ie, through collective effects), but due to
a different assignment of 'background' and resonance contributions
in the photoabsorption data. When moving into the time-like dilepton
regime (at small three-momentum), the resonance contributions (which
are larger in the spectral function approach, most importantly 
the $N(1520)$) are relatively much more enhanced than the structureless
background (which is larger in the chiral reduction approach), 
being essentially a kinematic effect.  
This is further supported by the fact that in both approaches the direct 
photon spectra actually agree quite well when integrated 
over the temperature and density history of central S+Au collisions. 
Thus a careful separation of background
and resonance contributions in the experimental photoabsorption 
data should provide the key for a more quantitative discrimination 
of the effects in the time-like regime.  

From a theoretical point of view, the most distinctive feature  
between the many-body and the chiral reduction treatment lies in 
the fate  of the $\rho$ resonance peak: in the master formula framework 
in-medium corrections are obtained through additive terms in 
a  temperature and density expansion, which have almost no impact on the 
free $\rho$ peak, whereas  the resummation of
large imaginary parts in the many-body treatment of the 
$\rho$ propagator induces a marked depletion. Model-independent 
results for the in-medium vector and axialvector correlators indeed 
require that, due to their mixing, strength should be removed 
from both the $\rho$ and the $a_1$ poles, at least to lowest order in 
temperature and in the chiral limit~\cite{DEI90} (cf.~Sect.~\ref{sec_vamix}). 
The phenomenological many-body
calculations seem to comply with this feature, although their theoretical  
relation to chiral symmetry is not always obvious. 

In finite-temperature
calculations based on chiral $\pi$-$\rho$-$a_1$ Lagrangians the mixing 
theorem has been shown to be satisfied~\cite{LSY95}, being 
mainly realized through  a reduction of the $\gamma$-$\rho$ vector dominance
coupling $g_{\rho\gamma}(T)$ due to a finite-temperature pion tadpole loop.
However, corresponding results for dilepton production rates cannot
account for the low-mass enhancement observed at the 
CERN-SpS~\cite{Koch96}. One may 
raise the question whether phenomenological calculations at finite
temperature {\em and} density are still compatible with the data 
when including a suppression of the VDM coupling. 
Since $[g_{\rho\gamma}(T)/g_{\rho\gamma}(0)]^2$ decreases to only
about 80\% at the relevant temperatures of 
$T\simeq 150$~MeV (using the physical pion mass rather than the chiral limit), 
the answer is that it will only mildly affect the many-body results 
(in fact, a more complete calculation should also include  
a (moderate) finite-temperature softening of the single-pion 
dispersion relation, which re-generates a small enhancement 
below the free $\rho$ mass). 

Another, more practical, issue concerns the 
role of a finite pion chemical potential in the evolution of 
(ultra-) relativistic heavy-ion collisions. Although it should have 
only moderate influence on the {\em shape} of the dilepton spectra, 
it certainly has a severe impact on the {\em total yield} of dilepton
pairs originating from $\pi\pi$ annihilation in the interacting
fireball (being proportional to the square of the pion density). 
Both microscopic transport and hydrodynamical calculations
correctly reproduce the total number of pions at freezeout, but the 
former seem to imply a finite $\mu_\pi$ (not present in current
hydrodynamical analyses), resulting in a larger dilepton signal by 
a factor of 2--3. 
The introduction of a  $\mu_\pi>0$ in a thermal fireball calculation 
has been shown to essentially resolve this discrepancy~\cite{RW99}. 
This needs to be confirmed in a full hydrodynamic treatment. 

Finally we repeat that there is currently no theoretical explanation 
available for the strong enhancement in the low-mass dilepton 
spectra measured by the DLS collaboration~\cite{DLS2} in heavy-ion 
collisions at relativistic projectile energies (1--2~AGeV). Both 
the dropping-mass scenario~\cite{BK98} and the broadening of the 
in-medium $\rho$ spectral function~\cite{BCRW}
 fall short of the data by a factor 
of 2--3 at invariant masses around 0.4~GeV.  
Hopes to resolve this puzzle reside on the upcoming  precision 
measurments to be performed with the HADES detector at SIS (GSI) 
at similar bombarding energies.

\chapter{Conclusions}

The investigation of hadron properties in a hot and dense environment
as produced in energetic collisions of heavy nuclei represents one of the 
main frontiers in modern nuclear physics. In particular, it is directly
related to the approach towards the QCD phase transition, which
constitutes the 'Holy Grail' of the ultrarelativistic heavy-ion
initiative. In the present article we have 
tried to review a very active subfield of this research program, 
namely the major theoretical accomplishments that
have been achieved in connection with low-mass dilepton production 
over the past five years or so. Of course, the strong interest in this 
rapidly developing field is largely fueled by the exciting data that 
our experimental colleagues have obtained despite the notorious  
difficulties in extracting these observables.  

The nature of dilepton final states mediated by electromagnetic
currents immediately attaches to the vector mesons as the key
objects for gaining direct non-trivial information on in-medium
effects.  In the low-mass region which we have focused on, these
are the $\rho$,  $\omega$ and $\phi$ mesons, with the $\rho$ meson 
playing the dominant role since it has the shortest lifetime
and the largest dilepton decay width.
The now widely accepted viewpoint is that the main nonperturbative 
feature of low-energy strong interactions  -- the spontaneous breakdown 
of the global chiral symmetry in the fundamental QCD Lagrangian --
is not only responsible for the build-up of the 'constituent' quark masses
but governs the appearance of the low-energy hadron spectrum altogether.  
Hence the approach to chiral restoration in hot/dense matter is
intimately related to changing in-medium properties of light hadrons.
Here the only strict prediction from QCD is that 
the spectral distributions of 'chiral partners' have  
to become identical (or 'degenerate'). 
This is encoded in 'Weinberg sum rules' which 
(in the chiral limit) relate the energy-integrated difference
of the vector and axialvector correlators to the weak pion-decay constant, 
$f_\pi$, one of the order parameters for chiral symmetry restoration. 
Medium effects of the $\rho$ meson therefore necessarily have to be put into
context with those of its chiral partner, the $a_1$ meson,
as well as the pion. {\em How} the 'degeneration' is  realized 
in nature is far from obvious and marks the 
central question to be answered in our context.  
For the $\omega$ meson the problem is further complicated by the fact 
that it couples to three-pion (or $\pi$-$\rho$) states through the 
Wess-Zumino term and hence the anomaly structure of QCD.
Also the assessment of in-medium properties of the $\phi$ meson is hampered
by the fact that the current mass of the $s$-quark is quite large so that 
arguments based on the chiral limit are not as stringent as in the $u$-$d$
sector.   

Based on general properties of QCD we have started our 
discussion with a focus on model-independent 
approaches. To lowest order in temperature (and in the chiral limit), 
soft pion theorems imply the leading temperature effect to be a mere 
mixing between vector and axialvector correlators with no 
medium effects in the correlators themselves. 
QCD sum rules relate physical vacuum correlators to the various condensates
thus providing a direct link between physically observed hadrons and
the underlying QCD vacuum structure.
It has now become clear that, when applied within a hadronic medium, they
have limited predictive power for in-medium spectral distributions since
they only provide a band of allowed 
combinations of large/small masses with corresponding 
large/small widths of the light vector mesons. 
  
More specific predictions can be obtained from hadronic model
Lagrangians which have been applied at various levels of approximations. 
Mean-field treatments typically focus on the in-medium behavior
of the masses, and here results from chiral Lagrangians including
vector mesons as gauge bosons (when applied at finite temperature) 
seem to allow for both an increasing
$\rho$-meson mass -- becoming degenerate with the $a_1$ -- as well as a 
decreasing mass as conjectured in the famous Brown-Rho paper based on scale
invariance arguments. On the other hand, several independent many-body
calculations of in-medium vector meson spectral functions come to the 
consensus that multiple interactions in hadronic matter inevitably induce 
a broadening of the spectral distributions. Nevertheless, certain 
$S$-wave scattering processes -- most importantly resonant 
$\rho N\to N(1520)$ excitations -- can be associated with a 
reduction of the $\rho$/$\omega$ mass.  Towards higher densities, however, 
the widths of the spectral functions (most prominently for the $\rho$ meson)
increase to such an extent that the entire resonance 
structures are 'melted' into an essentially structureless continuum. 
At the same time the real parts in the propagators also become
very flat so that the concept of a single mass ceases to be  
meaningful. 

We have argued that a scenario of melting resonances has in fact the very 
appealing feature to establish a continuous link ('duality') between
hadron- and quark-gluon-based calculations of the vector correlator, 
in the following sense: in vacuum the 'duality-threshold', \ie, the 
invariant mass where the total cross section $\sigma(e^+e^-\to hadrons)$ 
empirically starts to follow the perturbative QCD prediction for 
$q \bar q$ production is located at about $M\simeq 1.5$~GeV. Resonance 
formation at lower invariant masses is an inherently nonperturbative 
effect associated with spontaneous chiral symmetry breaking  which is a 
'large distance phenomenon'. It should not play any role at the 
small space-time distances probed beyond 1.5~GeV. 
Consequently, above this mass also the axialvector correlator, being
identical to the vector one, ought to be given by perturbative QCD 
(which is unfortunately not well-established experimentally).  
Medium modifications can be studied through dilepton production, the 
reverse process of $e^+e^-$ annihilation. We have pointed out that 
the finite-temperature 
mixing of vector and axialvector channels, which to lowest order  
arises from the coupling to pions of the heat bath, suffices to equalize
the hadronic and quark-antiquark description down to the $\phi$ meson
mass when extrapolated to temperatures where chiral symmetry is
restored. This feature is corroborated by model-independent
approaches, such as the chiral reduction formalism, as well as detailed
model calculations in  the region above the $\phi$ meson peak. It
can thus be stated that 'duality' of hadronic and quark descriptions
for the in-medium rates down to 1~GeV is well established for the case of 
vanishing baryon density and is likely to also hold for the latter case. 
The appealing physical interpretation is that, as electromagnetic probes 
couple to charges, in the vicinity of the phase boundary it becomes 
immaterial 
whether they reside in free quarks or in a large number of strongly
interacting hadrons. 

The continuation of the 'duality' argument to even 
lower invariant masses is not rigorously established at present, although
suggestive indications have emerged. It is clear that the 
lowest-order mixing does not affect dynamically generated (low-mass)  
resonance structures in the correlators. 
This is precisely where the many-body effects enter through 
a flattening of the resonance peaks, which now requires arbitrary 
orders in density introduced through resummations. Hadronic model calculations  
near the phase boundary show that the resulting dilepton rates in the 
$\rho$ region and somewhat below continue to match rather well the perturbative 
quark-antiquark rates. Deviations set in for $M\lsim 0.5$~GeV. At such low
invariant masses it is, however,  conceivable that 'soft' processes
could significantly alter the quark rates.    
From a more practical point of view, the melting of especially the 
$\rho$ resonance directly entails an enhanced dilepton production
{\em below} the free mass which has turned out to be compatible 
with current data from the SpS.  At the same time, to verify
the associated depletion of the in-medium signal in the free $\rho/\omega$ 
mass region of the spectra, it will be crucial
to discriminate free $\omega$ decays, occurring after freezeout,  
which is anticipated to be feasible with improved mass resolution
measurements at the CERN-SpS. Clearly, this applies equally to {\em both}
the commissioned low-energy run at 40~AGeV {\em and} additional 
future ones at the full SpS energy.

There remain a number of further problems which have to be resolved. 
First, it will be important to establish a more profound theoretical 
connection between the processes that reshape the vector ($\rho$)
and axialvector ($a_1$) spectral distributions. In particular one has 
to find reliable ways to perform a similarly advanced calculation 
for the in-medium properties of the $a_1$ as has been achieved for the 
$\rho$. Also, since (at comparable densities) the medium effects from 
baryonic matter seem to be more pronounced than at  
finite temperature, the fate of the (light) vector mesons in purely
mesonic matter near the phase boundary is not really settled.  
Here, the upcoming collider experiments at RHIC and LHC, where the 
meson-to-baryon ratios at midrapidity are expected to
increase by another substantial factor ($\sim$~5--10 at RHIC) as compared 
to current SpS conditions, will provide answers.

\vskip1cm
 
\centerline {\bf ACKNOWLEDGMENTS}
Many of our experimental and theoretical colleagues have integrally
contributed to the progress reported in this article. 
We especially thank our collaborators G.E. Brown, M. Buballa, M. Urban,   
G. Chanfray, W. Cassing, E.L. Bratkovskaya and C. Gale for the fruitful
joined efforts. 
We are also grateful for many productive conversations with  
P. Braun-Munzinger, A. Drees, J.W. Durso, M. Ericson, J. Friese, B. Friman, 
F. Karsch, F. Klingl, C.M. Ko, V. Koch, G.Q. Li, C. Louren\c{c}o, V. Metag, 
U. Mosel, W. N\"orenberg, M. Prakash, K. Redlich, M. Rho, A. Richter,   
P. Schuck, E.V. Shuryak, H. Sorge, H.J. Specht, J. Stachel, 
I. Tserruya, W. Weise, A. Wirzba and I. Zahed.  
We furthermore thank G.E. Brown, W. Cassing, C. Louren\c{c}o and B.J. Schaefer
for useful comments and a careful reading of the manuscript. 
We apologize to our colleagues whose work we did not appropriately discuss. 
One of us (RR) acknowledges support 
from the Alexander-von-Humboldt foundation through a 3-year  
Feodor-Lynen fellowship at Stony Brook during which most of his work 
on the presented topics has been performed. 
This work is supported in part by the National Science Foundation
under Grant No. NSF PHY98-00978, the U.S. Department of Energy 
under Grant No. DE-FG02-88ER40388, the BMBF and GSI Darmstadt.



\end{document}